\documentclass[twocolumn]{aastex631}

\usepackage{graphicx}
\usepackage{dcolumn}
\usepackage{bm}
\usepackage[nointegrals]{wasysym}
\usepackage{verbatim}
\usepackage{color}
\usepackage{amsmath}
\usepackage[utf8]{inputenc}
\usepackage[shortlabels]{enumitem}
\usepackage{wrapfig}
\usepackage{scalerel}
\usepackage[caption=false]{subfig}

\usepackage{listings}
\definecolor{background}{rgb}{0.95,0.95,0.92}
\definecolor{codeblack}{rgb}{0.1,0.1,0.1}
\definecolor{codegrey}{rgb}{0.3,0.3,0.3}
\definecolor{codered}{rgb}{0.6,0.1,0.1}
\definecolor{codegreen}{rgb}{0.1,0.6,0.1}
\definecolor{codeblue}{rgb}{0.1,0.1,0.6}

\newcommand\YAMLcolonstyle{\color{codered}\mdseries}
\newcommand\YAMLkeystyle{\color{codeblack}\bfseries}
\newcommand\YAMLvaluestyle{\color{codeblue}\mdseries}

\makeatletter

\newcommand\language@yaml{yaml}

\expandafter\expandafter\expandafter\lstdefinelanguage
\expandafter{\language@yaml}
{
  basicstyle=\YAMLkeystyle\ttfamily\footnotesize,
  keywords={true,false,null,y,n},
  keywordstyle=\color{codeblack}\bfseries,
  sensitive=false,
  comment=[l]{\#},
  morecomment=[s]{/*}{*/},
  commentstyle=\color{codegreen},
  stringstyle=\YAMLvaluestyle,
  moredelim=[l][\color{codered}]{\&},
  moredelim=[l][\color{codeblue}]{*},
  moredelim=**[il][\YAMLcolonstyle{:}\YAMLvaluestyle]{:},   
  morestring=[b]',
  morestring=[b]",
  literate =    {---}{{\ProcessThreeDashes}}3
                {>}{{\textcolor{codered}\textgreater}}1     
                {|}{{\textcolor{codered}\textbar}}1 
                {\ -\ }{{\mdseries\ -\ }}3,
  backgroundcolor=\color{background},
  numberstyle=\tiny\ttfamily\color{codegrey},
  numbers=left,
  numbersep=5pt,
  breaklines=true
}

\lst@AddToHook{EveryLine}{\ifx\lst@language\language@yaml\YAMLkeystyle\fi}
\makeatother

\newcommand\ProcessThreeDashes{\llap{\color{cyan}\mdseries-{-}-}}

\newcommand{\ba}{\begin{eqnarray}}
\newcommand{\ea}{\end{eqnarray}}

\newcommand{\ombh}{\Omega_{b} h^2}
\newcommand{\omegam}{\Omega_{m}}

\newcommand{\mnu}{\sum m_{\nu}}
\newcommand{\neff}{N_{\rm eff}}
\newcommand{\yhe}{Y_{\rm He}}

\newcommand{\lcdm}{$\Lambda\rm{CDM}$}
\newcommand{\Planck}{{\it{Planck}}}
\newcommand{\WMAP}{{\it{WMAP}}}
\newcommand{\act}{\textsf{ACT}}
\newcommand{\pact}{\textsf{P-ACT}}
\newcommand{\pactlb}{\textsf{P-ACT-LB}}
\newcommand{\pactlbb}{\textsf{P-ACT-LB$_{\textsf{BOSS}}$}}
\newcommand{\wactlbb}{\textsf{W-ACT-LB$_{\textsf{BOSS}}$}}
\newcommand{\pactlbD}{\textsf{P-ACT-LB-D}}
\newcommand{\pactlbHe}{\textsf{P-ACT-LB-He}}
\newcommand{\pactlbDHe}{\textsf{P-ACT-LB-D-He}}
\newcommand{\pactlbbicep}{\textsf{P-ACT-LB-BK18}}
\newcommand{\pactlbs}{\textsf{P-ACT-LBS}}
\newcommand{\pacts}{\textsf{P-ACT-S}}
\newcommand{\wact}{\textsf{W-ACT}}
\newcommand{\wactlb}{\textsf{W-ACT-LB}}
\newcommand{\plb}{\textsf{Planck-LB}}
\newcommand{\plbs}{\textsf{Planck-LBS}}
\newcommand{\DHe}{\textsf{D-He}}

\newcommand{\MeV}{\textrm{MeV}}

\newcommand{\Geff}{G_\mathrm{eff}}

\newcommand{\subsubsubsection}[1]{\vspace{-0.5cm}\begin{center}\item{}\emph{#1}\end{center}}
 
\usepackage{aas_macros}
\begin{document}
\setcounter{tocdepth}{4}

\title{The Atacama Cosmology Telescope: DR6 Constraints on Extended Cosmological Models}
 \shorttitle{}
  \shortauthors{}
\begin{abstract}
We use new cosmic microwave background (CMB) primary temperature and polarization anisotropy measurements from the Atacama Cosmology Telescope (ACT) Data Release 6 (DR6) to test foundational assumptions of the standard cosmological model, \lcdm, and set constraints on extensions to it. We derive constraints from the ACT DR6 power spectra alone, as well as in combination with legacy data from the \Planck\ mission. To break geometric degeneracies, we include ACT and \Planck\ CMB lensing data and baryon acoustic oscillation data from DESI Year-1.  To test the dependence of our results on non-ACT data, we also explore combinations replacing \Planck\ with \WMAP\ and DESI with BOSS, and further add supernovae measurements from Pantheon+ for models that affect the late-time expansion history.  We verify the near-scale-invariance (running of the spectral index $d n_s/d\ln k = 0.0062 \pm 0.0052$) and adiabaticity of the primordial perturbations. Neutrino properties are consistent with Standard Model predictions: we find no evidence for new light, relativistic species that are free-streaming ($\neff = 2.86 \pm 0.13$, which combined with astrophysical measurements of primordial helium and deuterium abundances becomes $\neff = 2.89 \pm 0.11$), for non-zero neutrino masses ($\sum m_\nu < 0.089$~eV at 95\% CL), or for neutrino self-interactions. We also find no evidence for self-interacting dark radiation ($N_{\rm idr} < 0.134$), or for early-universe variation of fundamental constants, including the fine-structure constant ($\alpha_{\rm EM}/\alpha_{\rm EM,0} = 1.0043 \pm 0.0017$) and the electron mass ($m_e / m_{e,0} = 1.0063 \pm 0.0056$).  Our data are consistent with standard big bang nucleosynthesis (we find $Y_{p} = 0.2312 \pm 0.0092$), the \emph{COBE/FIRAS}-inferred CMB temperature (we find $T_{\rm CMB} = 2.698 \pm 0.016$~K), a dark matter component that is collisionless and with only a small fraction allowed as axion-like particles, a cosmological constant ($w = -0.986 \pm 0.025$), and the late-time growth rate predicted by general relativity ($\gamma = 0.663 \pm 0.052$).  We find no statistically significant preference for a departure from the baseline \lcdm\ model.  In fits to models invoking early dark energy, primordial magnetic fields, or an arbitrary modified recombination history, we find $H_0 = 69.9^{+0.8}_{-1.5}$, $69.1 \pm 0.5$, or $69.6 \pm 1.0$~km/s/Mpc, respectively; using BOSS instead of DESI BAO data reduces the central values of these constraints by 1--1.5~km/s/Mpc while only slightly increasing the error bars.  In general, models introduced to increase the Hubble constant or to decrease the amplitude of density fluctuations inferred from the primary CMB are not favored over \lcdm\ by our data.
\vspace{0.7cm}
\end{abstract}

\suppressAffiliations
\author[0000-0003-0837-0068]{Erminia~Calabrese}\thanks{Corresponding author: Erminia Calabrese, \url{calabresee@cardiff.ac.uk}}\affiliation{School of Physics and Astronomy, Cardiff University, The Parade, Cardiff, Wales, UK CF24 3AA}
\author[0000-0002-9539-0835]{J.~Colin~Hill}\thanks{Corresponding author: Colin Hill, \url{jch2200@columbia.edu}} \affiliation{Department of Physics, Columbia University, New York, NY 10027, USA} \affiliation{Flatiron Institute, 162 5th Avenue, New York, NY 10010 USA}
\author[0000-0002-9429-0015]{Hidde~T.~Jense} \affiliation{School of Physics and Astronomy, Cardiff University, The Parade, Cardiff, Wales, UK CF24 3AA}
\author[0000-0002-2613-2445]{Adrien~La~Posta} \affiliation{Department of Physics, University of Oxford, Keble Road, Oxford, UK OX1 3RH}
\author[0000-0003-3230-4589]{Irene~Abril-Cabezas} \affiliation{DAMTP, Centre for Mathematical Sciences, University of Cambridge, Wilberforce Road, Cambridge CB3 OWA, UK} \affiliation{Kavli Institute for Cosmology Cambridge, Madingley Road, Cambridge CB3 0HA, UK}
\author[0000-0002-2147-2248]{Graeme~E.~Addison} \affiliation{Dept. of Physics and Astronomy, The Johns Hopkins University, 3400 N. Charles St., Baltimore, MD, USA 21218-2686}
\author[0000-0002-5127-0401]{Peter~A.~R.~Ade} \affiliation{School of Physics and Astronomy, Cardiff University, The Parade, Cardiff, Wales, UK CF24 3AA}
\author[0000-0002-1035-1854]{Simone~Aiola} \affiliation{Flatiron Institute, 162 5th Avenue, New York, NY 10010 USA} \affiliation{Joseph Henry Laboratories of Physics, Jadwin Hall, Princeton University, Princeton, NJ, USA 08544}
\author{Tommy~Alford} \affiliation{Department of Physics, University of Chicago, Chicago, IL 60637, USA}
\author[0000-0002-4598-9719]{David~Alonso} \affiliation{Department of Physics, University of Oxford, Keble Road, Oxford, UK OX1 3RH}
% ...

\date[]{\emph{Affiliations can be found at the end of the document}}

\tableofcontents
\vspace{0.4cm}

\section{Introduction}\label{sec:intro}

The $\Lambda$ cold dark matter ($\Lambda$CDM) cosmological model has emerged as the standard model of cosmology over the past quarter-century, undergirded by precision measurements of the cosmic microwave background (CMB) primary anisotropy power spectra in both temperature and polarization~\citep[e.g.,][and upcoming South Pole Telescope results]{wmap_spergel_2003,bennett/etal:2013,Planck_2018_params,rosenberg:2022,hillipop2024,Choi2020,Bakenhol_SPT,2025arXiv250106890C}, the expansion history of the universe as probed by baryon acoustic oscillations (BAO; ~\citealp{Eisenstein2005, BOSS2013, DESI-BAO-III}) and Type Ia supernovae measurements (SNIa;~\citealp{Riess1998, Perlmutter1999,Union3,Rubin2023_Union3,Pantheon,Pantheon+,DES_SN}), the growth of structure from gravitational lensing and galaxy clustering measurements~\citep[e.g.,][]{2004ApJ...606..702T, Smith2007,Das2011,vanEngelen2012,Planck_2018_lensing, Carron_pr4_lensing, SPT_lensing, Qu_dr6_lensing, SPT3G2024, DES_lensing, Amon_2022, Secco_2022, Heymans_2021,Asgari_2021, More_2023, Miyatake_2023, Sugiyama_2023, Dalal_2023,Li_2023,DES_KIDS_shear_2023, BOSS-LRG, eBOSS-LRG, PhilcoxIvanov2022, DESI_fullshape_2024}, and a wide array of additional probes.  However, there is strong motivation to further test the model and its underlying ingredients, particularly given our lack of microphysical knowledge of the dark sector.  The CMB is a uniquely powerful probe of extensions to $\Lambda$CDM~\citep[e.g.,][]{Planck_2018_params,Aiola2020,Balkenhol_SPT_ext}, both because of the theoretical accuracy with which new signals can be predicted in the CMB and because the CMB is sensitive to weakly-coupled new physics that is often otherwise difficult to probe.

In this paper, we stress-test \lcdm\ using a state-of-the-art CMB dataset built from the new Atacama Cosmology Telescope (ACT) Data Release 6 (DR6) measurements of the small-scale CMB temperature and polarization power spectra. A companion paper presents these data and combines them with the \Planck\ mission legacy data~\citep{Planck_2018_overview, Planck_2018_likelihood} to form the most statistically constraining CMB power spectrum dataset assembled to date~\citep{dr6ps}. To validate constraints derived from the combination of ACT and \Planck\ data, we also perform analyses combining ACT with legacy data from the \WMAP\ mission~\citep{bennett/etal:2013,WMAP9_Hinshaw}.  To assist in parameter degeneracy-breaking, we further add gravitational lensing measurements of the CMB from ACT DR6 and \Planck~\citep{Qu_dr6_lensing, Madhavacheril_dr6_lensing,Carron_pr4_lensing}, as well as BAO distance measurements from the Dark Energy Spectroscopic Instrument (DESI;~\citealp{DESI-BAO-III,DESI_BAO_Lya}) or the Baryon Oscillation Spectroscopic Survey (BOSS;~\citealp{BOSS-LRG,eBOSS-LRG}), and from the Pantheon+ supernovae compilation~\citep{Pantheon+}.  Further low-redshift data are used in some analyses as well, where significant additional constraining power can be gained.

We test the cosmological model via both single-parameter extensions of \lcdm\ and by relaxing its fundamental assumptions --- 
for example, considering variations in the underlying particle physics, energy densities of various components, and gravitational and non-gravitational interactions between them.  We constrain new physics operating at energy scales ranging from the inflationary epoch to the recombination epoch to the late-time universe, including models that have been constructed with the aim of increasing the value of the Hubble constant or decreasing the amplitude of late-time density fluctuations inferred from the primary CMB, as well as models motivated by more fundamental considerations in particle physics, such as the existence of new light species in the early universe.

This work builds on --- and extends --- previous cosmological explorations performed with ACT CMB power spectrum measurements~\citep{Dunkley2011,Sievers2013,Louis2017,Aiola2020,Thiele2021,Hill_ACT_EDE,An_ACT_DM,Li_ACT_DMB,Kreisch_ACT_SIN}. The new ACT DR6 data provide higher sensitivity over a broad range of angular scales, allowing us to access potential signals that would previously have been hidden in the noise. In some cases (depending on the model extension), the ACT DR6 sensitivity is comparable to that achieved by the \Planck\ legacy dataset~\citep{Planck_2018_params} and serves as a useful cross-check of the CMB response to a specific cosmological model. In other cases, because ACT DR6 provides constraining power in a different region of the power spectrum compared to \Planck\ (particularly in polarization), the joint fit to both datasets surpasses bounds from \Planck\ alone and thus represents a new state of the art. 

\begin{figure*}[hp]
	\centering
 \includegraphics[width=\textwidth]{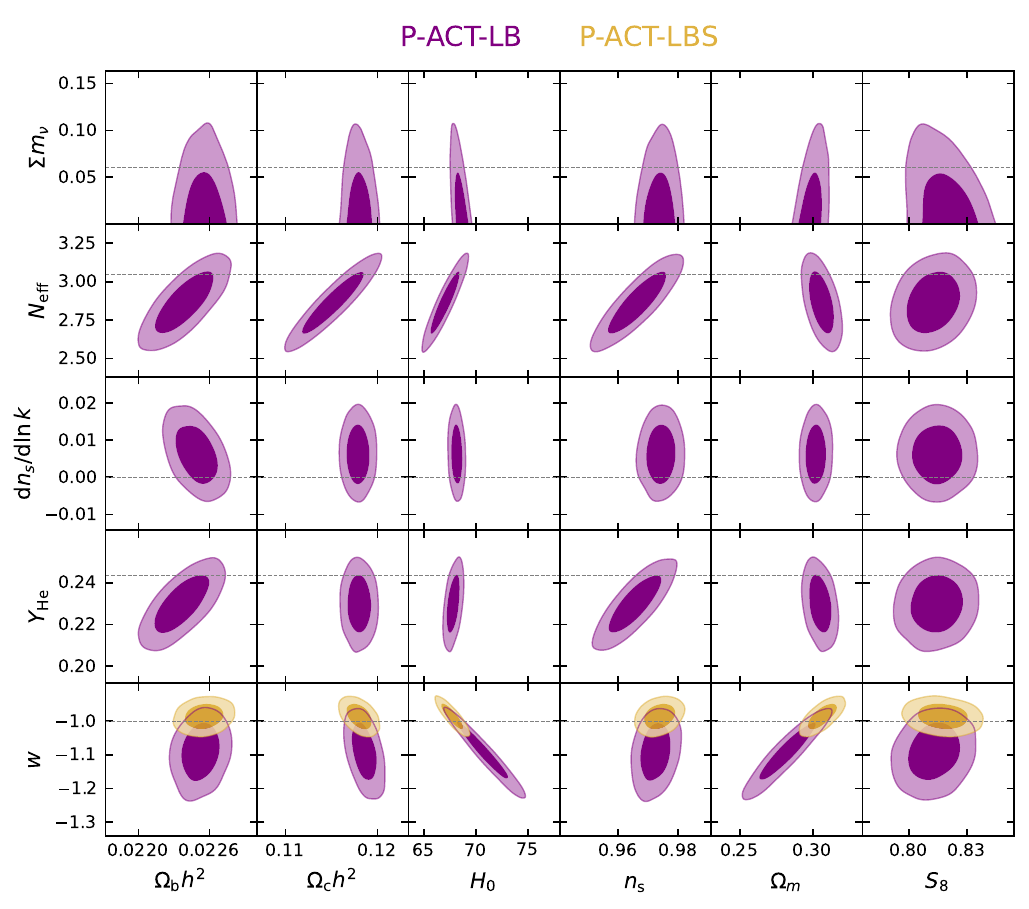}
	\caption{Constraints on single-parameter extensions to \lcdm\ (vertical axis) versus key \lcdm\ parameters (horizontal axis) from the combination of the new ACT DR6 power spectra with \Planck\ legacy CMB spectra, CMB lensing from \Planck\ and ACT, and BAO from DESI Year-1 data (purple; labeled \pactlb, as defined in~\S\ref{sec:glossary}).  We further add SNIa data from Pantheon+ (gold; with label \pactlbs) for models that affect the late-time expansion history. The rows span over models varying neutrino physics (the summed mass and number of neutrinos with $\mnu$ and $\neff$ respectively, \S\ref{sec:neutrinos}), variation of the primordial scalar perturbation spectral index with scale ($dn_s/d\ln k$, \S\ref{sec:running}), the abundance of primordial helium ($Y_{\rm He}$, \S\ref{sec:bbn}), and the dark energy equation of state ($w$, \S\ref{sec:de}). These are shown across columns against \lcdm\ parameters quantifying the baryon, cold dark matter, and total matter densities ($\Omega_b h^2$ and $\Omega_c h^2$ and $\Omega_m$, respectively), the Hubble constant ($H_0$) in km/s/Mpc, the spectral index of primordial scalar perturbations ($n_s$), and the amplitude of density fluctuations ($S_8$). The contours show confidence levels at 68\% and 95\% (dark and light shade, respectively). The dashed gray lines mark the canonical values expected for these parameters in the standard models of cosmology and particle physics.} 
	\label{fig:lcdm_single_extensions_pactlb}
\end{figure*}

\begin{figure*}[hp!]
	\centering
 \includegraphics[width=\textwidth]{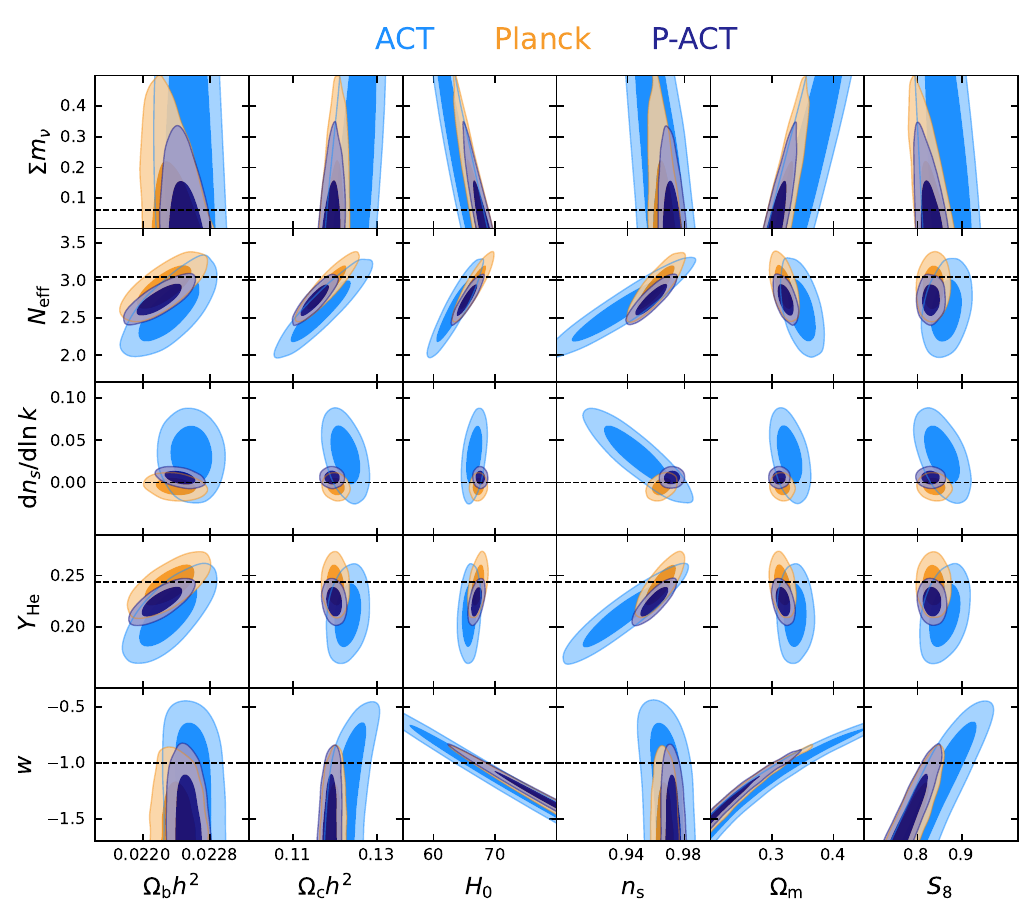}
	\caption{Similar to Fig.~\ref{fig:lcdm_single_extensions_pactlb}, but highlighting the specific constraints from the ACT DR6 CMB power spectra (light blue) compared to those from the \Planck\ legacy CMB power spectra (orange) and those from a joint analysis of the two datasets (\pact, navy blue, as defined in~\S\ref{subsec:cmb_spec}), which breaks degeneracies and tightens the constraints.}
	\label{fig:lcdm_single_extensions_pact}
\end{figure*}

Figure~\ref{fig:lcdm_single_extensions_pactlb} highlights new leading results from ACT DR6 combined with other datasets on a wide range of benchmark single-parameter \lcdm\ extensions that are studied in detail later in this paper.  This figure is not exhaustive and shows only a small fraction of the extended models analyzed in this work.  Figure~\ref{fig:lcdm_single_extensions_pact} isolates the specific contributions from the new ACT DR6 data to the primary CMB-derived constraints, by comparing constraints from ACT alone, \Planck\ alone, and their combination.  It is evident that the new ACT DR6 data have reached a level of precision competitive with that of \Planck; furthermore, due to the complementarity of the two datasets, their joint analysis yields significant gains over the sensitivity of \Planck\ alone, as will be explored throughout this paper.

Across all model extensions studied in this work, we find no preference for departures from the baseline \lcdm\ model. Accordingly, we set new limits on key fundamental physics parameters and theoretical models that deviate from the standard scenario, such as models featuring new particles or fields in the pre-recombination universe. Models introduced to increase the CMB-inferred Hubble constant or to decrease the CMB-inferred amplitude of density fluctuations are not favored by our data.

This paper is part of a suite of ACT DR6 papers, with companion papers describing the maps \citep[][N25 hereafter]{dr6maps}, and power spectra, likelihood, and baseline \lcdm\ parameter constraints \citep[][L25 hereafter]{dr6ps}. Other papers report the power spectrum covariance matrix estimation \citep{atkins/etal:2024}, beam measurements and modeling \citep{beams_inprep}, and foreground modeling for the power spectrum \citep{dr6fg}. The broad set of ACT DR6 papers is summarized in N25 and also includes noise simulations \citep{atkins/etal:2023}, CMB lensing maps and interpretation \citep{Qu_dr6_lensing,Madhavacheril_dr6_lensing,MacCrann_2024}, component-separated CMB temperature, CMB E-mode, and Compton $y$-maps \citep{act_ymap:2024}, studies of millimeter transients \citep{act_transient:2023}, and upcoming cluster and source catalogs.

The remainder of this paper is organized as follows. In~\S\ref{sec:data} we summarize the main datasets used in our analysis. The theoretical framework and assumptions of our analyses are covered in \S\ref{sec:method}, along with an overview of the various computational tools employed. \S\ref{sec:inflation}-\S\ref{sec:late-time} contain the main set of constraints on numerous extensions of the standard cosmology, covering primordial perturbations and inflation constraints (\S\ref{sec:inflation}), modifications of physics prior to and during recombination (\S\ref{sec:recomb}), properties of known and hypothetical fundamental particles (\S\ref{sec:astroparticle}), and modifications of gravity or other physics impacting cosmic evolution at late times (\S\ref{sec:late-time}). We discuss model consistency and the resulting impact on cosmological parameter concordance in \S\ref{sec:consistency_concordance}. In \S\ref{sec:conclusions} we provide a brief summary, highlighting the consistency of the ACT and \Planck\ CMB power spectra with the standard \lcdm\ cosmological model. A set of Appendices provides further technical details.

\section{Summary of data}\label{sec:data}

\subsection{CMB power spectra}\label{subsec:cmb_spec}

The ACT DR6 power spectra described in L25 are derived from maps made from five years of observations collected during 2017--2022, using detector arrays sensitive to three frequency bands: f090 (${77-112}$~GHz), f150 (${124-172}$~GHz), and f220 (${182-277}$~GHz). The maps are described in N25. The array-band combinations and the multipole ranges that pass a comprehensive battery of null tests form the nominal DR6 dataset, comprising temperature-temperature power spectra (TT), temperature-E mode polarization power spectra (TE), and polarization-polarization power spectra (EE). These power spectra have white noise levels that improve over those of \Planck\ by roughly a factor of three in polarization and a factor of two in temperature, with multi-frequency spectra measured over the multipole range ${600<\ell<8500}$ and the CMB signal extracted in the range ${600<\ell<6500}$. The spectra and their covariance matrix are used as inputs to a multi-frequency likelihood, \texttt{MFLike}, and a CMB-only (foreground-marginalized) likelihood, \texttt{ACT-lite}, also described in L25.

To leverage the full multipole range accessible with the CMB for cosmological analyses, we combine ACT with satellite data. A minimal addition used in all analyses of the ACT primary CMB (labeled \act\ throughout) is a \Planck\ measurement of the optical depth to reionization from the low-multipole EE power spectrum ($\ell<30$) \href{https://web.fe.infn.it/~pagano/low_ell_datasets/sroll2/}{\texttt{Sroll2}} likelihood~\citep{pagano/etal:2020}. In many cases we compress the information of {\texttt{Sroll2}} into a Gaussian prior on the optical depth, $\tau=0.0566 \pm 0.0058$; for models that include parameters degenerate with primordial power spectrum parameters, we use the full likelihood shape.  As described in L25, we use a baseline CMB combination, labeled \pact, which extends the combined dataset further with the inclusion of \Planck\ data on large-to-intermediate scales --- at $\ell<1000$ in TT and $\ell<600$ in TE/EE, truncating the multipole range of the {\texttt{plik\_lite}} likelihood~\citep{Planck_2018_likelihood} and generating a ``\Planck$_{\rm cut}$'' likelihood. This combination is built to increase constraining power while minimizing the overlap between the two experiments and allowing us to neglect their covariance.  In cases where it is useful to cross-check the results with a \Planck-independent CMB combination, we replace \Planck\ with \WMAP\ data from the final 9-year release~\citep{bennett/etal:2013,WMAP9_Hinshaw}, with this combination labeled \wact. In this case, we truncate the low-$\ell$ \WMAP\ polarization likelihood, replacing it with the \Planck\ \texttt{Sroll2} likelihood.  We use a Python implementation of the \WMAP\ likelihood, \href{https://github.com/HTJense/pyWMAP}{\texttt{pyWMAP}}, which retains only the data in temperature (on all scales) and at $\ell>23$ in polarization.

\subsection{CMB lensing}\label{subsec:cmb_lens}

Incorporating CMB lensing data into our cosmological constraints provides valuable complementary information to that obtained from the primary CMB power spectra. Gravitational lensing of the CMB probes large-scale structure across a wide range of cosmic history, with a broad peak at $z \approx 1$--$2$ and a tail extending to high redshift. 

The ACT DR6 CMB lensing release provides the most precise detection of CMB lensing to date, with a $43\sigma$~\citep{Qu_dr6_lensing,Madhavacheril_dr6_lensing} measurement of the lensing power spectrum. Lensing is robustly measured across multipoles ${40 < L < 763}$, with extensive tests confirming stability against systematics and foreground contamination~\citep{MacCrann_2024}. The associated likelihood conservatively uses only this baseline range to minimize potential systematic impacts.

The \Planck\ PR4 dataset provides a CMB lensing measurement comparable to ACT DR6 in signal-to-noise, achieving 42$\sigma$ using the reprocessed NPIPE maps across multipole range ${8 < L < 400}$~\citep{Carron_pr4_lensing}. Combining \Planck\ with ACT DR6 yields a state-of-the-art lensing power spectrum, with ACT adding high-precision data on smaller scales, $L > 400$.

Our analysis uses a Gaussian likelihood framework to combine the ACT DR6 and \Planck\ PR4 bandpowers, which appropriately accounts for the small correlation between the two datasets~\citep{Qu_dr6_lensing,Madhavacheril_dr6_lensing}. The effective signal-to-noise ratio of the combined ACT DR6 + \Planck\ NPIPE lensing spectrum, accounting for their joint covariance, corresponds to 58$\sigma$.  Perturbative adjustments are also applied to correct for the weak dependence of the measurements on cosmological assumptions, following methods outlined in \citet{2016A&A...594A..15P,Madhavacheril_dr6_lensing}.\footnote{Although the ACT DR6 CMB lensing likelihood corrections have not been recomputed based on the improved knowledge of ACT maps and power spectra presented in this new suite of papers, we do not expect this to impact our results.  We explicitly verify that even omitting these corrections altogether has negligible impact on our inferred parameter constraints.}

\subsection{BAO}\label{subsec:bao}

BAO data measure the acoustic scale at $\sim 150$ Mpc in the clustering of galaxies in the late-time, $z \lesssim 4$, universe. This feature allows one to constrain distance ratios parallel and perpendicular to the line of sight as a function of redshift.  The BAO feature along the line-of-sight direction measures $D_H(z)/r_d$, the inverse-ratio between the sound horizon at the baryon drag epoch, $r_d$, and $D_H(z) \equiv c/H(z)$; when combined with a calibration of $r_d$ (e.g., from CMB data), this allows a measurement of the Hubble parameter at redshift $z$.  The BAO feature in the angular correlation function of galaxies at redshift $z$ measures $D_M(z)/r_d$, the ratio between the comoving angular diameter distance and $r_d$, which thus allows inference of $D_M(z)$ when combined with a calibration of $r_d$.  These quantities are often combined to report measurements of the angle-averaged distance $D_V(z)/r_d$, with $D_V(z)=(zD_M(z)^2D_H(z))^{1/3}$.
This has a strong dependence on the matter density and other parameters affecting the expansion history of the universe~\citep[see, e.g.,][]{EISENSTEIN2005360,2010deot.book..246B,2013PhR...530...87W}. Because the main signature is generated during a phase of linear evolution of matter density perturbations, BAO measurements have become the most common cosmological dataset used to break geometric degeneracies in CMB analyses. 

Previous analyses used a compilation of BAO data at different redshifts from multiple surveys, such as the Baryon Oscillation Spectroscopic Survey (BOSS)~\citep{BOSS-LRG} and the 6dF Galaxy Redshift Survey~\citep{6dF}.  Here, we use the recent DESI Year-1 observations of the BAO feature in galaxy, quasar, and Lyman-$\alpha$ forest tracers~\citep{DESI-BAO-III,DESI_BAO_Lya,DESI-BAO-VI} as our baseline BAO combination.\footnote{The impact of the more recent DESI Year-3 observations is discussed in~Appendix~\ref{app:newDESI_tau}.}
This DESI dataset spans ${0.1<z<4.2}$ and includes twelve total data points, of which ten are pairs of $D_H(z)/r_d$ and $D_M(z)/r_d$, and two are combined $D_V(z)/r_d$ measurements, one each at the lowest and highest redshifts.  To ensure that our results are not solely driven by DESI, and in light of some $2.5\sigma$ deviations between the DESI luminous red galaxy (LRG) data points and previous measurements at the same redshifts, we also consider analyses with DESI replaced by BOSS/eBOSS BAO data, including both BOSS DR12 LRGs~\citep{BOSS-LRG} and
eBOSS DR16 LRGs~\citep{eBOSS-LRG}.  We primarily perform such analyses for models in which BAO data have a significant impact on the parameter constraints.

\subsection{SNIa}\label{subsec:SNIa}
Type Ia supernovae (SNIa) are powerful probes of cosmological distances in the modern universe (the last ten billion years). Using SNIa anchored to other cosmological distance indicators, supernovae constrain the luminosity distance across a range of redshifts; even in the absence of absolute calibration, SNIa precisely constrain the relative expansion history of the universe at late times. 
SNIa are thus sensitive probes of the matter density and the equation of state of dark energy, as well as the spatial curvature when analyzed in tandem with the CMB. To maximize redshift coverage and ensure consistency of calibration of the SNIa, surveys from different groups are combined and re-calibrated, and released with a likelihood including systematic errors from fitting a model of supernova brightness, calibration offsets, and telescope systematics~\citep[e.g.,][]{Union3,Rubin2023_Union3,DES_SN,Pantheon}. Here we use the latest SNIa compilation from Pantheon+~\citep{Pantheon+, Panthoen+cosmology}, which brings together 18 samples comprising 1550 spectroscopically-confirmed SNIa spanning ${0.001<z<2.26}$. We primarily include Pantheon+ as an additional dataset when exploring models that affect the expansion history of the late-time universe, and do not explore other SNIa compilations.

\subsection{Glossary of data combinations and other additional datasets}
\label{sec:glossary}

In this paper, we analyze the CMB primary anisotropy datasets introduced in \S\ref{subsec:cmb_spec} alone or combined with CMB lensing from \S\ref{subsec:cmb_lens}, BAO data from \S\ref{subsec:bao}, and SNIa data from \S\ref{subsec:SNIa}. We report the resulting constraints with the label conventions for different dataset combinations summarized in Table~\ref{tab:labels}. We present the constraints obtained with the most constraining data combination as our baseline results: depending on the model, this combination is either \pactlb\ or \pactlbs. We discuss and compare these results with those obtained from the primary CMB anisotropies --- this is slightly different from the approach taken by the \Planck\ collaboration, which focused primarily on comparisons between a joint CMB primary-anisotropy and lensing result versus results from the CMB combined with large-scale-structure (LSS) probes.  We justify our ability to combine these data in~\S\ref{subsec:consistency}, showing that the best-fit \lcdm\ model to \pact\ CMB data gives excellent predictions for the low-redshift measurements, and furthermore that the \lcdm\ model gives an excellent joint fit to all data.

\begin{table}[t]
    \centering
    \begin{tabular}{c|l}
        \hline
        \textsf{Planck} & \emph{Planck}$^{\rm TT/TE/EE}$ + Sroll2 \\
        \hline
        \act & ACT$^{\rm TT/TE/EE}$ + Sroll2 \\
        \hline
        \pact & ACT$^{\rm TT/TE/EE}$ + \Planck$_{\rm cut}^{\rm TT/TE/EE}$ + Sroll2\\
         \hline
        \wact & ACT$^{\rm TT/TE/EE}$ + WMAP$^{\rm TT/TE/EE}$ + Sroll2\\
        \hline
        \multicolumn{1}{l}{followed by}\\
        \hline
        \textsf{-LB} & when adding CMB lensing and BAO \\
        \hline
        \textsf{-LS} & when adding CMB lensing and SNIa \\
        \hline
        \textsf{-LBS} & when adding CMB lensing, BAO, and SNIa\\
        \hline 
        \hline  
    \end{tabular}
    \caption{Dictionary listing the main dataset combinations analyzed in this work.  Here, CMB lensing always refers to ACT DR6+\Planck\ NPIPE lensing; BAO refers to the DESI Year-1 release unless explicitly stated otherwise (e.g., ``\textsf{B$_{\textsf{BOSS}}$}'' referring to BOSS BAO); and SNIa refers to Pantheon+. All \textsf{Planck} results have been re-run to include an updated version of the \texttt{Sroll2} likelihood compared to \citet{Planck_2018_params}.}
    \label{tab:labels}
    \vspace{-0.3cm}
\end{table}

Depending on the physical signature probed by each model, additional astrophysical and cosmological measurements (beyond CMB lensing, BAO, and SNIa data) can help to further tighten the parameter constraints. These are folded into specific analyses as described in each section of interest below.

As shown in L25, our baseline \lcdm\ results confirm a discrepancy between the value of the Hubble constant, $H_0$, derived from the primary CMB and some measurements of this quantity from the local universe --- see \cite{Freedman_review,2024ARA&A..62..287V} for relevant reviews.  Specifically, the constraint on the Hubble constant using distance ladder methods with Cepheid-calibrated SNIa, from the SH0ES collaboration, $H_0=73.17 \pm 0.86$~km/s/Mpc~\citep{Riess_SHOES,Breuval:2024lsv}, is the most precise local measurement and also the most discrepant with the CMB estimate (L25). The latest value obtained by the CCHP program using tip of the red giant branch stars (TRGB) to calibrate SNIa distances is $H_0=70.39 \pm 1.94$~km/s/Mpc~\citep{H0Freedman,Freedman_CCHP}, which is consistent with the CMB value (and with the SH0ES value), but less precise than the SH0ES constraint. Further recent results from direct measurements can be found in \cite{2024ARA&A..62..287V}.  As a baseline choice, we do not combine our data with local measurements of $H_0$. However, we discuss in detail cases where a given model can accommodate a larger value of the Hubble constant compared to \lcdm\ or where there are important parameter degeneracies that impact $H_0$.

Similarly, as shown in L25, our baseline \lcdm\ results confirm the value of the fluctuation-amplitude parameter $S_8 \equiv \sigma_8 \sqrt{\Omega_m/0.3}$ (obtained from a combination of the amplitude of matter fluctuations on scales of ${8~h^{-1}}$~Mpc, $\sigma_8$, and the total matter density, $\omegam$) found in previous \Planck\ analyses, which lies $2$--$3\sigma$ higher than values found in some weak lensing and galaxy clustering studies (see, e.g.,~\citealp{Madhavacheril_dr6_lensing} for a collection of recent $S_8$ results from low-redshift probes).  We emphasize that ACT and \Planck\ CMB lensing data show no evidence of a low $S_8$ value~\citep{Madhavacheril_dr6_lensing,Planck_2018_lensing}.\footnote{The results from the KiDS Legacy data~\citep{2025arXiv250319442S,2025arXiv250319441W}, which appeared after the first version of this paper was submitted, show full consistency between the $S_8$ value inferred from cosmic shear and that measured from the CMB.} As with $H_0$, as a baseline choice, we do not combine our data with low-redshift measurements of $S_8$. This choice is also motivated by the fact that some extended models alter the shape and/or redshift evolution of the matter power spectrum, and thus a dedicated reanalysis of the relevant data within the context of each model would be necessary in order to derive valid constraints.  Nevertheless, for some models it is appropriate to consider external priors on $S_8$ from low-redshift data; where this is useful and allowed by the parameter posteriors, we include in our analysis a prior from the joint analysis of DES and KIDS weak lensing data, ${S_8 = 0.797 \pm 0.0155}$~\citep{DES_KIDS_shear_2023}.\footnote{This is a symmetrized Gaussian approximation of the fixed-neutrino-mass result from Table 4 of~\citet{DES_KIDS_shear_2023}.}


\section{Analysis methodology}\label{sec:method}

We obtain cosmological parameter constraints using the ACT DR6 multi-frequency or CMB-only likelihood\footnote{As shown in L25, these two likelihoods yield cosmological parameters that agree within $0.1\sigma$.} coupled to \href{https://cobaya.readthedocs.io/en/latest/}{\texttt{Cobaya}}~\citep{Cobaya}, which itself is coupled to the Einstein-Boltzmann codes  \href{https://camb.readthedocs.io/en/latest/}{\texttt{camb}}~\citep{Lewis_camb} or \href{https://lesgourg.github.io/class_public/class.html}{\texttt{class}}~\citep{Lesgourgues_class,CLASS_2011}, or to \href{https://alessiospuriomancini.github.io/cosmopower/}{\texttt{CosmoPower}} emulators of these codes~\citep{Spurio_Mancini_cosmopower,Bolliet_emulators,Jense_emulators,Qu:2024lpx}, to compute the lensed theoretical CMB power spectra at high precision, as described in Appendix~\ref{app:theory}. All codes exploring extended models requiring modifications to \texttt{camb} and \texttt{class} are benchmarked against the respective \texttt{camb} and \texttt{class} baseline \lcdm\ results. In Table~\ref{tab:models} in Appendix~\ref{app:theory}, we provide a summary of the theory code and likelihood that are used for each model.

Unless explicitly mentioned otherwise, we compute all theory predictions using \href{https://cosmo.nyu.edu/yacine/hyrec/hyrec.html}{\texttt{HyRec}}~\citep{Hyrec} or \href{https://www.jb.man.ac.uk/~jchluba/Science/CosmoRec/Welcome.html}{\texttt{CosmoRec}}~\citep{Cosmorec} (in \texttt{class} and \texttt{camb}, respectively), rather than \texttt{Recfast}~\citep{Recfast}, to obtain higher precision for recombination physics.\footnote{The Recfast ``fudge parameters'' were tuned to provide sufficient accuracy for \Planck, but are no longer sufficient for ACT DR6 sensitivity.  We test the latest versions of \texttt{HyRec} and \texttt{CosmoRec} to ensure that they provide accurate recombination calculations when we compute the DR6 likelihood evaluation across the full parameter space, while also cross-checking the two codes against each other. We use \texttt{CosmoRec} in {\tt camb} and {\tt HyRec} in {\tt class} because the newest \texttt{CosmoRec} version is not yet available within \texttt{class}.} 
As a baseline choice, we compute Big Bang Nucleosynthesis (BBN) predictions using calculations of the primordial helium abundance from \href{https://www2.iap.fr/users/pitrou/primat.htm}{\texttt{PRIMAT}}~\citep{2018PhR...754....1P}, but in~\S\ref{sec:bbn} we explore other options and discuss in detail the impact of different helium and deuterium abundance calculations.\footnote{Differences in the exact helium abundance predicted by the most recent BBN codes have a negligible impact on the CMB primary anisotropy spectra and therefore on cosmological parameters inferred from them. We also note that \texttt{PRIMAT}-based BBN data were not present in \texttt{class} when our analyses were performed (although they now are), so we manually imported a {\tt PRIMAT}-based table from \texttt{camb} to use in our {\tt class} calculations.  We take care to account for a small difference in the $\neff$ value assumed in the {\tt PRIMAT} BBN table (3.044) and that assumed in the {\tt class} BBN module (3.046).} We also use the latest version of \texttt{HMcode}~\citep{Mead2020} (rather than \texttt{Halofit}, \citealp{Halofit,Takahashi2012}) for modeling non-linear corrections to the matter power spectrum, unless explicitly stated otherwise.  At ACT DR6 precision, the impact of non-linear corrections to CMB lensing are non-negligible even in the primary CMB power spectra~\citep{McCarthy_2022}.  We adopt the dark-matter-only \texttt{HMcode} model for the non-linear matter power spectrum, with no baryonic feedback corrections.\footnote{At the time this work was performed, the latest \texttt{HMcode} model was not yet implemented in the main branch of \texttt{class}; we thank J.~Lesgourgues for providing an updated version with this model implemented (developed from {\tt class v3.2.2}). As of publication, the latest \texttt{HMcode} model is now available in the main public branch of \texttt{class}.}  We note that \texttt{HMCode} is not guaranteed to give accurate results when applied to models that alter late-time growth (e.g., models with interactions between dark matter and dark energy in the late universe, as in \S\ref{sec:dedm}); in such cases, we use custom alternative models or restrict to linear scales.  The vast majority of the new-physics models studied in this work alter physics in the early universe, and hence their impact on structure formation is generally captured in a change to the linear matter power spectrum at the onset of structure formation, which in most cases remains close to that in \lcdm.  These modeling choices are validated in Appendix~\ref{app:theory}.

Our parameter set includes the six basic $\Lambda$CDM cosmological parameters: the physical baryon and cold dark matter densities, $\Omega_b h^2$ and  $\Omega_c h^2$, the optical depth due to reionization, $\tau$, the amplitude and spectral index of the power spectrum of primordial adiabatic scalar perturbations, $A_s$ and $n_s$, both defined at a pivot scale ${k_*=0.05~{\rm Mpc}^{-1}}$,\footnote{This pivot scale is the default choice in many CMB analyses, used by \Planck, ACT, SPT, and other experiments. In some cases, following e.g., \WMAP, \Planck, and BICEP, results are also shown for $k_*=0.002$~Mpc$^{-1}$ --- when for example it is important to look at larger scales. This is explicitly mentioned in the relevant sections.} and a parameter that sets the absolute distance scale: (i) in \texttt{camb}, an approximation to the angular scale of the acoustic horizon at decoupling, $\theta_{\rm MC}$; (ii) in \texttt{class}, the angular size of the sound horizon at decoupling, $\theta_s$; or (iii) the Hubble constant, $H_0$ in km/s/Mpc (note that the exact definition of the angular scale varies between \texttt{camb} and \texttt{class} --- see, e.g., the discussion in~\citealp{Bolliet_emulators}).\footnote{$H_0$ is expressed throughout this paper in km/s/Mpc; units are generally omitted for brevity.}  We adopt broad, flat, uninformative priors on all \lcdm\ parameters in every analysis in this paper (except when imposing an \texttt{Sroll2}-informed prior on $\tau$ in some ACT-only fits).  Priors on extended-model parameters are described in each section, and are generally chosen to be uninformative as well.  From the sampled parameters, we obtain several derived parameters, including the rms amplitude of linear density fluctuations at $z=0$ on $8~h^{-1}\rm{Mpc}$ scales, $\sigma_8$, the matter density fraction $\Omega_m$, and $S_8 \equiv \sigma_8 \sqrt{\Omega_m/0.3}$.
For models in which neutrino physics is not varied, we fix neutrino properties to comprise one massive and two massless particles, with total mass ${\mnu = 0.06}$~eV and effective number of neutrino species $\neff=3.044$.  Following \Planck\ and the ACT DR6 CMB lensing analyses, we assume three massive eigenstates (zero massless) when analyzing models where the sum of the neutrino masses is a free parameter. Additional, model-specific parameters are added to the base \lcdm\ set as explained in each subsection below. 

Extra parameters are included in our theoretical model for astrophysical foregrounds, passband uncertainties, calibration, and polarization efficiency factors.  These parameters are varied as described in L25 when using the \texttt{MFLike} likelihood, or reduced to a minimum set of two varying parameters for overall calibration and polarization efficiency when using the \texttt{ACT-lite} likelihood (also described in L25).

MCMC chains are run with theory predictions computed up to $\ell_{\rm max} = 9000$ and with the Gelman-Rubin convergence parameter, $R-1$, reaching values $\sim O(10^{-2})$ and in nearly all cases smaller than 0.01. Parameter posteriors and statistical results are obtained with \href{https://getdist.readthedocs.io/en/latest/}{\texttt{GetDist}}~\citep{Getdist}.  Marginalized confidence intervals (for two-tailed limits) are calculated with the standard credible-interval approach, i.e., the interval between the two points with highest equal marginalized probability density.

Model comparisons are performed with respect to the benchmark \lcdm\ results presented in L25.  We determine the \emph{maximum a posteriori} (MAP) point in parameter space for each model (note that the ACT DR6 likelihood includes informative priors on some nuisance parameters, and thus the MAP is used, rather than the maximum-likelihood point).  The MAP for each model is then compared to the best-fit $\Lambda$CDM result for the relevant dataset combination.  We take care to discard contributions to the effective MAP $\chi^2$ value arising from uninformative priors --- the informative priors placed on nuisance parameters are always identical, and, as noted above, their contributions to the overall $\chi^2$ are accounted for in the best-fit model determination.  For reference, the best-fit \pact\ $\Lambda$CDM model yields MAP $\chi^2_{\rm \Lambda CDM, \texttt{MFLike}}=2180.5$ and $\chi^2_{\rm \Lambda CDM, \texttt{ACT-lite}} = 781.9$ for the full \pact\ dataset (see also L25).

\begin{figure}[t!]
	\centering
  \includegraphics[width=\columnwidth]{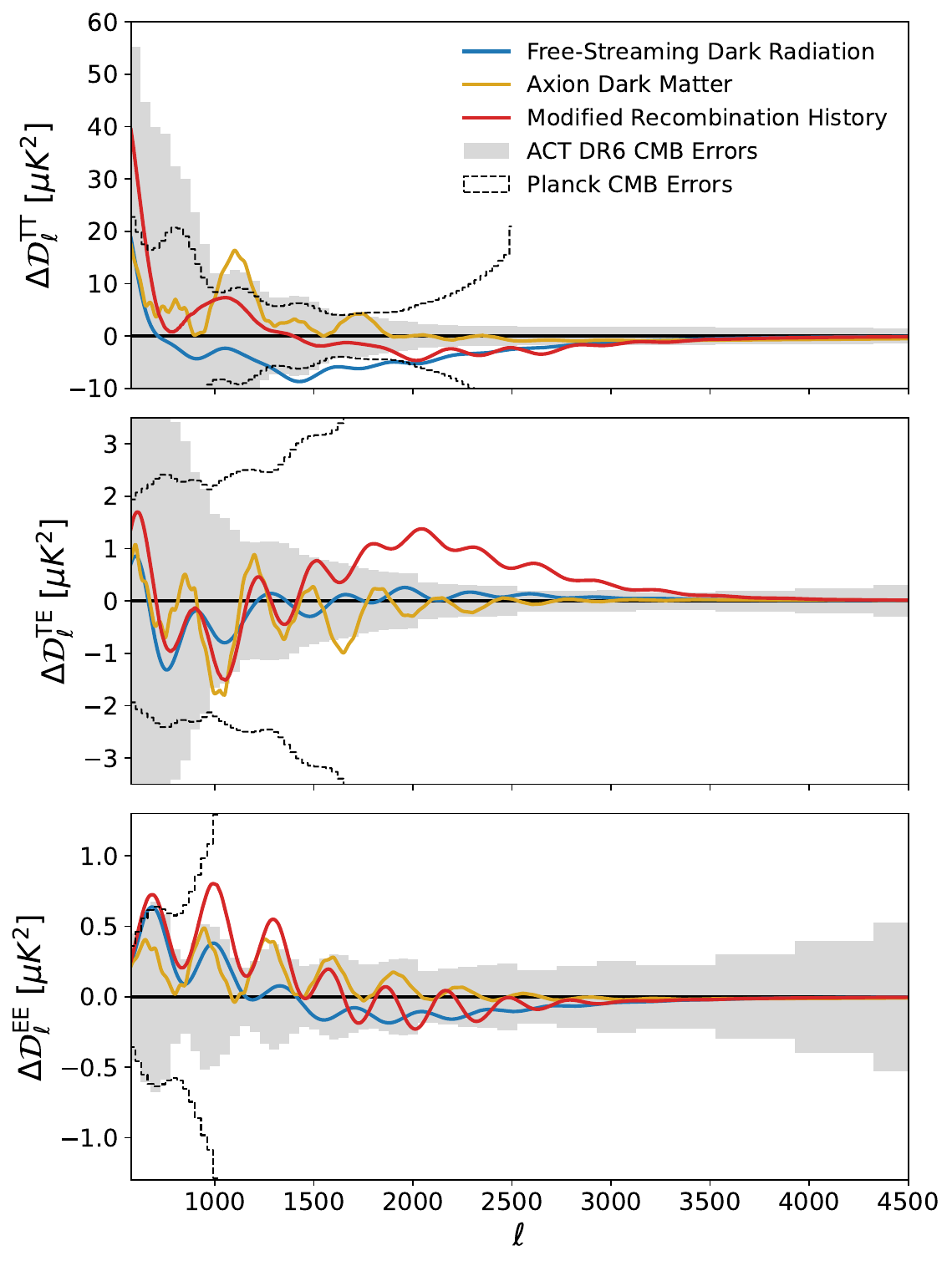}
	\caption{Models of free-streaming dark radiation (blue), axion dark matter (gold), and modified recombination history (red) that are allowed by \Planck\ data show differences in TT, TE, and EE compared to the baseline ACT \lcdm\ best-fit model (black line, L25), which are detectable by the new ACT data at high significance. We show theory differences for non-standard cosmologies consistent with existing \Planck\ constraints relative to ACT's \lcdm\ model, and compared to the sensitivity of ACT and \Planck\ through their CMB-only bandpower errors. For free-streaming dark radiation we consider a model allowing for an effective number of relativistic species $\neff=3.4$ (see~\S\ref{sec:neutrinos});$^a$ for axion dark matter we take a model with 5\% of the total dark matter density in the form of an axion of mass $m_{\rm ax}=10^{-26}\,\mathrm{eV}$ (see~\S\ref{sec:axions}); for modified recombination history we show a model with a non-standard ionization fraction between redshift $z=1000$ and $1500$ (see~\S\ref{sec:modrec}). 
    }
    \footnotetext{A self-interacting dark radiation model (\S\ref{sec:si_dr}) that fits \Planck\ with excess radiation would look similar to the model shown here with a positive excess in $\neff$.}
	\label{fig:spec}
\end{figure}

Constraints on parameters of the baseline \lcdm\ model are presented in L25.  In the following, we report constraints on a large suite of extended cosmological models, organized into four topical sections. In Fig.~\ref{fig:spec}, we illustrate with a few examples how ACT data push constraints on these models beyond what has been achieved by \Planck. Some models exhibit features on small scales ($\ell \gtrsim 2000$) not measured by \Planck. In other cases, the integrated sensitivity over an extended range of multipoles on small scales, and the sensitive ACT DR6 measurements of TE and EE at intermediate scales, rule out models at high significance that would be allowed within the \Planck\ bounds. 

We also discuss how ACT DR6 improves on previous ACT releases, in particular for models that showed a hint of a preference over \lcdm\ in the previous ACT DR4 dataset~\citep{Hill_ACT_EDE, Kreisch_ACT_SIN}. Additional comparisons with the DR4 results are shown and discussed in Appendix~\ref{app:dr4dr6comp} and Appendix~\ref{app:ede_full_constraints}. 

\section{Primordial perturbations and inflation}\label{sec:inflation}

\subsection{Scale invariance of scalar perturbations} \label{sec:running}

One key prediction of the simplest single-field slow-roll inflation models is an almost scale-invariant spectrum of primordial scalar perturbations --- the scalar spectral index $n_s$ being close to, but crucially different from (and usually less than) unity (see, e.g.,~\citealp{Mukhanov:1981xt,1992PhR...215..203M,2000cils.book.....L,Peiris_2003,Mukhanov_2007}). Analyses of \Planck\ CMB data have confirmed this prediction to high precision and shown its robustness across models~\citep{Planck_2018_params, Planck2018_inflation}. The new constraints on $n_s$ from ACT DR6 are presented and discussed in L25. Here, to test the inflation slow-roll approximation further, we take the common approach of expanding the power-law form of the primordial power spectrum of scalar curvature perturbations around a pivot scale $k_*$~\citep{1995PhRvD..52.1739K}
\begin{equation}
    \mathcal{P}_{\mathcal{R}}(k)=A_s\Big(\frac{k}{k_*}\Big)^{n_s-1 \ + \ (1/2)\ (\mathrm{d} n_s/\mathrm{d} \ln k)\ \ln(k/k_*)} \,,
    \label{eq:pks}
\end{equation}
and constrain the running of the spectral index, $\mathrm{d} n_s/\mathrm{d} \ln k$ --- i.e., the variation of $n_s$ as function of scale $k$, evaluated at ${k_*=0.05~{\rm Mpc}^{-1}}$. This is a single-parameter extension to \lcdm, varied in the range $[-0.2,0.2]$. 

\begin{figure}[t!]
	\centering
  \includegraphics[width=\columnwidth]{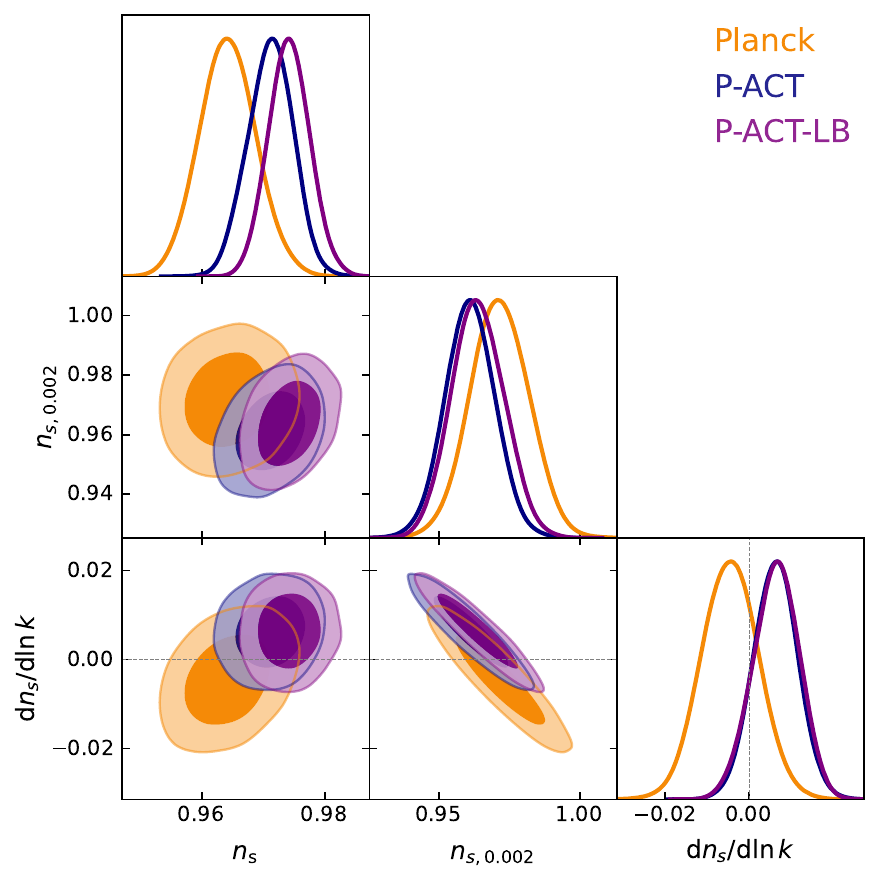}
	\caption{Constraints on parameters quantifying the scale invariance of primordial scalar perturbations: the scalar spectral index $n_s$ and its running $\mathrm{d} n_s/\mathrm{d} \ln k$ at $k_* = 0.05 \, \mathrm{Mpc}^{-1}$. Projections at larger scales (to facilitate comparisons with the \Planck\ results) are shown with $n_s$ at ${k_* = 0.002}\, \mathrm{Mpc}^{-1}$ computed as a derived parameter. The new ACT CMB spectra move the \Planck-alone constraint (orange) towards higher values of both $n_s$ and $\mathrm{d} n_s/\mathrm{d} \ln k$, as shown by \pact\ (navy). The addition of lensing and BAO data shown in \pactlb\ (purple) only shifts $n_s$ slightly, leaving the running constraint unchanged. In all cases, we find consistency with the \lcdm\ expectation (marked with gray dashed lines).}
	\label{fig:nrun}
\end{figure}

\begin{figure}[t!]
	\centering
  \includegraphics[width=\columnwidth]{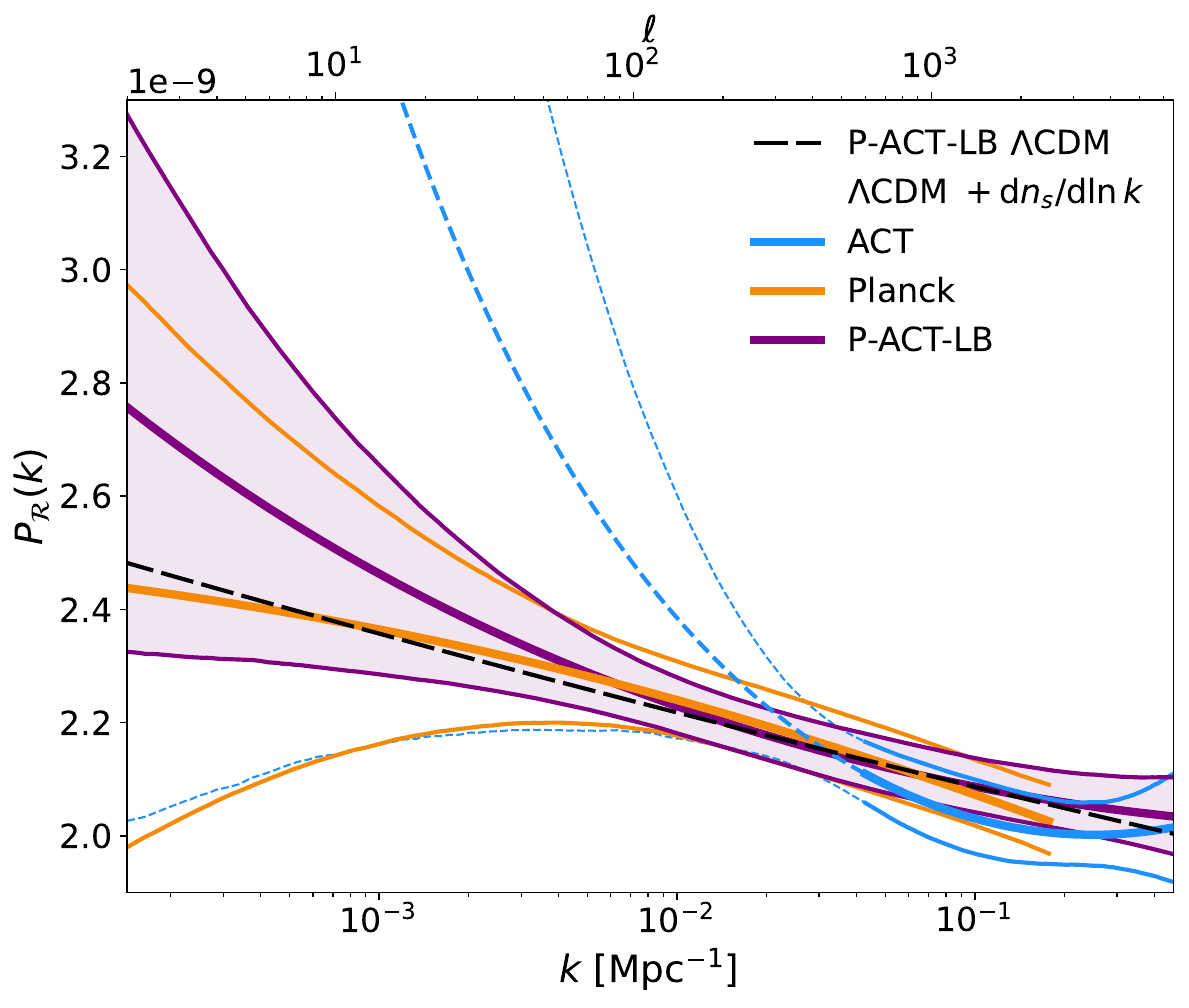}
	\caption{Inferred constraints on the primordial power spectrum of scalar curvature perturbations from \act\ (blue), \Planck\ (orange), and \pactlb\ (purple), assuming a primordial power spectrum  parameterized by an amplitude $A_s$, spectral index $n_s$, and running of the spectral index $\mathrm{d} n_s/\mathrm{d} \ln k$, each defined at a pivot scale $k_* = 0.05 \, \mathrm{Mpc}^{-1}$. The constraints shown are the 95\% CL allowed by the parametric constraints in Fig.~\ref{fig:nrun}, with the thick line being the mean and the thin lines the 95\% limits. Each data combination is only plotted along the range of scales probed by the experiment, i.e., \Planck\ is shown for $2 \le \ell \le 2508$, and \act\ is shown with solid blue lines in the range $600 \le \ell \le 6500$ probed directly by DR6 and with dashed blue lines extrapolating this down to $\ell = 2-32$ probed by the inclusion of the \textsf{Sroll2} data. The results are consistent with a vanishing running, as shown by the \pactlb\ \lcdm\ best-fit (long-dashed black line).}
	\label{fig:pknrun}
\end{figure}

This parameter was found to be consistent with zero by \Planck, with ${\mathrm{d} n_s/\mathrm{d} \ln k = -0.0041 \pm 0.0067}$ from combining \Planck\ CMB, lensing, and BAO data~\citep{Planck_2018_params,Planck2018_inflation}. With the addition of the new ACT DR6 spectra we confirm a vanishing running of the spectral index and tighten the error bar, finding 
\begin{eqnarray}
    \mathrm{d} n_s/\mathrm{d} \ln k &=& 0.0060\pm0.0055 \quad (68\%, \pact), \nonumber \\
    &=& 0.0062 \pm 0.0052 \quad (68\%, \pactlb).
\end{eqnarray}
Lacking the measurement of the first acoustic peak, and with both parameters acting to balance the tilt of the spectrum at small scales, ACT alone provides a looser bound on the combined $\mathrm{d} n_s/\mathrm{d} \ln k-n_s$ space compared to \Planck, as shown in Fig.~\ref{fig:lcdm_single_extensions_pact}, giving $\mathrm{d} n_s/\mathrm{d} \ln k=0.034 \pm 0.022 \ (68\%, \act)$.  With the \pact\ combination, $n_s$ is firmly constrained and the error bar on the running is tightened by $\approx 20\%$. Adding additional datasets has a marginal impact here, with the constraints stabilizing around a vanishing running. Figure~\ref{fig:nrun} shows the combined measurement of $\mathrm{d} n_s/\mathrm{d} \ln k-n_s$ and projections to \Planck's large scales, at $k_*=0.002~\rm{Mpc^{-1}}$. In Fig.~\ref{fig:pknrun} we report the same measurements in terms of the scalar primordial power spectrum using \act, \Planck, and \pactlb. We infer $\mathcal{P}_{\mathcal{R}}(k)$ (using Eq.~\ref{eq:pks}) by post-processing the chains using the amplitude, spectral index, and running of the spectral index of our runs, and computing the 95\% two-tailed confidence interval.
All three datasets measure a similar amplitude and spectral index at the pivot scale, but the constraints from \Planck\ prefer a slight negative running of the spectral index (due to needing to reconcile constraints from large and small scales --- see the discussion in \citealp{Planck_2018_params}), indicated by the concave shape of the mean of the constraints, while \act\ and the combination \pactlb\ mildly prefer a slight positive running of the spectral index, changing the overall shape of $\mathcal{P}_{\mathcal{R}}(k)$ to convex.

Our constraints on $\mathrm{d} n_s/\mathrm{d} \ln k$ disfavor the moderate evidence for negative running of the spectral index seen in combined fits to Lyman-$\alpha$ forest and \Planck\ data, ${\mathrm{d} n_s/\mathrm{d} \ln k = -0.010 \pm 0.004}$~\citep{eBOSS_Lya_2020}. From CMB data alone (\pact), we exclude $\mathrm{d} n_s/\mathrm{d} \ln k = -0.010$ at over $3\sigma$ significance.

\subsection{Primordial power spectrum}\label{sec:pk}

In order to explore a broader range of deviations from a simple power-law primordial adiabatic power spectrum than those captured solely by the running of the scalar spectral index, we consider a more model-independent approach.  

We reconstruct the primordial power spectrum of the scalar perturbations for wavenumber bins centered at $k = 10^{-4}, 10^{-3.5}, 10^{-3}, 10^{-2.5}$ and then 26 equally-spaced logarithmic bins from ${0.011\lesssim k ~/\mathrm{Mpc^{-1}}\lesssim 0.43}$, where $k_{i+1} \simeq 1.16k_i$ \citep[following previous similar analyses in][]{2003MNRAS.342L..72B, 2011JCAP...08..031G, 2012ApJ...749...90H, 2013PhRvD..87h3526A, 2014JCAP...01..025H, 2014PhRvD..89j3502D, 2016PhRvD..93b3504M, 2016JCAP...09..009H, 2017PhRvD..96h3526O, Planck2018_inflation,2025arXiv250310609R}.  
Given the degeneracy between the primordial power and the optical depth (often described in terms of the ${A_s-\tau}$ degeneracy), we sample the value of $e^{-2\tau}P_{\mathcal{R}}(k_{i})$ for each $i^{\rm th}$ bin. We use the cubic spline interpolation method implemented within \texttt{camb} to build the initial power spectrum from our binned values. For $k$ values below our minimum-assumed wavenumber $k_\mathrm{min}$, we set the amplitude of the primordial power spectrum to $P_{\mathcal{R}}(k \leq k_\mathrm{min}) = P_{\mathcal{R}}(k_\mathrm{min})$. This approach removes the scalar spectral amplitude $A_s$ and the scalar index $n_s$ as sampled parameters, but adds the amplitudes within the 30 different bins as described above, leading to 28 additional parameters/degrees of freedom in this model as compared to \lcdm. The prior ranges that we adopt for each $k$ bin are given in Table~\ref{table:binned priors} in Appendix~\ref{app:pkcompilation}, and we use the \texttt{ACT-lite} likelihood to sample the extended parameter space.

\begin{figure}
	\centering
\includegraphics[width=\columnwidth]{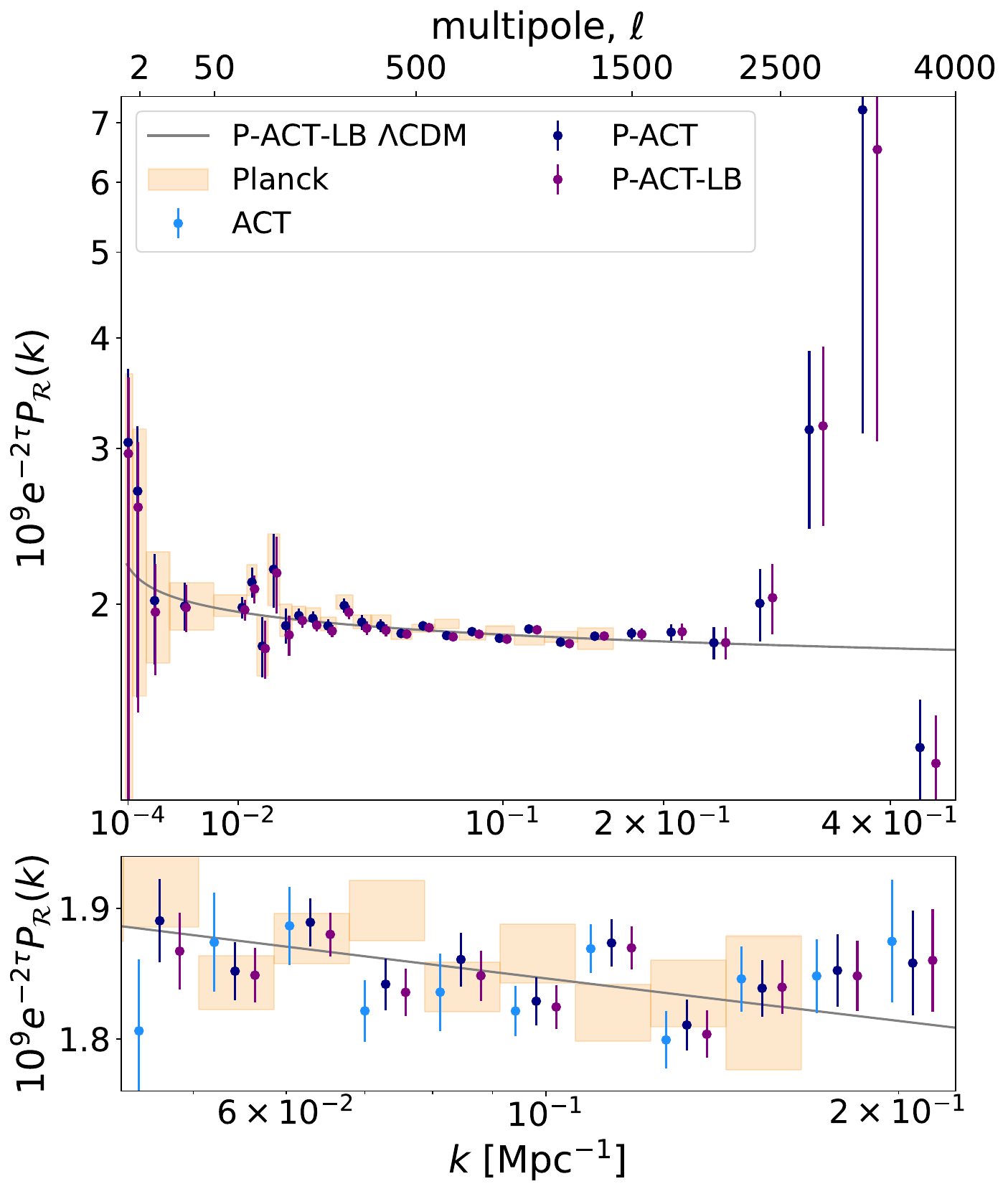}
	\caption{Binned reconstruction of the dimensionless primordial curvature power spectrum with 68\% CL errors for different datasets. The new constraints are centered on the \pact\ measurement; the \act\ (\pactlb) wavenumbers on the $x$-axis are shifted to the left (right) by 4\% in order to more easily see the values/errors for each dataset combination. \Planck\ is plotted as shaded boxes for comparison up to ${k=0.15~\mathrm{Mpc}^{-1}}$ --- beyond this wavenumber, \Planck\ alone gives unconstrained limits with the posteriors filling the whole prior range. The top panel shows the reconstruction over the full range of wavenumbers considered and highlights the improvement achieved with the addition of the ACT data (navy versus orange). For this panel, the $x$-axis is scaled as $k^{0.5}$ in order to best show the small scales. The bottom panel zooms into the region where ACT alone (light blue) becomes comparable and then overtakes \Planck\ in constraining power. Adding lensing and BAO data (purple) has minimal impact. The binned measurement is consistent with the \pactlb\ best-fit $\Lambda$CDM power-law spectrum shown in gray. }
	\label{fig:binned Pk}
\end{figure}

\begin{figure}
	\centering
\includegraphics[width=0.9\columnwidth]{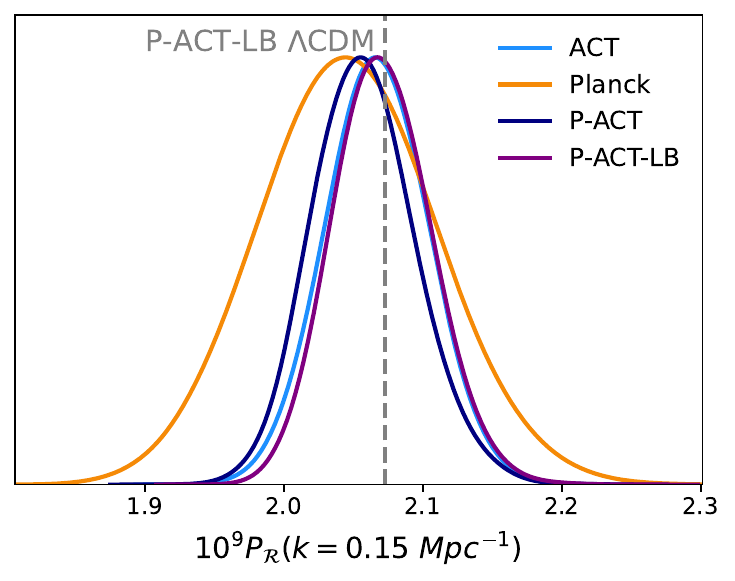}
	\caption{Marginalized posterior for the power at wavenumber ${k_\star = 0.15~{\rm Mpc^{-1}}}$, showing the improvement brought by \act\ over \Planck\ and the limit achieved with \pactlb. The results are consistent with the \pactlb\ best-fit $\Lambda$CDM power-law model (dashed gray line). \Planck\ cannot constrain the model above $k_\star$, while \act\ continues to measure the amplitude of primordial power with a two-sided posterior up to ${k=0.32~\mathrm{Mpc}^{-1}}$.}
	\label{fig:binned Pk 1D}
\end{figure}

In Fig.~\ref{fig:binned Pk} we present constraints on the binned primordial power spectrum from \act, \Planck, \pact, and \pactlb, compared to the \pactlb\ best-fit $\Lambda$CDM power law. The constraints from \pact\ are improved over the \Planck-alone measurements wherever \act\ data are included (for $\ell \gtrsim 850$), with the ratio of \pact\ to \Planck\ error bars ranging between 0.95 and 0.43 for wavenumbers between ${0.063< k~\mathrm{Mpc^{-1}}< 0.15}$ or (${880 \lesssim \ell \lesssim 2140}$). This range is chosen as ${k=0.063~\mathrm{Mpc}^{-1}}$ is the first wavenumber where \pact\ significantly improves over \Planck\ and ${k_\star=0.15~\mathrm{Mpc}^{-1}}$ is the highest wavenumber where \Planck\ has a two-sided marginalized posterior (rather than an upper limit). The correlations across bins can be very high and present either between direct neighbors or between more distant bins. In particular, we find correlations at the level of $\sim 30-40\%$, with a significant increase around $k=0.01$~Mpc$^{-1}$ where bins 6--9 are strongly anti-correlated reaching a level of 90\%, and above $k=0.1$~Mpc$^{-1}$ where correlations increase again to reach $50-60\%$ at small scales.

In Fig.~\ref{fig:binned Pk 1D}, we illustrate the improvement in constraining power for ${k_\star=0.15~\mathrm{Mpc}^{-1}}$. Above that wavenumber, the \Planck\ constraints essentially fill the assumed prior range, while \act\ constrains the amplitude of primordial power with a two-sided marginalized posterior up to ${k=0.32~\mathrm{Mpc}^{-1}}$ and with a one-sided 95\% CL to ${k=0.43~\mathrm{Mpc}^{-1}}$. The improvements come from the \act\ sensitivity to smaller angular scales ($\ell > 2500).$ In particular, at $k_\star$, we obtain a fractional error of 2.8\% and 1.4\% on $e^{-2\tau}P_{\mathcal{R}}(k)$ for \Planck\ alone and ACT alone, respectively, and a fractional error of
\begin{eqnarray}
\frac{\sigma(e^{-2\tau}P_{\mathcal{R}}(k_\star))}{e^{-2\tau}P_{\mathcal{R}}(k_\star)} & = & 1.2\% \ \ (\pact), \nonumber \\
 & = & 1.1\% \ \ (\pactlb),
\end{eqnarray}
with the individual fits consistent with $\Lambda$CDM. Our new constraints improve the range of $k$ over which there are significant constraints by about a factor of three compared to \Planck.

We map this measurement onto the linear matter power spectrum and compare with other measurements in Appendix~\ref{app:pkcompilation}.\\

\subsection{Isocurvature perturbations}\label{sec:isocurvature}

Standard single-field inflation predicts purely adiabatic primordial perturbations, in the sense that all perturbations arise from a single degree of freedom (the local time at which inflation ends).  This generally implies that fractional perturbations in the number density of all species are equal on super-horizon scales ($\delta n_i/n_i=\delta n_j/n_j$).  In contrast, isocurvature perturbations --- characterized by variations in the relative number densities or velocities of different species --- can arise in a variety of early-universe scenarios, particularly if there are additional fields present around the time of inflation~\citep{Linde:1985yf, Mollerach:1989hu, Polarski:1994rz, Garcia-Bellido:1995hsq, Linde:1996gt, Lyth:2001nq, Kawasaki:2001in, Lyth:2002my}. Whereas current cosmological observations are consistent with purely adiabatic perturbations, and a pure isocurvature perturbation has been excluded since early measurements of the large-scale CMB anisotropies~\citep{Stompor:1995py, Enqvist:2000hp, Enqvist:2001fu}, it is possible to have models with mixed adiabatic and isocurvature modes, as constrained in previous CMB anisotropy analyses~\citep[e.g.,][]{Bucher2004,Moodley:2004nz,Bean2006,Planck_2013_params,Planck:2015sxf,Planck2018_inflation}. In this section, we place constraints on a range of scenarios consisting of a mixture of an adiabatic mode with a single (possibly correlated) isocurvature mode. We refer the reader to Table~\ref{table:isocurvature_constraints} in Appendix~\ref{app:isocurvature_full_constraints} for a detailed compilation of the isocurvature constraints derived in this work.

There are four possible non-decaying isocurvature modes: cold dark matter density (CDI), baryon density (BDI), neutrino density (NDI), and neutrino velocity (NVI)~\citep{Bucher:1999re}. At linear order, BDI and CDI modes have an indistinguishable impact on the CMB power, differing only in amplitude; therefore, we do not analyze BDI modes directly.\footnote{Our CDI constraints can be converted to BDI constraints by applying the appropriate factors of $\Omega_c$ and $\Omega_b$, as discussed in~\cite{Planck2018_inflation}.}  Additionally, we do not consider NVI because it is less theoretically motivated than CDI and NDI. Consequently, we focus on two isocurvature modes: $\mathcal{I}_{\rm CDI}$ and $\mathcal{I}_{\rm NDI}$.

The primordial curvature and isocurvature fluctuations are described by the curvature power spectrum, $\mathcal{P}_{\mathcal{R}\mathcal{R}}(k)$, the isocurvature power spectrum, $\mathcal{P}_{\mathcal{I}\mathcal{I}}(k)$, and their cross-power spectrum, $\mathcal{P}_{\mathcal{R}\mathcal{I}}(k)$. Following \emph{Planck} analyses of isocurvature~\citep{Planck_2013_params, Planck:2015sxf, Planck2018_inflation}, we assume power-law primordial power spectra defined in terms of their amplitudes at two scales, $k_1=0.002~{\rm Mpc}^{-1}$ and $k_2=0.1~{\rm Mpc}^{-1}$, hence
\begin{align}\label{eq:scalar_osc_bispectrum}
\begin{split}
    \mathcal{P}_{ab}(k)=\exp\bigg[&\left(\frac{\ln(k)-\ln(k_2)}{\ln(k_1)-\ln(k_2)}\right)\ln\left(\mathcal{P}_{ab}^{(1)}\right)\\
    &+\left(\frac{\ln(k)-\ln(k_1)}{\ln(k_2)-\ln(k_1)}\right)\ln\left(\mathcal{P}_{ab}^{(2)}\right)\bigg]
\end{split}
\end{align}
where $\mathcal{P}_{ab}^{(i)}\equiv \mathcal{P}_{ab}(k_i)$ and $a,b\in\{\mathcal{R},\mathcal{I}\}$. The mixed adiabatic and isocurvature models are characterized by four new parameters: $\mathcal{P}_{\mathcal{R}\mathcal{I}}^{(1)}$, $\mathcal{P}_{\mathcal{R}\mathcal{I}}^{(2)}$, $\mathcal{P}_{\mathcal{I}\mathcal{I}}^{(1)}$, $\mathcal{P}_{\mathcal{I}\mathcal{I}}^{(2)}$. 

For computational reasons, we focus on models where the adiabatic and isocurvature perturbations are either uncorrelated or fully (anti)-correlated, and thus $\mathcal{P}_{\mathcal{R}\mathcal{I}}^{(i)}$ is a derived parameter. These are the most well-motivated isocurvature scenarios.  We sample the adiabatic amplitudes $\mathcal{P}_{\mathcal{R}\mathcal{R}}^{(1)}$ and $\mathcal{P}_{\mathcal{R}\mathcal{R}}^{(2)}$ assuming uniform priors between $[15, 40]\times 10^{-10}$ as a replacement of the amplitude and scalar spectral index of the baseline \lcdm\ parameter set. For the isocurvature amplitudes, we sample $\mathcal{P}_{\mathcal{I}\mathcal{I}}^{(1)}$ and $\mathcal{P}_{\mathcal{I}\mathcal{I}}^{(2)}$ (for models where $n_{\mathcal{I}\mathcal{I}}$ is not fixed) assuming uniform priors between $[0,100]\times 10^{-10}$. We sample the amplitudes at two scales instead of sampling a single amplitude and a spectral index to mitigate prior-volume effects that can arise if the data have no preference for isocurvature modes~\citep{Moodley:2004nz}.\footnote{We use \texttt{class} with the settings described in Appendix~\ref{app:theory} to compute theoretical predictions for isocurvature models. Additionally, we set \texttt{hmcode\_kmax\_extra}=100 to address issues in the \texttt{HMCode} non-linear power spectrum calculation for blue-tilted isocurvature perturbations. Despite this, the non-linear evaluation can still fail for a small fraction ($<1$\%) of samples in the \emph{Planck}-only uncorrelated CDI constraints with a free spectral index. We exclude this subset of samples from our analysis, slightly affecting \emph{Planck}-alone constraints for this model. The issues with the non-linear computation do not affect any of our constraints including ACT data, which already exclude such blue-tilted models.}

\subsubsection{Uncorrelated models}

We first present constraints on an uncorrelated ($\mathcal{P}_{\mathcal{R}\mathcal{I}}^{(i)}=0$) mixture of adiabatic perturbations with a single, possibly scale-invariant, CDI or NDI mode. Scale-invariant CDI perturbations can arise if there are axions or axion-like particles present during inflation~\citep{Turner:1983sj, Axenides:1983hj, Seckel:1985tj, Turner:1990uz, Linde:1991km, Beltran:2006sq, Hertzberg:2008wr}. Axions can also produce uncorrelated CDI with a blue spectral index~\citep{Kasuya:2009up}.

For scale-invariant CDI and NDI modes, we vary $\mathcal{P}_{\mathcal{I}\mathcal{I}}^{(1)}$ and fix $\mathcal{P}_{\mathcal{I}\mathcal{I}}^{(2)}=\mathcal{P}_{\mathcal{I}\mathcal{I}}^{(1)}$.  The \pact\ constraint on the amplitude of scale-invariant CDI perturbations is $10^{10}\mathcal{P}_{\mathcal{I}\mathcal{I}}^{(1)}<1.1$ (95\% CL). This is slightly weaker than the \emph{Planck}-only bound of $10^{10}\mathcal{P}_{\mathcal{I}\mathcal{I}}^{(1)}<0.9$ (95\% CL), due to the positive correlation between the adiabatic spectral index ($n_{\mathcal{R}\mathcal{R}}$) and the CDI amplitude. For NDI, the baseline \pact\ constraint on the isocurvature amplitude is $10^{10}\mathcal{P}_{\mathcal{I}\mathcal{I}}^{(1)}<1.8$ at (95\% CL), which is 15\% tighter than the \emph{Planck}-only bound, $10^{10}\mathcal{P}_{\mathcal{I}\mathcal{I}}^{(1)}<2.1$ (95\% CL). Neither of these constraints changes appreciably between \pact\ and \pactlb. 

\begin{figure}[!t]
\centering
\includegraphics[width=\columnwidth]{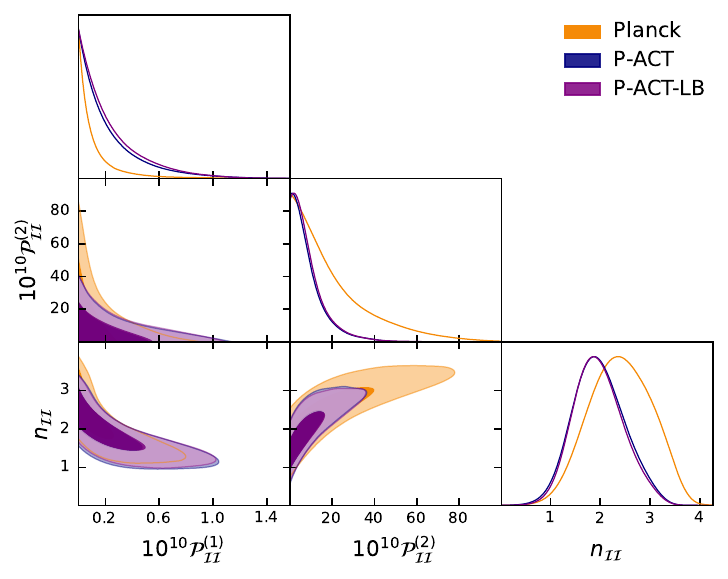}
\caption{Constraints on the primordial isocurvature power spectrum parameters for a mixed adiabatic and uncorrelated CDI mode with a power-law power spectrum for several dataset combinations.} \label{fig:cdi_posterior}
\end{figure}

Figure~\ref{fig:cdi_posterior} shows the marginalized posterior distributions for the uncorrelated CDI model with a free spectral index. Combining ACT and \emph{Planck}, we find
\begin{equation}
\left.
 \begin{aligned}
10^{10}\mathcal{P}_{\mathcal{I}\mathcal{I}}^{(1)} &< 0.7 \\ 
10^{10}\mathcal{P}_{\mathcal{I}\mathcal{I}}^{(2)} &< 26
\end{aligned}
\quad \right\} \mbox{\textrm{\; (\textsf{one-tail}\ 95\%, \pact).}}
\end{equation}
The \pact\ constraint on $\mathcal{P}_{\mathcal{I}\mathcal{I}}^{(2)}$ is a significant improvement over the \emph{Planck}-only constraint of $10^{10}\mathcal{P}_{\mathcal{I}\mathcal{I}}^{(2)} < 59$. Interestingly, the \pact\ constraint on $\mathcal{P}_{\mathcal{I}\mathcal{I}}^{(1)}$ is weaker than that from the \emph{Planck} dataset alone. This is a consequence of the anti-correlation between $\mathcal{P}_{\mathcal{I}\mathcal{I}}^{(1)}$ and $\mathcal{P}_{\mathcal{I}\mathcal{I}}^{(2)}$, assuming a power-law spectrum. In other words, the large values of $\mathcal{P}_{\mathcal{I}\mathcal{I}}^{(2)}$ that are consistent with the \emph{Planck} data, but not the \pact\ combination, drive the \emph{Planck}-only marginalized posterior on $\mathcal{P}_{\mathcal{I}\mathcal{I}}^{(1)}$ towards zero. The addition of CMB lensing and BAO data does not significantly change the \pact\ constraints on this model.

\begin{figure}[!t]
\centering
\includegraphics[width=\columnwidth]{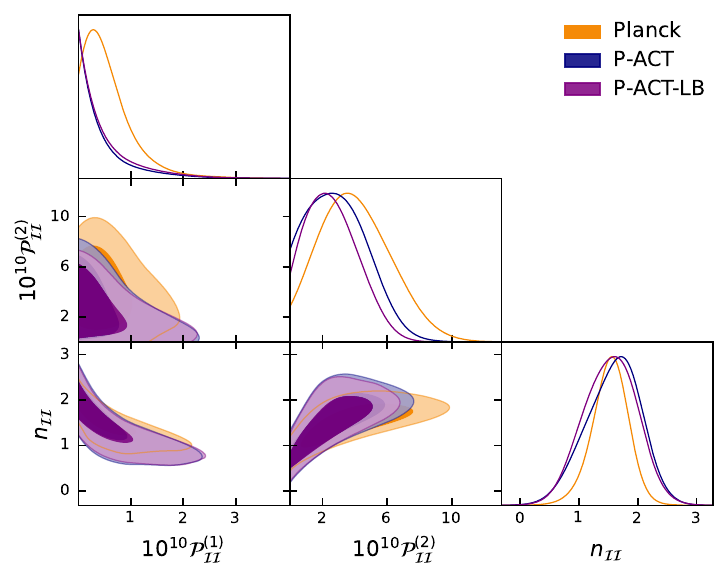}
\caption{Same as Fig.~\ref{fig:cdi_posterior} for a primordial isocurvature power spectrum with a mixed adiabatic and uncorrelated NDI mode.} \label{fig:ndi_posterior}
\end{figure}

Figure~\ref{fig:ndi_posterior} presents the marginalized posteriors on the isocurvature amplitudes for an uncorrelated mixture of adiabatic and NDI modes assuming a free spectral index. Combining ACT and \emph{Planck}, we find
\begin{equation}
\left.
 \begin{aligned}
10^{10}\mathcal{P}_{\mathcal{I}\mathcal{I}}^{(1)} &< 1.6 \\ 
10^{10}\mathcal{P}_{\mathcal{I}\mathcal{I}}^{(2)} &< 6.1 
\end{aligned}
\quad \right\} \mbox{\textrm{\; (\textsf{one-tail}\ 95\%, \pact).}}
\end{equation}
Interestingly, although the \emph{Planck} data alone showed a small, statistically insignificant peak in the posterior at non-zero values of $\mathcal{P}_{\mathcal{I}\mathcal{I}}^{(1)}$ and $\mathcal{P}_{\mathcal{I}\mathcal{I}}^{(2)}$, this feature disappears when ACT DR6 data are included. Finally, including CMB lensing and BAO data further improves the constraint on the amplitude of NDI perturbations at $k=0.1~{\rm Mpc}^{-1}$ to $10^{10}\mathcal{P}_{\mathcal{I}\mathcal{I}}^{(2)}<5.4$. In this case, the improvement from CMB lensing and BAO is predominately due to the anti-correlation between $n_{\mathcal{R}\mathcal{R}}$ and  $10^{10}\mathcal{P}_{\mathcal{I}\mathcal{I}}^{(2)}$, with \pactlb\ preferring a slightly larger value of $n_{\mathcal{R}\mathcal{R}}$ than \pact.

In summary, we find no evidence of uncorrelated CDI or NDI. Nevertheless, including ACT DR6 data leads to stringent bounds on the amplitude of small-scale isocurvature perturbations, highlighting the complementarity of large- and small-scale CMB observations.

\subsubsection{Correlated models}

We also present constraints on a correlated mixture of adiabatic and CDI perturbations. We consider the situation where the adiabatic and CDI modes are fully correlated, $\mathcal{P}_{\mathcal{R}\mathcal{I}}(k)=\sqrt{\mathcal{P}_{\mathcal{R}\mathcal{R}}(k)\mathcal{P}_{\mathcal{I}\mathcal{I}}(k)}$, as well as the fully anti-correlated scenario ($\mathcal{P}_{\mathcal{R}\mathcal{I}}(k)=-\sqrt{\mathcal{P}_{\mathcal{R}\mathcal{R}}(k)\mathcal{P}_{\mathcal{I}\mathcal{I}}(k)}$). In both cases, we vary only $\mathcal{P}_{\mathcal{I}\mathcal{I}}^{(1)}$ and fix $\mathcal{P}_{\mathcal{I}\mathcal{I}}^{(2)}$ assuming $n_{\mathcal{I}\mathcal{I}}=n_{\mathcal{R}\mathcal{R}}$. While the high-resolution ACT DR6 spectra are unlikely to significantly improve upon the \emph{Planck}-only constraints on these models as the transfer function for scale-invariant CDI perturbations is significantly suppressed at high multipoles, we explore correlated CDI models because they are theoretically well-motivated. For example, mixed adiabatic and CDI perturbations with large (anti)-correlations arise in the curvaton scenario~\citep{Mollerach:1989hu, Moroi:2001ct, Lyth:2001nq, Bartolo:2002vf, Lyth:2002my}. 

For the fully correlated model, the \pact\ dataset constrains the isocurvature amplitude to $10^{10}\mathcal{P}_{\mathcal{I}\mathcal{I}}^{(1)} < 0.025$ (95\% CL). Including CMB lensing and BAO data slightly relaxes this bound to $10^{10}\mathcal{P}_{\mathcal{I}\mathcal{I}}^{(1)} < 0.031$. For the fully anti-correlated model, \pact\ provides $10^{10}\mathcal{P}_{\mathcal{I}\mathcal{I}}^{(1)} < 0.027$, improving modestly on the \emph{Planck}-only limit ($<0.031$). \pactlb\ leads to $10^{10}\mathcal{P}_{\mathcal{I}\mathcal{I}}^{(1)} < 0.015$ at 95\% confidence.

Overall, we find no evidence of primordial isocurvature perturbations.

\subsection{Tensor modes and constraints on inflation models}\label{sec:tensors} 

Tensor perturbations, for example in the form of primordial gravitational waves predicted by inflation models (see, e.g.,~\citealp{2016ARA&A..54..227K} for a review), add power on very large scales in the CMB T/E angular power spectra and generate a unique large-scale B-mode polarization signal not measured by ACT. However, the full behavior of the primordial perturbations depends on the ratio/co-existence of tensor and scalar modes, with the latter measured by ACT small-scale data via $n_s$. Here, we study how the updated measurement of $n_s$ presented in L25 affects constraints on inflation models.

The modeling of tensor perturbations is similar to that used for scalars, $\mathcal{P}_t(k) = A_t(k/k_*)^{n_t}$, and is usually quantified by the tensor-to-scalar ratio parameter, $r=\mathcal{P}_t(k_*)/\mathcal{P}_{\mathcal{R}}(k_*)$.  Here we set $n_t$ via the inflationary consistency relation, $n_t = -r/8$.  Constraints on $r$ and $n_s$ can be directly related to the potential slow-roll parameters of inflation, $\epsilon_V$ and  $\eta_V$, which are determined by the shape of the inflaton potential, as
\begin{align}
    \epsilon_V &\equiv \frac{{M_{Pl}}^2}{2} \left( \frac{V_{,\phi}}{V} \right)^2, \nonumber \\
    \eta_V &\equiv {M_{Pl}}^2 \left( \frac{V_{,\phi\phi}}{V} \right) \,,
\end{align}
where $M_{Pl}$ is the Planck mass, $V(\phi)$ is the inflaton potential, and $V_{,\phi}$ and $V_{,\phi\phi}$ are its first and second derivatives. These parameters respectively describe the steepness and curvature of the inflaton potential, and they relate to the power spectrum parameters as
\begin{align}
    r &= 16 \epsilon_V, \nonumber \\
    n_s - 1 &= 2 \eta_V - 6 \epsilon_V \,.
\end{align}

\begin{figure}[t!]
	\centering
\includegraphics[width=\columnwidth]{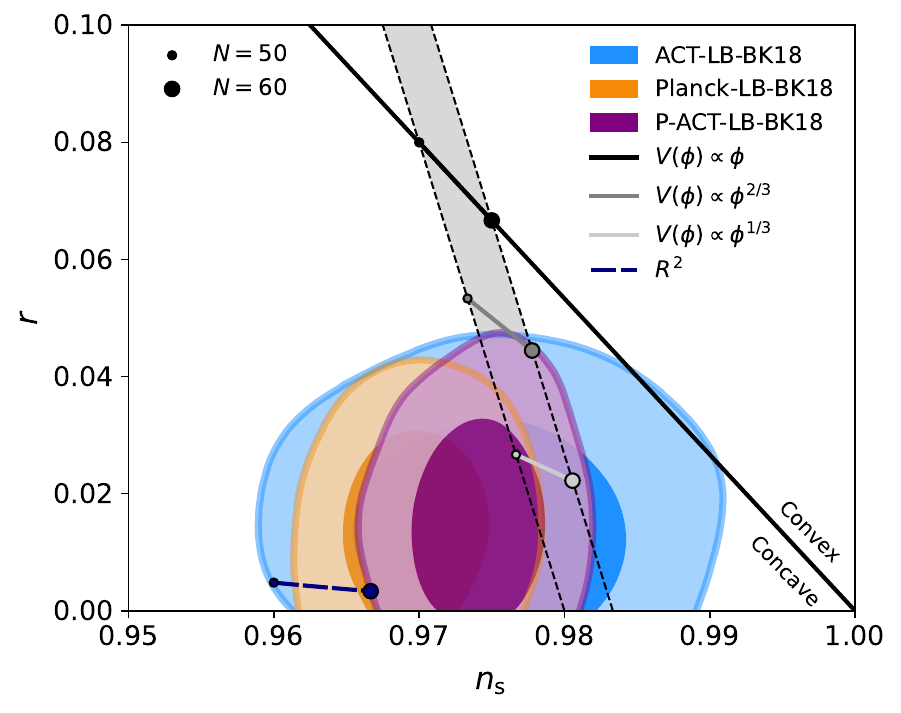}
	\caption{Constraints on the scalar and tensor primordial power spectra at $k_* = 0.05 \, \mathrm{Mpc}^{-1}$, shown in the $r-n_s$ parameter space. The constraints on $r$ are driven by the BK18 data, while the constraints on $n_s$ are driven by \Planck\ (orange), \act\ (blue), or \pact\ (purple). The combined dataset also includes CMB lensing and BAO in all cases. The various circles and solid lines in between the gray band show predictions for different power-law potentials with the number of e-folds of inflation $50<N<60$, while the solid black line shows the separation between convex- and concave-shaped potentials. The Starobinsky $R^2$ model is also shown for 50-60 e-folds (dashed navy line); the \pactlb\ measurement of $n_s$ disfavors this model at $\gtrsim 2\sigma$, for $50<N<60$.}
	\label{fig:tensors}
\end{figure}
\begin{figure}[t!]
	\centering
  \includegraphics[width=\columnwidth]{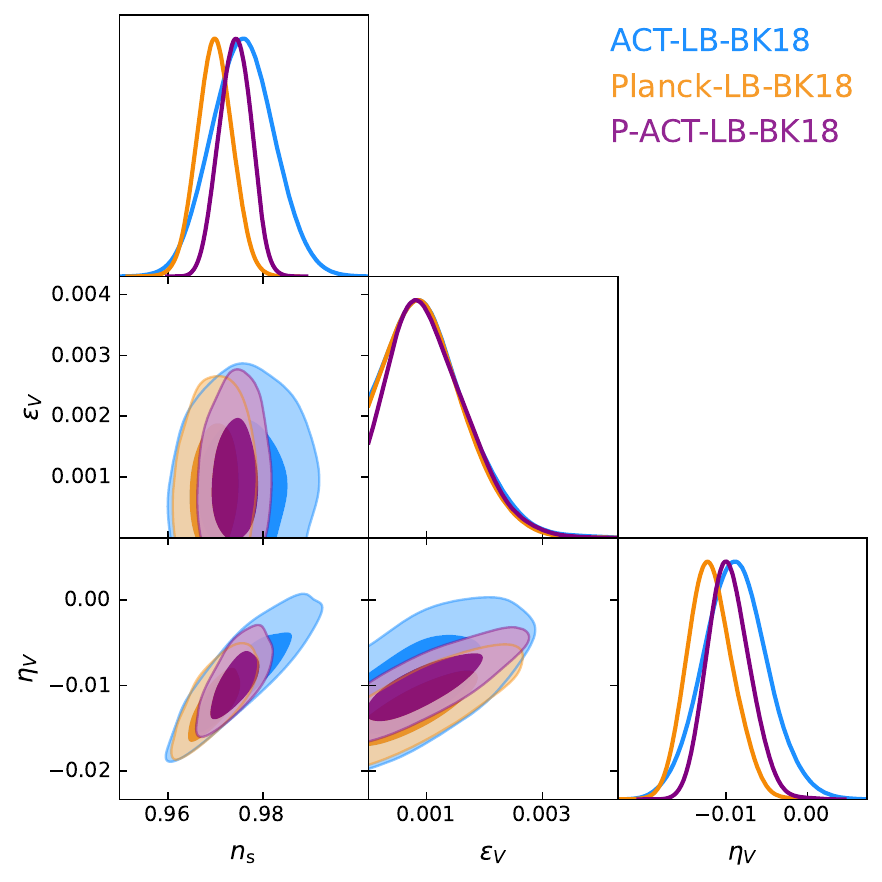}
	\caption{Constraints on the slow-roll parameters $\epsilon_V$ and $\eta_V$. The first parameter is driven by the BK18 limit on the tensor-to-scalar ratio, while $\eta_V$ tracks the contribution to $n_s$ from ACT (blue), \Planck\ (orange), and \pactlb\ (purple).}
	\label{fig:slow-roll}
\end{figure}

To derive new results in the $r-n_s$ plane presented in Fig.~\ref{fig:tensors}, we vary $r$ in the range $[0,1]$ and we include B-mode measurements from the BICEP and Keck telescopes at the South Pole, which have jointly accumulated more than fifteen years of observations of CMB polarization at degree scales. Their measurements of CMB B-modes are the most precise to date in the field~\citep[BK18]{BK18} and completely drive the upper limit on $r$, which (combined with $n_s$ from \Planck) strongly excludes monomial inflation models with convex potentials. The inclusion of \act\ data shows a preference for a slightly higher value of $n_s$, indicated by the purple contour shifting to the right of the orange contour in Fig.~\ref{fig:tensors}, and gives $r<0.038$ (95\%, \pactlbbicep). As a consequence, \pact\ can accommodate power-law inflation models with slightly higher power-law indices than those preferred by \Planck\ alone (assuming 50--60 e-folds of inflation), i.e., power-law models that are closer to a linear potential and hence possess a second derivative closer to zero. On the contrary, the Starobinsky $R^2$ inflation model~\citep{1980PhLB...91...99S,1992JETPL..55..489S}, which predicts lower values of $n_s$ for the standard range of 50--60 e-folds,\footnote{The number of e-folds depends on the (highly uncertain) post-inflation reheating history; see, e.g., the discussion in Sec.~2.4--2.5 of~\cite{CMB_S4_Science_Book}.} is more disfavored and lies on the $2\sigma$ boundary of \pactlb.

The constraints on the slow-roll parameters are shown in Fig.~\ref{fig:slow-roll}, with BK18 determining $\epsilon_V$ and ACT adding to the constraint on $\eta_V$. The slight preference for a higher $n_s$ value implies a value of $\eta_V$ closer to zero,
\begin{eqnarray}
    \eta_V &=& (-95^{+23}_{-30}) \times 10^{-4} \ (68\%, \pactlbbicep) \,,
\end{eqnarray}
and thus implies a preference for a slightly less concave inflaton potential.

The inclusion of running of the spectral index of scalar perturbations does not significantly change these results.\\

\section{Pre- and modified-recombination physics}\label{sec:recomb}
\subsection{Early dark energy}
\label{sec:ede} 

The early dark energy (EDE) extension to $\Lambda$CDM introduces a new field that acts to accelerate the expansion of the universe prior to recombination, reducing the sound horizon at last scattering and thus increasing the CMB-inferred Hubble constant~\citep{Poulin2019,Lin:2019qug,Agrawal:2019lmo}.  Here, we consider an axion-like EDE model that has been widely studied in the literature, which is described by the potential~\citep{Poulin2019,Smith2019}
\begin{equation}
    V(\phi)=m^2f^2(1-\cos(\phi/f))^n \, ,
\end{equation}
where $m$ is the mass of the field, $f$ is the axion decay constant, and $n$ is a power-law index. While $n=1$ is excluded on phenomenological grounds --- as the EDE would act as an additional contribution to dark matter at late times --- values of $n \geq 2$ constitute viable models. Here, we consider as baseline the $n=3$ model, as in previous literature (e.g.,~\citealp{Hill_2020, Ivanov2020, DAmico2020, LaPosta2022, Hill_ACT_EDE}), and discuss the $n=2$ case below and in Appendix~\ref{app:ede_full_constraints}. 

Following previous works \citep{Poulin2019, Smith2019, Hill_ACT_EDE}, we adopt a phenomenological parametrization that introduces three additional parameters beyond those in $\Lambda$CDM: $f_{\rm EDE}$, $z_c$, and $\theta_i$, where  
\begin{equation}
    f_{\mathrm{EDE}} = 8\pi G \rho_{\mathrm{EDE}}(z_c)/(3H^2(z_c)) 
\end{equation}
is the maximum fractional contribution of EDE to the cosmic energy budget, which is reached at ``critical redshift'' $z_c$, and the initial field displacement $\theta_i=\phi_i/f$. The energy density $\rho_{\mathrm{EDE}}$ is given by
\begin{equation}
    \rho_{\mathrm{EDE}}(z_c) = \frac{1}{2}\dot{\phi} (z_c)^2 + V(\phi(z_c))
\end{equation}
and the field $\phi$ evolves by the Klein-Gordon equation (dots here denote derivatives with respect to cosmic time)
\begin{equation}
    \label{eq.KleinGordon}
    \ddot{\phi} + 3H \dot{\phi} + \frac{dV}{d\phi}=0 \, .
\end{equation}
Initially the field is frozen due to Hubble friction. Around redshift $z_c$ when $H \sim m$, the field begins to roll. 

Several previous studies have constrained this model with various datasets (see, e.g.,~\citealp{Kamionkowski:2022,Poulin2023EDEreview,McDonough2024} for reviews). \cite{Hill_2020} found that \emph{Planck} alone has no preference for EDE, with $f_{\mathrm{EDE}}<0.087$ (95\% CL) and $H_0=68.3 \pm 1.0$ (68\% CL); these \Planck-derived bounds were further tightened using NPIPE data in~\cite{McDonough2024,Efstathiou2024EDE}.  However, the ACT DR4 data showed a hint of a preference for the EDE model over $\Lambda$CDM at $2$--$3\sigma$ significance~\citep{Hill_ACT_EDE}.  The combination of ACT DR4, large-scale \emph{Planck} TT, \emph{Planck} CMB lensing, and BOSS BAO data yielded $f_{\mathrm{EDE}}=0.091^{+0.020}_{-0.036}$ (68\% CL) and $H_0=70.9^{+1.0}_{-2.0}$~km/s/Mpc (68\% CL)~\citep{Hill_ACT_EDE} (see also \citealp{2021PhRvD.104l3550P,LaPosta2022,2022PhRvD.106d3526S,2024PhRvD.109j3506S}).  However, the statistical weight of \Planck\ was sufficiently large compared to DR4 that this mild preference disappeared when combining ACT DR4 with the full \emph{Planck} data.

We re-evaluate here the preference for this model with the new ACT DR6 power spectra.  Note that, in general, models with a significant EDE fraction leave imprints in the CMB power spectra on modes that are within (or entering) the horizon at redshifts around $z_c$.  While some of these imprints can be absorbed by changes in other cosmological parameters, particularly $\Omega_c h^2$ and $n_s$~\citep{Hill_2020}, high-resolution polarization data are expected to break these degeneracies and yield strong sensitivity to EDE, should it be present~\citep{Smith2019,Hill_ACT_EDE,2024PhRvD.109j3506S}.  For models with $z_c \sim z_{\rm eq}$, unique features are imprinted in CMB power spectra at ${\ell \sim 400-800}$, as well as broader imprints extending to smaller angular scales.  Indeed, the mild hint of EDE in the ACT DR4 analysis was largely driven by a fluctuation in the EE power spectrum at $\ell \sim 500$ and a broad trend in the joint ACT and \Planck\ TE power spectrum~\citep{Hill_ACT_EDE}.  Our analysis of the new ACT DR6 spectra is a high-precision test as to whether these features were the first hints of a real signal, or simply a statistical fluctuation.

We sample the model parameters with uniform priors $f_{\mathrm{EDE}}{\in [0.001,0.5]}$, $\log_{10}(z_c){\in [3.0,4.3]}$, and $\theta_i{\in [0.1,3.1]}$. We compute theoretical predictions using the EDE model implementation in {\tt class}, which itself was merged from the modified version {\tt class\_ede}~\citep{Hill_2020},\footnote{\url{https://github.com/mwt5345/class_ede}} as well as using {\tt CosmoPower}-based emulators of the {\tt class} EDE predictions (described in~\citealp{Qu:2024lpx}).  One important detail to note is that these emulators were trained on theory predictions computed assuming three massive neutrino species, which is different from our baseline convention in this paper of assuming one massive neutrino.  This difference leads to very small shifts in the $H_0$ posteriors ($<0.1$ km/s/Mpc), but otherwise leaves the parameter constraints unaffected, as we verify using full {\tt class} runs. For calculations in the ${n=2}$ EDE model, we compute theoretical predictions using the implementation in {\tt camb}, as we find that calculations for this model are more numerically stable in {\tt camb} than in {\tt class} (for some extreme $n=2$ scenarios, we find that {\tt class} does not run successfully).  As a test of the theory codes used for the EDE model, we verify that the $\chi^2_{\rm ACT}$ values computed for a benchmark $n=3$ EDE model with {\tt camb} and {\tt class} agree with one another to within 0.06, nearly as good as the \lcdm\ agreement for these codes (see Appendix~\ref{app:theory}).

Figures~\ref{fig:ede_base_triangle} and~\ref{fig:ede_kitchen_sink} show the marginalized posterior distributions for key parameters in the $n=3$ EDE model, for different dataset combinations.  With the new ACT DR6 spectra, we find 
\begin{eqnarray}
    f_{\mathrm{EDE}} &<& 0.088 \quad (95\%, \act),\label{eq.ede_f_act} \nonumber\\
    f_{\mathrm{EDE}} &<& 0.12 \quad \ \ (95\%, \pact), \label{eq.ede_f_pact}
\end{eqnarray}
and 
\begin{eqnarray}
    H_0 &=& 67.5^{+0.9}_{-1.7} \quad (68\%, \act), \label{eq.ede_h0_act} \nonumber\\
    H_0 &=& 69.3^{+0.9}_{-1.5}
 \quad (68\%, \pact) \label{eq.ede_h0_pact}\,, 
\end{eqnarray}
with the \wact\ case also shown in the figures to allow a \Planck-independent assessment of the EDE constraints.  We find that \act\ alone shows essentially no shift in $H_0$ within the EDE model, compared to \lcdm\ --- the hint of non-zero EDE seen in the DR4 data is not seen in DR6 (comparisons between the DR4 and DR6 EDE constraints can be found in Appendix~\ref{app:ede_full_constraints}).  When combining with \Planck, the $H_0$ posterior shifts upward, but when combining with \WMAP, this is significantly lessened: we find $H_0 = 68.1^{+0.8}_{-1.6} \,\, (68\%, \wact)$.  As discussed below, the improvement in quality of fit in all cases is not statistically significant. When we also include CMB lensing and BAO data, we obtain
\begin{eqnarray}
    f_{\mathrm{EDE}} &<& 0.12 \quad \ \ \quad (95\%, \pactlb) \label{eq.ede_f_pactlb} \nonumber\\ 
    H_0 &=& 69.9^{+0.8}_{-1.5} 
 \quad (68\%, \pactlb) \label{eq.ede_h0_pactlb} \,,
\end{eqnarray}
with \wactlb\ giving similar results. For comparison, $H_0 = 68.2 \pm 0.4$~km/s/Mpc in \lcdm\ for \pactlb\ (L25).  
We also consider substituting BOSS BAO for DESI BAO data, which yields $f_{\rm EDE} < 0.10$ (95\%, \pactlbb) and $H_0=69.2^{+0.7}_{-1.3}$~km/s/Mpc (68\%, \pactlbb), slightly decreasing $H_0$ and tightening the bound on $f_{\rm EDE}$ compared to \pactlb.\footnote{This is consistent with the results of~\citet{Qu:2024lpx}, who found that the shift in the best-fit $\Omega_m$ between BOSS and DESI BAO data led to a weaker upper limit on EDE from DESI in combination with data from \Planck\ and other probes.} We further verify that these results are essentially unchanged with the inclusion of SNIa data in the analysis.

\begin{figure}[t]
    \centering
    \includegraphics[width=\columnwidth]{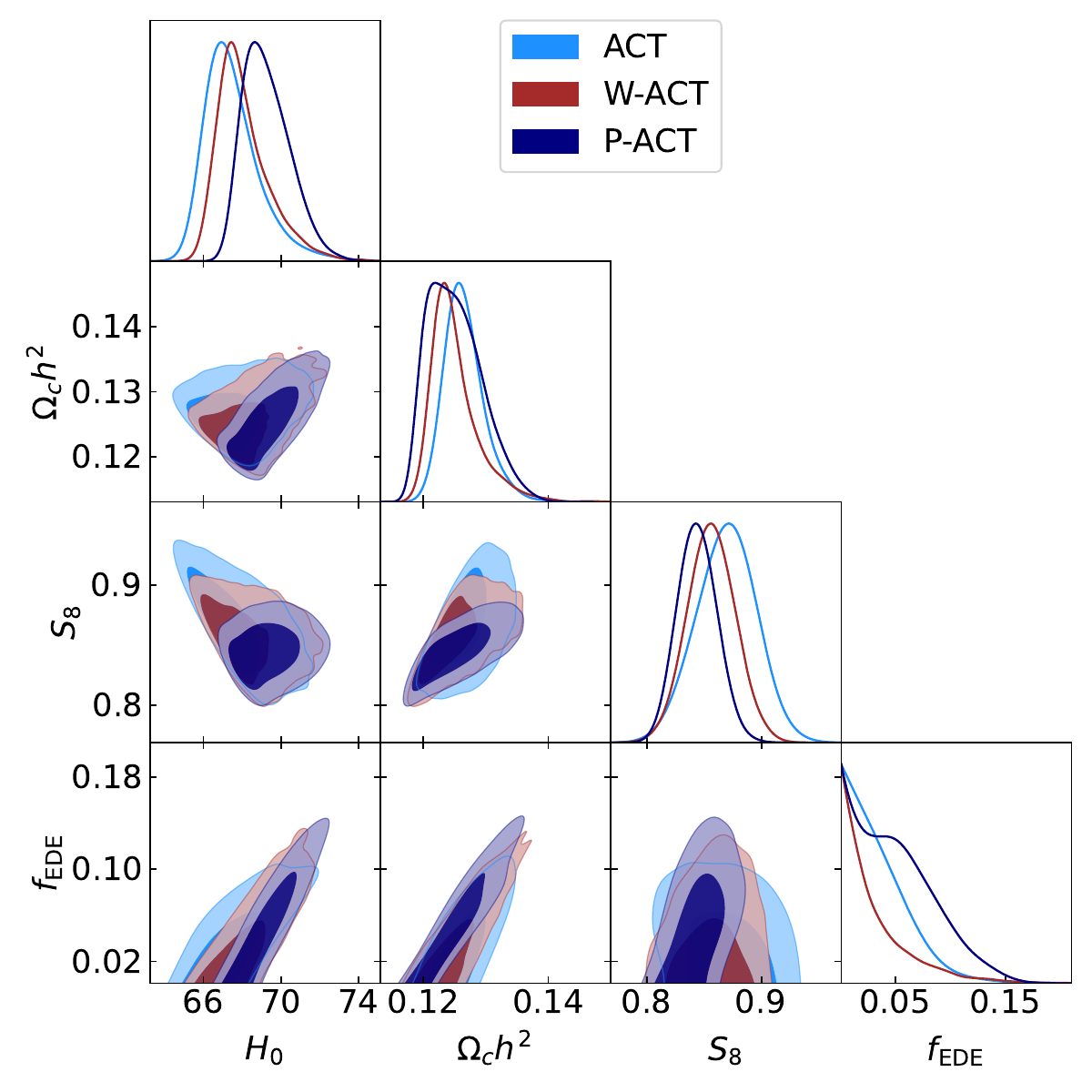}
    \vspace{-0.6cm}\caption{Marginalized contours at 68\% and 95\% confidence for the EDE fraction showing (well-known) positive correlations with $H_0$, $\Omega_c h^2$, and $S_8$ for different dataset combinations (\act\ in blue, \wact\ in brown, and \pact\ in navy).  The small bump in the marginalized $f_{\rm EDE}$ posterior for \pact\ is real and arises from the mild EDE improvement in $\chi^2$ over $\Lambda$CDM in this model --- see Table~\ref{table:ede_chi2}.  }
    \label{fig:ede_base_triangle}
\end{figure}

\begin{figure}[t]
    \centering
    \includegraphics[width=\columnwidth]{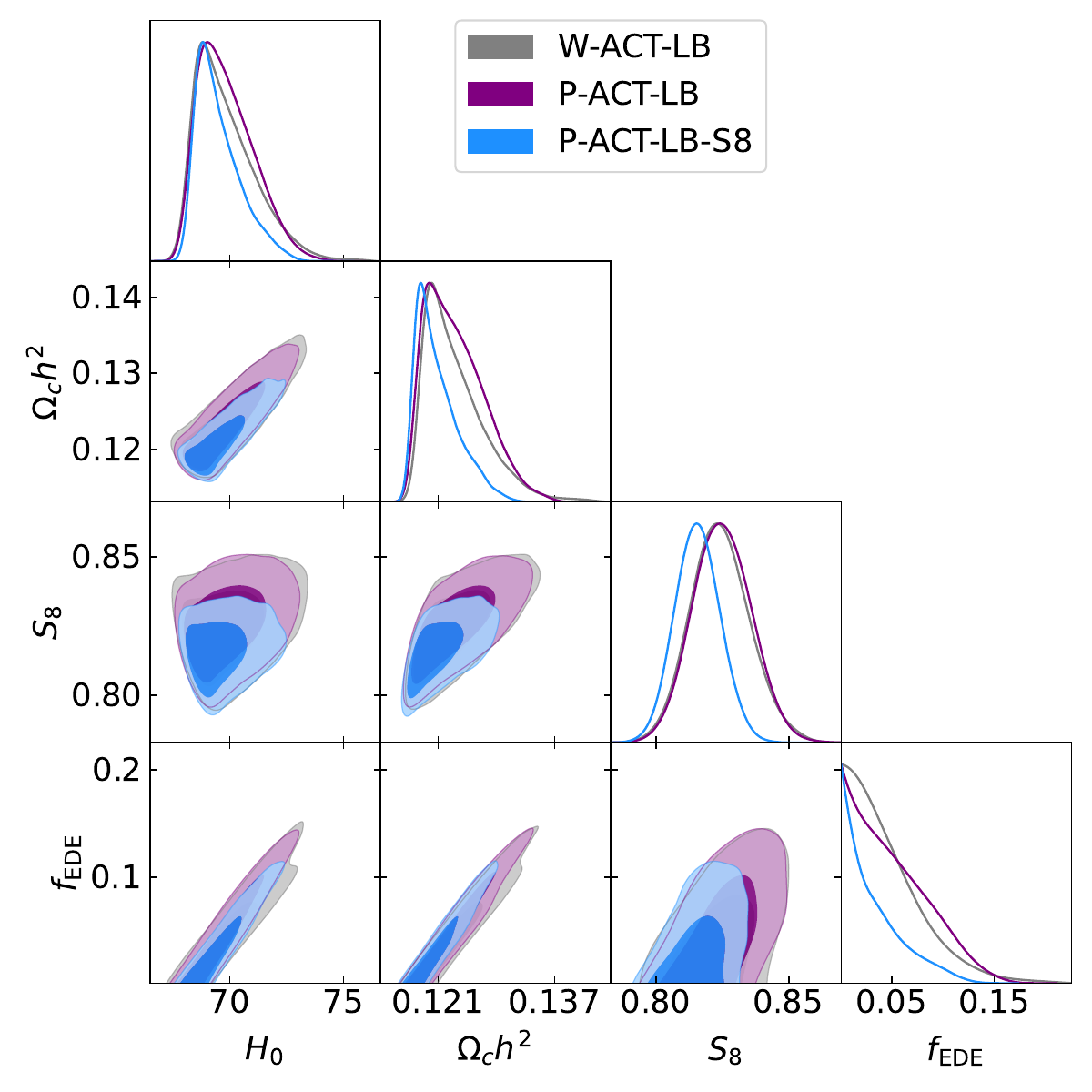}
    \caption{Same as Fig.~\ref{fig:ede_base_triangle}, tightening the limits with additional datasets (\wactlb\ in gray, \pactlb\ in purple, and \pactlb \textsf{-S8} in blue). } 
    \label{fig:ede_kitchen_sink}
\end{figure}

We also obtain constraints for \pactlb \textsf{-S8},\footnote{Note that we obtain $S_8 = 0.825\pm 0.011$ for \pactlb; this $S_8$ constraint agrees with that from~\cite{DES_KIDS_shear_2023} at $1.5\sigma$.} which includes an external prior on $S_8$ from the joint analysis of DES-Y3 and KiDS-1000 cosmic shear data~\citep{DES_KIDS_shear_2023}, as described in \S\ref{sec:glossary}. \cite{Hill_2020}, who also placed constraints on EDE, validated that a simple Gaussian prior on $S_8$ is a sufficiently accurate approximation to the inclusion of a full weak lensing likelihood in such joint analyses with CMB data (see their Appendix B). The inclusion of $S_8$ significantly tightens the upper limit on EDE (see Fig.~\ref{fig:ede_kitchen_sink}), as positive $f_{\rm EDE}$ values increase this parameter above its \lcdm\ value (the late-time matter power spectrum amplitude is increased due to increases in $\Omega_c h^2$ and $n_s$ in the CMB fit), which is in the opposite direction to the preference in the DES and KiDS data~\citep{Hill_2020}.

\begin{table}[t]
\centering
\begin{tabular}{c|c|c|c|c|c}
\hline
 & $\Delta\chi^2$ & Pref. in $\sigma$ & $H_0^{\rm (EDE)}$ & $f_{\rm EDE}$ & $\log_{10}{z_c}$ \\ \hline \hline
\act & $\approx 0.0$  & 0.0 &66.5 &0.012 & 3.00\\ \hline 
\wact &1.9  &0.5 &69.9 &0.089 &3.55 \\ \hline 
\pact &  $4.3$ & $1.2$ &70.4 &0.091 &3.56 \\ \hline
\wactlb &$2.9$  &$0.8$  &70.2 &0.070 &3.52 \\ \hline 
\pactlb & $6.6$ & $1.7$ &71.2 &0.093 &3.56 \\ \hline
\end{tabular}
  \caption{The  $\Delta\chi^2=\chi^2_{\Lambda\mathrm{CDM}}-\chi^2_{\mathrm{EDE}}$ from the \texttt{MFLike} likelihood MAP points for the $n=3$ EDE model  compared to $\Lambda$CDM for each dataset combination, and preference (in units of $\sigma$) for EDE over \lcdm\ using the likelihood-ratio test statistic. The values for $H_0$, $f_{\rm EDE}$, and $\log_{10}{z_c}$ in the MAP EDE model are also reported. The data show no significant preference for non-zero EDE. For ACT alone, the MAP $\chi^2$ for EDE is indistinguishable from that for $\Lambda$CDM within our numerical precision, indicating that adding EDE parameters does not improve the fit at all in this case. } 
\label{table:ede_chi2}
\end{table}

Switching now to the $n=2$ EDE model, for ACT alone, we obtain $f_{\rm EDE}< 0.091$ (95\%, \act) with $H_0=67.5^{+0.9}_{-1.5}$~km/s/Mpc. Adding \emph{Planck} yields $f_{\rm EDE}< 0.11$ (95\%, \pact) with $H_0=69.3^{+1.0}_{-1.5}$~km/s/Mpc. Finally, adding CMB lensing and BAO data yields $f_{\rm EDE}< 0.11$ (95\%, \pactlb) with $H_0=70.1^{+0.9}_{-1.5}$~km/s/Mpc. These results are similar to the $n=3$ case.  A full comparison between the two models is discussed in Appendix~\ref{app:ede_full_constraints}.

Some past studies (e.g., \citealp{Herold:2021, Herold:2022, Poulin2023EDEreview, Efstathiou2024EDE}) have discussed ``projection effects,'' where, in the $\Lambda$CDM limit ($f_{\rm EDE} \rightarrow 0$), the additional EDE-specific parameters ($z_c$ and $\theta_i$) have no impact. These can bias the Bayesian posterior for $f_{\rm EDE}$. Additionally, the exact choice of the $z_c$ prior range can have a significant effect on the final constraints in this model, as large values of $f_{\rm EDE}$ are allowed if $z_c$ is sufficiently high that no dynamical effects are imprinted on the CMB (i.e., the EDE is irrelevant by the recombination epoch, \citealp{LaPosta2022}). Finally, the lower bound of $f_{\mathrm{EDE}}=0$ only allows the model fit to increase $H_0$ or leave it unchanged, not decrease it. Therefore, to evaluate whether the data have a meaningful preference for the EDE model, it is useful to compare the effective $\chi^2$ value of the MAP EDE model to that of the $\Lambda$CDM model, rather than to solely inspect the Bayesian posteriors, as was done in \cite{Hill_ACT_EDE}. To maximize the posterior, we use both \texttt{Cobaya} and \texttt{PROSPECT}, a profile likelihood code that can also be used for accurate global optimization \citep{Holm:2023uwa}.\footnote{For the posterior maximization procedure, we use the full Boltzmann code \texttt{class} rather than the emulators, to obtain high numerical precision.} We then take the global MAP from the different minimizer runs.

Performing this exercise, we find no preference for the EDE model from the DR6 data alone or combined with other datasets, as shown in Table~\ref{table:ede_chi2} for the $n=3$ EDE model.  For \act\ alone, we do not find an MAP point for the $n=3$ EDE model with lower $\chi^2$ than that found for \lcdm, i.e., the additional parameters of the EDE model provide no meaningful improvement in goodness-of-fit to the ACT data within our numerical precision.  The MAP model is nearly indistinguishable from \lcdm, with $f_{\rm EDE}=0.012$ (very close to the \lcdm\ edge of the parameter space) and $\log_{10}{z_c}=3.00$, the lower edge of our prior, where the impact of EDE on the CMB is minimal.  In the other cases, we note that in the EDE model there is a narrow degeneracy direction in the parameter space, where parameters may shift while having very similar effective $\chi^2$ values.  Nevertheless, in none of the cases do we find the improvement in goodness-of-fit over \lcdm\ significant. 

Using the $\Delta \chi^2$ values, we compute the equivalent preference in $\sigma$ using the likelihood-ratio test as in~\cite{Hill_ACT_EDE}.\footnote{Because the $\Lambda$CDM model is a subset of the EDE model (with $f_{\rm EDE} = 0$), the $\Delta \chi^2$ between the best-fit EDE and best-fit $\Lambda$CDM models follows a chi-squared distribution with three degrees of freedom, corresponding to the three additional free parameters of the EDE model.}  We find that in all cases the preference for EDE over \lcdm\ is $< 2\sigma$. The $\Delta \chi^2$ values for the $n=2$ case, shown in Appendix~\ref{app:ede_full_constraints}, yield the same conclusion.

To further circumvent the problem of projection effects, some past works have used a profile likelihood approach (see, e.g., \citealp{Lewis:2002ah,Audren2013} for more detail). However, the $\Delta \chi^2$ values that we find between the EDE and $\Lambda$CDM MAP models are insignificant; as such, we see no evidence of preference for the EDE model in either a Bayesian or frequentist framework. Thus, we conclude that the significant computational expense of a full profile likelihood analysis is not justified, and is unlikely to affect the interpretation of these results.

\subsection{Varying fundamental constants}\label{sec:varconst} 

The dynamics of recombination depends critically on the values of fundamental constants during the decoupling epoch, including the fine-structure constant $\alpha_{\rm EM}$ and the electron mass $m_e$.  High-precision CMB observations thus allow for constraints on possible variations of these constants over vast scales in both distance and time (see, e.g.,~\citealp{SekiguchiTakahashi2021,HartChluba2020,HartChluba2018,2015A&A...580A..22P,Menegoni2012,Scoccola2009,Martins2004,Rocha2004,Battye2001,Avelino2001,Avelino2000,Kaplinghat1999,Hannestad1999}).  Furthermore, models featuring fundamental constant variations are amongst the few that have had some phenomenological success in increasing the CMB-inferred Hubble constant~\citep{KnoxMillea2020,SekiguchiTakahashi2021,H0_olympics_2022,2024JCAP...04..059K}.  However, we emphasize that these models are not physical, in the sense that the fundamental constant variation is not driven dynamically by a new field or interaction; we simply assume that $\alpha_{\rm EM}$ or $m_e$ can take a different value in the early universe than it does today, which opens degeneracies in the CMB.  Recently,~\cite{Baryakhtar2024} have found that a first-principles implementation of varying-constant models can yield rather different conclusions than crude phenomenological models, due to (for example) the impact of perturbations in the field driving the constant-variation, as well as its contribution to the cosmic energy budget. However, for simplicity, here we stick to a phenomenological approach in which $\alpha_{\rm EM}$ or $m_e$ are simply allowed to vary, thus testing their values at the CMB epoch as compared to today.

We use the new ACT DR6 power spectra to place updated bounds on possible variation of $\alpha_{\rm EM}$ and $m_e$ (considered individually).  For simplicity, we assume that the value of $\alpha_{\rm EM}$ (or $m_e$) undergoes an instantaneous, step-function transition well after recombination is completed, but well before the reionization epoch (specifically, we choose $z=50$ for the redshift of this transition).  The value of the constant after the transition is fixed to the laboratory-measured value~\citep{ParticleDataGroup:2024cfk}, while the value before the transition is taken to be a free parameter.  The relevant physical effects on the recombination dynamics are treated via the implementation in {\tt HyRec}~\citep{Hyrec}, as incorporated in {\tt class} (note that {\tt CosmoRec}~\citep{Cosmorec} includes a similarly accurate treatment of these effects).  The dominant physical effects are due to changes in the Thomson scattering cross-section, with $\sigma_{\rm T} \propto \alpha_{\rm EM}^2 m_e^{-2}$, and changes in the energy levels of atomic hydrogen, with $E \propto \alpha_{\rm EM}^2 m_e$.  Many additional, subtle effects arise due to the non-equilibrium nature of cosmological recombination --- see, e.g.,~\cite{HartChluba2018,Chluba_AliHaimoud2016,2015A&A...580A..22P} for a thorough discussion.  In general, variations of $\alpha_{\rm EM}$ or $m_e$ change the timing of recombination, with higher values of these constants associated with earlier recombination. Thus, such variations change the physical scales imprinted in the CMB power spectrum, including the damping scale. The new ACT DR6 spectra allow tests of these effects in a qualitatively new regime, deep into the damping tail in TT and across a wide range of scales in TE and EE.

As pointed out in~\cite{SekiguchiTakahashi2021}, a joint variation of $m_e$ and the spatial curvature $\Omega_k$ opens up a particularly significant degeneracy with $H_0$ and other cosmological parameters. In principle, such a model has sufficient freedom to accommodate an increased value of $H_0$ while providing a good fit to not only CMB data, but also late-time measurements of the distance-redshift relation from BAO and SNIa data.  Here, we revisit this scenario. In addition, we also consider a model in which both $\alpha_{\rm EM}$ and $\Omega_k$ are allowed to vary.

\subsubsection{Electron mass}\label{sec:var_me}

We parametrize a possible deviation of the electron mass from its present-day value by introducing a free parameter $m_e/m_{e,0}$, where $m_e$ ($m_{e,0}$) is the mass at $z>50$ ($z<50$, including today).  We adopt a flat, uninformative prior: $m_e/m_{e,0} \in \left[0.3, 1.7\right]$.
Combining ACT and \emph{Planck}, we find
\begin{eqnarray}
    m_e/m_{e,0} & = & 0.856^{+0.060}_{-0.073} \quad (68\%, \, \pact), 
\end{eqnarray}
whereas \emph{Planck} alone yields $m_e/m_{e,0} = 0.880^{+0.068}_{-0.080}$.  While the \pact\ result lies $2.4\sigma$ below unity, we find that this is driven mostly by the \emph{Planck} data: ACT alone yields $m_e/m_{e,0} = 1.02^{+0.11}_{-0.15}$, while combining ACT with \emph{WMAP} yields
\begin{eqnarray}
    m_e/m_{e,0} & = & 0.921\pm 0.083 \quad (68\%, \, \wact). 
\end{eqnarray}
Interpreting the \pact\ result as a statistical fluctuation, we further include CMB lensing and DESI BAO data to obtain
\begin{equation}
\left.
 \begin{aligned}
m_e/m_{e,0} & = & 1.0096 \pm 0.0060 \\ 
H_0 & = & 69.8 \pm 1.1
\end{aligned}
\quad \right\} \mbox{\textrm{\; (68\%, \pactlb).}}
\end{equation}
Further adding SNIa data slightly tightens the constraints while moving them closer to \lcdm, giving $m_e/m_{e,0} = 1.0063 \pm 0.0056$ and $H_0 = 69.1\pm 1.0$ (68\%, \pactlbs).

These bounds represent the tightest constraints to date on the value of the electron mass at the recombination epoch, now approaching the half-percent level. Here, the BAO data are crucial in breaking severe geometric degeneracies that arise in the primary CMB when the electron mass is allowed to vary (see, e.g.,~\citealp{HartChluba2018} and~\citealp{SekiguchiTakahashi2021} for detailed discussion). These severe degeneracies also allow much more freedom in the Hubble constant, which increases compared to the value found in \lcdm\ when fitting these datasets. Nevertheless, we emphasize again that this model is not physical: no dynamical field has been introduced that drives the evolution of $m_e$.\\

\subsubsection{Electron mass and spatial curvature}\label{sec:var_me_ok}

\cite{SekiguchiTakahashi2021} noted that variation of the electron mass has the unique property of allowing multiple scales in the CMB to be preserved (in combination with other \lcdm\ parameters changing), including the acoustic scale and the damping scale. However, this variation will change the late-time expansion history, which is why the inclusion of BAO data plays a key role in the constraints above.  This observation motivates further extending the model to allow freedom to adjust the late-time distance-redshift relation.  \cite{SekiguchiTakahashi2021} considered scenarios in which either the spatial curvature or the dark energy equation of state were allowed to vary in addition to $m_e$. They found spatial curvature to yield surprising success in accommodating significant variation of $m_e$ (and hence $H_0$) in fits to \Planck\ CMB and BOSS BAO data, a result that was further validated in~\cite{H0_olympics_2022}.

Here, we revisit this model with the new ACT DR6 spectra. We adopt a uniform prior $\Omega_k \in \left[-0.6, 0.6\right]$ and the same prior on $m_e$ used above. We compute theoretical predictions using {\tt class}.

\begin{figure}[tp]
  \includegraphics[width=\columnwidth]{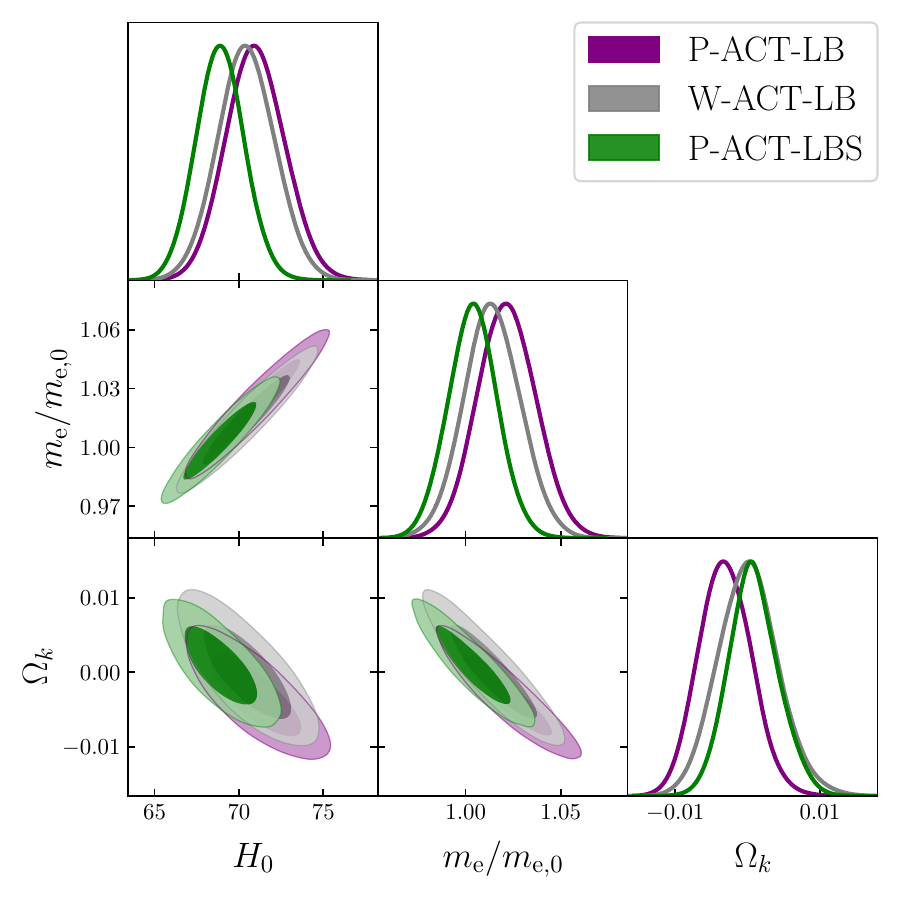}
 \vspace{-0.5cm} \caption{Constraints on a (non-physical) model with time variation of the electron mass and non-zero spatial curvature.  This model opens up significant degeneracies in the CMB (and BAO) data.  A joint fit to \pactlb\ yields a higher value of $H_0$ than in \lcdm, but with a much larger error bar; the same behavior is seen for \wactlb, indicating that the results are not specific to \Planck.  However, inclusion of SNIa data (\pactlbs, green) pushes the results toward \lcdm, disfavoring this approach for increasing the cosmologically-inferred value of $H_0$.}
  \label{fig:me_Omegak_H0}
\end{figure}

We find that this two-parameter extension of \lcdm\ is highly unconstrained in fits to CMB data alone.  We thus focus solely on joint analyses of CMB and late-time data here, with results shown in Fig.~\ref{fig:me_Omegak_H0}. Our baseline dataset combination yields
\begin{equation}
\left.
 \begin{aligned}
    m_e/m_{e,0} & = & 1.022\pm 0.016 \\
    \Omega_k & = & -0.0031\pm 0.0037 \\ 
    H_0 & = & 71.0\pm 1.7 
\end{aligned}
\quad \right\} \mbox{\textrm{\; (68\%, \pactlb).}}
\end{equation}
While neither non-standard $m_e$ or non-flat $\Omega_k$ are detected on their own in this fit, their combined impact allows $H_0$ to take on higher values, consistent with earlier results from \cite{SekiguchiTakahashi2021,H0_olympics_2022}.  To assess the additional constraining power contributed by ACT, note that the error bar on $H_0$ in our \pactlb\ analysis is roughly $20\%$ smaller than that found for the same model using \Planck\ and BOSS BAO data~\citep[$H_0 = 69.29 \pm 2.11$,][]{H0_olympics_2022}. We verify that these results are not driven by the high CMB lensing amplitude in \Planck\ by replacing \Planck\ with \WMAP\ in the dataset combination, finding
\begin{equation}
\left.
 \begin{aligned}
    m_e/m_{e,0} & = & 1.014 \pm 0.015 \\
    \Omega_k & = & 0.0001 \pm 0.0042 \\
    H_0 & = & 70.4 \pm 1.7
\end{aligned}
\quad \right\} \mbox{\textrm{\; (68\%, \wactlb).}}
\end{equation}

The parameter posteriors in these fits are significantly broadened compared to those obtained in \lcdm\ (cf.~${H_0 = 68.22 \pm 0.36}$ for \pactlb\ in L25) or \lcdm+$\Omega_k$. For comparison, we find $\Omega_k = 0.0019 \pm 0.0015$ for \pactlb\ in \S\ref{sec:curv}; this bound is weakened by more than a factor of two in the model studied here. Although the central values of some parameters shift compared to \lcdm, the significance of the shifts is counteracted by the increased error bars, and we find that this model does not yield a significant improvement in goodness-of-fit over \lcdm. For the MAP model obtained for \pactlb\ (\wactlb), we find $\Delta \chi^2 = -2.4$ ($-1.5$) with respect to \lcdm, with two additional free parameters.  These values correspond to a $1.0\sigma$ or $0.7\sigma$ preference over \lcdm, respectively, neither of which is significant.

It is also worth noting that the DESI BAO data play an important role in the central value of the $H_0$ posterior here (and in other models studied in this work). Due to the computational expense of running MCMC chains for this model (which is extremely slow to evaluate in {\tt class}), we use importance-sampling with an approximate version of the BOSS BAO likelihoods to assess how the results would shift if DESI were replaced with BOSS. We find that the central value of the $H_0$ posterior shifts downward by 1--1.5 km/s/Mpc, while the electron mass and spatial curvature shift closer to their \lcdm\ values. The error bars on all quantities increase only slightly. Thus, we conclude that the central values of the above parameter constraints are partially driven by the DESI BAO data.

Furthermore, we find that a joint analysis with Pantheon+ SNIa data pulls all parameters toward their \lcdm\ values, while tightening the error bars compared to the \pactlb\ results:
\begin{equation}
\left.
 \begin{aligned}
    m_e/m_{e,0} & = & 1.004 \pm 0.013 \\
    \Omega_k & = & 0.0008^{+0.0032}_{-0.0036} \\ 
    H_0 & = & 68.9\pm 1.4 
\end{aligned}
\quad \right\} \mbox{\textrm{\; (68\%, \pactlbs).}}
\end{equation}
Notably, the $H_0$ value in this fit is fully consistent with its \lcdm\ value.  From the joint analysis of \pactlbs\ and the insignificant improvement in quality of fit to \pactlb, we conclude that there is no evidence for a scenario with spatial curvature and a non-standard electron mass at recombination.  The possibility of obtaining a high value of $H_0$ from cosmological data in this model is significantly reduced by these new results.

\subsubsection{Fine-structure constant}\label{sec:var_a}

As done above for the electron mass, we parametrize possible deviation of the fine-structure constant from its present-day value by introducing a free parameter $\alpha_{\rm EM}/\alpha_{\rm EM,0}$, where $\alpha_{\rm EM}$ ($\alpha_{\rm EM,0}$) is the value at ${z>50}$ (${z<50}$, including today). We adopt a flat, uninformative prior: $\alpha_{\rm EM}/\alpha_{\rm EM,0} \in \left[0.6, 1.4\right]$.  
Combining ACT and \emph{Planck}, we find
\begin{eqnarray}
    \alpha_{\rm EM}/\alpha_{\rm EM,0} & = & 1.0037\pm 0.0017 \quad (68\%, \, \pact), 
\end{eqnarray}
representing a 30\% reduction in the error bar compared to that from \emph{Planck} alone, which yields $\alpha_{\rm EM}/\alpha_{\rm EM,0} = 1.0005\pm 0.0025$ (consistent with \citealp{HartChluba2020}). ACT alone yields $\alpha_{\rm EM}/\alpha_{\rm EM,0} = 1.0046\pm 0.0035$.

Further including CMB lensing and DESI BAO data yields
\begin{eqnarray}
    \alpha_{\rm EM}/\alpha_{\rm EM,0} & = & 1.0043\pm 0.0017 \,\, (68\%, \, \pactlb), \ \ 
\end{eqnarray}
thus fluctuating ${\approx 2.5\sigma}$ above unity.  Inclusion of SNIa data has negligible effect on this result.  
As higher $\alpha_{\rm EM}$ corresponds to earlier recombination, this result also yields a moderate increase in the Hubble constant, with $H_0 = 69.27 \pm 0.54$ km/s/Mpc. Interpreting this result as a statistical fluctuation, these bounds represent the tightest constraints to date on the value of the fine-structure constant at the recombination epoch, approaching the per-mille level.

\subsubsection{Fine-structure constant and spatial curvature}\label{sec:var_a_ok}

Unlike the electron mass, variation of the fine-structure constant (in combination with other parameters) can not simultaneously preserve all of the relevant physical scales in the CMB power spectrum; this is largely why the constraints on $\alpha_{\rm EM}$ above are tighter than those on $m_e$.  This is also why the Hubble constant cannot increase as significantly when varying $\alpha_{\rm EM}$ as when varying $m_e$~\citep{SekiguchiTakahashi2021}.  For completeness, and to better understand the results found above for the $m_e$+$\Omega_k$ model, here we consider extending the varying-$\alpha_{\rm EM}$ model to also allow the spatial curvature to vary.  This allows further freedom in the late-time expansion history, allowing wider range of parameter variations to be accommodated in fits to CMB and BAO data.

As above, we compute theoretical predictions for this model using {\tt class}.  We adopt an uninformative prior $\Omega_k \in \left[-0.6, 0.6\right]$, while allowing $\alpha_{\rm EM}$ to vary as in our previous analysis.

As expected, we find that this two-parameter extension is far less degenerate than the $m_e$+$\Omega_k$ model studied above.  Focusing for simplicity solely on the \pactlb\ dataset combination, we find
\begin{equation}
\left.
 \begin{aligned}
    \alpha_{\rm EM}/\alpha_{\rm EM,0} & = & 1.0041\pm 0.0017 \\
    \Omega_k & = & 0.0007\pm 0.0015 \\
    H_0 & = & 69.4\pm 0.6 
\end{aligned}
\quad \right\} \mbox{\textrm{\; (68\%, \pactlb).}}
\end{equation}

Comparing to the results above, it is evident that the constraint on $\alpha_{\rm EM}$ is essentially unchanged by the opening of the spatial curvature, which itself is also tightly constrained near zero.  Likewise, $H_0$ hardly changes, with the analogous result for fixed spatial curvature yielding $H_0 = 69.3 \pm 0.5$~km/s/Mpc.

Contrasting the varying-$\alpha_{\rm EM}$ results with the varying-electron-mass case, we find that the degeneracy structure of the two models is remarkably different, with the electron mass possessing much more flexibility in accommodating a wide range of parameter space than the fine-structure constant, consistent with~\cite{HartChluba2018,HartChluba2020,SekiguchiTakahashi2021}.\footnote{A similar result is found for the small-scale baryon clumping scenario studied in \S\ref{sec:pmf} --- allowing $\Omega_k$ to vary also has little degeneracy with the clumping parameter constraint.}  Nevertheless, we note again that all of the varying-constant models studied here are not physical, but rather simply crude phenomenological approximations.  These results may be a useful guide in constructing physical models that can achieve similar flexibility in accommodating a wide range of data. \\

\subsection{Primordial magnetic fields}\label{sec:pmf} 

The existence of primordial magnetic fields (PMFs) is a compelling possibility. Such PMFs could cause inhomogeneities in the baryon distribution around recombination.  Thus, the PMF model is an example of a slightly more generic scenario known as ``baryon clumping''~\citep{Jedamzik2011,Jedamzik2019}. Primordial magnetic fields with a blue-tilted power spectrum can naturally have kpc-scale correlation lengths. Once the photon gas dynamically decouples from the baryon fluid on small scales, the magnetic force causes efficient growth of baryon density perturbations to potentially $\mathcal{O}(1)$ contrasts. These kpc-scale perturbations are not directly resolvable in CMB observations, but they cause accelerated recombination due to the quadratic source term in the equation describing the recombination rate~\citep{1968ApJ...153....1P}. The corresponding decrease in the sound horizon could then partially reconcile CMB-based determinations of the Hubble constant with local universe measurements~\citep{Jedamzik2020}. As magnetic fields are part of the standard model and their generation during early-universe phase transitions is conceivable, PMFs (or baryon clumping models in general) are a well-motivated scenario to potentially increase the CMB-inferred Hubble constant.\footnote{Note that effects due to Faraday rotation~\citep[e.g.,][]{2022PhRvD.105f3536C} are not considered here and are expected to yield weaker constraints on such models~\citep{2019PhRvD.100b3507P}.}

Baryon clumping induces changes in the Silk damping tail in addition to the leading-order geometric effect described above. Two counteracting effects are at play~\citep{Jedamzik2011}: shortened photon diffusion due to accelerated hydrogen recombination suppresses Silk damping, but the previously accelerated helium recombination decreases the ionization fraction and thus increases the damping. These two effects do not generally cancel, and empirically the first one appears to be dominant~\citep[e.g.,][]{Thiele2021}. Thus, the high-resolution ACT DR6 data with good signal-to-noise in the damping tail enable stringent constraints on baryon clumping scenarios.

Theoretical modeling of PMF-induced baryon clumping is non-trivial. We adopt a simplified model used in earlier works~\citep{Jedamzik2020,Thiele2021,Rashkovetskyi2021,Galli2022,Jedamzik2023}. The baryon density distribution function is sampled by three zones in each of which the baryon density is taken as a constant. Taking the zones as indexed by $i=1,2,3$, each zone carries a volume fraction $f_i$ and a baryon density $\Delta_i \equiv \rho_{b,i} / \langle \rho_b \rangle$. Volume and mass conservation imply the constraints $\sum f_i = \sum f_i \Delta_i = 1$. For each zone, the recombination history is computed separately and the thermal history is then taken as the volume-weighted average over the three zones. This treatment neglects Ly$\alpha$ photon mixing~\citep{Jedamzik2023a}.

Within the three-zone model, we choose to work with the ``M1'' scenario of~\cite{Jedamzik2020}: $f_2=1/3$, $\Delta_1=0.1$, $\Delta_2=1$. After applying the constraints described above, the model thus has only a single free parameter, the variance $b$ of the small-scale baryon density fluctuations
\begin{equation}
    1 + b \equiv \sum_i f_i \Delta_i^2\,.
\end{equation}

Recent perturbative and simulation-based works have suggested refinements of this simplified treatment~\citep{Lee2021,Jedamzik2023a}.
\cite{Jedamzik2025} recently presented constraints on primordial magnetic fields using a simulation-informed model for the recombination history. 
Where comparisons can be made between this method and our three-zone model, there is a remarkably close agreement between the two approaches. In particular, the posteriors for $H_0$ in~\cite{Jedamzik2025} are similar to results found for the three zone-model.

In the absence of a more accurate publicly available model and in order to facilitate comparison with previous constraints, we choose to continue working with the three-zone model. However, we stress that our constraints are to be interpreted in the context of this model and further work will be needed to evaluate the viability of baryon clumping scenarios more generally. Nevertheless, previous work has indicated that increasing the flexibility of the model does not substantially alter the constraint on $b$ or $H_0$~\citep{Thiele2021}.

\begin{figure}[tp]
   \includegraphics[width=\columnwidth]{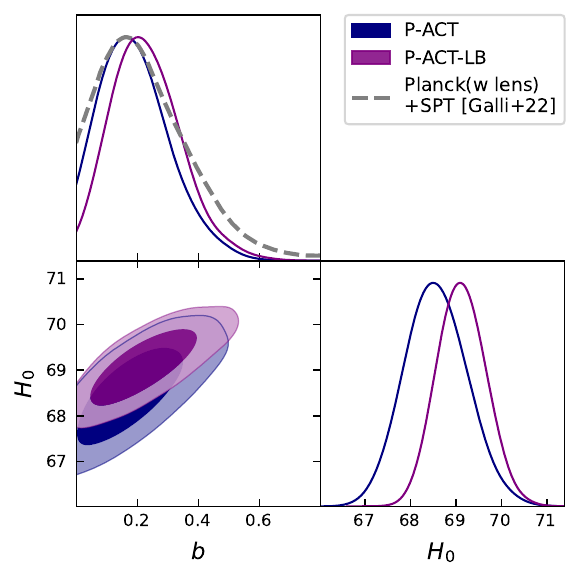}
   \vspace{-0.6cm}
  \caption{
    Constraints on the variance $b$ of the small-scale baryon density distribution at recombination from \pact\ (navy) and \pactlb\ (purple), compared with the latest results from \Planck\ (including CMB lensing) combined with SPT small-scale polarization (gray dashed line).
    Primordial magnetic fields would induce baryon clumping on small scales, and hence $b>0$.  No evidence of clumping is seen in our analysis.
  }
  \label{fig:pmfs}
\end{figure}

Figure~\ref{fig:pmfs} shows our constraints on the clumping parameter and the Hubble constant,
from primary CMB data alone and in combination with CMB lensing and BAO data. We find that the constraint on $b$ is tightened compared to previous analyses:
\begin{eqnarray}
    b \, &<& \, 0.41 \ (95\%, \pact), \nonumber\\
      \, &<& \, 0.44 \ (95\%, \pactlb) \,.
\end{eqnarray}
For comparison, the \Planck\ primary CMB alone constrains $b<0.51$, and the combination of \Planck\ (including CMB lensing) plus SPT-3G gives results consistent with our \pact\ measurement~\citep{Galli2022}.  If we combine all three datasets (ACT, \Planck, and the public SPT-3G data), there is only a marginal improvement on \pact. 

Note that the posteriors including ACT data peak at non-zero $b$; thus the improvement in constraining power is larger than what is suggested by comparing the upper limits.
As quantified by the standard deviation in $b$, adding \act\ improves upon the \Planck-only constraint by about $30\%$.
We find no detection of baryon clumping and hence further restrict this model's ability to increase the CMB-inferred $H_0$ value: we find $H_0=68.6 \pm 0.7$~km/s/Mpc and $H_0=69.1 \pm 0.5$~km/s/Mpc for \pact\ and \pactlb\ in this model, respectively.

In contrast to the seemingly similar model with varying electron mass, we find that the inclusion of spatial curvature as a free parameter does not appreciably change the $H_0$ posterior in the \pactlb\ data combination (see \S\ref{sec:var_me_ok} for contrasting results in the varying-electron-mass case).  The non-trivial small-scale effects in the baryon clumping model constrain $b$ beyond the information contained in the geometric sound-horizon modification.

\subsection{Temperature of the CMB}\label{sec:tcmb} 

Several studies \cite[e.g.,][]{Ivanov:2020mfr,Wen:2020txi,Hill:2023wda} have explored the possibility of increasing the CMB-anisotropy-inferred value of the Hubble constant by changing the CMB monopole temperature, $T_\mathrm{CMB}$. In particular, \cite{Ivanov:2020mfr} highlighted a strong negative degeneracy in the $H_0$-$T_\mathrm{CMB}$ plane when the monopole temperature is left free in analyses of \Planck\ data. Setting a SH0ES prior on $H_0$ would then yield a temperature measurement $3\sigma$ lower than the combined measurement of $T_\mathrm{CMB}=2.72548 \pm 0.00057$~K~\citep{Fixsen2009} from \textit{COBE/FIRAS} and other data.\footnote{Note that the error bar on $T_{\rm CMB}$ from~\cite{Fixsen2009} is sufficiently small that this value is taken as a fixed constant in analyses of \lcdm\ (and the extended models studied in this paper).} 

Obtaining a higher value of $H_0$ via a decrease in the monopole CMB temperature is difficult.  Models with some level of post-recombination reheating can have an impact by allowing a lower value of the monopole temperature in the early universe (e.g.,~\citealp{Hill:2023wda} propose to convert a fraction of DM into photons during the dark ages).  However, although such a process is straightforward to implement in a phenomenological way, it is hard to find a well-grounded physical mechanism to motivate it such that the blackbody spectrum is preserved \citep[see, e.g.,][]{Chluba:2014wda}. 

\begin{figure}[!t]
\centering
\includegraphics[width=\columnwidth]{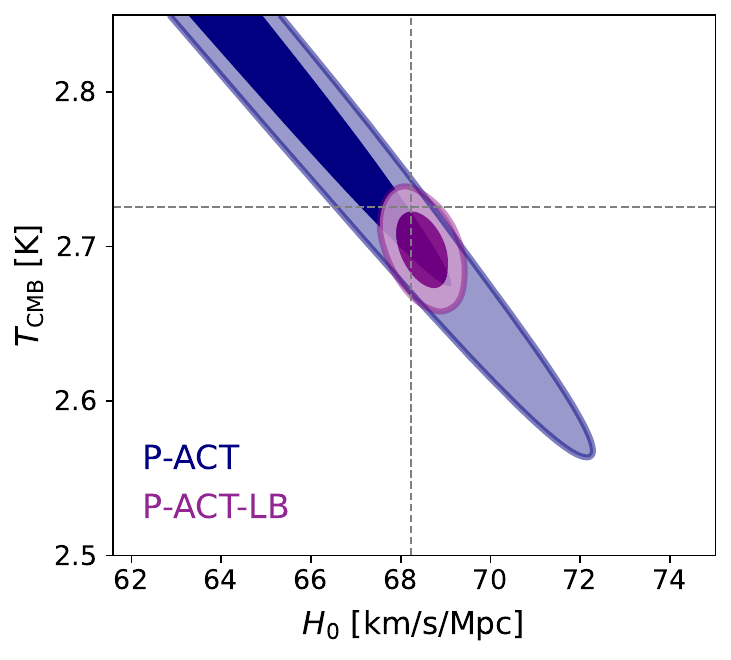}
\vspace{-0.5cm}
\caption{Constraints on the CMB monopole temperature ($T_\mathrm{CMB}$) and $H_0$. The \pactlb\ combination (purple) effectively breaks the degeneracy between these two parameters present in CMB-only analyses (navy) and gives a tight measurement of both parameters, similar to that achieved with CMB data from \Planck\ in previous literature.  The results are consistent with the $\Lambda$CDM case, in which $T_\mathrm{CMB}$ is fixed to the \textit{COBE/FIRAS} value (gray dashed lines).}
\label{fig:Tcmb}
\end{figure}

Nevertheless, these studies have led to the realization that current CMB anisotropy data in combination with BAO data provide, on their own, a powerful probe of the amount of radiation in the universe.\footnote{Note that it is important to self-consistently vary the calibration of the CMB anisotropy data in such analyses, as explained in~\cite{Ivanov:2020mfr}.}  A single-parameter extension to the $\Lambda$CDM model, where $T_\mathrm{CMB}$ is left free, is well constrained. \cite{Ivanov:2020mfr} combine \textit{Planck} 2018 CMB anisotropy and lensing data with BOSS DR12 BAO data \citep{eBOSS:2017cqx} and find $T_\mathrm{CMB}=2.706^{+0.019}
_{-0.020}$~K (68\% CL). Updating this \Planck\ result including BAO from DESI, we find a slightly tighter constraint, $T_\mathrm{CMB}=2.696\pm 0.017~\mathrm{K}$. With the addition of the new ACT DR6 spectra, we find a similar constraint
\begin{eqnarray}
    T_\mathrm{CMB} &= 2.698\pm 0.016\,\mathrm{K} \,\, (68\%, \pactlb).
\end{eqnarray}
This $0.6\%$ measurement of $T_\mathrm{CMB}$ is consistent with the direct \emph{COBE/FIRAS} monopole measurement.  The corresponding constraint on $H_0$ is consistent with the result in \lcdm, as shown in Fig.~\ref{fig:Tcmb}. \\

\subsection{Modified recombination history}\label{sec:modrec}

In this subsection, we present a model-independent reconstruction of the cosmological recombination history, including ACT DR6 data.  
Modifications to the standard recombination history have been extensively studied, both as a powerful test of the standard cosmological model and as a potential avenue to increase the value of the Hubble constant inferred from CMB observations~\citep[e.g.,][]{Lee:2022gzh, Lynch:2024gmp, Lynch:2024_desi, Pogosian:2024ykm, Mirpoorian:2024fka}. To investigate such modifications, we use the \texttt{ModRec} model developed in~\cite{Lynch:2024gmp}, which parametrizes departures in the ionization fraction,  $X_{\rm e}(z) \equiv n_{\rm e}(z)/n_{\rm H}(z)$, from the standard recombination history, where $n_{\rm e}(z)$ and $n_{\rm H}(z)=n_{\rm HI}(z)+n_{\rm HII}(z)$ are the number densities of free electrons and hydrogen nuclei, respectively. While phenomenological, the flexible parametrization employed by \texttt{ModRec} effectively captures a broad range of physical scenarios, including the effects of primordial magnetic fields, time-varying electron mass, and dark matter annihilation~\citep{Lynch:2024gmp}, all of which are also studied separately elsewhere in this paper.

Following~\cite{Lynch:2024gmp}, we parameterize departures from the standard ionization fraction $\Delta X_{\rm e}(z)=X_{\rm e}(z)-X_{\rm e}^{\rm std}(z)$ at seven control points spanning ${z = 533}$ to $1600$, leaving reionization and helium recombination unchanged. These control points are interpolated using a cubic spline that enforces physical bounds, $0\leq X_{\rm e}(z)\leq X_{\rm e}^{\rm max}$, where $X_{\rm e}^{\rm max}$ is the maximum allowed ionization fraction. We compute all theoretical predictions using the \texttt{class} implementation of \texttt{ModRec}, with precision parameters and settings described in Appendix~\ref{app:theory}. Our control points and their priors are identical to those used in~\citet{Lynch:2024_desi,Lynch:2024gmp}. 

We present constraints on the \texttt{ModRec} model using the \plb\ and \pactlb\ datasets --- we do not show or discuss discuss \pact\ alone as BAO data are essential for breaking degeneracies between the modified recombination parameters and standard cosmological parameters~\citep{Lynch:2024gmp}. We show results for both DESI and BOSS BAO datasets, as the choice of BAO data impacts the reconstructed $X_e(z)$ due to differences in the inferred values of $r_dH_0$ and $\Omega_m$ between BOSS and DESI measurements, assuming $\Lambda$CDM~\citep{Lynch:2024_desi, Pogosian:2024ykm}.

\begin{figure}[!t]
\centering
\includegraphics[width=\columnwidth]{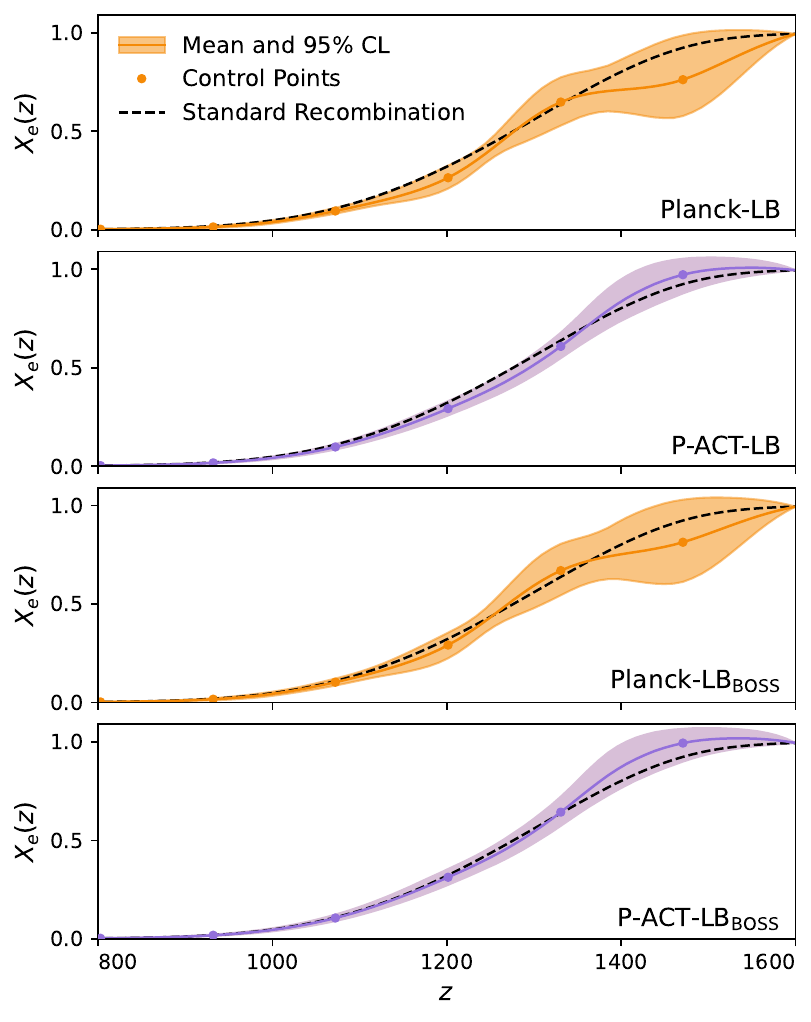}
\vspace{-0.6cm}\caption{Reconstructed ionization fraction as a function of redshift using the \texttt{ModRec} model. The shaded region shows the mean and 95\% two-tailed CL. The circles indicate the mean ionization fraction at the control points used in the reconstruction. The top (bottom) two panels use DESI (BOSS) BAO; the orange and purple colors distinguish between using \Planck-alone or \pact\ CMB data. The black dashed line shows the evolution of the ionization fraction for the standard recombination scenario assuming the \pactlb\ best-fit $\Lambda$CDM parameters. We see no evidence of a deviation from the standard recombination history.
} \label{fig:modrec_Xe_recon}
\end{figure}

Figure~\ref{fig:modrec_Xe_recon} shows the mean and 95\% two-tailed CL~constraints on the reconstructed ionization fraction as a function of redshift. Including ACT DR6 data significantly improves the constraints on $X_e(z)$, with a factor of two reduction in the uncertainty of the highest redshift ($z\approx 1467$) control point.\footnote{The modified recombination history cosmology in Fig.~\ref{fig:spec} is obtained by selecting a sample from the \plb~chain that has at least a 20\% difference in the highest-redshift ionization fraction relative to the standard recombination history and a $\Delta \chi^2<1$ compared to the chain’s best-fit cosmology.} This improvement arises because modifications to the ionization fraction impact the polarization and damping tail of the CMB~\citep{Hadzhiyska:2018mwh, Lynch:2024gmp}. The reconstructed recombination histories are consistent with the standard recombination scenario (black dashed lines).\footnote{The standard recombination history is computed using the best-fit $\Lambda$CDM parameters from the \pactlb\ dataset. Using the best-fit parameters from other datasets considered in this work has negligible impact on the predicted $X_e(z)$.}

Notably, the mean ionization fraction from the \pactlbb\ reconstruction is nearly indistinguishable from the standard recombination scenario across all redshifts analyzed, with all sampled control points agreeing with the standard recombination history within $1\sigma$, except for the highest redshift control point ($z= 1467$), where the ACT data favor a slightly higher ($1.5\sigma$) ionization fraction than the standard model. This is driven by ACT's mild preference for less damping compared to \emph{Planck}, as a larger value of $X_e$ at high redshifts leads to a decrease in damping~\citep{Lynch:2024gmp}. This mild preference for a higher $X_e$ at high redshifts is in the opposite direction of what is generally required by modified recombination models that aim to increase the value of $H_0$ inferred from the primary CMB. Finally, with respect to the MAP $\Lambda$CDM model, we find $\Delta \chi^2=-11.6$ ($-8.4$) for the \pactlb\ (\pactlbb) datasets,  for a model with seven additional free parameters. These values correspond to a preference of $1.6\sigma$ and $1.0\sigma$ over $\Lambda$CDM, respectively, and hence are not statistically significant. We present the full set of constraints on the control points in Appendix~\ref{app:modrec_full_constraints}.

In Fig.~\ref{fig:modrec_H0} we present the marginalized constraints on $H_0$ from the \texttt{ModRec} analysis. Including ACT data shifts $H_0$ closer to its $\Lambda$CDM value for both BAO datasets, consistent with our findings that ACT data support the standard recombination history. We find
\begin{eqnarray}
    H_0&=&69.6\pm 1.0 \ (68\%, \pactlb), \nonumber\\
   &=&68.1\pm 1.1~(68\%, \pactlbb).
\end{eqnarray}
For comparison, $H_0=70.1\pm 1.1$ ($68.6\pm 1.2$) km/s/Mpc for the $\textsf{Planck-LB}$ ($\textsf{Planck-LB}_{\textsf{BOSS}}$) dataset combinations. Additionally, we find that the choice of BAO data can lead to a shift of $\sim 1.5$ km/s/Mpc in the posterior mean of $H_0$. These findings are consistent with previous studies of the \texttt{ModRec} model~\citep{Lynch:2024_desi,Lynch:2024gmp} and are representative of a more general trend where the $H_0$ value inferred from models that modify the (pre)-recombination era physics depends sensitively on the BAO data. 

\begin{figure}[!t]
\centering
\includegraphics[width=\columnwidth]{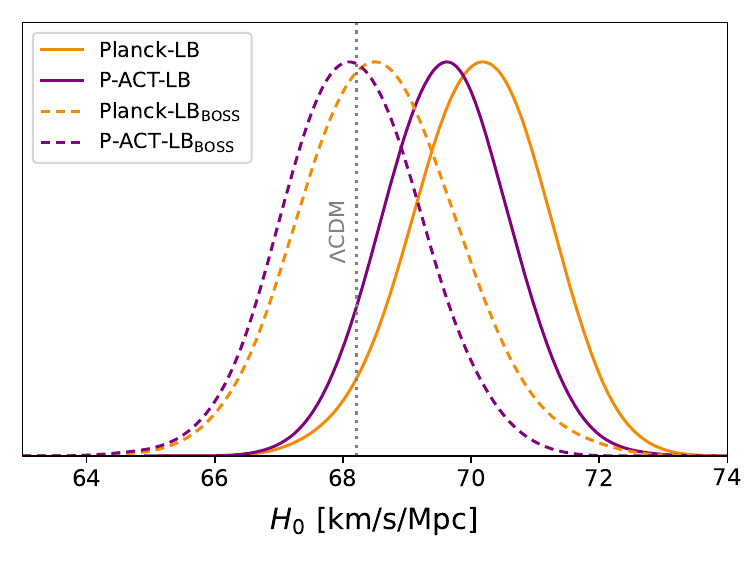}
\vspace{-0.6cm}\caption{Marginalized constraints on $H_0$ from the \texttt{ModRec} analysis for the \pactlb\ (purple) and \plb\ (orange) datasets. The solid (dashed) lines use DESI (BOSS) BAO. The vertical gray line denotes the best-fit $\Lambda$CDM value of $H_0$ from the \pactlb\ (DESI) dataset.
} \label{fig:modrec_H0}
\end{figure}

Finally, given that~\cite{Lynch:2024gmp} found that \texttt{ModRec} constraints on $\Omega_m$ including DESI BAO data prefer a lower value than recent $\Omega_m$ constraints from supernova surveys, we comment on the $\Omega_m$ constraints for the \texttt{ModRec} model. For \pactlb\ ($\textsf{Planck-LB}$), we find $\Omega_m=0.296\pm 0.007$ ($0.295\pm 0.007$). Therefore, including ACT data has a negligible impact on the inferred value of $\Omega_m$ for the \texttt{ModRec} model.

\section{Particle astrophysics}\label{sec:astroparticle}

\subsection{Neutrinos}\label{sec:neutrinos} 

The CMB is sensitive to a cosmological background of neutrinos produced in the early universe that contribute a significant fraction ($\sim41\%$) of the radiation density. In the Standard Model (SM) of particle physics, these neutrinos are expected to be three flavors of massless active particles that interact weakly with other SM particles but, as explained below, they are known to have non-zero masses.

\subsubsection{Number and mass}\label{sec:neu_n_m}

The exact number of neutrinos and their total mass are two key parameters of the cosmological model, and can point to beyond-Standard-Model (BSM) physics. 

\vspace{-0.12cm}
\subsubsubsection{Neutrino number} 
\vspace{-0.12cm}
The effective number of relativistic species quantifies the contribution to the radiation density of neutrinos, or other light particles, during their relativistic phase (including during the BBN epoch and the CMB decoupling epoch), $\neff \equiv (8/7) (11/4)^{4/3} \rho_\nu / \rho_\gamma$, where $\rho_\nu$ and $\rho_\gamma$ are respectively the neutrino and photon energy densities. Three neutrino species undergoing a non-instantaneous decoupling from the primordial plasma correspond to $\neff=3.044$~\citep{Neff3044,2020JCAP...12..015F,2020JCAP...08..012A,2024JCAP...06..032D}, accounting for energy-dependent interactions in the early universe. Numbers greater than this value might indicate the existence of additional light particles or extra free-streaming (dark) radiation --- as postulated by many BSM theories~\citep[e.g.,][]{PhysRevLett.40.223,2013arXiv1311.0029E,2013JHEP...12..058B,2012arXiv1204.5379A,2013PhRvL.110x1301W,2015PhRvD..91l5018K,2016PhRvL.117q1301B}. Values of $\neff$ smaller than $3.044$ are also possible in scenarios where radiation is affected not by the existence of extra particles but by non-standard properties of neutrinos or changes to the standard thermal history. In low-reheating temperature scenarios, for example, the late-time entropy production associated with the decay of a massive particle might delay the beginning of the radiation-dominated era down to a temperature of a few MeV, thus preventing the full thermalization of neutrinos \citep[see, e.g.,][]{Kawasaki:1999na,Kawasaki:2000en,Giudice:2000ex,2002PhRvD..66j3508T,Ichikawa:2005vw,Ichikawa:2006vm,DeBernardis:2008zz,deSalas:2015glj,Hasegawa:2019jsa,Barbieri:2025moq}. Entropy production occurring after neutrino decoupling might also reduce $\neff$ below its standard value by increasing the photon temperature relative to neutrinos (see, e.g., \citealp{2013PhRvD..87j3517S}, and also \citealp{Cadamuro:2011fd}, for a scenario involving the decay of an axion-like particle). A reduction of the effective number of neutrino species might also be the signature of sterile neutrinos with strong self-interactions \citep{Mirizzi:2014ama}. 

The CMB is able to constrain $\neff$ because its value affects the expansion rate of the universe, especially during the radiation-dominated phase, thereby altering the expansion history just before recombination and the predicted abundances of primordial light elements~\citep{Bashinsky&Seljak,Hou2013,Abazajian2015,Pan2016,Baumann2016}. At the perturbative level, $\neff$ alters the damping tail (high-$\ell$ region) of the TT/TE/EE spectra, both because the change to the expansion history alters the timescale for diffusion damping and because the free-streaming nature of the radiation damps the growth of perturbations, with the latter also inducing a characteristic phase shift in the acoustic peaks~\citep{Bashinsky&Seljak}. Combining the \Planck\ legacy CMB with \Planck\ CMB lensing and BOSS BAO, the neutrino number is measured to be $\neff=2.99\pm0.17$ at 68\% CL~\citep{Planck_2018_params}, or $\neff=3.06\pm 0.17$ at 68\% CL when we evaluate this estimate using \plb. Here, and as baseline throughout this paper, we assume a fixed total neutrino mass of 0.06~eV carried by a single massive species.

\begin{figure}[t!]
	\centering
    \includegraphics[width=\columnwidth]{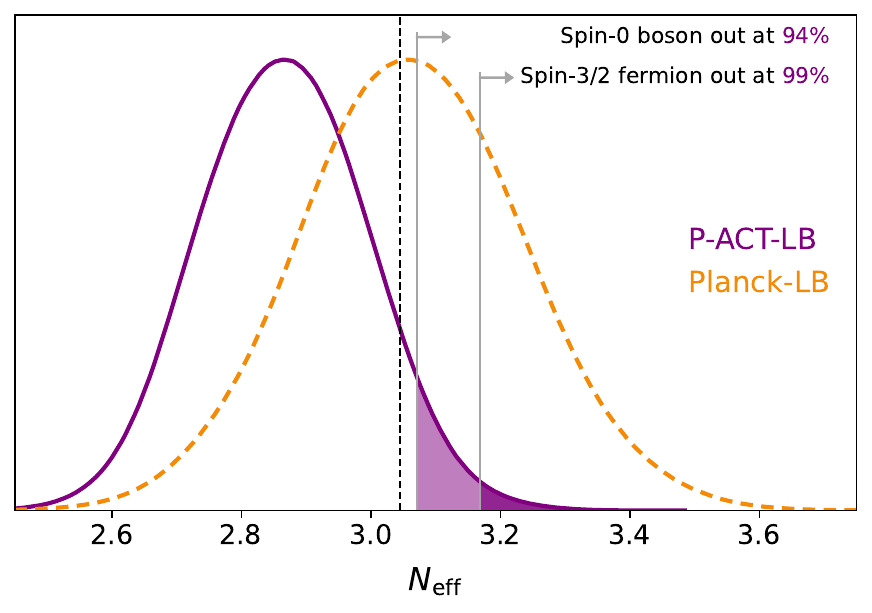}
	\vspace{-0.5cm}\caption{Constraints on the effective number of relativistic species, $\neff$. With the addition of the ACT DR6 spectra (purple) the constraint is more precise than the measurement from \Planck\ combined with CMB lensing and BAO (orange) by ${\sim25\%}$. The constraint is consistent with the SM expectation of $\neff=3.044$ for three light active neutrinos marked by the dashed vertical line. The purple bands show the region excluding a minimum contribution in extra radiation from another thermalized relativistic species, finding $\neff=3.044+0.027$ from a spin-0 boson at the 94\% confidence border (light purple), and the most stringent bound on the existence of another species, disfavoring $\neff=3.044+0.095$ from a spin-3/2 fermion at 99\% confidence (dark purple) --- see also Fig.~\ref{fig:nnu_dec}.}
	\label{fig:nnu_1D}
\end{figure}

With the new ACT DR6 spectra we find $\neff = 2.60^{+0.21}_{-0.29}$ at 68\% CL,  combining into
\begin{eqnarray}
    \neff &=& 2.73 \pm 0.14 \quad (68\%, \pact), \nonumber \\
     &=& 2.86 \pm 0.13 \quad (68\%, \pactlb), 
\end{eqnarray}
and giving
\begin{equation}
    \neff < 3.08 \quad (\textsf{one-tail}\ 95\%, \pactlb) .
\end{equation} 
This measurement is consistent with SM predictions for three relativistic neutrinos and the existence of another thermalized particle (irrespective of its spin or decoupling time) lies at the 94\% confidence border, as calculated from the likelihood of having a minimal additional contribution of 0.027 from a scalar boson (see below) beyond the expected value of 3.044, as shown in Fig.~\ref{fig:nnu_1D}.  Higher-spin particles are disfavored at higher significance.  A strong contribution to this new limit comes from the peak of the ACT $\neff$ distribution shifting to lower values compared to the \Planck\ measurement because, as shown in Fig.~\ref{fig:spec}, the new ACT data prefer somewhat less damping than the \Planck\ best-fit model. 

While our constraints are consistent with the SM expectation --- and the shift is not statistically significant enough to draw strong conclusions --- it is interesting to consider physical models that could yield values of ${\neff < 3.044}$.  Recalling that $\neff$ depends on the neutrino-to-photon energy density ratio, a simple possibility to decrease $\neff$ is an injection of energy into the photon-baryon plasma after neutrino decoupling, but before last scattering (note that this would thus lead to different values for $\neff$ at the BBN and CMB epochs).  Taking the central value of our \pact\ (\pactlb) posterior, the photon energy density would need to be increased by 12\% (6\%) between neutrino decoupling and last scattering.  If this energy injection takes place between redshifts ${5 \times 10^4 < z < 2 \times 10^6}$, a $\mu$ distortion will be produced in the CMB monopole energy spectrum; if it takes place at lower redshifts ${z < 5 \times 10^4}$, a $y$-type distortion will be produced.  Using standard formulae for the distortion amplitudes (e.g.,~\citealp{Chluba2012}), the \pact\ (\pactlb) energy-injection values would yield ${\mu = 0.16}$ or ${y = 0.03}$ ($\mu = 0.09$ or ${y = 0.02}$).  Such large spectral distortions are strongly excluded already by the \emph{COBE/FIRAS} data~\citep{Fixsen1996,Bianchini2022}.\footnote{The \emph{COBE/FIRAS} spectral distortion bounds are sufficiently tight that only very small increases in the photon energy density are allowed during the $\mu$ and $y$ epochs, corresponding roughly to $\Delta \neff = -0.0002$ (for both the $\mu$ and $y$ limits).}  Nevertheless, a low $\neff$ value could still be obtained via energy injection into the photon bath at ${2 \times 10^6 < z < 10^9}$, by cooling the neutrinos via a BSM interaction, or by other, more exotic mechanisms as discussed above.

\begin{figure}[t!]
	\centering
\includegraphics[width=\columnwidth]{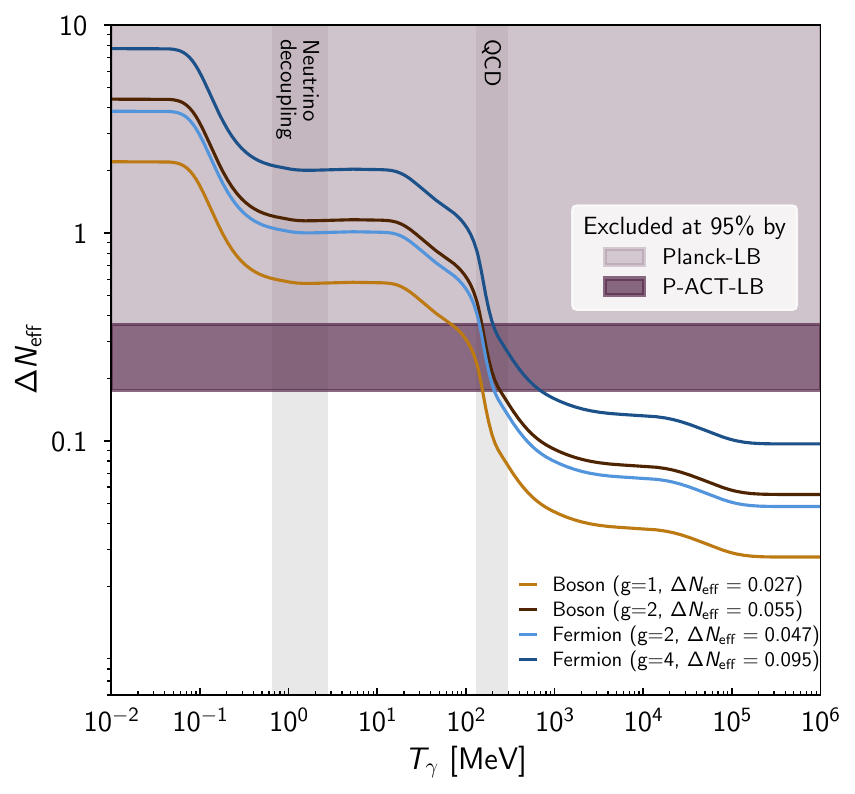}
	\vspace{-0.6cm}\caption{Excess in the contribution to radiation, $\Delta \neff$, caused by the existence of additional relativistic species with $g$ independent spin states $s$ (with $g=2s+1$) as a function of the temperature at which they decouple from the thermal bath. Particles decoupling after the QCD phase transition are excluded by our constraints; in particular, light particles of spin 0 and 1/2 must decouple at temperatures $\gtrsim 200$ MeV and all light particles of spin 3/2 must decouple at temperatures $\gtrsim 1$ GeV. The figure is adapted from the versions in~\citet{Planck_2018_params} and ~\citet{Simons_Observatory} but, differently from those versions, this figure is based on the fit of a positive excess rather than from the tail of the full $\neff$ posterior, i.e., we assume the existence of another species in addition to the three SM neutrinos (varying only $\Delta \neff >0$).}
	\label{fig:nnu_dec}
\end{figure}

If we instead start by assuming the existence of the three SM neutrinos and add new, specific light relic particles with mass $\lesssim$ eV --- i.e., only looking at regions of parameter space with $\neff>3.044$ --- then we can set limits on the new species' nature by exploiting the fact that the excess in $\neff$ that they generate depends on the spin of the particle and on the temperature at which they decoupled from the thermal bath in the early universe, as shown in Fig.~\ref{fig:nnu_dec} (with predictions from~\citealp{2016arXiv160607494B} for different species as adopted in previous CMB literature, including~\citealp{Planck_2015_params, Planck_2018_params,CMB_S4_Science_Book, Simons_Observatory}). We do this by fitting for the positive $\Delta \neff$ that such new species would contribute to $\neff$. We find $\Delta \neff< 0.32$ at 95\% CL from \act, which becomes 
\begin{eqnarray}
    \Delta \neff &<& 0.14 \quad (95\%, \pact), \nonumber \\
    &<& 0.17 \quad (95\%, \pactlb),
\end{eqnarray}
where lensing and BAO data relax the bound by moving the overall $\neff$ posterior to higher values as seen previously. The same limit using \plb\ is $\Delta \neff<0.36$ at 95\% confidence. Our \pactlb\ combination excludes at 95\% CL any light particle of spin 0 or 1/2 that decoupled after the start of the QCD phase transition at temperatures $\lesssim 200$ MeV and all light particles with spin 3/2 that decoupled at temperatures $\lesssim 1$ GeV. These are the strongest constraints to date on the possible existence of light, weakly coupled particles in the early universe.

As shown in Figs.~\ref{fig:lcdm_single_extensions_pactlb} and~\ref{fig:lcdm_single_extensions_pact}, the region of parameter space preferred by ACT is in the low $H_0$ corner of the ${\neff-H_0}$ correlation --- we find ${H_0=67.00\pm 0.91}$~km/s/Mpc (68\%, \pactlb) when varying $\neff$. This is driven by ACT's preference for values of $\neff$ below 3.044, while values higher than the SM expectation would be necessary to increase the expansion rate.  Therefore, this model extension is no longer viable to move the CMB-inferred Hubble constant toward the local SH0ES estimate. On the contrary, the \pactlb\ constraint is well within the CCHP measurement of $H_0$. If we import this latter measurement as a prior on $H_0$, we find a small pull of the central value of $\neff$ to higher values and negligible tightening of the error bar.

Including also a prior on $S_8$ (as described in~\S\ref{sec:glossary}) leaves the constraint essentially unchanged. In fact, including all state-of-the-art non-CMB data has limited impact on the precision of this measurement, which is dominated by \pact.

\begin{figure}[t!]
	\centering
\includegraphics[width=\columnwidth]{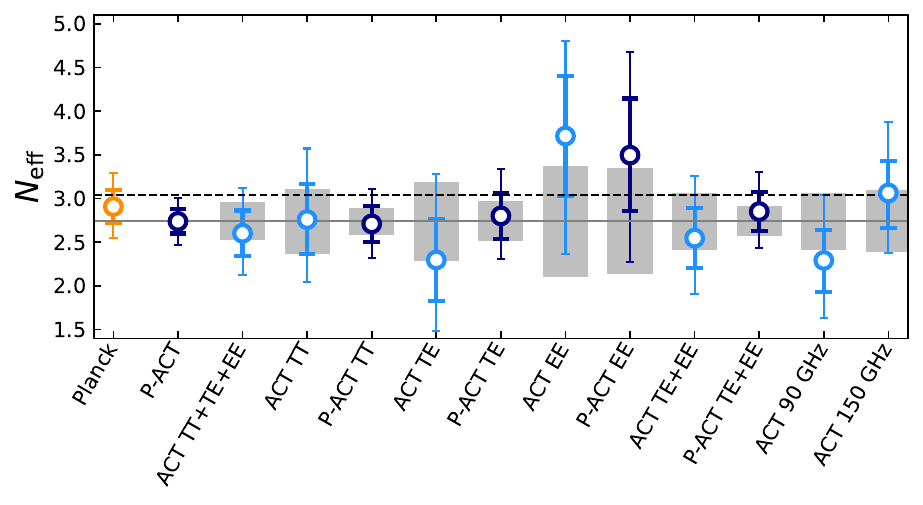}
	\caption{Robustness tests of the ACT measurement of $\neff$. The \pact\ value used as a reference and marked with the solid gray line is compared with the \Planck\ estimate (orange), with individual TT, TE, and EE probes within \pact\ (navy points) and with different subsets of ACT data (light blue points). All measurements are reported with $1$ and $2\sigma$ uncertainties (thicker and thinner colored error bars, respectively) and the SM prediction of $\neff=3.044$ is shown with the dashed black horizontal line. The gray bands show the expected statistical fluctuation incurred when looking only at a subset of the data --- calculated as the $\sigma$ difference between the full dataset and the nested subset of data used for each point~\citep{2020MNRAS.499.3410G}.}
	\label{fig:neff_whisker}
\end{figure}

$\neff$ is often used as a proxy parameter to explore the sensitivity of small-scale CMB data to physics that impacts the damping tail; conclusions drawn for $\neff$ in fact speak more broadly about patterns in --- and robustness of --- results from small scales. As such, we perform a number of tests to look at the stability of the $\neff$ measurement. In Fig.~\ref{fig:neff_whisker}, we report the variation in the constraint when comparing our baseline \act\ and \pact\ datasets with only a subset of the data.\footnote{To allow fair comparisons across values, all these runs are performed with the ACT multi-frequency likelihood; the ACT TE- and EE-only runs allow the polarization efficiencies to vary freely within a 20\% uniform prior.} The results are stable across probes, frequencies, and combinations with \Planck. The overall constraint is dominated by the ACT TT and TE spectra.\\

\vspace{-0.12cm}
\subsubsubsection{Neutrino mass}
\vspace{-0.12cm}
The total mass of all neutrino species is parametrized by $\mnu$. Measurements of flavor oscillations between neutrinos have provided evidence for a non-zero mass~\citep{Super-Kamiokande_mnu,SNO_mnu} and set a lower limit on the value of $\mnu$. Depending on the hierarchy (ordering) between mass eigenstates, ${\mnu \gtrsim 60}$~meV for the normal hierarchy and ${\mnu \gtrsim 100}$~meV for the inverted hierarchy~\citep[e.g.,][]{ParticleDataGroup:2024cfk}. The absolute value of the total mass is still unknown but, given that massive neutrinos with masses at sub-eV scales cannot be explained with the mass-generation mechanisms of the SM of particle physics, we already know that new physics is at play. Cosmology and laboratory experiments are highly complementary probes of the neutrino mass scale, being sensitive to different combinations of individual neutrino masses and mixing parameters, and relying on different assumptions in the derivation of the constraints~\citep[see, e.g.,][]{Gerbino:2022nvz}.  

Neutrino mass impacts the growth of structures in the universe after the neutrinos become non-relativistic~\citep[e.g.,][]{2006PhR...429..307L,2012arXiv1212.6154L,2011APh....35..177A,2011ARNPS..61...69W,2017FrP.....5...70G,2019BAAS...51c..64D}, suppressing growth at the level of a few percent. This in turn affects the amplitude of density perturbations measurable in the lensing of the CMB temperature and polarization anisotropies. \plb\ gives ${\mnu<0.077}$~eV at 95\% confidence~\citep{DESI-BAO-VI}, with a significant contribution from the CMB lensing data. With the new ACT DR6 spectra, assuming three massive neutrino species, we find
\begin{eqnarray}
    \mnu &<& 0.089~\rm{eV} \quad (95\%, \pactlb), \nonumber \\
    &<& 0.088~\rm{eV} \quad (95\%, \wactlb).
\end{eqnarray}
The addition of ACT DR6 to \Planck\ leaves the bounds almost unchanged. However, the neutrino mass constraints from \Planck\ correlate with the high fluctuation of the lensing-like smearing effect in the \Planck\ CMB spectra (measured by the $A_{\rm lens}$ parameter, \citealp{Calabrese2008}) so the ACT dataset serves here as a useful cross-check of the result. The impact of extra lensing is mitigated in the \pact\ combination by the cut in $\ell$ that we introduce in \Planck\ temperature and is completely avoided in the \wact\ combination --- since, as shown in L25, ACT DR6 measures $A_{\rm lens}$ consistent with \lcdm\ predictions. Once CMB lensing and BAO data are included, the CMB power spectra contribution to the bound is sub-dominant and the same across experiments. The conclusion that extra lensing-like smearing in the power spectra is not the primary contributor to the very tight bounds on $\mnu$ was also reached by \cite{Green&Meyers24}, who delve into all the different pieces of data contributing to a neutrino mass constraint (see also~\citealp{2024arXiv241205451G}).

\begin{figure}[t!]
	\centering
\includegraphics[width=\columnwidth]{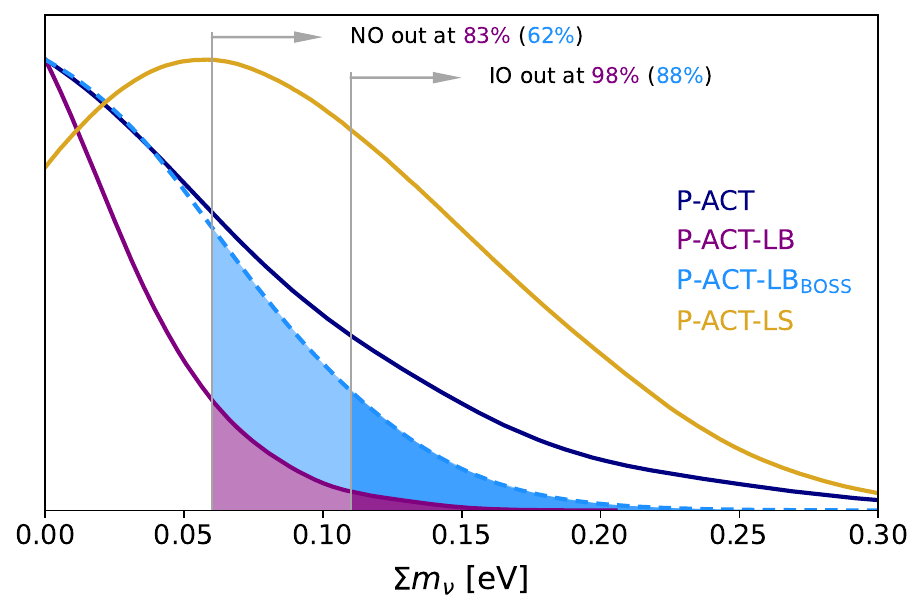}
    \caption{Upper limits on the sum of the neutrino masses, $\Sigma m_\nu$, within the \lcdm\ model. The gray lines mark the lower bounds coming from neutrino oscillation experiments, assuming a Normal Ordering (NO, ${\Sigma m_\nu > 0.06 }~\mathrm{eV}$) or Inverted Ordering (IO, ${\Sigma m_\nu > 0.10}~\mathrm{eV}$), respectively, and the colored bands show the exclusion regions coming from cosmology for each scenario. The most stringent bound is given by \pactlb\ (purple), with a significant contribution coming from DESI BAO. For this data combination, and within the assumption of the \lcdm\ model, the IO scenario is disfavored at 98\% confidence. The limits get more relaxed but still competitive when replacing DESI BAO with BOSS/eBOSS BAO (light blue) or when using SNIa (gold).}
	\label{fig:mnu}
\end{figure}

\begin{figure}[tp!]
	\centering
\includegraphics[width=\columnwidth]{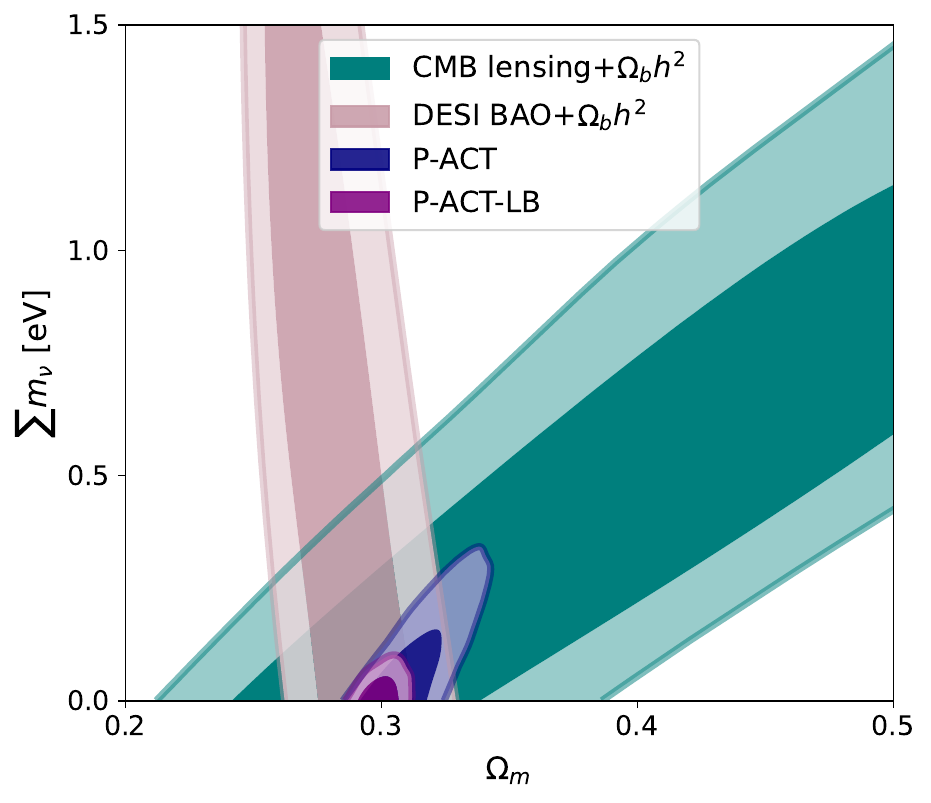}
	\vspace{-0.6cm}\caption{2D marginalized posteriors at 68\% and 95\% CL highlighting different degeneracy directions between the neutrino mass and the matter density for different cosmological probes. The BAO data (pink) are mostly insensitive to how matter is distributed across different components and only bound the total matter density. CMB lensing (teal) breaks this degeneracy, separating out the neutrinos from other matter components, once the baryon density is anchored by BBN constraints. Together with CMB anisotropies (purple), the combination of these data sets a stringent limit on the sum of the neutrino masses.}
	\label{fig:mnu_omegam}
\end{figure}

As shown in Figs.~\ref{fig:mnu} and~\ref{fig:mnu_omegam}, neutrino mass measurements from cosmology need different datasets to come together to disentangle the effect of neutrinos on the geometry of the universe and growth of structure from the late-time expansion effects~\cite[for a recent review see, e.g.,][]{2024arXiv241000090L}. In particular, DESI's constraint on the total matter fraction $\Omega_m$, through measuring the BAO and inferring cosmic distances, plays a large role in these neutrino mass limits, as shown in Fig.~\ref{fig:mnu_omegam}. The CMB lensing measurement then breaks the degeneracy between the total density in matter and in neutrinos, due to the unique effect of the neutrino on suppressing the clustering at smaller scales. The addition of ACT DR6 spectra confirms the tight limit that \cite{DESI-BAO-VI} found with \Planck. The combination of \pactlb\ disfavors at $\sim98\%$ confidence the inverted ordering scenario. This conclusion holds also for an independent-of-\Planck\ combination, with \wactlb\ giving very similar results.

Given the strong impact on these constraints from BAO, we explore other dataset combinations that can help the CMB for neutrino mass constraints. 
We replace DESI with BOSS BAO and obtain a bound of
\begin{eqnarray}
    \mnu&<& 0.13~\rm{eV} \ (95\%,\pactlbb),
\end{eqnarray}
which is unchanged when replacing \Planck\ with \WMAP. 
We also exclude BAO completely and instead exploit the information on the matter density coming from SNIa measurements, finding
\begin{equation}
        \mnu< 0.23~\rm{eV} \ (95\%,\textsf{P-ACT-LS}).
\end{equation}
In both cases, the limits become weaker and raise the likelihood of the inverted ordering (see Fig.~\ref{fig:mnu}). The broadening of the bound in \textsf{P-ACT-LS} is due to the preference of a larger value of the matter density exhibited by the supernovae measurements, as described in~\cite{2024arXiv241000090L}.

\begin{figure}[t!]
	\centering
 \includegraphics[width=\columnwidth]{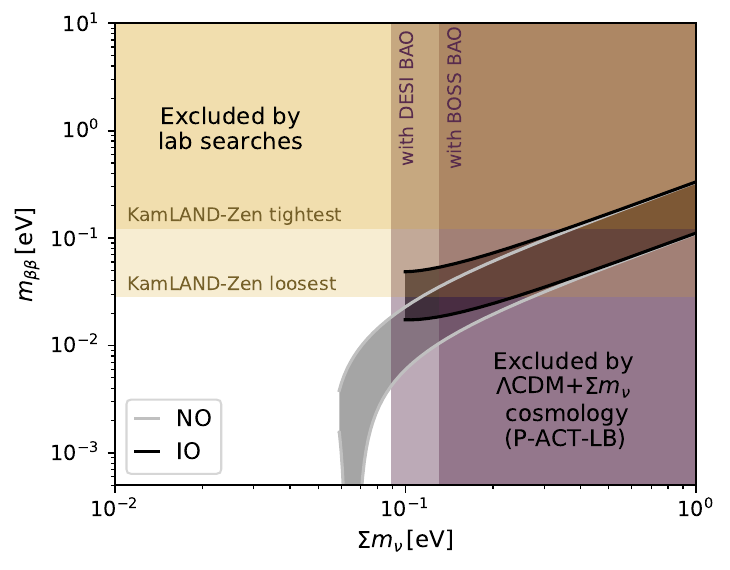}
	\vspace{-0.6cm}\caption{Neutrino oscillation experiments require that the neutrino mass scale parameters live in one of the two bands for normal ordering (NO, grey) or inverted ordering (IO, black), with the width of the bands reflecting the lack of knowledge of the Majorana phases~\citep{2020PTEP.2020h3C01P}. Cosmology helps distinguish between the two scenarios, disfavoring at 95\% CL the vertical band in purple, within the \lcdm\ model.  The dark purple region is disfavored when including BOSS BAO data, while when using DESI BAO data more parameter space is excluded, as shown by the light purple band.  Neutrinoless double-beta decay results at 90\% CL from KamLAND-Zen~\citep{KamLAND-Zen:2024eml} exclude the horizontal gold bands (these account for the uncertainty in the calculation of the nuclear matrix elements when translating the bounds on the xenon half life from KamLAND-Zen to a bound on the effective Majorana mass $m_{\beta\beta}$). The still viable parameter space is within the colored bands for the two neutrino mass orderings. The new cosmology results presented here reduce the allowed parameter space compared to $0\nu2\beta$ by excluding within the \lcdm+$\mnu$ model the inverted ordering region and a large fraction of the normal ordering region. This result however strongly depends on the use of DESI BAO data; with BOSS BAO the limits are more relaxed.}
	\label{fig:mnu_pp}
\end{figure}

In Fig.~\ref{fig:mnu_pp}, we compare our \pactlb\ bounds with constraints on the effective Majorana mass $m_{\beta\beta}$ from the neutrinoless double-beta decay ($0\nu2\beta$) experiment KamLAND-Zen~\citep{KamLAND-Zen:2024eml} (comparable constraints --- not shown here --- have also been obtained by the GERDA collaboration,~\citealp{GERDA:2020xhi}). The viable parameter space, represented by the two colored bands for the two neutrino mass orderings, is cut by the combination of cosmology and $0\nu2\beta$ results within \lcdm. Assuming that the neutrino mass mechanism leads to the $0\nu2\beta$ process (which thus assumes neutrinos are Majorana particles), KamLAND-Zen is able to exclude almost entirely the inverted ordering region. Cosmology further excludes the still-allowed IO region and a large fraction of the normal ordering. This sets a challenging target for next-generation $0\nu2\beta$ searches to probe the region of $m_{\beta\beta}<18\,\mathrm{meV}$ (the lowest value for the Majorana effective mass in the inverted ordering when considering measurements from neutrino oscillation experiments, i.e., the lowest value at the bottom of the black IO band in Fig.~\ref{fig:mnu_pp}). However, as discussed in, e.g.,~\cite{DESI-BAO-VI,2024arXiv240902295S}, these limits are model-dependent: for example, extended models with time-varying dark energy will relax the bounds on neutrino mass and bring the IO back into the allowed region. Cosmological bounds also assume that neutrinos are long-lived particles with a Fermi-Dirac distribution function. Models that deviate from these assumptions, such as scenarios featuring decaying neutrinos or neutrinos with average momenta larger than that of a thermal distribution~\citep[e.g.,][]{2016PhRvD..93k3002D,2019PhRvD..99b3501L,2019JCAP...04..049O,2020JHEP...04..020C,2021PhRvD.103d3519C,2021JCAP...03..087B,2022JCAP...02..037A,2022PhRvD.105f3501A,2022arXiv220309075Z,2022JHEP...08..076A}, would likely still be compatible with masses in the IO band.

The constraints on both $\neff$ and $\mnu$ remain stable when varying both parameters at the same time, with
\begin{equation}
\left.
 \begin{aligned}
\neff &= 2.85 \pm 0.25 \\ 
\mnu & < 0.073~{\rm eV}
\end{aligned}
\quad \right\} \mbox{\textrm{\; (95\%, \pactlb)}}
\end{equation}
and little correlation in the joint parameter region. 

While the constraint on $\neff$ agrees with SM predictions, the most probable value for $\mnu$ from cosmology is currently zero, in contrast with oscillation measurements. If this difference were to become significant with more precise data, it would imply a tension between our constraints on neutrino properties in the early universe versus neutrino measurements in the laboratory. Fortunately, with the increasing sensitivity of future CMB and BAO measurements, we can hope to determine the sum of neutrino masses to 0.02--0.03~eV precision~\citep[e.g.,][]{CMB_S4_Science_Book,Simons_Observatory,2015AAS...22533605E,Enhanced_SO_2025}. \\

\subsubsection{Neutrino self-interactions} \label{sec:neu_si}

Neutrino self-interactions can arise in a class of BSM models, for example in models where Majorana neutrinos interact through a (pseudo)scalar, the Majoron $\phi$, related to the breaking of lepton number and to the generation of small neutrino masses \citep{Chikashige:1980qk,Chikashige:1980ui,Schechter:1981cv}. In a wide class of models, the neutrino-Majoron coupling constant $g$, controlling the strength of neutrino self-interactions, is proportional to $m_\nu/v_L$, where $m_\nu$ is the neutrino mass and $v_L$ is the scale of the new physics that breaks lepton number. In general, a signature of neutrino self-interactions in cosmological observations might point to BSM physics. Here, we study the CMB phenomenology related to neutrino self-interactions arising from neutrino--neutrino scattering processes, which is mostly independent of the spin and parity properties of the mediator. The symbol $\phi$ in the following will denote a generic mediator, including the Majoron, unless otherwise stated.
We consider interactions among the three active neutrinos that do not couple different mass eigenstates, and with eigenstate-independent strength (in other words, the matrix of couplings is proportional to the identity matrix in the mass basis).

Self-interactions affect the evolution of perturbations by making neutrinos behave, at certain times, as a collisional fluid and not as free-streaming particles. Acoustic oscillations will then propagate in the neutrino fluid and enhance, through gravity, the amplitude of photon fluctuations. Collisions will also cancel the characteristic phase shift imprinted on photon and matter perturbations by neutrino free streaming. 
These effects are relevant during the radiation-dominated era, when neutrinos provide a non-negligible contribution to the total energy density, and are imprinted at the time when perturbation modes re-enter the horizon. The range of scales affected thus also depends on the energy dependence of the scattering rate, which simplifies in the two limiting scenarios considered in the following: a very light (``massless'') or very heavy mediator. 

\vspace{-0.12cm}
\subsubsubsection{Heavy mediator}
\vspace{-0.12cm}
We consider first the case where the mass $m_\phi$ of the mediator is much larger than the neutrino temperature at all times directly probed by CMB anisotropies \citep{Cyr-Racine:2013jua}. In this case, the neutrino interaction is effectively a four-fermion vertex controlled by a dimensional coupling constant ${G_{\rm eff} = g^2/m_\phi^2}$, i.e., the effective low-energy Lagrangian is 
${\mathcal{L}_\mathrm{eff} = G_{\rm eff}\bar\nu \nu \bar\nu \nu}$.
This is analogous to the low-energy behavior of standard weak interactions, just with $G_{\rm eff}$ taking the place of the Fermi constant $G_\mathrm{F}
\simeq 1.17\times 10^{-5}\,\,\mathrm{GeV}^{-2}$. 
This low-energy limit does not depend on the nature of the mediator, so the analysis here naturally encompasses both the scalar and vector cases. The Boltzmann hierarchy for neutrino perturbations including the collision term $\mathcal{L}_\mathrm{eff}$ generated by the interaction has been derived by \citet{Kreisch:2019yzn}.

In this scenario, neutrino free-streaming does not start at the time of weak decoupling, but is instead delayed until $T_{\mathrm{fs}}=T_{\nu,\mathrm{dec}} (G_F/G_{\rm eff})^{2/3}$, where $T_{\nu,\mathrm{dec}}\simeq 1\,\rm{MeV}$ is the neutrino decoupling temperature.
Neutrino self-interactions through a heavy mediator leave an imprint at angular scales $\theta\lesssim\theta_\mathrm{fs}$ (assuming $\theta_\mathrm{fs}<\theta_\mathrm{eq}$), where $\theta_\mathrm{fs}$ is the scale entering the horizon at $T=T_\mathrm{fs}$.

Previous analyses have shown that CMB and BAO data are compatible with, and in some cases prefer, neutrino self-interactions with $G_{\rm eff}\gg G_F$ \citep{Cyr-Racine:2013jua,Archidiacono:2013dua,Lancaster:2017ksf,Oldengott:2017fhy,Park:2019ibn,Kreisch:2019yzn,Barenboim:2019tux,Brinckmann:2020bcn,Das:2020xke,Mazumdar:2020ibx,RoyChoudhury:2020dmd,2025arXiv250310485P}. In fact, the posterior for $G_{\rm eff}$ has been found to be bimodal, with probability being concentrated in two distinct regions: a moderately interacting (MI$\nu$) mode, compatible with no self-interactions, and a strongly interacting (SI$\nu$) mode. The analysis of ACT DR4 data showed a slight preference for neutrino self-interactions at the $2-3\sigma$ level, finding $G_{\rm eff}\lesssim 10^{-3}\,\rm{MeV}^{-2}$ for MI$\nu$, and $G_{\rm eff}\simeq 10^{-1.5}\,\rm{MeV}^{-2}$ for SI$\nu$ \citep{Kreisch_ACT_SIN}.

We start by considering a one-parameter extension of \lcdm, including $\Geff$ as an extra parameter and keeping fixed ${\sum m_\nu = 0.06\,\mathrm{eV}}$ and $\neff=3.044$. To check if the bimodal behavior persists with the ACT DR6 spectra, we split the parameter space in two regions, ${\Geff < 10^{-2}~\mathrm{MeV}^{-2}}$~(MI$\nu$) and ${\Geff> 10^{-2}~\mathrm{MeV}^{-2}}$~(SI$\nu$), and perform separate MCMC runs with flat priors on $\log_{10}\left(\Geff \, \MeV^2\right)$: ${-8 \leq \log_{10}\left(\Geff \, \MeV^2\right) \leq -2}$ (MI$\nu$) and ${-2 \leq \log_{10}\left(\Geff \, \MeV^2\right) \leq 0}$ (SI$\nu$). The logarithmic prior allows us to explore a wide range of values of $\Geff$, spanning several orders of magnitude.

\begin{figure}[t!]
	\centering
    \includegraphics[width=\columnwidth]{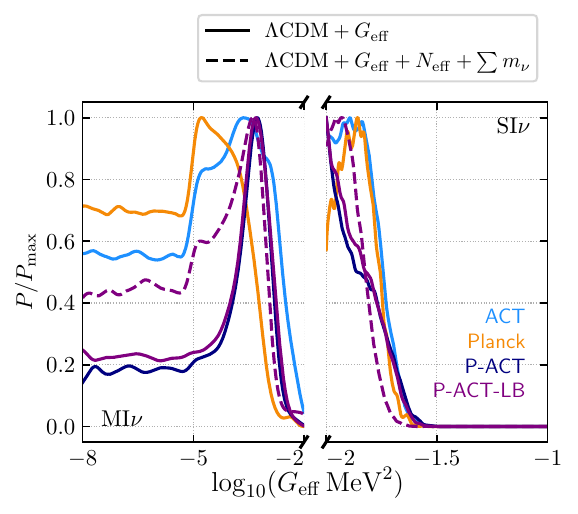}
	\vspace{-0.4cm}\caption{Posterior distribution of the neutrino self-interaction effective coupling constant, $G_\mathrm{eff}$, for various datasets within the ${\Lambda\text{CDM}+G_\mathrm{eff}}$ model (solid lines) and the more extended case of ${\Lambda\text{CDM}+G_\mathrm{eff}+\neff+\mnu}$ (dashed line). The posterior is presented in two distinct regions of parameter space, corresponding to the MI$\nu$ and SI$\nu$ modes, each independently normalized due to separate sampling. Analysis of the MAP models in the two regions shows that neutrino self-interactions are not preferred over \lcdm\ (corresponding to $G_\mathrm{eff}=0$), for any dataset combination. In the $\Lambda\text{CDM}+G_\mathrm{eff}$ model, for \pactlb\ we find the MI$\nu$ at $1.8\sigma$ and no peak for SI$\nu$; a similar behavior is also found for the more extended model. Minimal smoothing is applied to the plot to preserve the features of the distribution.}
	\label{fig:Geff_posterior}
\end{figure}

In the MI$\nu$ region, the \pact\ and \pactlb\ combinations are compatible with $\Geff=0$, but show a mild preference for ${\log_{10}\left(\Geff \, \MeV^2\right) \simeq -3}$. In contrast, in the SI$\nu$ region, the posteriors for the same dataset combinations peak at the lower edge of the prior range, i.e., at ${\log_{10}\left(\Geff \, \MeV^2\right) = -2.0}$.
We show the two posteriors in Fig.~\ref{fig:Geff_posterior}. Since we are performing separate parameter estimation runs for MI$\nu$ and SI$\nu$, we cannot use the distributions to assess the relative probability of the two scenarios. Instead we compare the effective $\chi^2$ of MAP models in the two regions to that of the \lcdm\ model, as reported in Table~\ref{table:selfnu_chi2_transposed}. 
We find that the SI$\nu$ model is never preferred over \lcdm. The MI$\nu$ model yields basically no improvement to the goodness-of-fit for the ACT DR6 data alone with respect to \lcdm, while it gives marginal improvement for \pact\ and \pactlb. Specifically, we find $\Delta\chi^2_{\mathrm{MI}\nu}= 2.9$ (\pact) and $\Delta\chi^2_{\mathrm{MI}\nu} = 3.1$ (\pactlb), corresponding respectively to a mild $1.7 \sigma$ and 1.8$\sigma$  preference over \lcdm, as computed using the likelihood-ratio test.

\begin{table}[t]
    \begin{tabular}{c|c|c|c}
    \hline
    & \act & \pact & \pactlb \\ \hline \hline
    $\Delta\chi^2_{\text{MI}\nu}$ & $-0.2$ & $2.9$ & $3.1$ \\ \hline
    $\Delta\chi^2_{\text{SI}\nu}$ & $-3.2$ & $-10.6$ & $-7.3$ \\ \hline
    $\sigma_{\text{MI}\nu}$ & -- & 1.7 & $1.8$\\ \hline
    $\sigma_{\text{SI}\nu}$ & -- & -- & -- \\ \hline
    \end{tabular}
    \caption{$\Delta\chi^2 \equiv \chi^2_{\Lambda\text{CDM}} - \chi^2_{\Lambda\text{CDM}+G_\text{eff}}$ from the MAP points of the MI$\nu$ and SI$\nu$ regions for different data combinations. When self-interacting neutrino models yield an improvement of the fit over \lcdm\ we also report the preference for the model in units of $\sigma$. We find no statistically significant preference for neutrino self-interactions.}
    \label{table:selfnu_chi2_transposed}
\end{table}

Since ${\Geff=0}$ is compatible with the data, the posterior for the MI$\nu$ scenario ideally extends with non-zero probability down to ${\log_{10}\left(\Geff \MeV^2\right)\to-\infty}$. The posterior is therefore ill-defined with a diverging integrated probability. This prevents the computation of meaningful Bayesian credible intervals. Previous analyses chose to arbitrarily cut the distribution at some lower bound and reported credible intervals depending on this choice. Here, we take a different approach; once the information about the position of the peaks has been acquired from the runs with a logarithmic prior, the proposal density of the MCMC can be correspondingly tuned and we repeat the runs using a uniform prior on $\Geff^2$, namely $ 0 \leq \Geff^2 \leq (10^{-2} \, \text{MeV}^{-2})^2 $, ensuring a proper posterior. We choose a uniform prior on $\Geff^2$ rather than on $\Geff$ because the former amounts to a uniform prior on the scattering rate $\Gamma\propto\Geff^2$.  Adopting this strategy, we find
\begin{equation}
    \Geff~(\textrm{MI}\nu)<{7.9 \times 10^{-3}} \, \text{MeV}^{-2}\ (95\%, \pactlb). 
\end{equation}

We repeat the analysis for a model in which only one of the three neutrino families is self-interacting. In this case, we find that both \pact\ and \pactlb\ yield a posterior with a well-defined peak in the SI region; in particular, ${\log_{10}\left(\Geff \, \MeV^2\right) = -1.33^{+0.21}_{-0.14}}$ (68\% \pactlb). The posterior in the MI region is instead qualitatively similar to the one found in the scenario with three interacting neutrinos. Both these single interacting-neutrino models are however not significantly preferred over \lcdm, with $\Delta\chi^2_{\mathrm{MI}\nu}= - 1.2$ and $\Delta\chi^2_{\mathrm{SI}\nu} = - 5.5$ (\pactlb), corresponding respectively to a 1.1$\sigma$ (MI$\nu$) and 2.3$\sigma$ (SI$\nu$) preference over \lcdm.

We further consider a 9-parameter extended model with three interacting neutrinos in which, in addition to the base \lcdm\ parameters and $\Geff$, also $\neff$ and $\mnu$ are allowed to vary. As in \citet{Kreisch_ACT_SIN}, $\neff$ is used to rescale the neutrino temperature. We find results similar to those for the \lcdm$+\Geff$ model with three interacting neutrinos (see Fig.~\ref{fig:Geff_posterior}). In particular, we do not find a peak in the posterior in the SI region, in contrast to the mild preference for such a model seen in the analysis of the ACT DR4 data~\citep{Kreisch_ACT_SIN}. The mild hint of SI$\nu$ in the DR4 analysis was largely driven by a high fluctuation in the EE power spectrum at intermediate scales, ${700<\ell<1000}$, which is no longer present in the DR6 spectra (see Appendix~\ref{app:dr4dr6comp}).

To summarize, we find that limits on interacting neutrino models are sensitive to the underlying assumptions used to describe the broader physics in the neutrino sector.
However, neutrino self-interactions are not detected in any scenario that we consider and not preferred over \lcdm.

This lack of evidence also limits the ability of this model to increase the value of the Hubble constant. For example, a heavy mediator in the MI$\nu$ scenario gives $H_0 = 68.2 \pm 0.4$~km/s/Mpc (\lcdm$+\Geff$) and $H_0 =67.5 \pm 1.0$~km/s/Mpc (\lcdm$+\Geff+\neff+\mnu$) using \pactlb. In general, we find no significant shift compared to \lcdm\ for the \lcdm$+\Geff$ model. In the extended case of \lcdm$+\Geff+\neff+\mnu$, the $H_0$ distribution shifts lower by ${\sim1}$~km/s/Mpc and has a twice larger uncertainty. This is driven by the ${\neff - H_0}$ correlation and by ACT's preference for $\neff<3.044$ found in \lcdm+$\neff$ (see~\S\ref{sec:neu_n_m}), which persists in this model. In the SI$\nu$ scenario, we find instead $H_0 = 69.0 \pm 0.4$~km/s/Mpc (\lcdm$+\Geff$) and $H_0 =67.0 \pm 0.9$~km/s/Mpc (\lcdm$+\Geff+\neff+\mnu$) for \pactlb. The ${\sim1}$~km/s/Mpc upward shift in $H_0$ in the \lcdm$+\Geff$ model is induced by the positive correlation between $\Geff$ and $\theta_s$. In the \lcdm$+\Geff+\neff+\mnu$ model, this effect is countered by a lower value of $\neff$, resulting in a lower value of $H_0$ with respect to the corresponding MI$\nu$ scenario. 

\begin{figure}[t!]
	\centering
\includegraphics[width=\columnwidth]{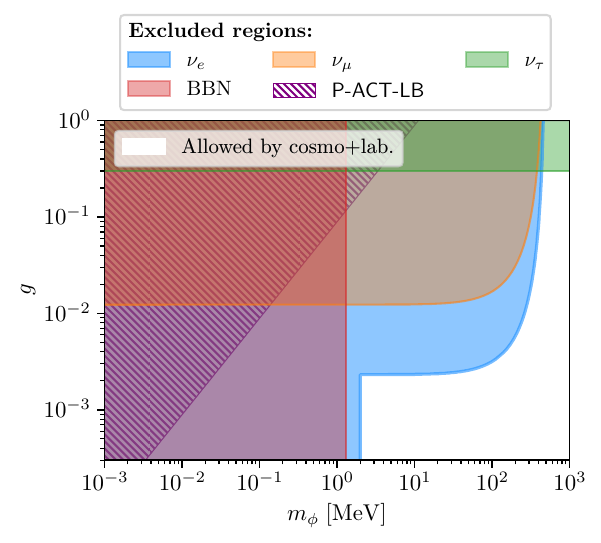} 
	\vspace{-0.6cm}\caption{Cosmological and laboratory limits on neutrino couplings with a new scalar $\phi$. The purple band with hatches in the upper left corner corresponds to the region excluded by \pactlb\ accounting for both MI and SI neutrinos. The other overlaid color bands are excluded by particle physics and BBN limits. The shaded blue region refers to scalars coupling to electron neutrinos and is excluded by neutrinoless double $\beta$ decay searches \citep{Blum:2018ljv,Kharusi:2021jez,NEMO-3:2013pwo,GERDA:2020xhi} and measurements of the ratio ${\mathrm{Br}(K^+\to e^+\nu_e)/\mathrm{Br}(K^+\to \mu^+\nu_\mu)}$ \citep{Lessa:2007up,Fiorini:2007zzc}. The orange shaded region is excluded by laboratory constraints on the decay ${K^+\to \mu^+\nu_\mu \phi}$ \citep{ParticleDataGroup:2024cfk} that apply to scalars coupling to $\mu$ neutrinos. The green shaded region is excluded by laboratory constraints on the decay ${\tau^-\to \ell_\alpha^+\bar\nu_\alpha\nu_\tau \phi}$ \citep{Lessa:2007up}, applying to scalars coupling to $\tau$ neutrinos. The red-shaded region is excluded by BBN constraints on $\neff$ \citep{Blinov:2019gcj}. Together, cosmology and laboratory measurements allow the parameter space remaining in the right bottom corner (in white).}
	\label{fig:massive_self_nu}
\end{figure}

Laboratory searches, especially double-$\beta$ decay experiments and observations of meson and lepton decays, severely constrain the coupling of neutrinos with a new light boson \citep{Blinov:2019gcj}. Taken at face value, the values of $\Geff$ in the SI$\nu$ region, as well as those corresponding to the MI$\nu$ peak, are excluded by these searches as shown in Fig.~\ref{fig:massive_self_nu}, unless the new scalar couples almost exclusively to $\tau$ neutrinos. 

\vspace{0.12cm}
\subsubsubsection{Light mediator}
\vspace{-0.12cm}
In this scenario the mediator mass is much smaller than the average neutrino momentum at all times of interest and the scattering rate $\Gamma\propto g^4 T$, so that the ratio between the scattering and Hubble rates increases with time. Neutrinos will then start free streaming at weak decoupling as usual, and become collisional again at later times \citep{Archidiacono:2013dua,Forastieri:2015paa,Forastieri:2019cuf}. 
The effects of collisions are confined to scales between $\theta_\mathrm{coll}$, the scale entering the horizon when neutrinos stop free streaming at late times, and $\theta_\mathrm{eq}$. These correspond to intermediate angular scales in the CMB power spectra (larger scales compared to those affected by a heavy mediator) and so we expect less contribution to this limit from ACT DR6.

We write the thermally averaged cross section $\langle \sigma v \rangle=\xi g^4/T^2 \equiv g_\mathrm{eff}^4/T^2$, where the coefficient $\xi$ depends on the details of the interaction, including the nature of the mediator. We consider a one-parameter extension of \lcdm, varying $g_\mathrm{eff}^4$ with a flat prior. As in the heavy mediator case, this is equivalent to a flat prior on the neutrino interaction rate $\Gamma\propto g^4$.  We follow the approach of \citet{Forastieri:2015paa,Forastieri:2019cuf} where the collision term is approximated using the thermally-averaged scattering rate in the relaxation time approximation. From \Planck\ CMB data, we find $g_\mathrm{eff}^4 < 1.5\times 10^{-27}$ at 95\% CL. The new ACT DR6 spectra alone give a limit about three times weaker, with $g_\mathrm{eff}^4 < 5.2 \times 10^{-27}$ at 95\% CL. Combining the two datasets gives a $\sim 20\%$ improvement on \Planck\ alone, with
\begin{eqnarray}
    g_\mathrm{eff}^4 &<& 1.2\times 10^{-27} \quad (95\%, \pact), \nonumber \\
    &<& 1.3\times 10^{-27} \quad (95\%,  \pactlb),
\end{eqnarray}
or ${|g_\mathrm{eff}| < 1.1\times 10^{-7}}$. 

\begin{figure}[t!]
	\centering
\includegraphics[width=\columnwidth]{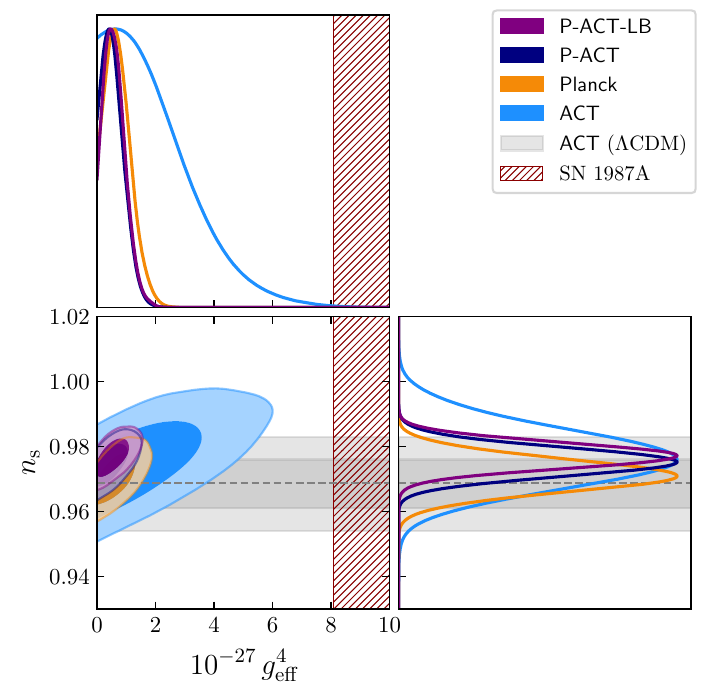}
	\vspace{-0.5cm}\caption{Constraints on the effective neutrino self-coupling, $g^4_\text{eff}$, in the light mediator limit from multiple data combinations. The hatched area is excluded by a luminosity argument applied to SN1987A~\citep{Kachelriess:2000qc}. The shaded gray region shows the 68\% and 95\% credible intervals for $n_s$ in the $\Lambda$CDM model (L25). The degeneracy with $n_s$ stems from the fact that large $g_\mathrm{eff}$ increases power at scales typically larger than the pivot scale $k_*=0.05\,\mathrm{Mpc}^{-1}$, while increasing the slope of the primordial spectrum reduces the power at those scales. The inclusion of ACT shifts the contours to larger values of $n_s$, consistent with what is reported in~\S\ref{sec:running}, and mildly tightens the bounds on $g_\mathrm{eff}$ compared to \Planck\ alone.}
	\label{fig:massless_self_nu}
\end{figure}

Neutrino couplings with a light scalar can be constrained from $0\nu2\beta$ experiments and from supernovae explosions. These constraints are expressed in terms of the couplings $g_{\alpha\beta}$ in the basis of the eigenstates of the weak interaction, with ${\alpha,\, \beta=e,\,\mu,\tau}$, as opposed to the mass basis used in our analysis. The two bases are related through the neutrino mixing matrix. Currently, EXO-200 provides the most stringent laboratory limits on the electron neutrino-Majoron coupling: ${|g_{ee}| < (0.4 - 0.9) \times 10^{-5}}$ \citep{Kharusi:2021jez}, where the range expresses the uncertainty due to nuclear matrix elements. Supernova data can also be used to constrain neutrino-Majoron couplings. In particular, scalar emission might overly shorten the observed neutrino signal from SN1987A. This luminosity argument excludes the range of couplings ${3\times 10^{-7} < |g_{\alpha\beta}| < 2\times 10^{-5}}$ \citep{Kachelriess:2000qc}. In most of the neutrino parameter space allowed by oscillation experiments, $\max{(|g_{\alpha\beta})|}\sim g_\mathrm{eff}$ and \pact\ yields the strongest constraints on neutrino interactions with a light (pseudo)scalar. In Fig.~\ref{fig:massless_self_nu}, we show the posterior distribution for $g_\mathrm{eff}^4$ for various data combinations, and compare with the limit from the luminosity argument. In the same figure we also show the mild correlation between $g_\mathrm{eff}$ and the scalar spectral index $n_s$, which we interpret as the result of the additional power at intermediate scales (i.e., on the left of the pivot scale $k_*$) for $g_\mathrm{eff}>0$.

\subsection{Helium and deuterium abundances} \label{sec:bbn} 

Light elements formed in the early universe during BBN, with abundances depending on the baryon-to-photon density ratio, $\eta_\textrm{b}\equiv n_\textrm{b}/n_\gamma$, and the number of relativistic degrees of freedom, $\neff$. Throughout this section we assume no leptonic asymmetry in the electron neutrino sector, i.e., a vanishing chemical potential of electron neutrinos. Given the photon temperature today \citep{Fixsen2009}, $\eta_\textrm{b}$ can be related to the physical baryon density $\omega_b\equiv \Omega_b h^2$. 

To compute primordial abundances of light elements as a function of $\ombh$ and $\neff$ we can use various computational codes, each built upon different underlying assumptions. In this paper we make use of the \texttt{PRIMAT} \citep{2018PhR...754....1P} and \texttt{PRyMordial} \citep{2024EPJC...84...86B} codes in three configurations:
\begin{enumerate}[(a)]
\itemsep0em
\item \texttt{PRIMAT\_2021}: this is the default BBN computation used in \texttt{camb} version \texttt{1.5}, but it was not present in the public version of \texttt{class} at the time our analyses were performed.\footnote{We manually import {\tt PRIMAT}-based tables from \texttt{camb} to use in \texttt{class}, as described in~\S\ref{sec:method}.  As of writing, {\tt PRIMAT} has recently been included in the public version of {\tt class}.} We use {\tt PRIMAT} as the baseline BBN code in all analyses presented in this paper. It implements tabulated reaction rates based on theoretical ab-initio calculations (see \citealp{2018PhR...754....1P} and references therein for details) and assumes a neutron lifetime value $\tau_\textrm{N}=(879.4 \pm 0.6)\,\,\textrm{s}$, i.e., the average provided in \texttt{PDG 2020} \citep{2020PTEP.2020h3C01P};
\item \texttt{PRIMAT\_2024}: uses the same rates as \texttt{PRIMAT\_2021}, with an updated value of $\tau_\textrm{N}=(878.4 \pm 0.5)\,\,\textrm{s}$ from \texttt{PDG 2024} \citep{ParticleDataGroup:2024cfk};
\item \texttt{PRyMordial}: uses a modified version of the \texttt{NACRE II} database \citep{2013NuPhA.918...61X} for the thermonuclear rates (see \citealp{2024EPJC...84...86B} and references therein for details) and the same value of the neutron lifetime used by \texttt{PRIMAT\_2024}, i.e., $\tau_\textrm{N}=(878.4 \pm 0.5)\,\,\textrm{s}$.
\end{enumerate}

These different BBN calculations return the same cosmological parameters and only affect the prediction of BBN abundances, so a distinction and detailed discussion is only required in this section.  We focus on BBN predictions for the primordial helium abundance, expressed as the nucleon number density fraction, $Y_p\equiv n_\textrm{He}/n_\textrm{b}$ --- rather than the mass fraction parameter $\yhe$ normally used in CMB calculations --- and the primordial deuterium abundance, $\textrm{D}/\textrm{H}\equiv\textrm{D}/\textrm{H}|_\textrm{p}\times 10^5$. Astrophysical determinations of the helium abundance are primarily derived from observations of low-metallicity extragalactic HII regions, whereas deuterium abundances are measured from high-redshift quasar absorption systems. The latest determinations are summarized in \texttt{PDG 2024} which also provides average estimates for both elements\footnote{Section 24.3 of \cite{ParticleDataGroup:2024cfk} details the averaging procedure. The \texttt{PDG 2024} average helium abundance does not incorporate the recent measurements from the EMPRESS survey performed on the Subaru telescope \citep{2022ApJ...941..167M}, which indicate a helium abundance, $Y_p = 0.2370 \pm 0.0034$, $\approx 2.4\sigma$ lower than the \texttt{PDG} value.}, namely $Y_p=0.245\pm0.003$ 
\citep{2021JCAP...03..027A,2022MNRAS.510..373A,2019ApJ...876...98V,2021MNRAS.505.3624V,2019MNRAS.487.3221F,2021MNRAS.502.3045K,2020ApJ...896...77H}
and ${\textrm{D}/\textrm{H}=2.547\pm0.029}$ \citep{2014ApJ...781...31C,2016ApJ...830..148C,2015MNRAS.447.2925R,2016MNRAS.458.2188B,2017MNRAS.468.3239R}.
We do not discuss here abundances of other light elements, such as tritium, $^3\textrm{He}$, and $^7\textrm{Li}$; these elements are more uncertain in both their determination and their interpretation \citep{2011ARNPS..61...47F, 2020ApJ...901..127I,ParticleDataGroup:2024cfk}.

\subsubsection{Abundances assuming standard BBN} 

\begin{figure}
\includegraphics[width=\columnwidth]{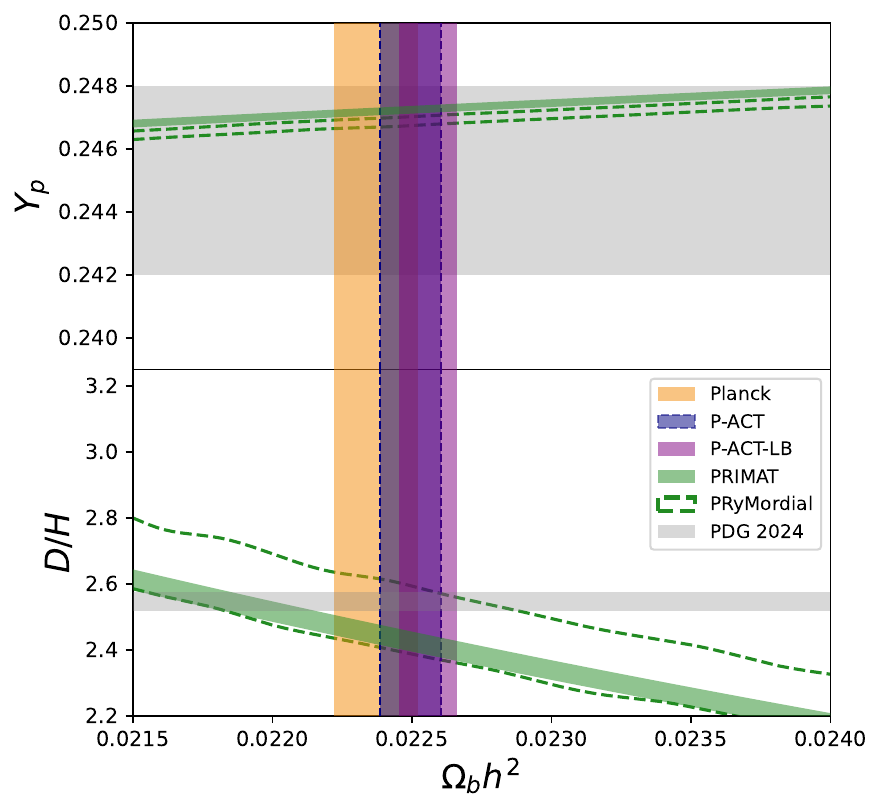}
    	\vspace{-0.6cm}\caption{Comparison of the constraints at 68\% confidence on the baryon density $\Omega_b h^2$ from \Planck\ (orange), \pact\ (navy), and \pactlb\ (purple), versus the measured primordial helium fraction $Y_p$ (top panel) and deuterium fraction $\mathrm{D}/\mathrm{H}$ (bottom panel). The green bands show the $1\sigma$ predicted relations between the parameters according to \texttt{PRIMAT} 2021 (filled) and \texttt{PRyMordial} (empty dashed). The gray bands are the average $68\%$ confidence levels of the astrophysical measurements of these fractions, according to \texttt{PDG 2024}~\citep{ParticleDataGroup:2024cfk}.}
	\label{fig:ydomb}
\end{figure}

Fixing the radiation density to its standard value (${\neff=3.044}$), the baryon density in the baseline \lcdm\ model can be directly related to the helium fraction. L25 shows that with the new ACT data the baryon density is measured to 0.5\% uncertainty; for the three configurations listed above, this leads to\footnote{Note that $Y_p$ is only logarithmically sensitive to the baryon density, which leads to the small error bars here.}
\begin{equation}
\left.
 \begin{aligned}
 \textrm{(a)}\;Y_p &= 0.24727 \\
 &\quad \pm 0.000046 \ (0.000131)\\
    \textrm{(b)}\;Y_p &= 0.24707 \\
    &\quad  \pm 0.000046 \ (0.000118)\\
    \textrm{(c)}\;Y_p &= 0.24691 \\
    &\quad \pm 0.000049 \ (0.000153)
\end{aligned}
\quad \right\} \mbox{\textrm{\; (68\%, \pactlb),}}
\end{equation}
where the first set of errors reflect only the uncertainty on $\ombh$ and the second in parentheses includes the theoretical BBN uncertainty added in quadrature. The shift between (a) and (b) is mainly driven by the different value adopted for the neutron lifetime, while the shift between (b) and (c) reflects the different treatment of the reaction rates and nuclear cross-sections between \texttt{PRIMAT} and \texttt{PRyMordial}; see \cite{2024JCAP...06..006S} for an exhaustive comparison. However, all these shifts are within the theoretical errors provided by the different codes, and encompassed by the \texttt{PDG} average of the observational bounds, which shows an order of magnitude larger error bar; see the upper panel of Fig.~\ref{fig:ydomb}.

We find that relaxing the assumptions on the radiation density by varying $\neff$ weakens the constraints on the helium fraction, substantially absorbing all the discrepancies between different BBN codes:
\begin{equation}
\left.
 \begin{aligned}
\label{eq:bbn_Yp_standardBBN_varyngneff}
    \textrm{(a)}\;Y_p &= 0.2448 \pm 0.0019 \\
    \textrm{(b)}\;Y_p &= 0.2446 \pm 0.0018 \\
    \textrm{(c)}\;Y_p &= 0.2445 \pm 0.0018
\end{aligned}
\quad \right\} \mbox{\textrm{\; (68\%, \pactlb).}}
\end{equation}
Here the error is dominated by the uncertainty on $\ombh$ and $\neff$, and the shift with respect to the $\neff=3.044$ case is driven by the shift of $\ombh$ between \lcdm\ and \lcdm+$\neff$ models.

Following the same approach, we can predict the deuterium abundance under the assumption of standard BBN. In the baseline \lcdm\ model the baryon density translates into 
\begin{equation}
\left.
 \begin{aligned}
 \label{eq:bbn_D_standardBBN}
    \textrm{(a)}\;\textrm{D}/\textrm{H} &= 2.415 \pm 0.019 \ (0.036) \\
    \textrm{(b)}\;\textrm{D}/\textrm{H} &= 2.413 \pm 0.019 \ (0.033)  \\
    \textrm{(c)}\;\textrm{D}/\textrm{H} &= 2.480 \pm 0.020 \ (0.104)
\end{aligned}
\quad \right\} \mbox{\textrm{\; (68\%, \pactlb).}}
\end{equation}

For deuterium, the neutron lifetime value has a negligible impact \citep{2022hnp..book..127G} and the shift is substantially caused by differences between \texttt{PRIMAT} and \texttt{PRyMordial} calculations, and mainly driven by the different choices made for the reaction rates. Theoretical errors also reflect the different choices made in the marginalization procedure by \texttt{PRIMAT} and \texttt{PRyMordial} --- see again~\cite{2024JCAP...06..006S} for an exhaustive comparison. When compared to direct measurements of the deuterium abundance, estimations based on \texttt{PRIMAT} exhibit a mild tension, ranging between 2 and 3$\sigma$. In contrast, the inferred values obtained using the \texttt{PRyMordial} computation are consistent with the \texttt{PDG} average, regardless of whether theoretical uncertainties are included; see the lower panel of Fig.~\ref{fig:ydomb}.

In the \lcdm+$\neff$ model, the scenario remains unchanged, with all constraints being shifted downward due to the higher baryon abundance found in this model
\begin{equation}
\left.
 \begin{aligned}
 \label{eq:bbn_D_standardBBN_varyngneff}
    \textrm{(a)}\;\textrm{D}/\textrm{H} &= 2.384 \pm 0.029\ (0.042) \\
    \textrm{(b)}\;\textrm{D}/\textrm{H} &= 2.383 \pm 0.029\ (0.040)\\
    \textrm{(c)}\;\textrm{D}/\textrm{H} &= 2.455 \pm 0.028\ (0.099)
\end{aligned}
\quad \right\} \mbox{\textrm{\; (68\%, \pactlb).}}
\end{equation}
Again, the deuterium abundance inferred assuming \texttt{PRyMordial} is consistent with astrophysical measurements while both \texttt{PRIMAT}-based estimates are in mild tension ($\sim 3 \sigma$).

\begin{figure}
\centering
\hspace{-0.7cm}
\includegraphics[width=\columnwidth]{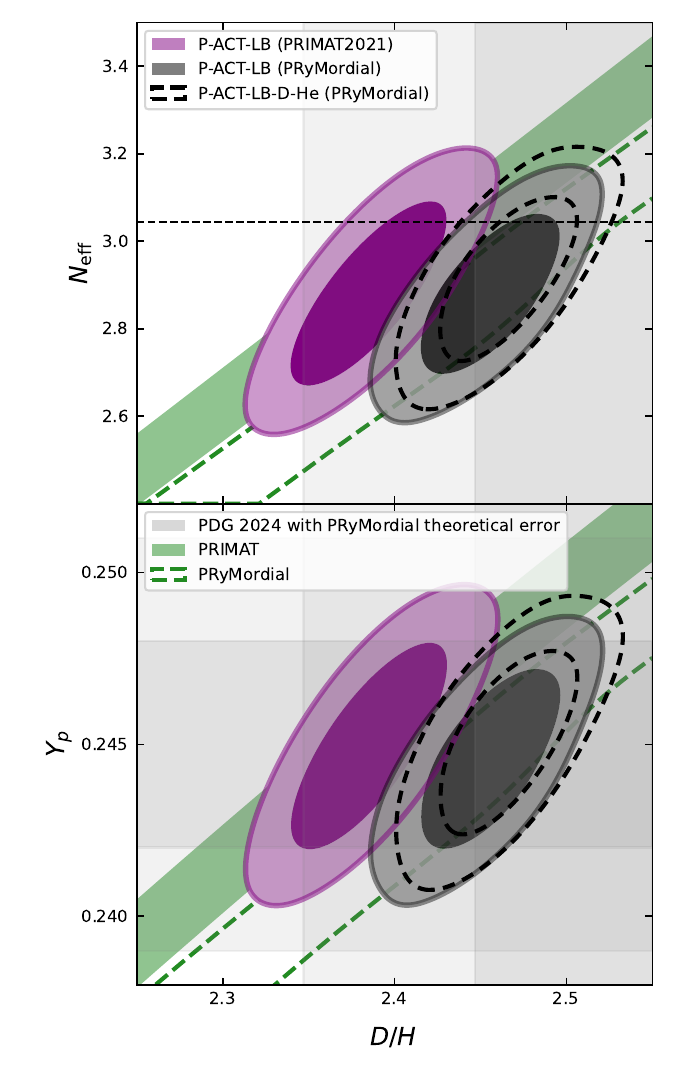}
\vspace{-0.4cm}
\caption{Constraints on the plane $\neff$--$\textrm{D}/\textrm{H}$ (upper panel) and $Y_p$--$\textrm{D}/\textrm{H}$ (lower panel) from $\pactlb$ assuming \texttt{PRIMAT} 2021 as BBN consistency (purple), $\pactlb$ assuming \texttt{PRyMordial} (filled black),  and $\pactlbDHe$ again assuming \texttt{PRyMordial} (dashed empty black). The green bands show predicted relations between the parameters according to \texttt{PRIMAT} 2021 (filled) and \texttt{PRyMordial} (empty dashed), assuming 1$\sigma$ error on $\ombh$ from $\pactlb$ in the \lcdm\ model. The gray bands show the \texttt{PDG} averages with 68\% and 95\% errors obtained combining in quadrature the observational error and the \texttt{PRyMordial} theoretical error.}
	\label{fig:yhedneff}
\end{figure}

Given these results, we combine direct deuterium measurements, accounting for both observational and theoretical uncertainties, with the posterior distribution of $(\ombh, \neff)$ only under the assumption of \texttt{PRyMordial} BBN calculations.  This approach yields joint CMB+BBN predictions. After marginalizing over $\ombh$, we obtain
\begin{equation}\label{eq:bbn_Neff_standardBBN_withD}
    \neff = 2.87 \pm 0.13 \quad (68\%, \pactlbD).
\end{equation}
The uncertainty on $\neff$ remains substantially unchanged compared to the limit from \pactlb~(\S\ref{sec:neutrinos}), as the theoretical error on deuterium dominates. The mean of the distribution is shifted towards slightly higher values, driven by the higher astrophysical measurement of the deuterium abundance as shown in Fig.~\ref{fig:yhedneff}.

We also expand this to include helium. In this case, the total uncertainty is dominated by the astrophysical measurement, which is only $50\%$ larger than the CMB determination (see Eq.~\ref{eq:bbn_Yp_standardBBN_varyngneff}). Combining the CMB data  with the \texttt{PDG} averages we get
\begin{equation}\label{eq:bbn_Neff_standardBBN_withDandHe}
    \neff = 2.89 \pm 0.11 \quad (68\%, \pactlbDHe),
\end{equation}
which represents the tightest limit on $\neff$ to date. 

By assuming the \texttt{PDG} averages for helium and deuterium abundances and employing \texttt{PRyMordial} as the BBN calculator, we can also derive a pure BBN constraint on $\neff$, finding $\neff = 2.93 \pm 0.21 \; (68\%,\DHe)$ after marginalizing over $\eta_\textrm{b}$. This estimate shows that the number of neutrinos at BBN and at the CMB recombination epoch are in very good agreement.

\subsubsection{Model-independent bounds on the helium fraction}

\begin{figure}
    \includegraphics[width=\columnwidth]{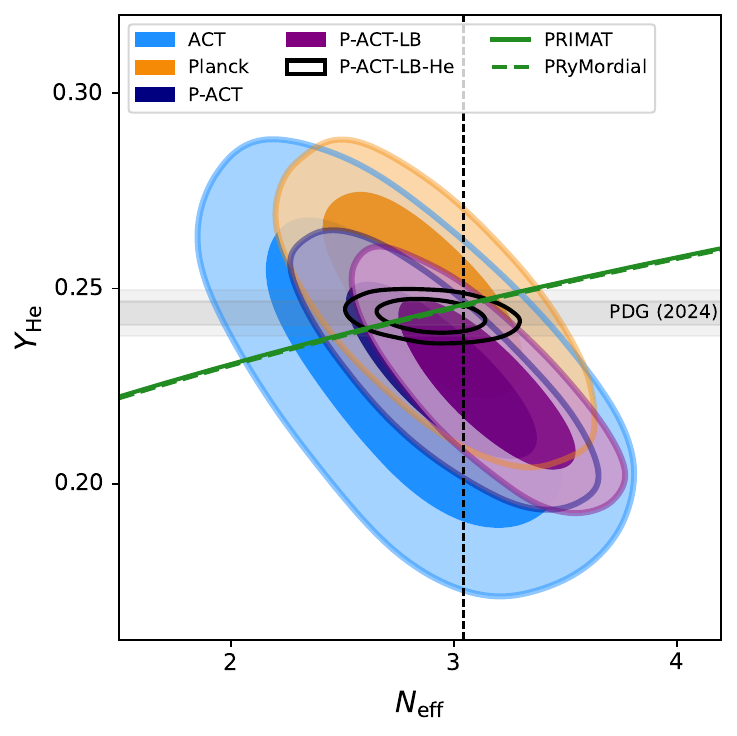}
    \vspace{-0.6cm}\caption{Constraints on the helium abundance, $Y_{\rm He}$, and number of effective neutrino species, $\neff$, with both parameters varying freely, for different data combinations. The horizontal bands are the \texttt{PDG} average of observational bounds with 68\% and 95\% errors. The green curves show the predicted relations between the parameters according to \texttt{PRIMAT} and \texttt{PRyMordial}, assuming the best-fit of \pactlb, and the dashed black line marks the SM prediction of $\neff=3.044$.}
	\label{fig:yheneff}
\end{figure}

Instead of deriving the primordial helium abundance from BBN codes using constraints on $(\ombh, \neff)$, we can measure it directly with the CMB. Variations in $\yhe$ alter the density of free electrons between helium and hydrogen recombination, thereby affecting the damping tail of the CMB anisotropies. By sampling the \lcdm+$\yhe$ model and converting the helium mass fraction into the helium nucleon fraction, we find 
\begin{equation}
    Y_p = 0.2312 \pm 0.0092 \quad (68\%, \pactlb),
\end{equation}
in good agreement with the astrophysical helium abundance measurements.  
Since both the helium abundance and the number of relativistic degrees of freedom influence the CMB damping tail, allowing both to vary simultaneously leads to weaker constraints
\begin{equation}
\left.
 \begin{aligned}
Y_p &= 0.227 \pm 0.014 \\ 
\neff &= 3.14 \pm 0.25
\end{aligned}
\quad \right\} \mbox{\textrm{\; (68\%, \pactlb).}}
\end{equation}
This is still consistent with direct helium abundance measurements and with the expected number of neutrinos, but with larger errors. The relatively low values of $Y_p$ in this case have the same origin as the low value of $\neff$ in the \lcdm+$\neff$ model, since lower values of both of these parameters reduce the damping, which is preferred by the ACT data. The joint constraints in the ${\neff-Y_{\rm He}}$ plane are shown in Fig.~\ref{fig:yheneff}, where the anti-correlation between these two parameters is clearly evident.

Considering the \texttt{PDG} collection of helium abundance measurements as an external dataset, it is possible to sample the \lcdm+$\neff$+$\yhe$ model imposing a prior on $Y_p$, which leads to
\begin{equation}
\left.
 \begin{aligned}
Y_p &= 0.2444 \pm 0.0029 \\ 
\neff &= 2.86 \pm 0.14
\end{aligned}
\quad \right\} \mbox{\textrm{\; (68\%, \pactlbHe),}}
\end{equation}
where the degeneracy between $\neff$ and $\yhe$ is efficiently broken by helium measurements (see Fig.~\ref{fig:yheneff}) and the retrieved $\neff$ constraint is consistent with the one obtained in the \lcdm+$\neff$ model.\\

\subsection{Axion-like particles}\label{sec:axions} 

We constrain contributions to the total dark-sector content from ultralight axion-like particles (with masses between ${10^{-28}~ < m_\mathrm{ax}~/{\rm eV}< 10^{-24}}$). This ultralight pseudo-scalar dark matter model is well-motivated by high-energy physics, as axions can generically arise from a variety of broken symmetries or compactified extra dimensions, the so-called ``axiverse'' \citep{Arvanitaki:2009fg,Duffy:2009ig}. These ultralight axions suppress structure formation below their de Broglie wavelength, which manifests on astrophysical scales. The axion Jeans scale $\lambda_J$, below which structure is smoothed out, depends on the axion mass: $\lambda_J \propto m_\mathrm{ax}^{-1/2}$ \citep{amendola/etal:2006, Duffy:2009ig, Arvanitaki:2009fg, park/etal:2012,hlozek/etal:2015,  Marsh:2015xka, hui/etal:2016, lague/etal:2021,ohare:2024}. The dynamics of the axion are described through the Klein-Gordon equation
\begin{equation}
    {\phi}^{\prime\prime} + 2 \mathcal{H} {\phi}^\prime + a^2 \frac{dV}{d\phi} = 0.
    \label{eq:axionEOM}
\end{equation}
Here $\phi$ is the axion field, with $\phi_0$ indicating its homogeneous component, and primes denote derivatives with respect to conformal time.  In the simplest case with a quadratic axion potential, this simplifies to 
\begin{equation}
    {\phi}^{\prime\prime} + 2 \mathcal{H} {\phi}^\prime + a^2 m_{\rm ax}^2\phi = 0,
    \label{eq:quadaxion}
\end{equation}
where $\mathcal{H}\equiv {a}^\prime/a = aH$ is the conformal Hubble parameter, which acts as a friction term in the oscillator equation governing the axion dynamics.  When the Hubble term drops (as the universe cools) to near the axion mass $m_{\rm ax} \simeq 3H(a_{\rm osc})$, the axion field begins to coherently oscillate, and begins to behave as a dark matter component (with $\rho_{\rm ax} \propto a^{-3}$), whereas it behaves as a frozen-in ``dark-energy-like'' component at earlier times. This links the axion mass directly to its observational impact on the CMB and large-scale structure clustering. We solve for the relic axion density by solving the field equations given that \begin{equation}
\label{eq.Omega_ax}
\Omega_{\rm ax} = \left[\frac{1}{2} \left( \frac{\phi_0^\prime}{a} \right)^2 + \frac{m_{\rm ax}^2}{2}\phi_0^2\right]_{m_{\rm ax} = 3H} \frac{a_{\rm osc}^3}{\rho_\mathrm{crit}} \,,
\end{equation}
where $\rho_\mathrm{crit}$ is the critical density at $z=0$.  We model the effects of axions via the \texttt{axionCAMB} Boltzmann solver,\footnote{\href{https://github.com/dgrin1/axionCAMB}{https://github.com/dgrin1/axionCAMB}} which has been used in several previous analyses of the impacts of axions on the CMB and LSS~\citep[e.g.,][]{hlozek/etal:2015,hlozek/etal:2018,rogers/etal:2023,lague/etal:2021}. There have been several improvements in the full Boltzmann solution \citep[e.g.,][]{liu/etal:2025} that are particularly relevant in extreme ranges of parameter space that we do not consider here. We leave a full comparison between codes for future work.

In order to correctly model the power on small scales \citep[and to not introduce spurious signals of axion physics, see][]{hlozek/etal:2017,dentler/etal:2022}, we use a modified halo model including mixed dark matter \citep{vogt/etal:2023}. The computational cost of both the halo model code and \texttt{axionCAMB} requires using an emulator for efficient computation. We employ the \texttt{axionEmu} emulator,\footnote{\href{https://github.com/keirkwame/axionEmu}{https://github.com/keirkwame/axionEmu}} which is based on a modified version of {\texttt{CosmoPower}}. 

\begin{figure}[t!]
\vspace{-0.7cm}\hspace{-0.6cm}\includegraphics[width=1.2\columnwidth]{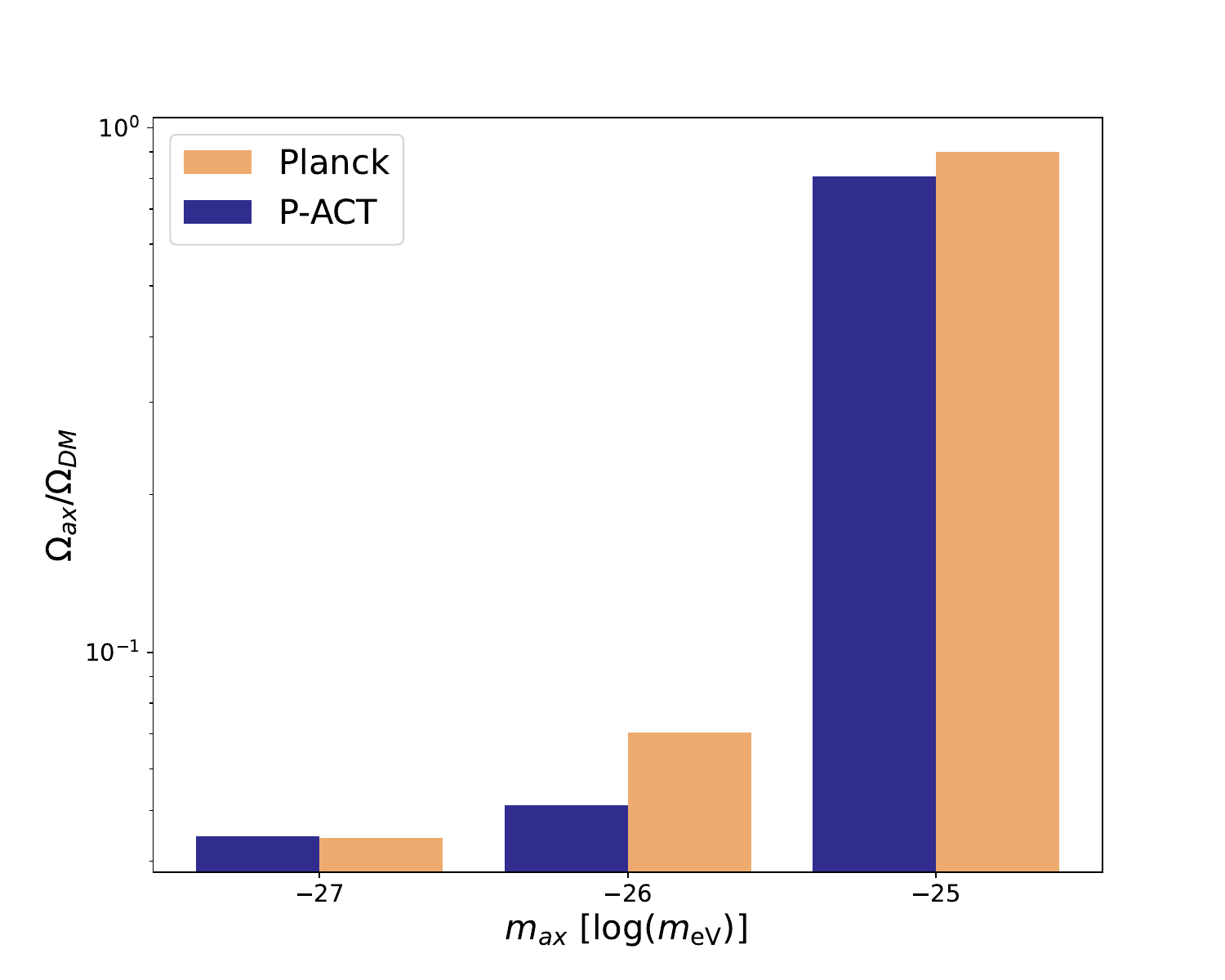}
    \vspace{-0.6cm}\caption{Constraints on the axion fraction using \pact\ (navy) compared to \Planck\ alone (orange). Combining ACT with \Planck\ significantly improves the constraints on the axion fraction compared to \Planck\ alone for the axion mass of $m_{\rm ax}=10^{-26}\,\mathrm{eV}.$ For lighter masses the axion starts to roll in its potential at later times, leading to changes on scales where \Planck\ has the majority of the constraining power. \label{fig:ULA}}
\end{figure}

We present our constraints on the allowed fraction of the dark matter density that is comprised of axions in Fig.~\ref{fig:ULA}. For the lightest axion species ${m_{\rm ax}\lesssim 10^{-28}~\mathrm{eV}}$ (not studied here, but see, e.g., \citealp{hlozek/etal:2015}) this represents a limit on how axions can constitute the dark energy.  For the ${m_{\rm ax}\gtrsim 10^{-27}~\mathrm{eV}}$ axions that we consider here, constraints on the axion fraction reflect a disentangling of axion effects from standard cold dark matter in large-scale clustering.

Using the primary CMB spectra from \Planck\ alone or \pact, we find
\begin{eqnarray}
m_{\rm ax} =10^{-26} \,\, \textrm{eV}: &\nonumber\\
\Omega_\mathrm{ax}/(\Omega_\mathrm{ax}+\Omega_c)
    &<&  0.070 \quad (95\%, \, \textsf{Planck})\nonumber\\
    &< & 0.052 \quad (95\%, \, \pact) .
\end{eqnarray} 
The upper limits for $m_{\rm ax} = 10^{-27}\,\, \mathrm{eV}$ are comparable between \Planck\ and \pact\ at $4.5\%$, while \pact\ reduces the allowed fraction of axion-like dark matter to roughly 5\% (95\% CL) for the mass $m_{\rm ax} =10^{-26}\, \mathrm{eV}$. For masses lighter than $m_{\rm ax} <10^{-26},$ the axion starts to roll in its potential at later times, leading to changes in the primary CMB spectra on scales where \Planck\ is most constraining. Some of the improvement in the constraints comes from the degeneracy between the axion density and the scalar spectral index. As the preferred value of $n_s$ slightly increases between \Planck\ and \pact, the limits on the axion density also tighten. The impact of increased sensitivity in the temperature and polarization power spectra is more pronounced for the CMB lensing deflection power spectrum, particularly for $m_{\rm ax} \geq 10^{-25}\, \mathrm{eV}$. A full presentation of axion constraints including the ACT DR6 lensing power spectrum, and detailed modeling of the nonlinear clustering that impacts the lensing deflection on small scales, will be provided in a future paper~\citep{Lague_inprep}.\\

\subsection{Dark matter}\label{sec:dm}
Many compelling dark matter (DM) models can be tested with cosmological probes, including scenarios where DM interacts with baryons or annihilates into photons as described in this section, and scenarios where DM interacts with dark radiation (\S\ref{sec:idr}) or with dark energy (\S\ref{sec:dedm}).

\subsubsection{DM-baryon interactions}\label{sec:dm-b}

Dark matter models beyond the standard paradigm of cold, collisionless DM (CDM) include scenarios in which DM interacts with baryons \citep{Snowmass_2013, battaglieri2017cosmic,  Gluscevic_2019,akerib2022snowmass2021}. As done in previous ACT analyses~\citep{Li_ACT_DMB}, we consider the specific case in which DM elastically scatters with protons \citep{Gluscevic_2018, Boddy_2018} through a contact interaction. These models were previously explored using both linear cosmology \citep{Boehm_2005,Sigurdson2004,Dvorkin2014,Gluscevic_2018, Boddy_2018, Boddy2018,Nguyen2021,Rogers_2022,Li_ACT_DMB,Slatyer_2018, Buen_Abad_2022, Hooper_2022, Becker_2021} and near-field cosmology \citep{Maamari_2021,Nadler_2019, Nadler_2021,Gluscevic_2019}. In a DM-baryon scattering scenario, the two corresponding cosmological fluids exchange heat and momentum, leading to collisional damping of small-scale perturbations in the early universe. The strength of the interaction is quantified with the momentum-transfer cross-section $\sigma_\mathrm{MT}=\sigma_0 (v/c)^n$, where $v$ is the relative particle velocity (sourced primarily by thermal velocities at the redshifts of interest), $n$ is the power-law index determined by the interaction theory, and the cross-section normalization $\sigma_0$ is a free parameter of the model.

We consider models with ${n=-4}$, 0, and 2 and fix the DM particle mass to $m_\chi=1$~MeV. We choose a sub-proton mass because direct detection experiments probe values of $m_\chi$ greater than 1~GeV, and constraints from $N_{\rm eff}$ exclude thermal-relic DM with masses lower than $\sim 1$~MeV~\citep{LEWIN199687,An_ACT_DM}. The choice of $m_\chi=1$~MeV allows us to explore the extreme end of what direct detection does not reach. We then choose to explore the set of $n$ values that correspond to the most salient DM-baryon scattering models: millicharged DM (${n=-4}$), a velocity-independent and spin-independent contact interaction (${n=0}$), and DM with a velocity-dependent contact interaction (${n=2}$) \citep{Dvorkin2014}. We use a modified version of {\tt class} \citep{Gluscevic_2018, Boddy2018} that incorporates the effect of DM-baryon scattering in the linear-theory Einstein-Boltzmann equations. We assume a broad, flat, positive prior on $\sigma_0$ for each value of $n$.  Each interaction model is thus a single-parameter extension of \lcdm, for a given $n$ and DM particle mass.

\begin{figure}[t!]
	\centering
\includegraphics[width=\columnwidth]{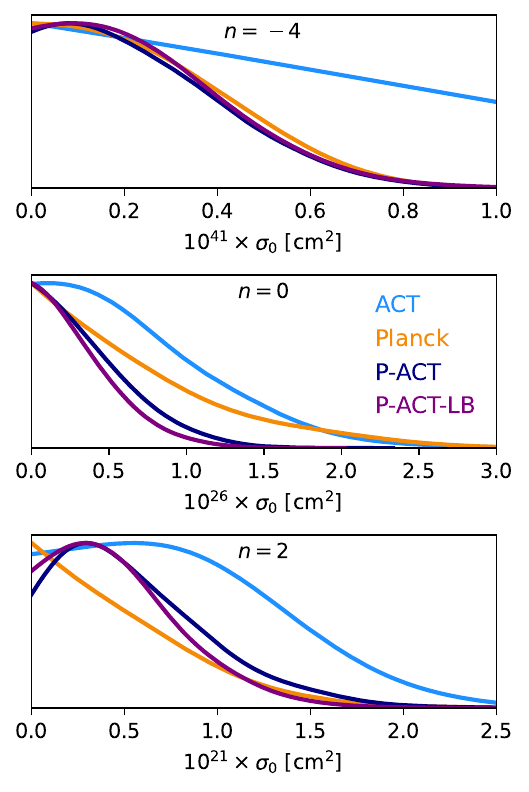}
    \vspace{-0.5cm}\caption{Constraints on the normalization of the momentum-transfer cross-section $\sigma_0$ for DM-proton elastic scattering are shown as 1D marginalized posterior probability distributions for three models: ${n=-4}$ (top), ${n=0}$ (middle), and ${n=2}$ (bottom). In the top and bottom cases, the addition of ACT does not alter the constraint from \Planck\ (compare navy and orange lines). In the case of $n=0$, ACT data (light blue) have similar constraining power to \Planck, leading to a factor of two improvement in the upper bound when the two data sets are analyzed jointly. In all cases, the addition of lensing and BAO data only marginally affects the bounds.}
	\label{fig:three_n_dm}
\end{figure}

We present constraints corresponding to the three interaction scenarios in Fig.~\ref{fig:three_n_dm}. For ${n=-4}$, we find that ACT data do not improve on \Planck\ constraints, giving $\sigma_0 ({n=-4}) < 6.1 \times 10^{-42}~\rm{cm}^2~(95\%, \pact)$; similarly, the addition of CMB lensing and BAO data has a negligible impact on the constraint. This is because the ${n=-4}$ model produces a scale-independent suppression at high $\ell$, and the high-multipole measurements from ACT do not add significantly to the signal-to-noise for this scenario.  
On the other hand, the $n=0$ model produces a strong $\ell$-dependent power suppression in both temperature and polarization~\citep[e.g.,][]{he2025boundsvelocitydependentdarkmatterbaryon}, and thus ACT alone has comparable constraining power to \Planck, leading to an upper bound of $\sigma_0 ({n=0}) < 1.7 \times 10^{-26}~\rm{cm}^2~(95\%, \act)$. A combined analysis of ACT and \Planck\ data leads to a factor of two improvement in the upper bounds,
\begin{eqnarray}
    \sigma_0 (n=0) & < & 9.9 \times 10^{-27} \,\, \rm{cm}^2 \ (95\%, \pact), \nonumber\\
    & < & 8.5 \times 10^{-27} \,\, \rm{cm}^2 \ (95\%, \pactlb). 
\end{eqnarray}
For ${n=2}$, ACT only marginally improves on \Planck\ constraints, giving $\sigma_0 ({n=2}) < 1.5 \times 10^{-21}~\rm{cm}^2$~(95\%, \pact). The addition of CMB lensing and BAO data again has a minimal impact, giving $\sigma_0 ({n=2}) < 1.2 \times 10^{-21}~\rm{cm}^2~(95\%, \pactlb)$. As ACT data are particularly sensitive to power suppression on small scales, the difference in constraining power for $n=0$ and $n=2$ stems from the baryon-loading-like effect occurring in the ${n=2}$ model, which leads to an enhancement of power at intermediate multipoles, rather than a strong suppression seen in the $n=0$ case~\citep{Boddy2018}.

The limits presented here are the tightest bounds on DM-baryon interactions obtained to date from linear cosmology. In particular, these bounds are consistent with and comparable to the results of joint analyses of \Planck, BOSS, and DES data \citep{he2023s8,he2025boundsvelocitydependentdarkmatterbaryon}. Meanwhile, the bounds obtained from highly nonlinear systems, such as the Milky Way satellite galaxy population, are stronger for all models that produce strong scale-dependent suppression (in this case, $n=0, 2$), as they lead to a notable decrement of small halos in the late universe \citep{Maamari_2021, Nadler_2019, 2024arXiv241003635N}. However, the non-linear modeling required for satellite analyses does not yet allow the exploration of the full cosmological parameter space, and the bounds obtained from near-field cosmology are thus not an apples-to-apples comparison to the present results. Moreover, mixed DM with a subcomponent interacting through velocity-independent ($n=0$) scattering was also found to yield a lower value of $S_8$, by suppressing the linear matter power spectrum on semi-nonlinear scales~\citep{he2023s8,he2025boundsvelocitydependentdarkmatterbaryon}. The ACT DR6 analysis results presented here are consistent with the preferred parameter space of interacting DM that reduces $S_8$ by a non-negligible amount, for the case where all DM interacts with protons. However, models that are most interesting in this context feature mixed DM and dedicated analyses of these scenarios are left for future work.

\subsubsection{DM annihilation} \label{sec:dm-ann}

If DM annihilates into Standard Model particles, energy released by the annihilation process is injected into the photon-baryon plasma around the time of recombination. This energy injection affects the CMB by altering the ionization history so as to broaden the width of the last-scattering surface and introducing a unique signature in matter clustering.  The main observational signatures in the CMB power spectra are: (i) damping of high-$\ell$ power in both temperature and polarization; (ii) enhancement of the polarization power spectrum at $\ell \lesssim 400$; (iii) shifts of the low-$\ell$ acoustic peaks in polarization (e.g.,~\citealp{2005PhRvD..72b3508P,PhysRevD.89.103508,2019JCAP...04..025G}).  The high-$\ell$ signature is degenerate with other parameters that alter the damping tail, such as $n_s$; thus, most of the constraining power beyond \Planck\ for this model is anticipated to come from large-scale polarization measurements.  Because of this, we do not expect major improvements in the constraints for this model from ACT DR6, given the scales probed by our data.

We follow earlier literature \citep{Finkbeiner_2012,PhysRevD.89.103508} and constrain the annihilation parameter
\begin{equation}
\label{eq.pann}
    p_\mathrm{ann} = f_\mathrm{eff} \frac{\langle \sigma v \rangle}{m_\chi}
\end{equation}
\noindent where $m_\chi$ is the DM particle mass, $\langle \sigma v \rangle$ is the thermally-averaged annihilation cross-section, and $f_\mathrm{eff}$ is the fraction of the energy due to annihilation that is injected into the plasma~\citep{Planck_2018_params}. The transferred energy fraction $f(z)$ is indeed redshift-dependent, but this redshift dependence only weakly impacts the CMB and can thus be approximated by a constant fraction $f(z) \approx f_\mathrm{eff}$ \citep{Slatyer_2016a,Slatyer_2016b}. Thus, $f_\mathrm{eff}$ corresponds to the energy transferred to the plasma around the redshifts at which the CMB is most sensitive to DM annihilation ($z\approx 500-600$) \citep{Finkbeiner_2012}. We assume that this energy from DM annihilation is transferred to the plasma immediately, known as the ``on-the-spot approximation" \citep{Planck_2018_params}.  We set a broad, flat prior on $p_{\rm ann}\in [0, 10^{-26}~{\rm cm^3/s/GeV}]$ and use the implementation of this model in the current standard version of \texttt{class}.

\begin{figure}[t!]
	\centering
    \includegraphics[width=\columnwidth]{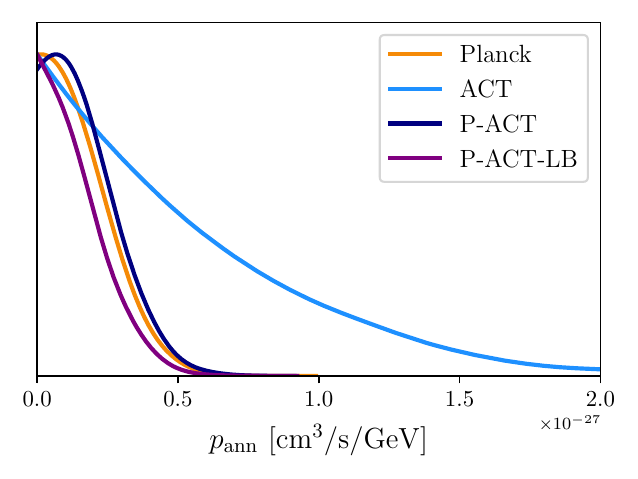}
    \vspace{-0.6cm}\caption{Constraints on DM annihilation (see Eq.~\eqref{eq.pann}). The posterior peak shifts slightly away from zero when combining ACT and \Planck\ data, leading to a slightly weaker upper limit than that from \Planck\ alone.  Including CMB lensing and DESI BAO data pushes the peak back to zero and yields an upper bound slightly tighter than that found with \Planck.}
	\label{fig:DMann}
\end{figure}

The results of this analysis are summarized in Fig.~\ref{fig:DMann}.  We find no evidence of DM annihilation from the new ACT DR6 spectra, with $p_{\rm{ann}} < 1.4 \times 10^{-27} \ \rm{cm^3/s/GeV} \ (95\%, \act)$.  When combining with \Planck, we find that the joint upper limit does not improve upon the \Planck-only constraint of $p_{\rm{ann}} < 3.9 \times 10^{-28} \ \rm{cm^3/s/GeV} \ (95\%)$,\footnote{We find a slightly weaker bound for \emph{Planck} alone than that in Eq.~(89a) of~\cite{Planck_2018_params}; this may be due to a slight shift of the central value arising from our use of {\tt Sroll2} for the low-$\ell$ EE data, which is particularly important in constraining this model.} but this is primarily due to the peak of the posterior for $p_{\rm ann}$ shifting slightly away from zero, as can be seen in Fig.~\ref{fig:DMann}.  The joint constraint gives $p_{\rm{ann}} < 4.1 \times 10^{-28} \ \rm{cm^3/s/GeV} \ (95\%, \pact)$.  The addition of CMB lensing and BAO data helps to tighten the constraints and pushes the peak of the posterior back to zero, yielding $p_{\rm{ann}} < 3.6 \times 10^{-28} \ \rm{cm^3/s/GeV} \ (95\%, \pactlb)$.  Further addition of SNIa data does not tighten this limit.  While the ACT DR6 data do not significantly improve constraints on $p_{\rm ann}$ beyond those from \Planck, future improvements from large-scale CMB polarization data are expected (see, e.g.,~\citealp{PhysRevD.89.103508,2019JCAP...04..025G}).

\subsection{Interacting dark radiation}\label{sec:idr}

\subsubsection{Self-interacting DR}\label{sec:si_dr}

A wide range of dark radiation (DR) models have been constructed, beyond the simple free-streaming case parameterized by $N_{\rm eff}$~\citep[e.g.,][]{Jeong_Takahashi_2013,Buen-Abad_2015,Cyr-Racine_2016,Lesgourgues_2016,Aloni_2022,Joseph_2023,Buen-Abad_2023,Rubira_2023,Schoneberg_2023,Zhou_Weiner_2024}.  These models generically involve self-interactions amongst the DR, interactions between the DR and (a subset of) the DM, or combinations thereof, potentially with non-trivial time-dependence (e.g., due to the temperature of the DR-DM sector falling below the mass of a massive mediator particle).  As a first step toward investigating these scenarios, we consider a simple model of (strongly) self-interacting DR (SIDR), for example due to a new gauge interaction in the dark sector.  At the background level, this model is identical to $N_{\rm eff}$, with a free parameter $N_{\rm idr} \geq 0$ describing the number of additional relativistic species (and hence the additional DR energy density).  However, the SIDR and free-streaming DR models differ at the perturbative level: the SIDR forms a perfect relativistic fluid with ${w = 1/3 = c_s^2}$, with interactions sufficiently strong that no anisotropic stress (or higher-order Boltzmann moments) can be supported.  Thus, the perturbative dynamics are characterized fully by the continuity and Euler equations.  Unlike free-streaming DR, SIDR can cluster on small scales, thus reducing the impact of Silk damping on the high-$\ell$ power spectra (at fixed DR energy density).  In addition, SIDR generates a smaller phase shift in the power spectra.  Thus, CMB fits to the SIDR model can accommodate larger amounts of DR than the free-streaming DR model, which can thus allow higher values of $H_0$~\citep[e.g.,][]{Aloni_2022,H0_olympics_2022,Allali_2024}.

We consider a simple SIDR model parameterized solely by $N_{\rm idr}$, with $N_{\rm idr} \geq 0$.  We assume no interactions between the SIDR and the DM.  We use the implementation of this model in {\tt class}\footnote{We set {\tt idm\_dr\_tight\_coupling\_trigger\_tau\_c\_over\_tau\_h = 0.005} and {\tt idm\_dr\_tight\_coupling\_trigger\_tau\_c\_over\_tau\_k = 0.008} to ensure high accuracy, identical to the values we use for the analogous {\tt class} accuracy parameters in the SM sector (see Appendix~\ref{app:theory}).} and we adopt a flat, uninformative prior $N_{\rm idr} \in [0,6]$.  We assume that the SM neutrino sector consists of one massive state carrying 0.06 eV and two massless states, with $N_{\rm eff} = 3.044$.

\begin{figure}[t!]
	\centering
    \includegraphics[width=\columnwidth]{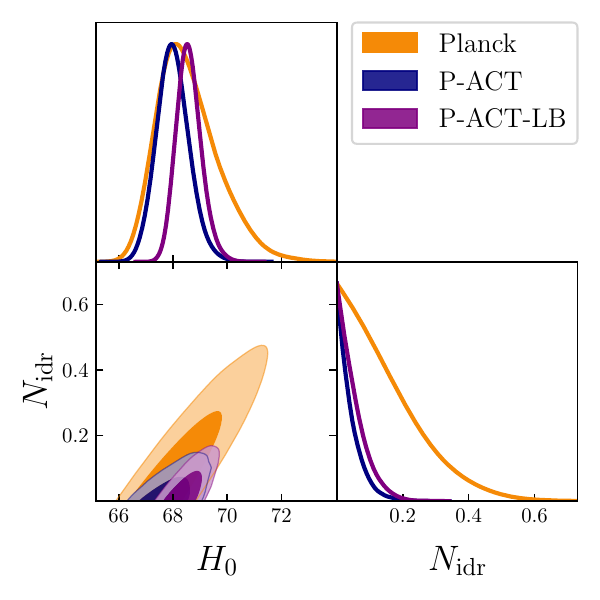}
    \vspace{-0.6cm}
    \caption{Constraints on the number of strongly self-interacting dark relativistic species, $N_{\rm idr}$. The addition of ACT DR6 spectra improves the constraint from \Planck\ by more than a factor of three (navy versus orange) and notably disfavors values of $H_0$ above 70~km/s/Mpc that are allowed by \Planck\ alone.  Inclusion of CMB lensing and DESI BAO data (purple) slightly weakens the SIDR upper limit due to small shifts in the best-fit model parameters, but nevertheless further tightens the $H_0$ posterior.  These are the tightest bounds on SIDR obtained to date.
    }
	\label{fig:SIDR}
\end{figure}

Constraints on this model are presented in Fig.~\ref{fig:SIDR}. 
From the ACT DR6 primary CMB power spectra alone, we find
\begin{eqnarray}
    N_{\rm idr} & < & 0.339 \quad \quad \quad \  (95\%, \, \act) \nonumber\\
    H_0 & = & 66.85^{+0.85}_{-1.0} \,\, \quad (68\%, \, \act).
\end{eqnarray}
Including \emph{Planck} primary CMB power spectra, we obtain
\begin{eqnarray}
    N_{\rm idr} & < & 0.114 \quad \quad \quad \ (95\%, \, \pact) \nonumber\\
    H_0 & = & 68.00^{+0.55}_{-0.63} \, \quad (68\%, \, \pact).
\end{eqnarray}
The joint \pact\ constraint represents more than a factor-of-three improvement over the constraint from \emph{Planck} alone, for which we find $N_{\rm idr} < 0.379$ at 95\% confidence.
The origin of this tight constraint is the lack of preference for excess high-$\ell$ damping in the DR6 power spectra; since SIDR can only increase damping (due to the physical bound $N_{\rm idr} > 0$), its existence is thus disfavored by the data.  From the primary CMB, we thus see no evidence of additional SIDR species, extending our results for the free-streaming DR case ($\neff$) to a very different physical regime.

Including additional low-redshift datasets does not significantly tighten the bound on $N_{\rm idr}$, but further tightens the error bar on $H_0$ and other cosmological parameters.  From the combination of ACT and \emph{Planck} CMB spectra, ACT and \emph{Planck} CMB lensing, and DESI BAO, we obtain
\begin{eqnarray}
    N_{\rm idr} & < & 0.134 \quad \quad \ \ (95\%, \, \pactlb) \nonumber\\
    H_0 & = & 68.59^{+0.41}_{-0.50} \, \ (68\%, \, \pactlb).
\end{eqnarray}
The upper limit on $N_{\rm idr}$ weakens slightly compared to that found for \pact\ above, due to small shifts in the best-fit parameter values.  Further including SNIa data in the analysis yields negligible changes to these constraints. 

The MAP SIDR model found for \pact, as well as that found for \pactlb, yields negligible improvement in the quality of fit over $\Lambda$CDM  --- in fact, we find that the MAP SIDR model lies at the \lcdm\ limit of the parameter space in both cases, i.e., with $N_{\rm idr} = 0$, indicating no preference for this model extension.  We emphasize again that this lack of preference arises from the lack of excess damping seen at high-$\ell$ in ACT: because SIDR can only increase damping over that in \lcdm, the model cannot accommodate decreased damping (unlike $N_{\rm eff}$, which can do so via a value of $N_{\rm eff} < 3.044$).

These are the tightest bounds on SIDR presented to date.  The ability of this model to increase the value of $H_0$ inferred from cosmological data is strongly limited by these observations, with the new ACT DR6 spectra playing a crucial role in significantly tightening the constraints compared to those from \emph{Planck}.

\subsubsection{Interacting DR-DM}\label{sec:i_dr_dm}

A wide range of dark sector models have been considered in the literature, featuring various types of interactions between dark radiation and dark matter species~\citep[e.g.,][]{Cyr-Racine_2016,Buen-Abad_2015,Lesgourgues_2016}.  As a first step in probing this space of models, here we consider the fiducial scenario studied in~\citet{Rubira_2023}, in which an SIDR fluid (identical to that studied above) interacts with all of the DM via an interaction with a momentum transfer rate $\Gamma(a) = \Gamma_{\rm 0,nadm} a^{-2}$ between the DM and the DR~\citep{Buen-Abad2018}.  Microphysically, such an interaction can be realized in dark-sector models featuring non-Abelian gauge bosons (hence the ``nadm'' subscript) that mediate interactions between Dirac fermion DM particles~\citep{Buen-Abad_2015,Lesgourgues_2016}.  In this particular model, $\Gamma$ has a redshift (or temperature) dependence that matches that of the Hubble rate during radiation domination --- see \cite{Buen-Abad_2015} for an explicit calculation of the momentum transfer rate in the ``nadm'' model. As a consequence, the interaction between the DR and the DM is relevant while modes covering a broad range of scales enter the horizon (essentially during all of radiation domination), which leads to a roughly scale-invariant suppression of power for such modes.  This makes the model a promising candidate to decrease the CMB-inferred value of $S_8$~\citep{Buen-Abad_2015,Rubira_2023}. The SIDR abundance (set by $N_{\rm idr}$) determines the scale where power is suppressed, while the interaction strength $\Gamma_{\rm 0,nadm}$ determines the amount of suppression~\citep{Buen-Abad_2015}.\footnote{Note that the IDR-IDM interaction studied here is not relevant in the late universe, and thus late-time growth is not affected; modifications to structure formation are captured by the change in the linear matter power spectrum.}

\begin{figure}[t!]
	\centering
    \includegraphics[width=\columnwidth]{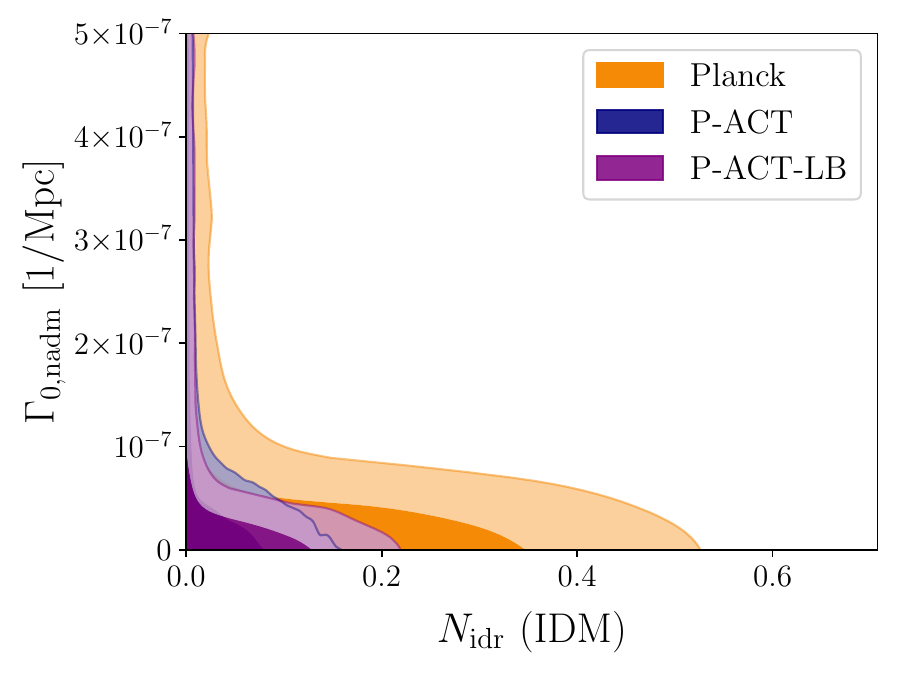}
	\vspace{-0.5cm}\caption{Constraints on the IDR-IDM model.  Here the interaction strength $\Gamma_{\rm 0,nadm}$ is given in 1/Mpc.  The inclusion of ACT DR6 spectra (navy) significantly improves the constraints from \Planck\ alone (orange). Inclusion of CMB lensing and DESI BAO data (purple) slightly weakens the $N_{\rm idr}$ upper limit due to small shifts in the best-fit model parameters, but further tightens the constraint on $\Gamma_{\rm 0,nadm}$.  The notation $N_{\rm idr}$ (IDM) indicates that the IDR in the model constrained here is interacting with the DM, unlike that in Fig.~\ref{fig:SIDR}.
    }\label{fig:SIDR_IDM}
\end{figure}

We use the implementation of this interacting DR -- interacting DM (IDR-IDM) model in {\tt class}~\citep{Lesgourgues_2016,Archidiacono2019}.  The model contains two additional free parameters beyond those of base \lcdm: $N_{\rm idr}$ (as in the SIDR model above) and $\Gamma_{\rm 0,nadm}$.  We adopt the same priors and model assumptions as for the SIDR scenario for all common parameters, while for $\Gamma_{\rm 0,nadm}$, we adopt an uninformative, uniform prior $\Gamma_{\rm 0,nadm} / ({\rm Mpc}^{-1}) \in \left[0.0, 5.0 \times 10^{-7}\right]$.

We find no preference for the IDR-IDM scenario in the ACT DR6 spectra.  Constraints on this model are shown in Fig.~\ref{fig:SIDR_IDM}. 
Joint analysis of ACT and \Planck\ yields
\begin{equation}
\left.
 \begin{aligned}
N_{\rm idr} &< 0.0977 \\ 
\Gamma_{\rm 0,nadm} & < 4.45 \times 10^{-7}~{\rm Mpc}^{-1}
\end{aligned}
\quad \right\} \mbox{\textrm{\; (95\%, \pact).}}
\end{equation}
This is a significant improvement over analogous results from \Planck\ alone, for which we find $N_{\rm idr} < 0.367$ at 95\% CL~and no bound on $\Gamma_{\rm 0,nadm}$ (at 95\% CL).  We find negligible reduction in $S_8$ in the \pact\ fit to the IDR-IDM model, with $S_8 = 0.814^{+0.023}_{-0.015}$ at 68\% CL.  Inclusion of CMB lensing and DESI BAO data in the combined dataset yields:
\begin{equation}
\left.
 \begin{aligned}
N_{\rm idr} &< 0.135 \\ 
\Gamma_{\rm 0,nadm} & < 4.09 \times 10^{-7}~{\rm Mpc}^{-1}
\end{aligned}
\quad \right\} \mbox{\textrm{\; (95\%, \pactlb).}}
\end{equation}
The bound on $S_8$ tightens considerably compared to \pact, with $S_8 = 0.808^{+0.013}_{-0.0095}$ at 68\% CL, further limiting this model's ability to yield low values of this parameter.

We note that Bayesian constraints on the IDR-IDM model can suffer from volume effects, since $\Gamma_{\rm 0,nadm}$ becomes degenerate in the $N_{\rm idr} \rightarrow 0$ limit.  We thus confirm that the best-fit IDR-IDM model to \pact\ or \pactlb\ does not yield a significantly better fit to these data than \lcdm.  In fact, as found for SIDR, we find that the MAP model in the IDR-IDM scenario for both \pact\ and \pactlb\ is indistinguishable from \lcdm\ within our numerical precision. Overall, we see no evidence of extended dark-sector physics in this analysis.\\

\section{Gravity and late-time physics}\label{sec:late-time}

\subsection{Geometry: spatial curvature}\label{sec:curv} 

\begin{figure}[t!]
	\centering
\includegraphics[width=\columnwidth]{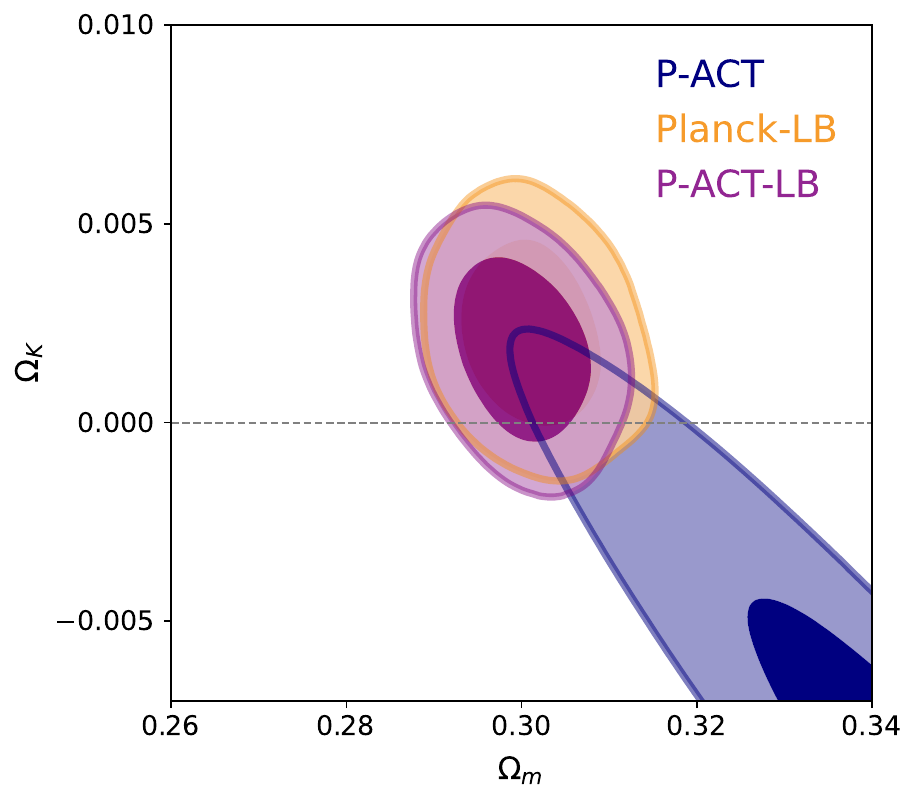}
	\vspace{-0.6cm}\caption{Curvature can be measured from the CMB alone due to lensing effects in the primary CMB; the new ACT DR6 spectra are consistent with a flat universe (dashed line) both alone and in combination with \Planck, as described in L25 and as shown by the navy contour. The \pactlb\ combination (purple) tightens the constraint around zero curvature, further breaking the degeneracy with the matter density and improving on \plb\ (orange).}
	\label{fig:omk}
\end{figure}

In the inflationary scenario, at the end of inflation the universe is predicted to emerge in a state very close to spatially flat~\citep{LINDE1982389,PhysRevLett.48.1220}. This hypothesis is tested empirically by allowing the curvature density parameter, $\Omega_{k}$, which is fixed to zero in the baseline \lcdm\ model, to be a free parameter.  As discussed in L25, the new ACT DR6 power spectra alone prefer a flat geometry, breaking degeneracies via lensing effects in the power spectra; in a joint fit with \Planck, ACT DR6 moves the \Planck\ contours toward vanishing curvature. When adding BAO and CMB lensing data to the primary CMB, the degeneracy with the matter density is more effectively broken, as shown in Fig.~\ref{fig:omk}, and the constraints tighten significantly, yielding 
\begin{equation}
    \Omega_{k} = 0.0019 \pm 0.0015\quad (68\%, \pactlb),
\end{equation}
similar to $\Omega_{k}=0.0022 \pm 0.0015$ (68\%, \plb). These limits derived from the combination of CMB and low-redshift data tightly constrain the universe's spatial geometry, and are fully consistent with spatial flatness.

This result can be converted into a new limit on the present-day radius of spatial curvature of the universe, $R_{ k}=c/H_0/\sqrt(|\Omega_{\rm tot}-1|)$ where $\Omega_{\rm tot}$ is the total cosmic energy density, such that $\Omega_{k} = 1-\Omega_{\rm tot}$.  With \pactlb\ we find at 95\% confidence $R_{k}>105$~Gpc or equivalently $R_{k}>343$ billion light-years for a closed universe ($\Omega_{\rm tot}>1$ or $\Omega_{k}<0$) and $R_{k}>66$~Gpc or $R_{k}>215$ billion light-years for an open universe ($\Omega_{\rm tot}<1$ or $\Omega_{k}>0$). 

\subsection{Late-time dark energy}\label{sec:de} 

Within the \lcdm\ model, the dark energy density is constant over time, described by a cosmological constant $\Lambda$ with an equation of state $w\equiv -1$ at all times. A common way of exploring the dark energy component is to relax the assumption on its equation of state, considering a time evolution of the pressure-density ratio $P/\rho = w(a)$ with~\citep{Chevallier2001,Linder_w}\footnote{See, e.g.,~\cite{Shlivko:2024llw} for consideration of time-evolving dark energy beyond the $(w_0,w_a)$ parametrization.}
\begin{equation}
    w(a) = w_0 + w_a(1 - a).
\end{equation}
This is equivalent to a cosmological constant $\Lambda$ when setting today's value of the equation of state to ${w_0=-1}$ and its time dependence to ${w_a=0}$. 
This expansion can then be used for explorations varying either only $w_0$ or both $w_0$ and $w_a$, as we do below. Opening these degrees of freedom in the dark energy component causes strong degeneracies between parameters when using only CMB data and therefore we mostly report results for combinations with CMB lensing and external probes of the expansion history, including both BAO and SNIa. In particular, as shown in~\cite{DESI-BAO-VI} in fits to models with additional degrees of freedom in the dark energy equation of state, SNIa are essential to break degeneracies with $H_0$ and the matter density. We do not expect significant improvements from the new ACT power spectra for late-time dark energy, but we report constraints here given interest in this model following the DESI Year-1 results.

When $w\equiv w_0$ is free to vary as a single-parameter extension to \lcdm\ (without time dependence --- ${w_a=0}$ --- and imposing a flat prior $w_0{\in [-3, 1]}$)\footnote{To get theory predictions for this model we use \texttt{camb} with the Parametrized Post-Friedmann approach~\citep{Fang2008} to compute dark energy perturbations (\texttt{dark\_energy\_model=DarkEnergyPPF}).}, we obtain the contours shown in Fig.~\ref{fig:lcdm_single_extensions_pactlb} for \pactlb. Adding also SNIa data, we find  
\begin{equation}
    w_0 = -0.986 \pm 0.025 \; (68\%, \pactlbs),
\end{equation}
in good agreement with the \lcdm\ expectation, and with the same error bar as previous measurements using CMB from \Planck\ ($w_0 = -0.997 \pm 0.025$, \citealp{DESI-BAO-VI}).

\begin{figure}[t!]
    \centering
\includegraphics[width=\columnwidth]{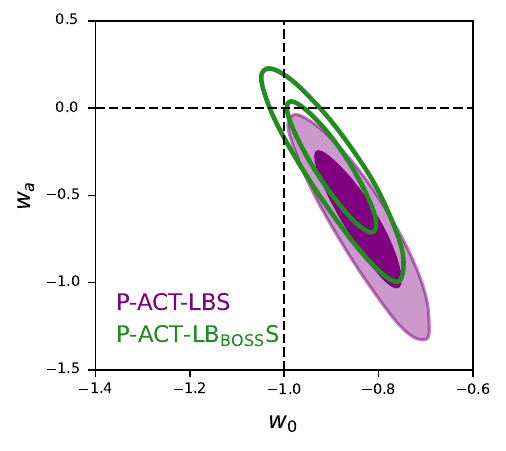}
    \vspace{-0.5cm}
    \caption{Constraints on the dark energy equation of state parameters, 
    varying both today's value, $w_0$,
    and its time variation, $w_a$. Similar to other studies, we find that DESI drives a preference for time-varying dark energy (compared to the dashed \lcdm\ line), which is relaxed when considering BOSS BAO instead (green contours). The CMB contribution to this measurement is sub-dominant, apart from breaking parameter degeneracies, with \textsf{Planck}, \wact, and \pact\ giving similar results.}
    \label{fig:wcdm_2d}
\end{figure}

When we expand the model further, with both $w_0$ and $w_a$ varying (with a flat prior $w_a \in [-3, 2]$), we find
\begin{equation}
\left.
 \begin{aligned}
    w_0 &= -0.837 \pm 0.061\\
    w_a &= -0.66^{+0.27}_{-0.24}
\end{aligned}
\quad \right\} \mbox{\textrm{\; (68\%, \pactlbs),}}
\end{equation}
with marginalized posteriors shown in Fig.~\ref{fig:wcdm_2d}. 
We see a moderate preference for the $w_0w_a\mathrm{CDM}$ model over \lcdm, similar to what was reported in~\cite{DESI-BAO-VI} using CMB data from \Planck\ ($w_0 = -0.827 \pm 0.063$, $w_a = -0.75^{+0.29}_{-0.25}$).\footnote{Quantitatively, comparing the MAP $w_0 w_a$CDM and \lcdm\  models, we find $\Delta \chi^2 = -7.0$ ($2.2\sigma$) for \pactlbs, while~\cite{DESI-BAO-VI} find $\Delta \chi^2 = -8.7$ ($2.5\sigma$) for \plbs.} This result is driven partially by the DESI data and partially by the SNIa data. Substituting the BAO dataset with BOSS BAO, the posteriors are consistent with the cosmological constant scenario. The specific CMB dataset used for this analysis has essentially no impact: \Planck\ alone, \pact, or \wact\ yield similar results.

With the combined \pactlbs\ dataset, we find a well-constrained value of $H_0=67.73\pm 0.68$~km/s/Mpc and $H_0=68.02\pm 0.72$~km/s/Mpc for the $w_0$ and ${w_0-w_a}$ extensions, respectively. Similarly to~\cite{DESI-BAO-VI}, we also find that SNIa reduce by a factor of $\sim 2$ the uncertainty on $H_0$ and shift the mean to lower values by $\sim 3$~km/s/Mpc (see Fig.~\ref{fig:lcdm_single_extensions_pactlb}).

\subsection{Interacting DE-DM}\label{sec:dedm}

An extension of $\Lambda$CDM with kinetic interactions between the DM and DE fluids was initially considered in~\cite{Simpson2010ScatteringOf}. This model introduces a coupling constant, $\Gamma_{\rm DMDE}$, describing pure momentum transfer between the DM and DE.  This coupling leaves the homogeneous background cosmological evolution unchanged, but alters the dark sector perturbation evolution equations as follows~\citep{Simpson2010ScatteringOf}:
\begin{align}
    \delta_{\mathrm{DM}}^{\prime} &= -\theta_{\mathrm{DM}} + 3\phi^{\prime},\label{eq:DM-dens-deriv}\\
    \delta_{\rm DE}^{\prime} & =-\left[(1+w)+9 \frac{\mathcal{H}^2}{k^2}\left(1-w^2\right)\right] \theta_{\rm DE}\nonumber \\ &\;\;\;\;+3(1+w) \phi^{\prime}-3 \mathcal{H}(1-w) \delta_{\rm DE}, \label{eq:DE-dens-deriv}\\
    \theta_{\mathrm{DM}}^{\prime} &= -\mathcal{H} \theta_{\mathrm{DM}}+k^2 \phi+\frac{a \Gamma_{\rm DMDE}}{\bar{\rho}_{\rm DM}}\Delta \theta, \label{eq:DM-vel-deriv}\\
    \theta_{\mathrm{DE}}^{\prime}= & 2 \mathcal{H} \theta_{\mathrm{DE}}+\frac{k^2}{\left(1+w\right)} \delta_{\mathrm{DE}} +k^2 \phi-\frac{a\Gamma_{\mathrm{DMDE}}}{(1+w)\bar{\rho}_{\rm DE}} \Delta \theta \label{eq:DE-vel-deriv},
\end{align}
where $w$ is the DE equation of state, $\delta\equiv\rho/\bar{\rho}-1$ is the energy density contrast, $\theta\equiv \nabla\cdot \mathbf{v}$ is the velocity divergence, $\Delta \theta \equiv \theta_{\rm DE}-\theta_{\rm DM}$,  $\mathcal H$ is the conformal Hubble factor ($\mathcal{H}=aH$), $\phi$ is the Newtonian gravitational potential, and ${}^\prime$ denotes differentiation with respect to conformal time.

This type of interaction leads to slower structure growth, and this idea was revisited in~\cite{Beltran2021ProbingElastic,Poulin2023Sigma-8,Beltran2025OnEvidence} as a potential way to reduce the value of $S_8$ inferred from CMB and LSS data. The coupling strength needed to appreciably reduce $S_8$ is around $\Gamma_{\rm DMDE}/(H_0 \rho_c) \sim 1$ (where $\rho_c$ is the critical density). While previous work assumed linear theory, \citet{Lague2024ConstraintsDMDE} developed a halo-model approach to characterize the behavior of DM with non-zero $\Gamma_{\rm DMDE}$ on semi-linear scales. This model is an adapted version of \texttt{HMcode} (still assuming dark matter only) and accounts for changes in halo formation resulting from DE-DM interactions using modified fitting functions for the critical and virial overdensities. A combination of CMB lensing from ACT DR6 and \Planck, BAO from SDSS/BOSS, and CMB primary anisotropy measurements from \textit{Planck} constrains the contribution of $\Gamma_{\rm DMDE}/H_0 \rho_c<2.76$ at the 95\% CL using this prescription~\citep{Lague2024ConstraintsDMDE}.

Here, we enlarge the dataset to include the new ACT DR6 primary CMB power spectra, and we replace the BAO likelihood with that of DESI.\footnote{We note here that the interacting DE-DM model is the only case in this paper for which we use the \emph{extended} setting for the ACT DR6 + \Planck\ CMB lensing data. To constrain $\Gamma_{\rm DMDE}$, we make use of the full CMB lensing likelihood up to $L=1250$~\citep{Madhavacheril_dr6_lensing,Qu_dr6_lensing}.} We study two configurations for this model. We first fix the dark energy equation of state to three different values of $w\in \{-0.999, -0.975, -0.95\}$, allowing $\Gamma_{\rm DMDE}$ to vary (the model is thus a two-parameter extension of $\Lambda$CDM with only one additional degree of freedom). We avoid $w=-1$ due to a divergence that arises in Eq.~\eqref{eq:DE-vel-deriv} in this case. We then also jointly sample both parameters, but we find that Metropolis-Hastings sampling performs poorly when varying both $w$ and $\Gamma_{\rm DMDE}$. This occurs because $w$ can take a very wide range of values when $\Gamma_{\rm DMDE}$ goes to $0$.\footnote{In this case, due to the computational expense in this model, we stop the chains at a Gelman-Rubin statistic of $R-1< 0.05$; thus, slight variations in the final bounds are expected.} We adopt a flat prior $\Gamma_{\rm DMDE}/(H_0 \rho_c) \in [10^{-8},50]$.  We compute theoretical predictions for this scenario using the modified version of {\tt class} from~\cite{Lague2024ConstraintsDMDE}.\footnote{\url{https://github.com/fmccarthy/Class_DMDE}}

\begin{figure}
    \centering
    \includegraphics[width=\linewidth]{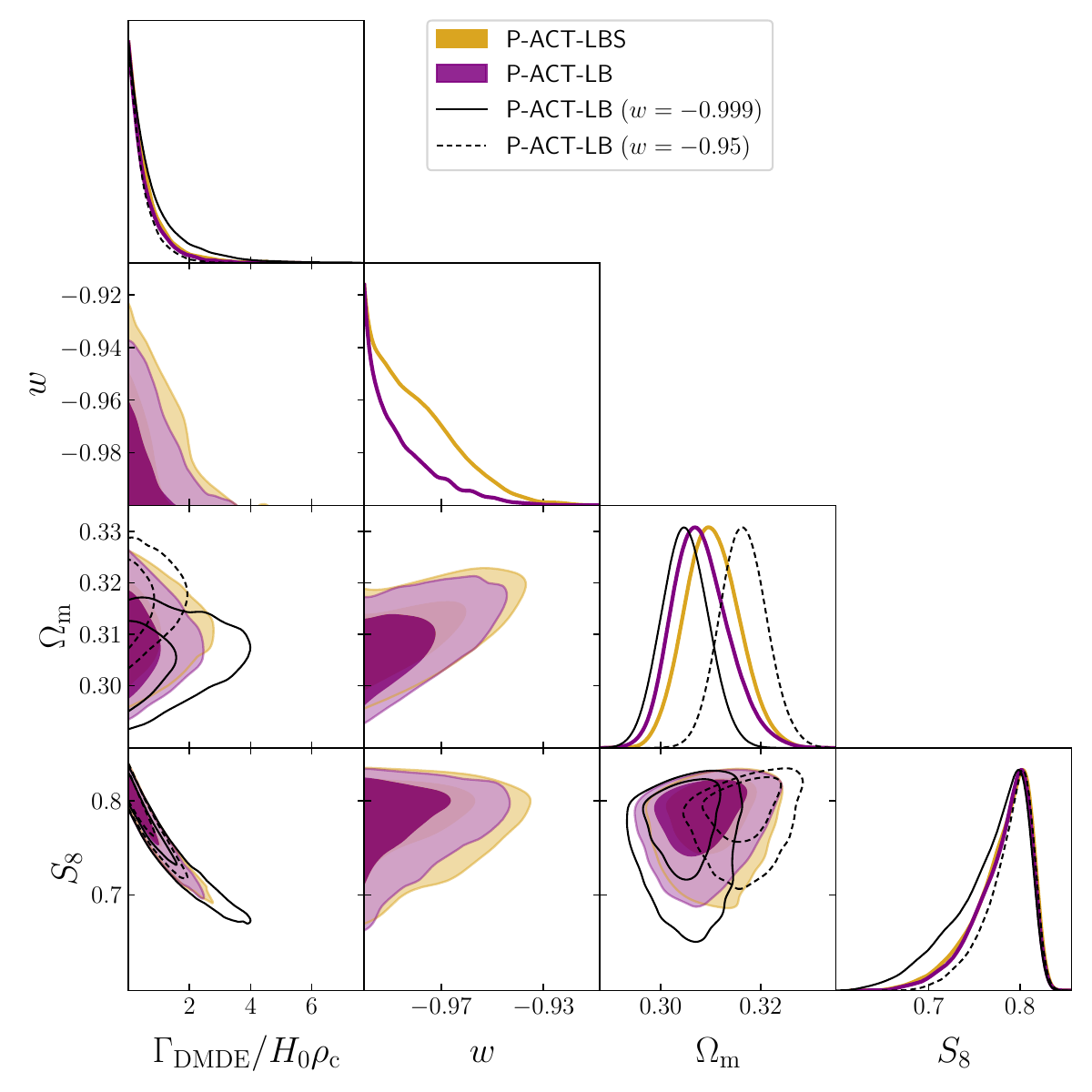}
    \caption{Constraints on the strength of the DM-DE interaction rate for the data combinations \textsf{P-ACT-LB} and \textsf{P-ACT-LBS}, for varying equation of state (filled contours) and two fixed values of $w$ (empty contours and black lines). The preferred parameters are in agreement with the \lcdm\ values of $w=-1$ and $\Gamma_{\rm DMDE} = 0$ (no coupling between DM and DE). }
    \label{fig:2w-Gamma-dmde}
\end{figure}

We show the parameter posterior distributions for a subset of the runs in Fig.~\ref{fig:2w-Gamma-dmde}. Given the large degeneracies present in this model when using only CMB data, we focus on results for the data combinations \pactlb\ and \pactlbs. For fixed values of $w$, we find that the limit on $\Gamma_\mathrm{DMDE}/(H_0\rho_\mathrm{c})$ gets tighter moving away from the $w=-1$ boundary: $\Gamma_\mathrm{DMDE}/(H_0\rho_\mathrm{c})_{w=-0.999} < 3.10$, $\Gamma_\mathrm{DMDE}/(H_0\rho_\mathrm{c})_{w=-0.975} < 1.64$, and $\Gamma_\mathrm{DMDE}/(H_0\rho_\mathrm{c})_{w=-0.95} < 1.64$ (all at 95\% CL from \pactlb). Similarly, the $S_8$ limit gets tighter, with the low-end tail of the distribution being more excluded (see Fig.~\ref{fig:2w-Gamma-dmde}).
When varying $w$ and $\Gamma_\mathrm{DMDE}/(H_0\rho_\mathrm{c})$ simultaneously, we find
\begin{align}
    &\Gamma_\mathrm{DMDE}/(H_0\rho_\mathrm{c}) < 1.9 \; (95\%, \pactlb), \nonumber\\
    &S_8 =0.781^{+0.04}_{-0.02} \quad \quad \quad (68\%, \pactlb),
\end{align}
and when including SNIa data from Pantheon+
\begin{align}
    &\Gamma_\mathrm{DMDE}/(H_0\rho_\mathrm{c}) < 2.1 \; (95\%, \pactlbs), \nonumber\\
    &S_8 =0.780^{+0.04}_{-0.02} \quad \quad \quad (68\%, \pactlbs).
\end{align}

We also perform the same analyses with BOSS BAO data used in place of DESI, finding similar constraints on $S_8$ (with error bars $10$-$30$\%~larger) and somewhat weaker bounds on $\Gamma_{\rm DMDE}$ (roughly $20$-$50$\% larger upper limits). We see no preference for this model over $\Lambda$CDM for any dataset combination considered here, with the best-fit model lying at the $\Gamma_{\rm DMDE} = 0$ edge of the parameter space (note that there are no projection effects in the Bayesian posteriors in the cases where we fix $w$ to specific values).  The inclusion of the ACT DR6 and DESI data tightens the upper limit on $\Gamma_{\rm DMDE}$ by 33\% compared to the bounds from~\cite{Lague2024ConstraintsDMDE} when varying both $w$ and $\Gamma_{\rm DMDE}$ using the \pactlb\ dataset. The improvement in constraining power comes mainly from using the DESI BAO dataset instead of BOSS and the improved bounds on $\Omega_{m}$ from the primary CMB in ACT DR6. The addition of SNIa data from Pantheon+ does not improve our bounds.

As observed in previous work, the coupling between the DM and DE leads to reduced values of $S_8$, significantly broadening the marginalized posterior for this parameter. When $-w$ approaches unity, the effects of the coupling are delayed to lower redshifts.  This makes CMB lensing, which is sensitive to redshifts around $z\sim 1-2$, less constraining for this model. For this reason, we find a more extended posterior for $\Gamma_{\rm DMDE}/(H_0\rho_{\rm{c}})$ when $w$ is near $-1$, as shown in Fig.~\ref{fig:2w-Gamma-dmde}. In general, we find that this model allows values of $S_8$ from \pactlb\ that are consistent with values found in galaxy weak lensing surveys, but we caution that a joint analysis with such datasets would require a full model for the nonlinear matter power spectrum in this scenario.\\

\subsection{Modified gravity}\label{sec:mg}

The growth rate of large-scale structure $f(a) \equiv d\ln D(a)/d\ln a$, where $D(a)$ is the linear growth factor and $a$ is the scale factor, can be accurately approximated in the matter- or dark energy-dominated eras as 
\begin{equation}
\label{eq.gamma}
    f(a) = \Omega_m^{\gamma}(a) \,,
\end{equation}
where $\gamma$ is the growth index, which takes a canonical value of 0.55 in general relativity (GR) with a flat $\Lambda$CDM cosmology~\citep{FRY1985211, Wang_1998, Linder_2005}. This simple model with values of $\gamma$ differing from 0.55 is part of a broader class of modified gravity theories, which generically predict modifications to the growth of structure. Various studies have considered and constrained different prescriptions for the growth and geometry~\citep[e.g.,][]{Wang:2007fsa, Guzzo:2008ac, Dossett:2010gq, Hudson2012, Rapetti:2012bu, Pouri:2014nta, Ruiz:2014hma, Bernal:2015zom, Johnson:2015aaa, Alam:2015rsa, Moresco:2017hwt, Basilakos:2019hlb, DES:2020iqt, Garcia-Garcia:2021unp, Ruiz-Zapatero:2021rzl, Ruiz-Zapatero:2022zpx, White:2021yvw, Andrade:2021njl, Chen:2022jzq, DES:2022ccp}. Some studies have found evidence for suppression of the growth of structure, while others have not. One of the most sensitive datasets for such theories is the combination $f\sigma_8(z)$ (the product of $f(z)$ and $\sigma_8(z)$) which is constrained by various redshift-space distortion (RSD) and peculiar velocity surveys.  RSD surveys measure the growth rate of structure via the imprint of peculiar velocities on the quadrupole of redshift-space galaxy clustering measurements~\citep[e.g.,][]{Kaiser1987,Hamilton1998,Percival2009,Song2009}. Surveys of the peculiar velocities of individual galaxies can be performed at low-to-moderate redshifts, when an estimate of the absolute distance to each galaxy is available (e.g., from Cepheid, TRGB, or SNIa observations), thus allowing the cosmological contribution to the observed redshift to be isolated from the peculiar-velocity contribution~\citep[e.g.,][]{Miller1992,Riess1998,Turnbull2012,Tully2016,Stahl2021}. CMB data also contribute constraining power via lensing effects~\citep[e.g.,][]{2009PhRvD..80j3516C}.

A recent analysis by \cite{Nguyen:2023fip} found moderate evidence of a potential deviation from the GR prediction for $\gamma$.  They considered measurements of peculiar velocities from \cite{Boruah:2019icj}, \cite{Huterer:2016uyq}, \cite{Turner:2022mla}, and \cite{Said:2020epb}, and measurements of RSD from \cite{Beutler2012}, \cite{Howlett:2014opa}, \cite{Blake2012}, \cite{Blake:2013nif}, \cite{Pezzotta:2016gbo}, \cite{eBOSS-LRG}, and \cite{Okumura:2015lvp}. Combining these with data from \emph{Planck}, they found evidence for suppression of the growth of structure at $3.7\sigma$ significance, with $\gamma=0.639^{+0.024}_{-0.025}$. This suppression both significantly reduced $S_8$ and eliminated \emph{Planck}'s (mild) preference for positive spatial curvature.

Here, we study the impact of including the new ACT DR6 spectra on these results, following the approach of \cite{Nguyen:2023fip}. Since Eq.~\eqref{eq.gamma} is only valid for sub-horizon perturbations, this approach leaves the unlensed primary CMB perturbations unchanged, but modifies the CMB lensing potential and other late-time observables self-consistently.  We use all measurements considered by \cite{Nguyen:2023fip} to form a $f\sigma_8$ likelihood, consisting of a simple Gaussian likelihood for each survey's constraint on $f\sigma_8$. DESI also recently released $f\sigma_8$ measurements~\citep{2024desi} which we do not include here to facilitate comparison with \cite{Nguyen:2023fip}, but we discuss them in~\S\ref{subsec:consistency}. In combination with ACT and other datasets, we use the $f\sigma_8$ likelihood to constrain the growth index $\gamma$ as a single-parameter extension to $\Lambda$CDM, varied with a uniform prior $\gamma \in \left[0,2\right]$. To perform the theoretical calculations, we use the modified version of {\tt camb} released by~\cite{Nguyen:2023fip}.\footnote{\url{https://github.com/MinhMPA/CAMB_GammaPrime_Growth}}

\begin{figure}[t]
    \centering
    \includegraphics[width=\columnwidth]{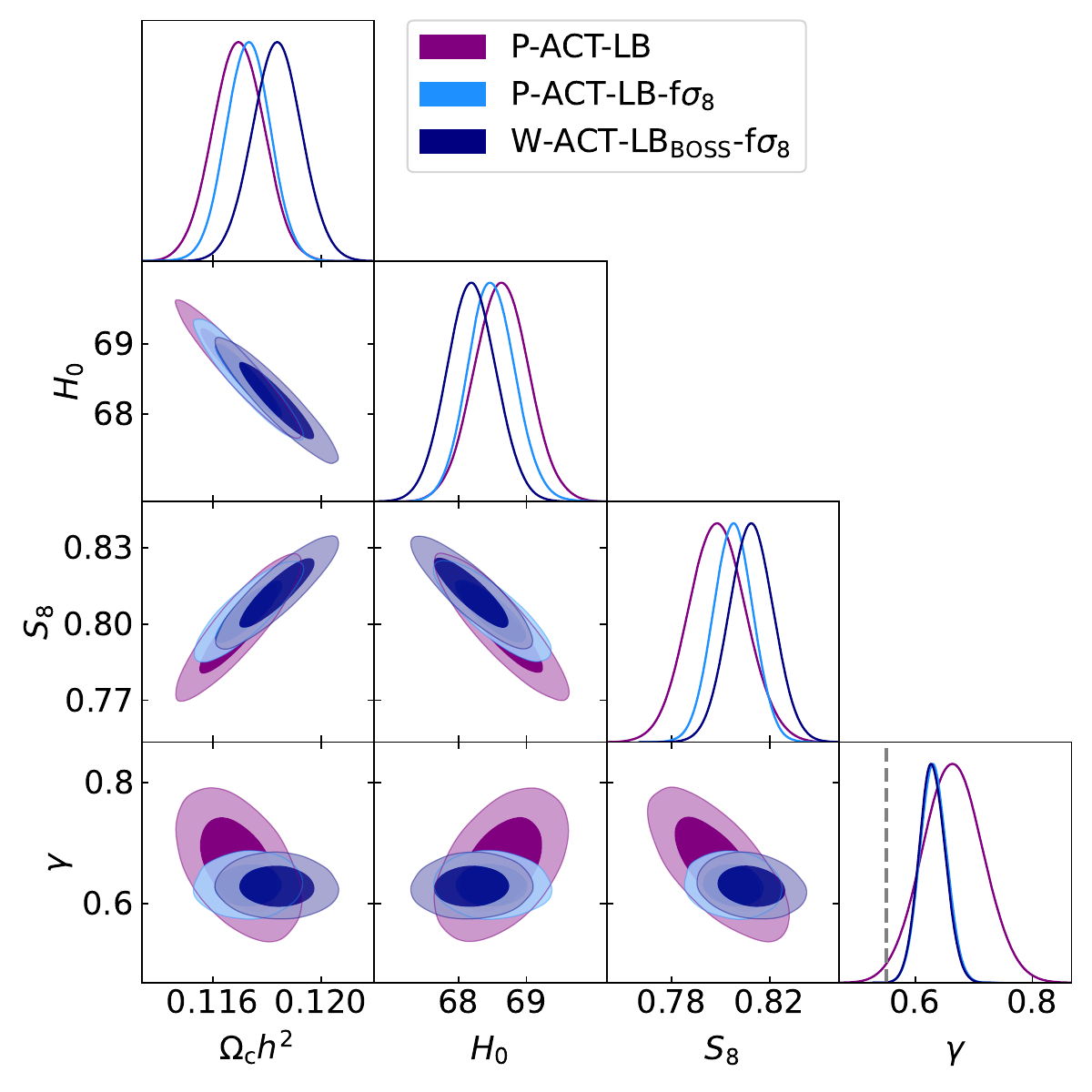}
    \vspace{-0.5cm}\caption{Constraints on a modified growth of structure model in which the growth index $\gamma$ is allowed to vary, for different dataset combinations (\pactlb\ in purple, \pactlb\textsf{-f$\sigma_8$}\ in light blue, and \wactlbb\textsf{-f$\sigma_8$}\ in navy). The addition of $f\sigma_8$ measurements effectively breaks degeneracies between $\gamma$ and other parameters, but moves the constraint away from the GR prediction of $\gamma=0.55$ (marked by the gray dashed line) at $>3\sigma$. The consistency between \pactlb\textsf{-f$\sigma_8$}\ and \wactlbb\textsf{-f$\sigma_8$}\ indicates that the $\gamma$ result is not driven strongly by the preference for high $A_{\mathrm{lens}}$ in \Planck, or by DESI BAO. 
    }
    \label{fig:gamma_base_triangle}
\end{figure}

With the new DR6 spectra, we find the growth index to be $\gamma = 0.560 \pm 0.110$, with $S_8=0.872 \pm 0.042~(68\%, \act)$.  Combining with \emph{Planck}, these limits tighten significantly, yielding $\gamma = 0.688 \pm 0.071$ and $S_8 = 0.812 \pm 0.017\; (68\%, \pact)$.  Further adding CMB lensing and DESI BAO data, we obtain 
\begin{equation}
\left.
 \begin{aligned}
\gamma &=& 0.663 \pm 0.052 \\
S_8 &=&  0.799 \pm 0.012\; 
\end{aligned}
\quad \right\} 
\mbox{\textrm{\; (68\%, \pactlb).}}
\end{equation}
This value of $\gamma$ lies $2.2\sigma$ above the GR prediction, but is nevertheless consistent with it (as shown in Fig.~\ref{fig:gamma_base_triangle}). 
When replacing the DESI BAO data with BOSS BAO data, we obtain $\gamma = 0.635 \pm 0.053$ and $S_8 = 0.812 \pm 0.013\; (68\%,\pactlbb)$, which is consistent with the \pactlb\ result but moves closer to the GR expectation. Including now the $f\sigma_8$ measurements, we obtain\footnote{We note that when including $f\sigma_8$ measurements, for the BOSS likelihood we use a likelihood that combines BAO and RSD data from BOSS DR12 and eBOSS DR16, as described in \cite{eBOSS:2020yzd}. When we refer to \pactlbb\textsf{-f$\sigma_8$} or \wactlbb\textsf{-f$\sigma_8$}, this is the BOSS likelihood that is used.}

\begin{equation}
\left.
 \begin{aligned}
\gamma  &=& 0.630\pm 0.023 \\ 
S_8  &=& 0.8050\pm 0.0081
\end{aligned}
\quad \right\} \mbox{\textrm{\; (68\%, \pactlb\textsf{-f$\sigma_8$}).}}
\end{equation}
Replacing DESI BAO data with BOSS has minimal impact, giving $\gamma=0.634 \pm 0.024$ and $S_8 =0.8121 \pm 0.0084$ (68\%, \pactlbb\textsf{-f$\sigma_8$}). The central values of $\gamma$ in these constraints deviate from GR at $3.5\sigma$.  The data driving this result are those from RSD and peculiar velocity surveys, but the central values are consistent with those found in our analyses without these data.

Note that values of $\gamma>0.55$ can improve concordance between CMB-derived and LSS-derived constraints on $S_8$, compared to those in \lcdm.  For example, our \pact\ result in the modified-growth model ($S_8 = 0.812 \pm 0.017$) is lower than that in \lcdm\ ($S_8 = 0.830 \pm 0.014$, L25).  We do not run constraints using non-CMB data on their own, but~\cite{Nguyen:2023fip} find that the combination of $f\sigma_8$, DES-Y1 3$\times$2-point, and pre-DESI BAO data yields $S_8 = 0.784^{+0.017}_{-0.016}$ in this model.  This agrees with the \pact\ $S_8$ result at $1.2\sigma$, whereas in \lcdm, the same comparison is discrepant at $3.0\sigma$.  The moderate tension is reduced in the modified-growth model both by $\sim1\sigma$ shifts in the central $S_8$ values and small increases in the error bars.

The effects of varying $\gamma$ on the lensed CMB power spectrum are similar to those from the phenomenological lensing amplitude parameter, $A_{\mathrm{lens}}$~\citep{Nguyen:2023fip}, for which \Planck\ has a moderate preference for a value $A_{\mathrm{lens}}>1$~\citep{Planck_2018_params}. To assess whether it is the \Planck\ $A_{\mathrm{lens}}$ behavior driving the evidence for a high growth index, we swap in \emph{WMAP} for \emph{Planck} data. For the CMB-only combination, we find $\gamma = 0.621 \pm 0.087$ and $S_8=0.840 \pm 0.028\; (68\%, \wact)$. Adding CMB lensing, BOSS BAO, and the $f\sigma_8$ data yields
\begin{equation}
\left.
 \begin{aligned}
\gamma  &=& 0.628\pm 0.022 \\ 
S_8  &=& 0.8122\pm 0.0091
\end{aligned}
\quad \right\} \mbox{\textrm{\; (68\%, \wactlbb\textsf{-f$\sigma_8$}),}}
\end{equation}
still finding a $3.5\sigma$ deviation from the GR value of $\gamma$. This shows that the current evidence for a non-standard growth index is not driven by \emph{Planck}'s preference for a high CMB lensing amplitude or by DESI BAO data. The precision on this result is driven strongly by the $f\sigma_8$ data. We discuss in more detail how this $f\sigma_8$ dataset compares with \lcdm\ in~\S\ref{subsec:consistency}. Any preference for this model appears to be largely driven by the $f\sigma_8$ data (in fact, by just two of these data points) rather than the ACT data or other CMB data, as discussed further in~\S\ref{subsec:consistency}.  \\

\section{Consistency with low-redshift data and impact on cosmological concordance} \label{sec:consistency_concordance}

\subsection{\lcdm\ consistency with low-redshift observations}\label{subsec:consistency}

Given our findings that no model is significantly preferred over \lcdm, we confirm the consistency of our baseline \lcdm\ model (described in L25) with the main low-redshift datasets used here in combination with \pact. The agreement between the predictions of the best-fit \pact\ \lcdm\ model and direct measurements of the CMB lensing power spectrum, BAO distances, and SNIa distances provides a powerful consistency test of the cosmological model and justifies our joint inclusion of these datasets in \lcdm\ and extended model fits.  

\begin{figure}[ht!]
    \centering
    \includegraphics[width=\columnwidth]
    {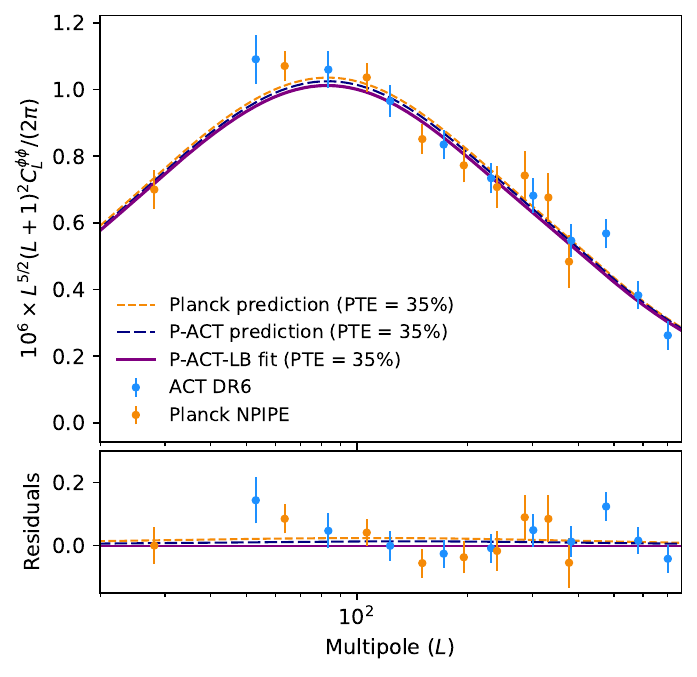}
    \includegraphics[width=\columnwidth]{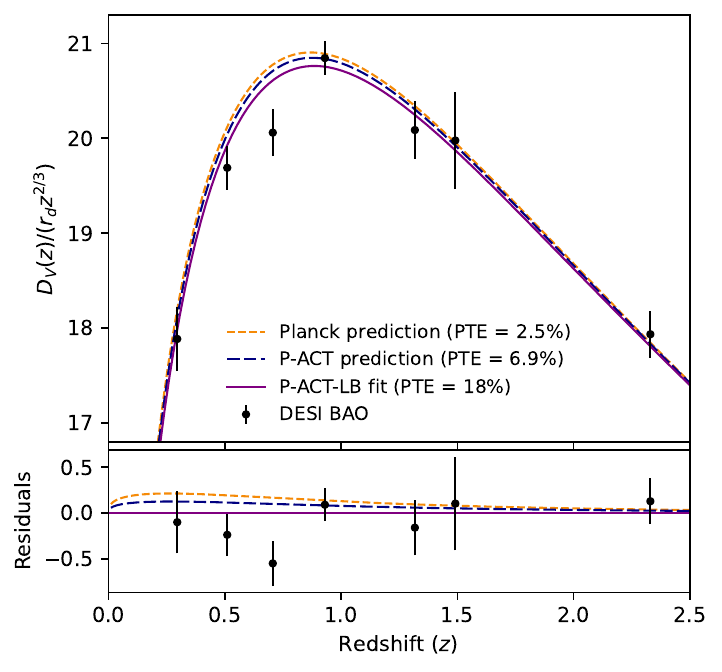}
    \caption{The ACT DR6 and \Planck\ NPIPE CMB lensing power spectra (top panel) and the DESI BAO distance-redshift relation (bottom panel) are compared to \lcdm\ predictions from primary CMB anisotropy data (\Planck\ in orange and \pact\ in navy). The joint \pactlb\ \lcdm\ best-fit model is shown in purple. For CMB lensing, we show only the bandpowers used in the baseline likelihood described in~\S\ref{subsec:cmb_lens}. The DESI dataset is described in~\S\ref{subsec:bao} and includes twelve total data points; of these twelve points, five pairs of $D_H(z) / r_d$ and $D_M(z)/r_d$ are combined into five different $D_V(z)/r_d$ measurements leading to a total of seven data points in this figure. The PTEs are calculated approximating the number of degrees of freedom to be equal to the number of data points and without propagating uncertainties in theory predictions.}
    \label{fig:lensing_bao_residuals}
\end{figure}

In Fig.~\ref{fig:lensing_bao_residuals}, we show predictions for the two main datasets we combine with: ACT DR6+\Planck\ NPIPE CMB lensing and DESI BAO. We find that the \pact\ \lcdm\ best-fit model predicts these observations well, and slightly improves the agreement with the DESI data, compared to that seen for the \lcdm\ model determined by \Planck\ alone. The same figure also shows the combined \pactlb\ \lcdm\ fit to all datasets, including residuals and the probability to exceed (PTE) for CMB lensing and BAO.\footnote{The three models give very similar PTEs for CMB lensing --- 34.7\%, 35.3\%, 35.3\% for \Planck, \pact, and \pactlb, respectively --- which round to the same number when reporting only two significant figures.} We find here and in L25 that the jointly-derived \lcdm\ model is a good fit to all datasets.  The residuals and best-fit PTE for \Planck\ and ACT primary CMB power spectra are discussed in L25. 

\begin{figure}[]
    \centering
\includegraphics[width=\columnwidth]{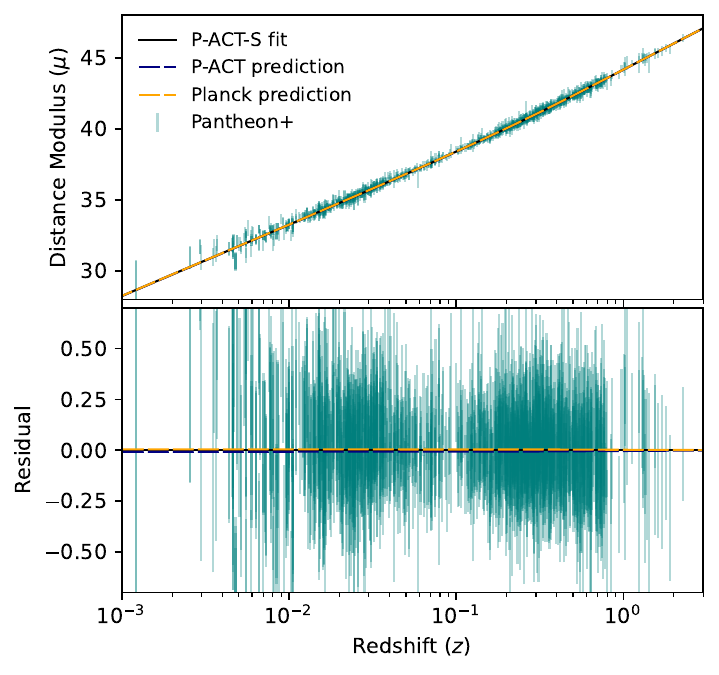}
    \vspace{-0.6cm}\caption{The SNIa Hubble diagram, showing the distance modulus $\mu$ as a function of redshift for the Pantheon+ compilation. The full SNIa sample is compared with the CMB anisotropy best-fit \lcdm\ model predictions (from either \Planck\ or \pact), and with the joint \pact\ CMB plus SNIa \lcdm\ best-fit (\pacts\ in black). The distance modulus has been calibrated relative to each \lcdm\ best-fit cosmology (rather than using SH0ES, as done in \citealp{Panthoen+cosmology}).} 
    \label{fig:SNIa_lcdm_comp}
\end{figure}

Figure~\ref{fig:SNIa_lcdm_comp} shows the equivalent comparison for the Pantheon+ SNIa compilation. We show the usual Hubble diagram of the distance modulus (corrected apparent magnitude) of all SNIa in the compilation as a function of redshift, and compare it with \lcdm\ predictions from the CMB and the joint best-fit model. Differently from \cite{Panthoen+cosmology}, we do not normalize the distance modulus using the SH0ES calibration. Instead, we separately obtain and correct for the absolute magnitude calibration from the \textsf{Planck} and \pact\ cosmology. The fit of the joint \pacts\ dataset marginalizes over this calibration while sampling the cosmological parameters. We find that all \lcdm\ solutions are in good agreement with the whole sample, recovering maximum-likelihood values similar to those reported in the Pantheon+ public chains.

\begin{figure}
    \centering
\includegraphics[width=\columnwidth]{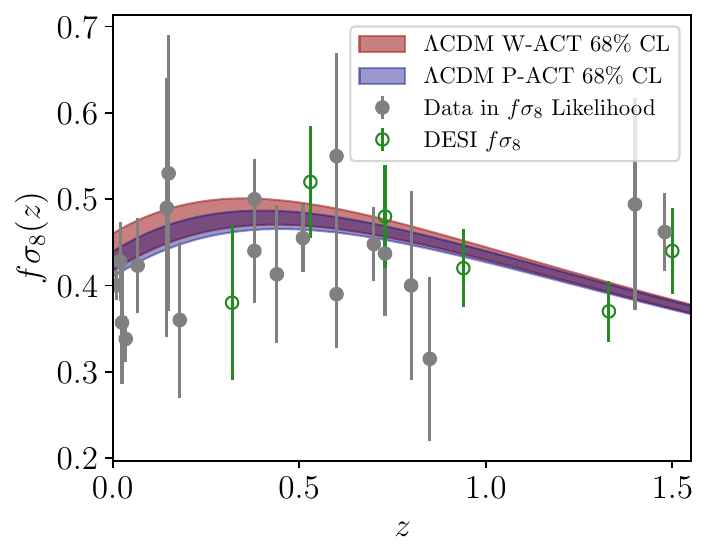}
    \vspace{-0.6cm}\caption{Measurements of $f\sigma_8$ used in our likelihood analysis of the modified gravity model (\S\ref{sec:mg}) with 1$\sigma$ error bars (gray). The navy band shows the theoretically-predicted $f\sigma_8(z)$ with the 68\% confidence interval for the best-fit \lcdm\ cosmology to \pact, with the brown band representing the same for \wact. We also show $f\sigma_8$ measurements from DESI~\citep{2024desi} in green, but note that these are not used in our $f\sigma_8$ likelihood analysis and are only shown for comparison.} 
    \label{fig:fsigma8_points}
\end{figure}

Figure~\ref{fig:fsigma8_points} shows the data used in the $f\sigma_8$ likelihood from \S\ref{sec:mg}, compared with the theoretical predictions in \lcdm\ for $f\sigma_8(z)$ from the 68\%~CL~constraints for \pact\ and \wact\ (L25). It appears that the hint of deviation from \lcdm\ found in \S\ref{sec:mg} is driven by low-$z$ points in the $f\sigma_8$ dataset that have small error bars and lie below the CMB expectation for $f\sigma_8(z)$. To assess the difference between the CMB-predicted $f\sigma_8(z)$ and the data points from RSD and peculiar velocity surveys, we compute the $\chi^2$ between them, finding $\chi^2=28.0$ with respect to the \pact\ prediction and $\chi^2=37.3$ with respect to the \wact\ prediction, both for 20 degrees of freedom, indicating some discordance (note, however, that these $\chi^2$ values do not propagate the theoretical uncertainty on the CMB predictions). These $\chi^2$ values decrease significantly if we omit two of the low-$z$ peculiar velocity data points that have the lowest $f\sigma_8$ values. These points are at redshifts $z=0.025$ (from \citealp{Turner:2022mla}, with $f\sigma_8=0.357 \pm 0.071$) and $z=0.035$ (from \citealp{Said:2020epb} with $f\sigma_8 = 0.338 \pm 0.027$). Omitting these points, we find $\chi^2=14.4$ for the \pact\ prediction and $\chi^2=19.1$ for \wact, both for 18 degrees of freedom, illustrating no tension. Because of the significant $\chi^2$ reduction that results from removing the lowest two redshift points, we conclude that deviations from $\Lambda$CDM when including $f\sigma_8$ data in \S\ref{sec:mg} are largely driven by those points. In theory, one could redo the analysis omitting those points from the likelihood. However, we do not perform this exercise here since it is not the ACT DR6 data driving these deviations.

Figure~\ref{fig:fsigma8_points} also shows recent $f\sigma_8$ measurements from DESI~\citep{2024desi}, though these points are not used in our joint likelihood analysis.  Assuming a diagonal covariance matrix, we find that the DESI measurements have $\chi^2=4.0$ with respect to the \pact\ \lcdm\ prediction and $\chi^2=4.2$ with respect to the \wact\ \lcdm\ prediction, both for 6 degrees of freedom. Thus, the DESI RSD data are in good agreement with $\Lambda$CDM, further suggesting that the two outlying $f\sigma_8$ points at very low-$z$ are not indicative of new physics.

\subsection{Cosmological concordance}\label{subsec:concordance}

Many of the models studied in this paper impact the value of the Hubble constant inferred from the primary CMB power spectra, and the precision of the $H_0$ measurement changes because of parameter degeneracies in these scenarios.  In some cases, such as in models with variations in the particle physics content and/or particle interactions, the degeneracies are due to changes in the ingredients of the universe affecting the expansion history at early times. In other cases, for example for the early dark energy model, the physics prior to or at recombination is changed significantly to reduce the sound horizon and this increases the Hubble constant at the expense of additional degrees of freedom in the cosmological model correlating with $H_0$. While no robust detection of any of these extended models has been made in previous CMB or LSS data, some of these models have previously had non-negligible success in shifting the CMB-derived $H_0$ value toward the local distance ladder estimate from SH0ES.  In this work, we have shown that this is generally no longer the case when including the new ACT data.  ACT DR6 contributes to limiting the allowed region of parameter space to $66.1<H_0<71.0$~km/s/Mpc within all of the models studied here.  This range is calculated as the minimum and maximum value at 68\% CL of all the \pactlb\ and \pactlbs\ results (whichever yields higher precision) from all models explored in this work; the neutrino self-interaction result from \S\ref{sec:neu_si} sets the lower end of the allowed range and the early dark energy result from \S\ref{sec:ede} sets the upper end.  This range of values is in agreement with the CCHP measurement~\citep{Freedman_CCHP}, but lies below the latest SH0ES measurement~\citep{Breuval:2024lsv}.

\begin{figure}[tp!]
    \centering
    \includegraphics[width=\columnwidth]{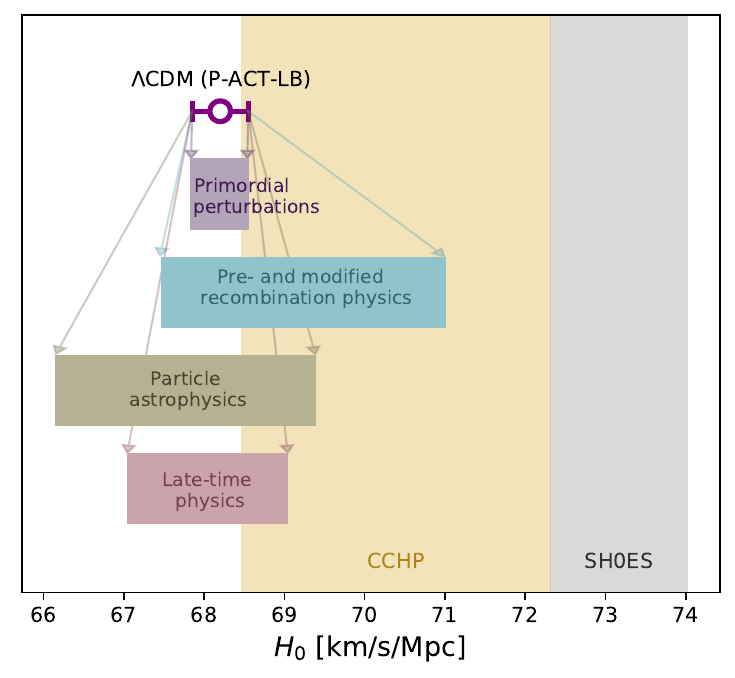}
    \caption{Inferred measurements of the Hubble constant across classes of models from \pact\ primary CMB combined with low-redshift data. Each box captures the range of $H_0$ allowed by all models explored in this paper within that class (\S\ref{sec:inflation}-\S\ref{sec:late-time}) --- drawing from the 68\% limits obtained with the most stringent data combination for that model, \pactlb\ or \pactlbs\ depending on the model (see also Fig.~\ref{fig:H0_1D}) --- and marks shifts and broadening with respect to the 68\% CL estimate in \lcdm\ from \pactlb\ (purple bar, L25). All estimates are statistically consistent with the CCHP measurement (68\%~CL from \citealp{Freedman_CCHP}, gold band), but no class of extensions fully meets the bounds from SH0ES (68\%~CL from \citealp{Breuval:2024lsv}, gray band). See also~\citealp{2024ARA&A..62..287V} for additional direct $H_0$ measurements.} 
    \label{fig:H0_variation}
\end{figure}

\begin{figure}[tp!]
    \centering
    \includegraphics[width=\columnwidth]{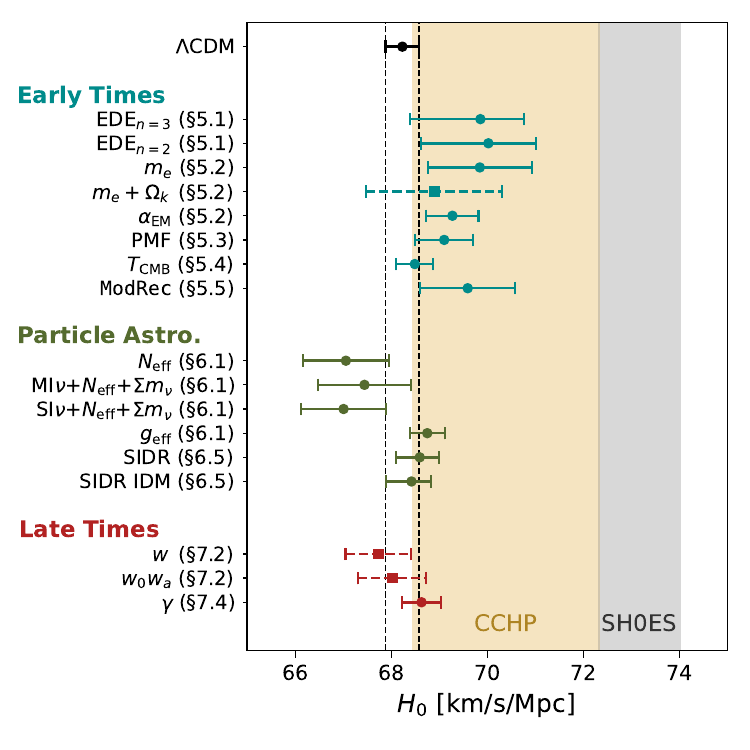}
    \vspace{-0.6cm}\caption{The three classes of models shown in Fig.~\ref{fig:H0_variation} that have an impact on the inferred value of $H_0$ --- changes in early-time physics before or at recombination (teal), new particle astrophysics (olive green), and new physics impacting the late-time expansion of the universe (dark red) ---  are broken down here into the constraints for specific models within each class.  Solid (dashed) bars are constraints at 68\% confidence derived from \pactlb\ (\pactlbs).} 
    \label{fig:H0_1D}
\end{figure}

\begin{figure}[hp!]
    \centering
    \includegraphics[width=0.93\columnwidth]{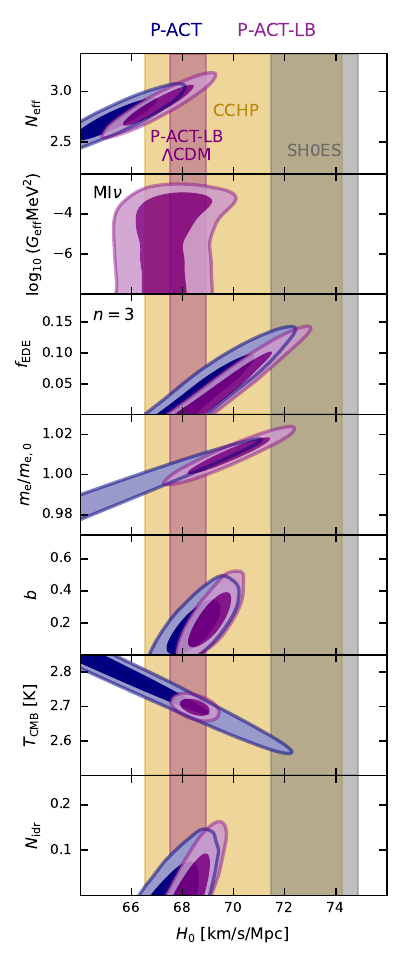} \vspace{-0.2cm}
    \caption{Constraints at 68\% and 95\% confidence on $H_0$ in select extended models from \pact\ (navy) and \pactlb\ (purple), with the vertical bands indicating the $2\sigma$ limits for the latter data combination within \lcdm. Degeneracies with extended-model parameters do not move the \pact\ or \pactlb\ limits into agreement with the SH0ES constraints ($2\sigma$ region from \citealp{Breuval:2024lsv} shown as a gray band), but the constraints are in agreement with the CCHP measurement ($2\sigma$ region from \citealp{Freedman_CCHP} shown as a gold band). Note that the MI$\nu$ model shown in the second panel also includes varying $N_{\rm eff}$ and $\sum m_\nu$.}
    \label{fig:H0-tension}
\end{figure}

These results are summarized in Fig.~\ref{fig:H0_variation}.  Figure~\ref{fig:H0_1D} provides more details on the full suite of $H_0$ constraints for the three classes of models that have a non-negligible impact on $H_0$.
Figure~\ref{fig:H0-tension} breaks this result further down into the specific behavior of $H_0$ versus key model parameters in many extended models of interest, and shows that both the CMB-only dataset, \pact, and its combination with CMB lensing and BAO, \pactlb, are inconsistent with the SH0ES estimate.  Because of correlations with the inferred value of the total matter density $\Omega_m$,\footnote{Note that the primary CMB most directly constrains a degenerate combination of $\Omega_m$ and $H_0$, roughly $\Omega_m h^3$, with the exact value of the exponent depending on the range of angular scales measured, due to the different physical effects at play~\citep[e.g.,][]{2002MNRAS.337.1068P,2019ApJ...871...77K}.} BAO data play an important role in determining the central value of $H_0$ in many extended models.  Notably, as shown in many of the analyses in \S\ref{sec:recomb}--\S\ref{sec:late-time}, using BOSS BAO instead of DESI data reduces the central value of $H_0$ by 1--1.5~km/s/Mpc --- hence moving constraints in the opposite direction of SH0ES --- while only slightly increasing the error bars.  Upcoming DESI data will be instrumental in further clarifying this situation.

From a more model-independent perspective, the consistency of the inferred physical matter density $\Omega_m h^2$ from \Planck\ and ACT (as well as their combination; L25) within \lcdm\ provides a strong indication that the \lcdm\ model accurately describes physics just prior to recombination.  As noted in~\cite{KnoxMillea2020}, the shape of the radiation-driving envelope in the CMB power spectra is directly tied to the physics operating around matter-radiation equality.  If new physics were present at this epoch, a generic expectation is that \lcdm-based inferences of $\Omega_m h^2$ should exhibit deviations when inferred on different angular scale ranges, due to different scales probing $H(z)$ at different times via radiation driving.  We see no such evidence of such deviations in $\Omega_m h^2$ (L25), thus providing a strong validation of the physics of \lcdm\ in this important redshift range.

\begin{figure}[hp!]
    \centering
\includegraphics[width=0.916\columnwidth]{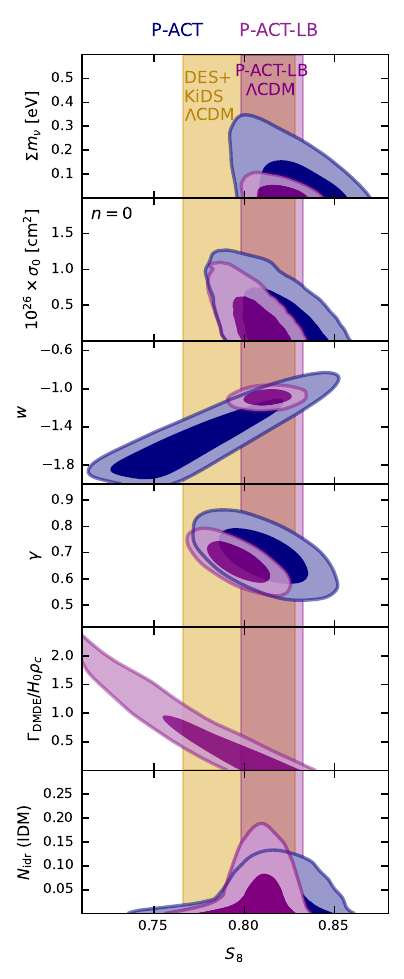} 
    \vspace{-0.25cm}
    \caption{Constraints at 68\% and 95\% confidence on $S_8$ in select extended models from \pact\ (navy) and \pactlb\ (purple) compared to the DES-Y3+KiDS cosmic shear limit ($2\sigma$ region from the \lcdm\ case with fixed neutrino mass from \citealp{DES_KIDS_shear_2023} shown as a gold band).  The \pactlb\ \lcdm\ constraint ($2\sigma$ purple vertical band) is already in statistical agreement with DES-Y3+KiDS (see L25). Degeneracies with extended-model parameters do not generally move the \pactlb\ central values to even lower $S_8$ values at the center of the DES-Y3+KiDS posterior.}
    \label{fig:S8-tension}
\end{figure}

We also find that models that alter the growth of structure are not preferred by our data.  In particular, we do not find evidence in any extended model for a significant shift in the CMB-inferred $S_8$ value toward lower late-universe estimates, for example from cosmic shear surveys, as summarized in Fig.~\ref{fig:S8-tension}. Our results generally lie on the upper end of the limits from DES+KiDS~\citep{DES_KIDS_shear_2023}, although it is important to note that the $S_8$ value in \lcdm\ is already consistent between \pactlb\ and this result.\footnote{The more recent KiDS Legacy measurement, $S_8=0.815^{+0.016}_{-0.021}$, shifts to even higher values and is almost perfectly centered on the \pactlb\ result.} Also in this case the matter density inferred via DESI BAO will contribute to the location of the maximum-likelihood point. For all models tested, using BOSS instead of DESI BAO data yields qualitatively similar conclusions.  In addition, it is important to note that the inference of $S_8$ itself from galaxy weak lensing or other late-time data will generally change in the presence of the new physics shown in these models (e.g., non-GR values of $\gamma$ or non-zero $\Gamma_{\rm DMDE}$), and thus one should take the comparison shown here with a grain of salt.  For a high-precision comparison, constraints from late-time data should be self-consistently derived in these extended models, which we leave to future, dedicated analyses.

\section{Summary}\label{sec:conclusions}

In this paper, we have used a state-of-the-art CMB primary anisotropy dataset built from new ACT DR6 power spectra combined with large-scale data from the \Planck\ satellite to set constraints on a large suite of extended cosmological models, both testing foundational assumptions made in the standard \lcdm\ model and constraining more complex scenarios.  We show that the new ACT DR6 data have reached a level of precision competitive with that of \Planck\ for the primary CMB contribution to the constraints, and because of the complementarity of the two datasets --- particularly at $\ell>2000$ in TT and on essentially all scales measured by ACT in TE and EE --- their joint analysis yields significant gains over the sensitivity of \Planck\ alone.  We further include ACT and \Planck\ CMB lensing data, BAO data from DESI and SNIa from Pantheon+ to obtain the tightest limits to date on many fundamental physics parameters and cosmological models.  To validate the results, we also explore combinations replacing \Planck\ with \WMAP\ large-scale CMB data, and DESI with BAO data from BOSS.  

Key takeaways from this work include:
\vspace{-0.09cm}
\begin{itemize}
\itemsep0em 
\item We test predictions for the early-universe inflationary epoch, verifying the near-scale-invariance and adiabaticity of the primordial scalar perturbations, and restricting the parameter space for inflation models;
\item We tightly bound the presence of early dark energy, or primordial magnetic fields that would impact the pre-recombination period, as well as variations in fundamental constants or in the monopole temperature of the CMB or the standard recombination history;
\item We measure the properties of neutrinos and find no evidence for new light, relativistic species that are free-streaming, no evidence for non-zero neutrino masses, and no evidence for neutrino self-interactions. We also set limits on primordial helium and deuterium abundances that are consistent with standard BBN predictions and with astrophysical measurements;
\item Dark matter is consistent with the standard collisionless CDM assumption, with no evidence for annihilation or interactions with baryons. We also limit the contribution to DM from ultralight axion-like particles to be no more than 5\% at axion masses of $10^{-26}$~eV; 
\item We find no evidence for interactions in the dark sector, either within a single dark radiation component, between dark radiation and dark matter components, or between dark matter and dark energy;
\item Dark energy and late-time gravity generally behave as expected in the standard model ($w=-1$ and $\gamma = 0.55$), although we find a moderate hint for time-evolving dark energy with \pactlbs, consistent with (and driven by the BAO and SNIa data from)~\cite{DESI-BAO-VI}.  Upcoming BAO and SNIa data are expected to clarify the robustness of this hint.
\end{itemize}

Assessing the preference for each extended model over \lcdm\ via the marginalized posterior for the extended-model parameter (in the case of 1D extensions) or via comparison of the MAP model to that in \lcdm\ (in the case of multi-parameter extensions), we find that no extended model is preferred over \lcdm.  This conclusion includes a wide range of extended models constructed with the aim of increasing the value of the Hubble constant or decreasing the amplitude of density fluctuations inferred from the primary CMB.  With these results providing stringent tests in new observational regimes, the footing of the standard model of cosmology is further solidified.

This work paves the way for further, higher-precision tests of the cosmological model in the coming decade with even more sensitive CMB polarization data, including that expected from the Simons Observatory~\citep{Simons_Observatory,Enhanced_SO_2025}, CMB-S4~\citep{CMB_S4_Science_Book,CMBS4DSR}, LiteBIRD~\citep{LiteBIRD}, and other experiments.  The noise levels of these surveys will push nearly an order of magnitude below those of ACT DR6 over the next 10-15 years, enabling precise searches for new physics, including not only the models studied in this work, but also new scenarios yet to be constructed.

Shortly after the appearance of this paper, new BAO measurements from three years of DESI observations and corresponding cosmological interpretation were presented in~\cite{DESI1_DR2,DESI-DR2}. In Appendix~\ref{app:newDESI_tau}, we present updated \lcdm\ consistency and evolving dark energy constraints from this new dataset, finding results similar to those presented in the main text with DESI Year-1 observations and consistent with those reported in \cite{2025arXiv250418464G}.  We also present updated constraints on a subset of the extended models studied earlier in the paper.  In addition, we show that relaxing the constraint on the optical depth leads to slightly reduced evidence for $w_0/w_a$ and more relaxed bounds on the neutrino mass.

\section*{ACKNOWLEDGEMENTS}
We are grateful to Julien Lesgourgues for assistance in providing the version of {\tt class} used throughout this work, to Antony Lewis for useful exchanges regarding {\tt camb}, and to Alessio Spurio Mancini for help with \texttt{CosmoPower}.  We also thank Jens Chluba and Yacine Ali-Haimoud for useful correspondence regarding {\tt CosmoRec} and {\tt HyRec}, and we thank Gabriel Lynch for useful correspondence regarding the Modified Recombination scenario, Henrique Rubira for useful correspondence regarding the IDR-IDM scenario, and Minh Nguyen for useful correspondence regarding the modified-growth-index scenario. 

Support for ACT was through the U.S.~National Science Foundation through awards AST-0408698, AST-0965625, and AST-1440226 for the ACT project, as well as awards PHY-0355328, PHY-0855887 and PHY-1214379. Funding was also provided by Princeton University, the University of Pennsylvania, and a Canada Foundation for Innovation (CFI) award to UBC. ACT operated in the Parque Astron\'omico Atacama in northern Chile under the auspices of the Agencia Nacional de Investigaci\'on y Desarrollo (ANID). The development of multichroic detectors and lenses was supported by NASA grants NNX13AE56G and NNX14AB58G. Detector research at NIST was supported by the NIST Innovations in Measurement Science program. Computing for ACT was performed using the Princeton Research Computing resources at Princeton University and the Niagara supercomputer at the SciNet HPC Consortium. SciNet is funded by the CFI under the auspices of Compute Canada, the Government of Ontario, the Ontario Research Fund–Research Excellence, and the University of Toronto. This research also used resources of the National Energy Research Scientific Computing Center (NERSC), a U.S. Department of Energy Office of Science User Facility located at Lawrence Berkeley National Laboratory, operated under Contract No. DE-AC02-05CH11231 using NERSC award HEP-ERCAPmp107 from 2021 to 2025. We thank the Republic of Chile for hosting ACT in the northern Atacama, and the local indigenous Licanantay communities whom we follow in observing and learning from the night sky.

Some cosmological analyses were performed on the Hawk high-performance computing cluster at the Advanced Research Computing at Cardiff (ARCCA). We made extensive use of computational resources at the University of Oxford Department of Physics, funded by the John Fell Oxford University Press Research Fund.

We are grateful to the NASA LAMBDA archive for hosting our data. This work uses data from the Planck satellite, based on observations obtained with Planck (http://www.esa.int/Planck), an ESA science mission with instruments and contributions directly funded by ESA Member States, NASA, and Canada. We acknowledge work done by the Simons Observatory collaboration in developing open-source software used in this paper.

IAC acknowledges support from Fundaci\'on Mauricio y Carlota Botton and the Cambridge International Trust.  SA, MH and DNS acknowledge the support of the Simons Foundation. ZA and JD acknowledge support from NSF grant AST-2108126. EC, IH and HJ acknowledge support from the Horizon 2020 ERC Starting Grant (Grant agreement No 849169). OD acknowledges support from a SNSF Eccellenza Professorial Fellowship (No. 186879). JD acknowledges support from a Royal Society Wolfson Visiting Fellowship and from the Kavli Institute for Cosmology Cambridge and the Institute of Astronomy, Cambridge. RD and CS acknowledge support from the Agencia Nacional de Investigaci\'on y Desarrollo (ANID) through Basal project FB210003. RD acknowledges support from ANID-QUIMAL 240004 and FONDEF ID21I10236. SG acknowledges support from STFC and UKRI (grant numbers ST/W002892/1 and ST/X006360/1). SJG acknowledges support from NSF grant AST-2307727 and acknowledges the Texas Advanced Computing Center (TACC) at the University of Texas at Austin for providing computational resources that have contributed to the research results reported within this paper. V.G. acknowledges the support from NASA through the Astrophysics Theory Program, Award Number 21-ATP21-0135, the National Science Foundation (NSF) CAREER Grant No. PHY2239205, and from the Research Corporation for Science Advancement under the Cottrell Scholar Program.This research was carried out in part at the Jet Propulsion Laboratory, California Institute of Technology, under a contract with the National Aeronautics and Space Administration (80NM0018D0004). JCH acknowledges support from NSF grant AST-2108536, the Sloan Foundation, and the Simons Foundation, and thanks the Scientific Computing Core staff at the Flatiron Institute for computational support. MHi acknowledges support from the National Research Foundation of South Africa (grant no. 137975). ADH acknowledges support from the Sutton Family Chair in Science, Christianity and Cultures, from the Faculty of Arts and Science, University of Toronto, and from the Natural Sciences and Engineering Research Council of Canada (NSERC) [RGPIN-2023-05014, DGECR-2023- 00180]. JPH (George A. and Margaret M. Downsbrough Professor of Astrophysics) acknowledges the Downsbrough heirs and the estate of George Atha Downsbrough for their support. JK, MM  and KPS acknowledge support from NSF grants AST-2307727 and  AST-2153201. ALP acknowledges support from a Science and Technology Facilities Council (STFC) Consolidated Grant (ST/W000903/1). ML acknowledges that IFAE is partially funded by the CERCA program of the Generalitat de Catalunya. SN acknowledges a grant from the Simons Foundation (CCA 918271, PBL). TN thanks support from JSPS KAKENHI Grant No. JP20H05859, No. JP22K03682, and World Premier International Research Center Initiative (WPI Initiative), MEXT, Japan. FN acknowledges funding from the European Union (ERC, POLOCALC, 101096035). LP acknowledges support from the Wilkinson and Misrahi Funds. KKR is supported by an Ernest Rutherford Fellowship from the UKRI Science and Technology Facilities Council (grant number ST/Z510191/1). NS acknowledges support from DOE award number DE-SC0025309. KMS acknowledges support from the NSF Graduate Research Fellowship Program under Grant No.~DGE 2036197, and acknowledges the use of computing resources from Columbia University's Shared Research Computing Facility project, which is supported by NIH Research Facility Improvement Grant 1G20RR030893-01, and associated funds from the New York State Empire State Development, Division of Science Technology and Innovation (NYSTAR) Contract C090171, both awarded April 15, 2010.
NB, MG, ML and LP acknowledge the financial support from the INFN InDark initiative and from the COSMOS network through the ASI (Italian Space Agency) Grants 2016-24-H.0 and 2016-24-H.1-2018. NB, MG and ML are funded by the European Union (ERC, RELiCS, project number 101116027). MG acknowledges support from the PRIN (Progetti di ricerca di Rilevante Interesse Nazionale) number 2022WJ9J33.

We gratefully acknowledge the many publicly available software packages that were essential for parts of this analysis. They include: camb~\citep{Lewis_camb}, class~\citep{CLASS_2011, Lesgourgues_class}, CosmoPower~\citep{Spurio_Mancini_cosmopower}, HyRec~\citep{Hyrec}, CosmoRec~\citep{Cosmorec}, HMcode~\citep{Mead2020}, PRIMAT~\citep{2018PhR...754....1P}, Cobaya~\citep{Cobaya}, PROSPECT~\citep{Holm:2023uwa}, GetDist~\citep{Getdist}, numpy~\citep{Numpy}, and matplotlib~\citep{Hunter:2007}.

\clearpage
\bibliography{msm}
\bibliographystyle{act_titles}
\clearpage

\appendix
\section{Theory specifications}\label{app:theory}

For the results presented in this paper, we use multiple Einstein-Boltzmann codes to calculate theory predictions. We explore and define settings for both \texttt{camb} and \texttt{class} to ensure that the uncertainty in the theory calculations is well below the statistical error bars of the DR6 power spectra on all scales. In particular, some of the settings defined for the \Planck\ baseline analyses require revision to give unbiased results with our new small-scale measurements, as pointed out in, e.g.,~\citealp{Hill_ACT_EDE, McCarthy_2022}. For this work, we revise precision parameters for the calculation of the lensed CMB spectra, as well as baseline choices for non-linear modeling of the matter power spectrum, effects of baryonic feedback in the non-linear matter power spectrum model, recombination calculations, and the choice of BBN calculations (the latter is described in detail in~\S\ref{sec:bbn}).  We obtain the baseline settings summarized in~\S\ref{sec:method} and reported in Figs.~\ref{snip:camb_accuracy} and \ref{snip:class_accuracy} below. 

To assess robustness of the theory specifications with respect to the data, we consider both cosmological parameter posteriors using different assumptions and single likelihood evaluations to study the impact on the $\chi^2$ of a reference best-fit model.  In general, we find that parameter posteriors are unbiased, but $\chi^2$ values can vary across the parameter space, for both \lcdm\ and extended models.  We summarize the results of these investigations here:
\vspace{-0.1cm}
\begin{itemize}
\itemsep0em
    \item \emph{Numerical accuracy settings---} Accuracy parameters for the calculations of the lensed CMB theory power spectra were studied in previous works and validated for sensitivity levels beyond that of ACT DR6~\citep[see, e.g.,][]{Bolliet_emulators, Jense_emulators}.
    
    \item \emph{Recombination---} For recombination modeling, we compare the (previously-used-as-default) \texttt{RecFast} code with the alternative \texttt{HyRec} and \texttt{CosmoRec} codes, both of which can be manually installed within \texttt{camb}.  When evaluating the DR6 CMB-only likelihood \texttt{ACT-lite} at the best-fit \lcdm\ point from L25, we find $\Delta \chi_{\rm Recfast-HyRec}^2 = 0.37$ and $\Delta \chi_{\rm Recfast-CosmoRec}^2 = 0.49$, and with \texttt{HyRec} and \texttt{CosmoRec} agreeing within $\Delta \chi_{\rm HyRec-CosmoRec}^2 = 0.12$. Moving away from the best-fit point, the differences increase and vary across parameter space, with \texttt{RecFast} calculations compared to either \texttt{HyRec} or \texttt{CosmoRec} reaching $\Delta \chi^2\sim O(100)$.  Looking at the contributions to this difference across multipoles, we find that it mostly comes from TT at $\ell \sim 2000$. No significant difference in parameter posteriors is found. Given that \texttt{Recfast} relies on fudge factors optimized for the \Planck\ sensitivity but not beyond, and given that \texttt{CosmoRec} and \texttt{HyRec} have been proven to be consistent for precise calculations of hydrogen recombination --- with \texttt{CosmoRec} implementing also more complex helium recombination \citep{2012MNRAS.423.3227C} --- we choose to use \texttt{CosmoRec} as our baseline for use with \texttt{camb}. Since \texttt{CosmoRec} is not available with \texttt{class}, we use \texttt{HyRec} for models using this code.
    
    \item \emph{Non-linear matter power spectrum---} Previous CMB analyses~\citep[e.g.,][]{Choi2020,Planck_2018_params} used \texttt{Halofit} or the 2016 version of \texttt{HMcode} to compute the non-linear matter power spectrum, which is necessary for modeling CMB lensing.  Here, we switch to the updated 2020 version of \texttt{HMcode}.  We adopt the dark-matter-only model with the default parameters.  At the \lcdm\ best-fit point, we find $\Delta \chi_{\rm Halofit-HMcode2020}^2 = 0.09$.
    
    \item \emph{Baryonic feedback in non-linear modeling---} To assess the impact of parameters describing baryonic feedback in \texttt{HMcode}, we run calculations with dark-matter-only versus dark-matter-plus-baryons, with the default settings for baryonic feedback.  At the \lcdm\ best-fit point, this gives $\Delta \chi_{\rm DM-DM+b}^2=0.20$. We also test the impact of the strength of baryonic feedback, comparing our baseline with DM+baryons and with baryons having \texttt{HMcode\_logT\_AGN=8.0} (rather than 7.8), thus corresponding to a stronger feedback model with larger suppression of the small-scale matter power spectrum at late times. This gives $\Delta \chi_{\rm DM-DM+b+f}^2=0.39$.  For future analyses with Simons Observatory and CMB-S4 data, this effect will likely need to be accounted for in the modeling (as found in \citealp{McCarthy_2022}).
\end{itemize}

Figures~\ref{snip:camb_accuracy} and \ref{snip:class_accuracy} give code snippets from our \texttt{Cobaya} yaml files summarizing the baseline analysis settings. These configurations give results within $\Delta \chi_{\texttt{camb}-\texttt{class}}^2\sim O(10^{-2})$.  Figure~\ref{fig:lcdm_benchmark} shows validation across Einstein-Boltzmann codes for our baseline \lcdm\ parameter constraints, including {\tt camb}, {\tt class}, and {\tt CosmoPower}-based emulators; similar agreement is found also for extended models that are accessible in the three codes.  In Table~\ref{tab:models} we list all the models explored in this work and enumerate the theory and likelihood codes used for each analysis.

\begin{figure} [h!]
    \begin{lstlisting}[language=yaml]
camb:
  extra_args:
    kmax: 10
    k_per_logint: 130
    nonlinear: true
    lens_potential_accuracy: 8
    lens_margin: 2050
    lAccuracyBoost: 1.2
    min_l_logl_sampling: 6000
    DoLateRadTruncation: false
    recombination_model: CosmoRec
    \end{lstlisting}
    \caption{Baseline settings used for \texttt{camb} theory calculations, updating the default assumptions of \texttt{camb v1.5}.  Note that \texttt{halofit\_version=mead2020} and BBN consistency from PRIMAT 2021 are already default settings in this \texttt{camb} version.}
    \label{snip:camb_accuracy}
\end{figure}

\begin{figure} [h]
    \begin{lstlisting}[language=yaml]
classy:
  extra_args:
    N_ncdm: 1
    m_ncdm: 0.06
    N_ur: 2.0308
    T_cmb: 2.7255
    YHe: BBN
    non_linear: hmcode
    hmcode_version: '2020'
    recombination: HyRec
    lensing: 'yes'
    output: lCl,tCl,pCl,mPk
    modes: s
    l_max_scalars: 9500
    delta_l_max: 1800
    P_k_max_h/Mpc: 100.
    l_logstep: 1.025
    l_linstep: 20
    perturbations_sampling_stepsize: 0.05
    l_switch_limber: 30.
    hyper_sampling_flat: 32.
    l_max_g: 40
    l_max_ur: 35
    l_max_pol_g: 60
    ur_fluid_approximation: 2
    ur_fluid_trigger_tau_over_tau_k: 130.
    radiation_streaming_approximation: 2
    radiation_streaming_trigger_tau_over_tau_k: 240.
    hyper_flat_approximation_nu: 7000.
    transfer_neglect_delta_k_S_t0: 0.17
    transfer_neglect_delta_k_S_t1: 0.05
    transfer_neglect_delta_k_S_t2: 0.17
    transfer_neglect_delta_k_S_e: 0.17
    accurate_lensing: 1
    start_small_k_at_tau_c_over_tau_h: 0.0004
    start_large_k_at_tau_h_over_tau_k: 0.05
    tight_coupling_trigger_tau_c_over_tau_h: 0.005
    tight_coupling_trigger_tau_c_over_tau_k: 0.008
    start_sources_at_tau_c_over_tau_h: 0.006
    l_max_ncdm: 30
    tol_ncdm_synchronous: 1.e-6
    \end{lstlisting}
    \caption{Baseline settings used for \texttt{class} theory calculations, updating the default assumptions of the public \texttt{class} version. In this work, we use a version of the code that has been updated to implement the latest {\tt HMcode-2020} model for the non-linear power spectrum, provided by J.~Lesgourgues (developed from {\tt class v3.2.2}).}
    \label{snip:class_accuracy}
\end{figure}

\begin{figure*}[htp]
	\centering
 \includegraphics[width=\textwidth]{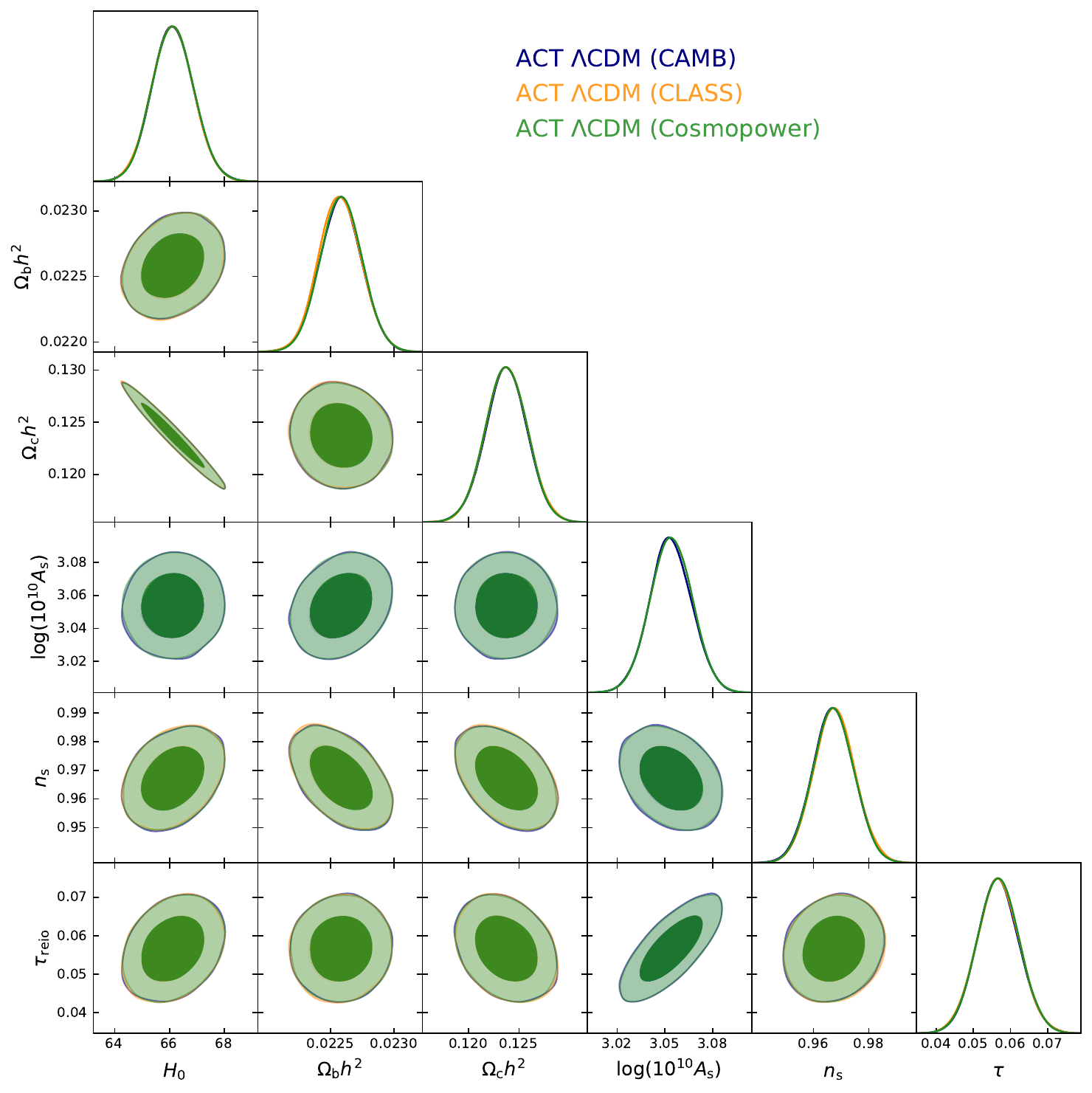}
	\caption{Comparison of baseline \lcdm\ parameter posteriors, validating different approaches to compute the lensed CMB theory --- using \texttt{camb} (blue), \texttt{CosmoPower} (green), or \texttt{class} (orange).}
	\label{fig:lcdm_benchmark}
\end{figure*}

\begin{wraptable}{l}{\textwidth}
\centering
\begin{tabular}{c|c|c|c}
\hline
\hline
{\bf Model}  & {\bf Section} & {\bf Theory Code} & {\bf Likelihood}\\
 \hline
 Running of scalar spectral index & \S\ref{sec:running} & \texttt{camb}& \texttt{MFLike}/\texttt{ACT-lite}\\
  \hline
 $P_{\mathcal{R}}(k)$& \S\ref{sec:pk} & \texttt{camb}& \texttt{ACT-lite}\\
 \hline
  Isocurvature perturbations & \S\ref{sec:isocurvature} & \texttt{class}& \texttt{MFLike}\\
  \hline
  Tensor modes & \S\ref{sec:tensors} & \texttt{camb}& \texttt{ACT-lite}\\
  \hline
  Early dark energy& \S\ref{sec:ede} & \texttt{class}/{\tt camb}/ & \texttt{MFLike}\\
&  & class emulators~\citep{Qu:2024lpx} & \\
 \hline
  Varying electron mass & \S\ref{sec:var_me} & \texttt{class}& \texttt{MFLike}\\
 \hline
Varying electron mass and curvature & \S\ref{sec:var_me_ok} & \texttt{class}& \texttt{MFLike}\\
 \hline
  Varying fine-structure constant & \S\ref{sec:var_a} &\texttt{class}& \texttt{MFLike}\\
 \hline
   Varying fine-structure constant and curvature & \S\ref{sec:var_a_ok} &\texttt{class}& \texttt{MFLike}\\
 \hline
 Primordial magnetic fields & \S\ref{sec:pmf}  & modified \texttt{class}& \texttt{MFLike}\\
 \hline
CMB temperature & \S\ref{sec:tcmb}  & \texttt{class}& \texttt{MFLike}\\
 \hline
 Modified recombination history & \S\ref{sec:modrec}  & modified \texttt{class}& \texttt{MFLike}\\
 \hline
 Neutrino number, $\neff$ & \S\ref{sec:neu_n_m}  & \texttt{camb}/& \texttt{MFLike}/\texttt{ACT-lite}\\
 & & camb emulators~\citep{Jense_emulators}& \\
 \hline
 Neutrino mass, $\mnu$ & \S\ref{sec:neu_n_m}  & \texttt{camb}/& \texttt{MFLike}/\texttt{ACT-lite}\\
&  & camb emulators~\citep{Jense_emulators}& \\
 \hline
 $\neff+\mnu$ & \S\ref{sec:neu_n_m}  & \texttt{camb}/& \texttt{ACT-lite}\\
&  & camb emulators~\citep{Jense_emulators}& \\
 \hline
 Neutrino self-interactions & \S\ref{sec:neu_si}  & modified \texttt{camb} & \texttt{MFLike}\\
 \hline
Helium and deuterium & \S\ref{sec:bbn}  &  \texttt{camb} & \texttt{MFLike}/\texttt{ACT-lite}\\
 \hline
Axion-like particles & \S\ref{sec:axions}  &  modified \texttt{camb}/ & \texttt{ACT-lite}\\
 & &  camb emulators & \\
 \hline
DM-baryon interactions & \S\ref{sec:dm-b}  &  modified \texttt{class} & \texttt{MFLike}\\
 \hline
DM annihilation & \S\ref{sec:dm-ann}  &  \texttt{class} & \texttt{MFLike}\\
\hline
Self-interacting DR & \S\ref{sec:si_dr}  & \texttt{class} & \texttt{MFLike}\\
\hline
Interacting DR-DM & \S\ref{sec:i_dr_dm}  & \texttt{class} & \texttt{MFLike}\\
 \hline
Spatial curvature& \S\ref{sec:curv}  & \texttt{camb}& \texttt{MFLike}/\texttt{ACT-lite}\\
 \hline
 Dark energy equation of state, $w$& \S\ref{sec:de}  & \texttt{camb}& \texttt{MFLike}/\texttt{ACT-lite}\\
 \hline
Dark energy equation of state, $w_0/w_{\rm a}$& \S\ref{sec:de}  & \texttt{camb}& \texttt{MFLike}\\
 \hline
Interacting DE-DM & \S\ref{sec:dedm}  &  modified \texttt{class} & \texttt{MFLike}\\
 \hline
Modified gravity & \S\ref{sec:mg} &  modified \texttt{camb} & \texttt{MFLike}\\
 \hline
\hline
\end{tabular}
\caption{Summary of models explored in this paper. For each case, we list the Einstein-Boltzmann code and likelihood that are used for each model, noting when chains have been run with more than one code (for robustness tests and reproducibility). The likelihood codes and the baseline \lcdm\ results are presented in L25.}\label{tab:models}
\end{wraptable}
\clearpage

\section{ACT DR6 versus DR4 cosmology} \label{app:dr4dr6comp}

\begin{wrapfigure}{r}{0.5\textwidth}
	\centering
   \includegraphics[width=0.48\textwidth]{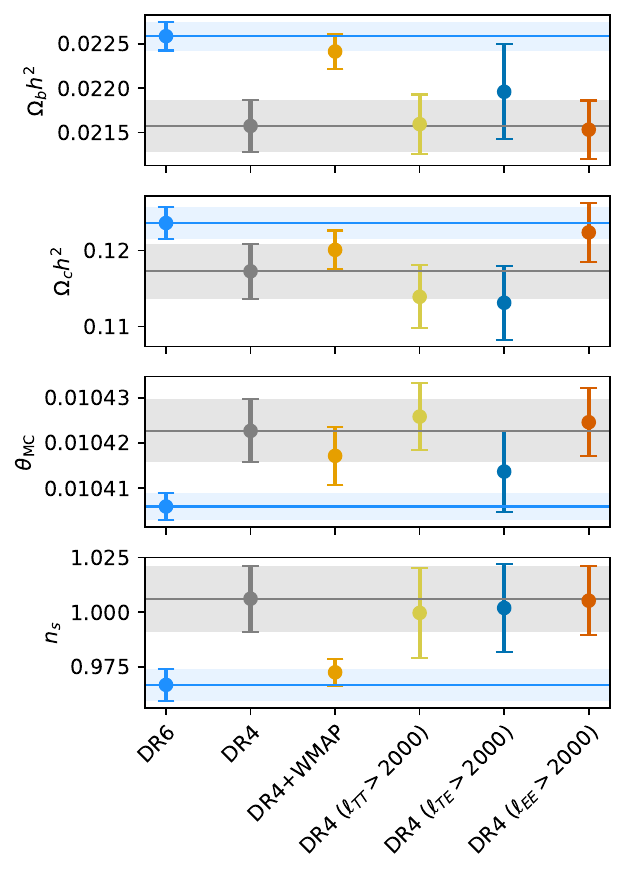}
	\caption{Comparison of \lcdm\ parameters (mean and $1\sigma$ error) estimated from different ACT datasets. The DR6 constraints (blue) are compared with various DR4 results from both ACT alone and ACT + \emph{WMAP},~\citep{Choi2020,Aiola2020}. In all cases we are combining with a measurement of the optical depth to reionization from \Planck\ {\tt Sroll2}. The DR6 estimates are in good agreement with the DR4 + \emph{WMAP} constraints, which comprised the baseline result presented in~\cite{Aiola2020}, but differ from the ACT-alone DR4 cosmology at $>2.8\sigma$, with $\Omega_b h^2$ driving most of this difference. The two datasets can be reconciled by discarding the DR4 TE measurements at $\ell<2000$, which could be impacted by beam modeling.}
	\label{fig:whisker_cosmology_dr4dr6}
\end{wrapfigure}

Here we compare results for the baseline \lcdm\ model derived in L25 from the DR6 maps with those from DR4 presented in~\cite{Choi2020,Aiola2020}. 
Figure~\ref{fig:whisker_cosmology_dr4dr6} shows the subset of \lcdm\ cosmological parameters measured by ACT for DR6 (L25) and DR4~\citep{Choi2020}, with the latter explored also in combination with \WMAP\ and for different subsets of the data. Considering the 4-dimensional space of these parameters, we find that the DR6 and DR4 ACT-alone parameters differ at the 2.8$\sigma$ level if we ignore correlations between the two datasets. This is therefore a lower limit on the difference if covariance were included. 
In particular, we find that the DR6 measurement of $\Omega_b h^2$ shifts upwards by $3\sigma$ compared to DR4. When ACT DR4 is combined with large-scale modes from \WMAP, which was the nominal combination in \cite{Aiola2020}, we find that DR6 agrees with the DR4 + \emph{WMAP} \lcdm\ best-fit model to within $1\sigma$.\footnote{Comparison of EDE constraints from DR4 and DR6 can be found in Appendix~\ref{app:ede_full_constraints}.}

\begin{figure*}[tp]
	\centering
	\includegraphics[width=0.495\textwidth]{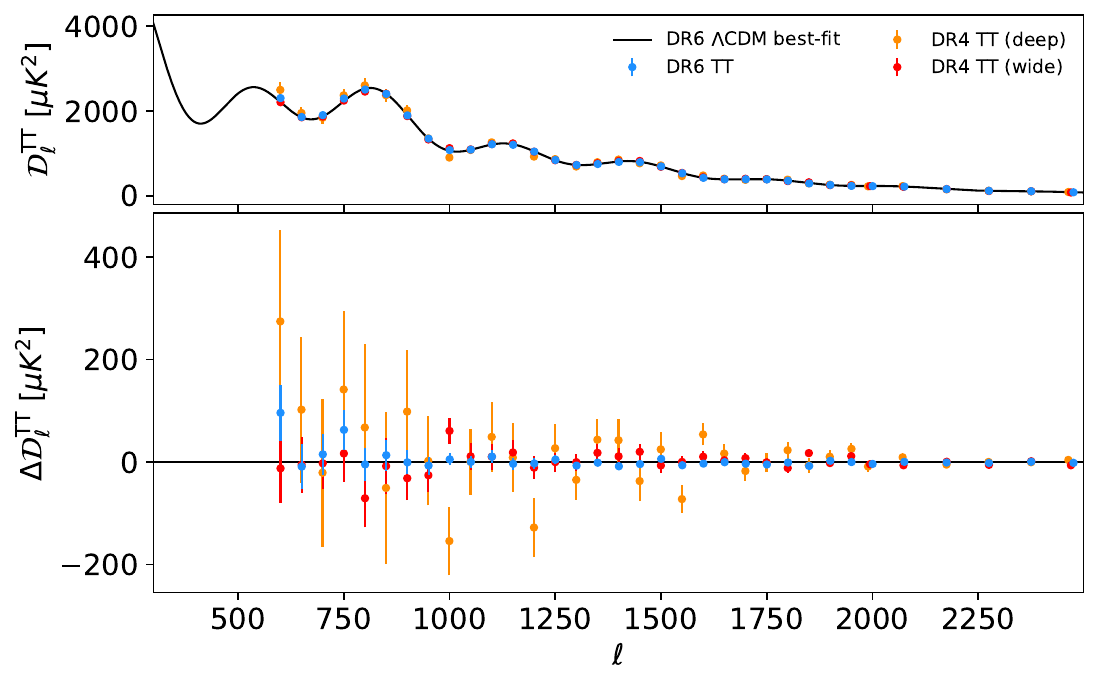}
    \includegraphics[width=0.48\textwidth]{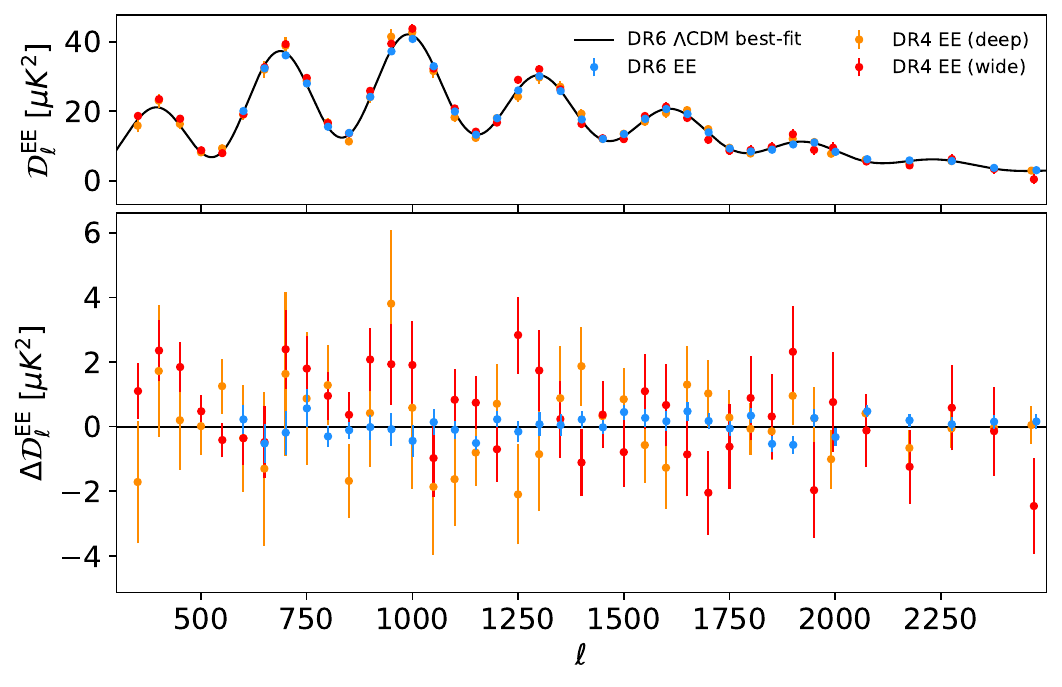}
	\centering
	\includegraphics[width=0.57\textwidth]{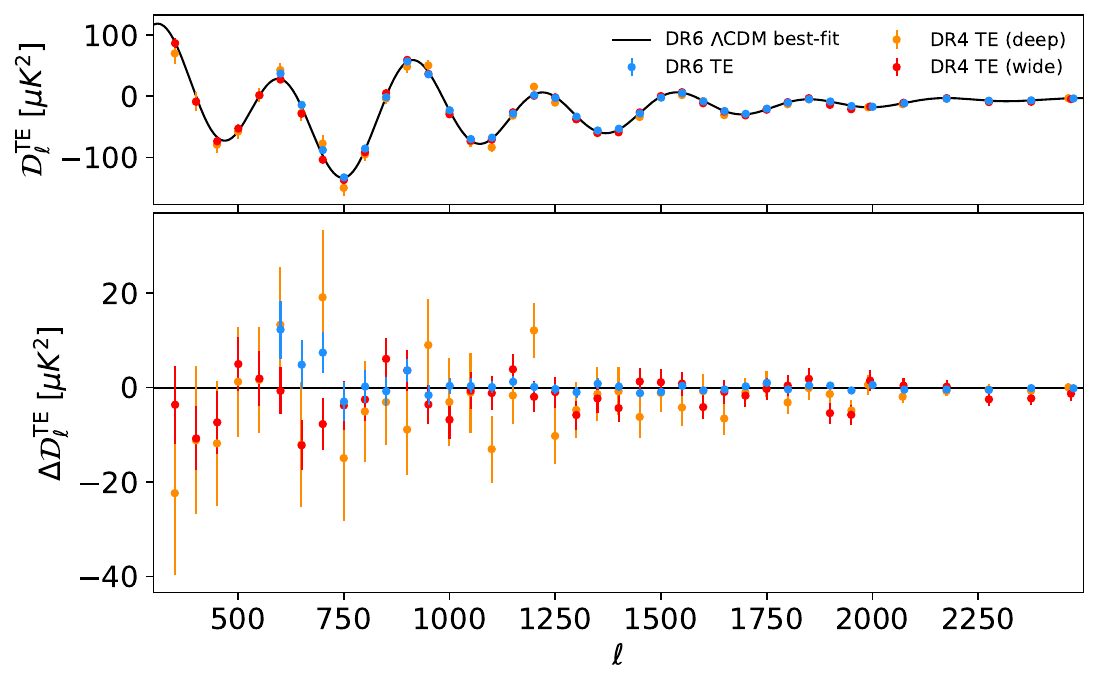}
    \includegraphics[width=0.35\textwidth]{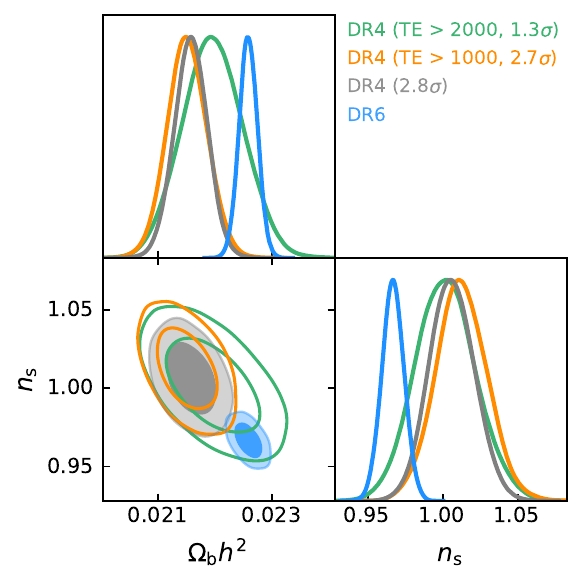}
    
	\caption{Comparison of the DR6 and DR4 CMB power spectra (TT in top left, EE in top right, and TE in bottom left) with respect to the DR6 \lcdm\ best-fit model (L25). The DR4 residuals in TE are mostly negative at $\ell<2000$, disfavoring the DR6 \lcdm\ cosmology. The bottom right panel shows the $\Omega_b h^2-n_s$ parameter space, which, as shown in~\citet{Aiola2020}, is very sensitive to the amplitude of the TE spectrum. Discarding the DR4 TE data at $\ell<2000$ brings the DR6 and DR4 contours in closer agreement.}
	\label{fig:residuals_te_omegabns_dr4dr6}
\end{figure*}

In terms of understanding the difference between the cosmology preferred by ACT-alone for DR4 and the best-fit model for DR6 (or DR4 + \emph{WMAP}), we note that \citet{Aiola2020} found that an artificial 5\% re-calibration of TE compared to TT (dividing the DR4 TE bandpowers by 1.05) would bring the ACT-alone DR4 parameters into better agreement with DR4 + \emph{WMAP} or \Planck. This re-calibration had the effect of moving parameters along the degeneracy direction for $\Omega_b h^2-n_s$ (see Fig.~14 of~\citealt{Aiola2020}). In comparing CMB power spectra for DR4 and DR6, shown in Fig.~\ref{fig:residuals_te_omegabns_dr4dr6}, we find good agreement in TT and EE, and note that the TE residuals at $\ell<2000$ with respect to the DR6 best-fit \lcdm\ model are predominantly negative, indicating less power in the DR4 spectra compared to DR6. The DR6 TE spectra are more consistent with the model preferred by DR4 + \emph{WMAP} and \Planck. 
Although we have not re-analyzed the DR4 data at this stage, we speculate that our improved modeling of temperature-to-polarization leakage between DR4 and DR6 could impact the TE measurement. During the DR6 beam calibration analysis, we determined that the DR4 leakage estimation method was insensitive to low-$\ell$ ($\ell < 2000$) leakage. Given that in DR6 we find significant leakage at these large angular scales, we have reason to speculate that the $\ell < 2000$ leakage in DR4 could have been underestimated, both in central value and in uncertainty. A rough estimate suggests that the $\Delta D_{\ell}^{\rm TE}$ from underestimating the leakage has an RMS value between $2$ and $4$~$\mu$K$^2$ for $350<\ell< 1000$ and between $0.2$ and $0.3$~$\mu$K$^2$ for $1000<\ell<2000$ --- more details are given in~\citealp{beams_inprep}. These numbers are estimated from the difference between the nominal DR6 leakage estimate and an estimate made using the DR4 leakage estimation method. Other factors that might have contributed to a difference in TE between DR6 and DR4 are improvements in the map-making procedure (see N25): in DR6 we have not subtracted an estimate of the pickup (due to the ground and potential other sources) during map-making and we have upgraded to a pointing matrix that uses bilinear interpolation instead of nearest-neighbor interpolation. Although it is not understood how these two changes could influence the TE spectrum, it has not been verified that these upgrades would leave the DR4 TE spectrum unchanged. Figure~\ref{fig:whisker_cosmology_dr4dr6} and the bottom right panel of Fig.~\ref{fig:residuals_te_omegabns_dr4dr6} also show that simply removing data in TE at $\ell<2000$ moves the DR4 limits into closer agreement with DR6. Cutting TT or EE at $\ell<2000$ (Fig.~\ref{fig:whisker_cosmology_dr4dr6}) does not yield the same agreement, expanding parameter degeneracies in different directions. 
This exploration of TT and EE subsets also tests other aspects noted in the DR4 analyses, including the impact of the DR4 lack of power in TT and the $\ell<1000$ region in EE where some deviations from \lcdm\ (e.g., early dark energy and self-interacting neutrinos) were moderately preferred.  

This assessment has been done assuming that the underlying model is \lcdm; we do not rule out the possibility that the difference between DR4 and DR6 is due to the true model not being \lcdm, as the two datasets do not fully overlap in angular scale, with DR4 having more statistical weight at smaller angular scales than DR6.

\clearpage

\section{Primordial matter power spectrum compilation}\label{app:pkcompilation}

\begin{wraptable}{r}{0.5\textwidth}
\centering
\begin{tabular}{c|c}
\hline
$k$ $\times 10^{2}$ [Mpc$^{-1}$] & $e^{-2\tau}P_{\mathcal{R}}(k)$ Prior Range $\times 10^{9}$\\ \hline \hline
0.01 &  0.00 -- 15.0 \\ \hline
0.03 &  0.10 -- 10.0 \\ \hline
0.10 &  0.30 -- 9.00 \\ \hline
0.32 &  0.50 -- 8.00 \\ \hline
1.06 &  1.40 -- 2.60 \\ \hline
1.23 &  1.40 -- 2.40 \\ \hline
1.43 &  1.40 -- 2.40 \\ \hline
1.66 &  1.60 -- 3.30 \\ \hline
1.92 &  1.60 -- 2.30 \\ \hline
2.23 &  1.60 -- 2.20 \\ \hline
2.59 &  1.60 -- 2.20 \\ \hline
3.00 &  1.60 -- 2.20 \\ \hline
3.48 &  1.68 -- 2.14 \\ \hline
4.03 &  1.75 -- 2.06 \\ \hline
4.68 &  1.50 -- 2.10 \\ \hline
5.42 &  1.50 -- 2.10 \\ \hline
6.29 &  1.80 -- 2.05 \\ \hline
7.29 &  1.70 -- 2.00 \\ \hline
8.45 &  1.75 -- 2.00 \\ \hline
9.81 &  1.76 -- 1.95 \\ \hline
11.4 &  1.70 -- 2.00 \\ \hline
13.2 &  1.70 -- 1.93 \\ \hline
15.3 &  1.65 -- 2.00 \\ \hline
17.7 &  1.51 -- 2.12 \\ \hline
20.6 &  1.32 -- 2.30 \\ \hline
23.8 &  1.04 -- 2.55 \\ \hline
27.7 &  0 -- 4.26 \\ \hline
32.1 &  0 -- 10.7 \\ \hline
37.2 &  0 -- 20.0 \\ \hline
43.1 &  0 -- 8.46 \\ \hline
\end{tabular}
  \caption{Central-bin wavenumber and prior ranges used for sampling $e^{-2\tau}P_{\mathcal{R}}(k)$ for each $k$ bin.}
\label{table:binned priors}
\end{wraptable}

The primordial power spectrum constraints presented in~\S\ref{sec:pk} are derived sampling the primordial power over 30 bins centered on specific $k$ wavenumbers, with individual flat uninformative priors as summarized in Table~\ref{table:binned priors}.

\begin{figure}
	\centering
\includegraphics[width=0.8\textwidth]{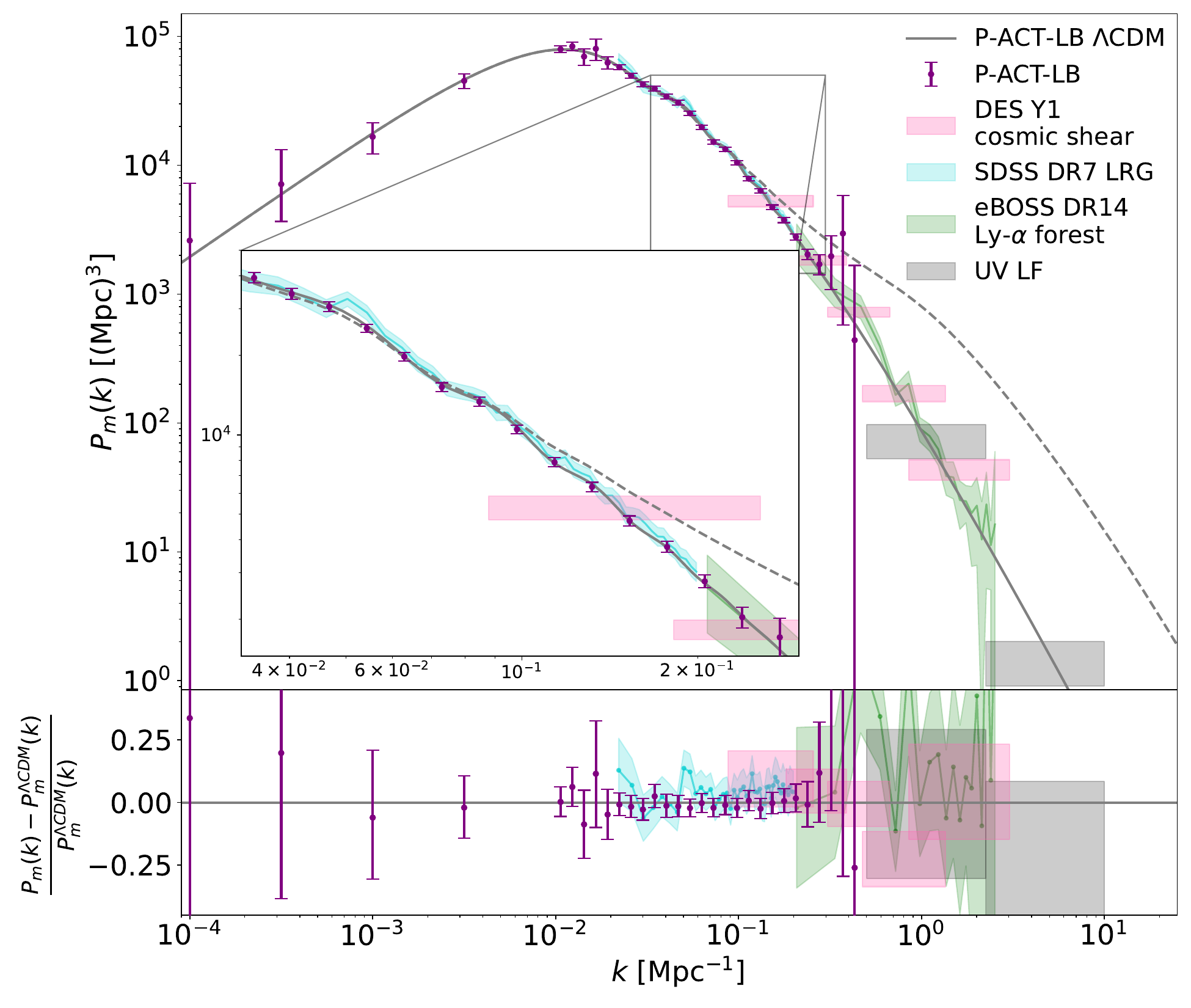}

	\caption{Constraints on the linear matter power spectrum.  The \pactlb\ best-fit \lcdm\ linear matter power spectrum prediction is shown as a solid gray line; the dashed gray line shows the non-linear power spectrum computed from this best-fit \lcdm\ model using {\tt HMcode}. 
    The extrapolation shown here includes propagation of the cosmological parameter uncertainties on the transfer function.
    Our \pactlb\ reconstruction of the binned linear $P_{\mathcal{R}}(k)$, presented in Fig.~\ref{fig:binned Pk} in~\S\ref{sec:pk}, is shown in purple. The \pact\ CMB dataset dominates this measurement. 
    Other constraints are shown from DES-Y1 cosmic shear~\citep{2018PhRvD..98d3528T}, SDSS luminous red galaxies~\citep{2010MNRAS.404...60R}, eBOSS Lyman-$\alpha$ forest~\citep{2018ApJS..235...42A,2019MNRAS.489.2247C}, and HST measurements of the UV galaxy luminosity function (UV LF) \citep{2022ApJ...928L..20S}, as labeled. This plot was made based on code from \cite{2022ApJ...928L..20S} and \cite{2019MNRAS.489.2247C}. The bottom panel shows the fractional residuals with respect to the \pactlb\ best-fit \lcdm\ linear power spectrum, with the y-axis optimized to highlight the scales more precisely measured.}
	\label{fig:matter power}
\end{figure}

The measurement across bins presented in~\S\ref{sec:pk} can be mapped onto the linear matter power spectrum through the matter transfer function $\mathcal{T}(k)\equiv T(k)/k^2$ via
\begin{align}
    P_m(k, z = 0) &= \frac{2\pi^2}{k^3} P_{\mathcal{R}}(k) \bigg(\frac{T(k)}{k^2}\bigg)^2k^4 \nonumber \\
    &= 2\pi^2 k P_{\mathcal{R}}(k) \mathcal{T}^2(k) \,,
\end{align}
where the dimensionless primordial power is converted to units of Mpc$^{3}$ through the $2\pi^2/k^3$ prefactor (note that $T(k)$ is dimensionless). As done in previous works, we use this relationship to show the CMB constraints on the primordial power spectrum alongside those from late-time probes such as galaxy surveys \citep[e.g.,][]{2002PhRvD..66j3508T, 2004ApJ...606..702T, 2012ApJ...749...90H, Planck_2018_overview, 2019MNRAS.489.2247C}. Our main results for the binned $P_{\mathcal{R}}(k)$ posterior distributions from \pactlb\ (Fig.~\ref{fig:binned Pk}) are projected onto the linear matter power spectrum in Fig.~\ref{fig:matter power}. We take the samples from our chains of the binned $P_{\mathcal{R}}(k)$ analysis using \pactlb\ and compute the linear matter power spectrum as a derived parameter in order to account for the uncertainties from the cosmology on the transfer function. From Fig.~\ref{fig:binned Pk} we note that \pact\ would give similar projections. We also show the \pactlb\ best-fit $\Lambda$CDM model for both the linear and non-linear matter power spectrum.\footnote{We use the \texttt{HMcode-2020}~\citep{Mead2020} dark-matter-only model of the non-linear power spectrum.}  
Other constraints are shown from the Dark Energy Survey (DES) \citep{2018PhRvD..98d3528T}, the Sloan Digital Sky Survey (SDSS) \citep{2010MNRAS.404...60R}, the extended Baryon Oscillation Spectroscopic Survey (eBOSS) \citep{2018ApJS..235...42A, 2019MNRAS.489.2247C}, and the Hubble Space Telescope (HST) measurements of the UV galaxy luminosity function (UV LF) \citep{2022ApJ...928L..20S}. Note that \pactlb\ detects non-zero power at $>95\%$ CL up to $k=0.43~\mathrm{Mpc}^{-1}$ whereas the \Planck-alone constraints cut off at $k=0.15~\mathrm{Mpc}^{-1}$. The bottom panel shows the residuals with respect to the \pactlb\ best-fit \lcdm\ linear power spectrum. At scales $k>0.1$~Mpc$^{-1}$, neighboring bins are more than 50\% correlated. The fact that the highest $k$ bins have mostly positive residuals is precisely why we find a slightly positive best-fit value for the running of the spectral index (in~\S\ref{sec:running}), although this is not statistically significant.

\clearpage
\section{Additional constraints on isocurvature}\label{app:isocurvature_full_constraints}

Table~\ref{table:isocurvature_constraints} presents the constraints on isocurvature models. In addition to the constraints on the amplitude of the primordial isocurvature power spectrum at two scales, we present constraints on the isocurvature power spectrum spectral index and the primordial isocurvature fraction,
\begin{equation}
\beta_{\rm iso}(k)\equiv \frac{\mathcal{P}_{\mathcal{I}\mathcal{I}}(k)}{\mathcal{P}_{\mathcal{R}\mathcal{R}}(k)+\mathcal{P}_{\mathcal{I}\mathcal{I}}(k)}.
\end{equation}
Overall, we find no evidence of isocurvature perturbations and impose stringent constraints on a range of isocurvature scenarios. 

\begin{table*}[h]
\centering
    \renewcommand{\arraystretch}{1.1}
    \begin{tabular}{l c c c c c}
    \hline
    \hline
     \multicolumn{1}{c}{Model and dataset}&  $~10^{10} P_{\mathcal{I}\mathcal{I}}^{(1)}~$    &   $~10^{10} P_{\mathcal{I}\mathcal{I}}^{(2)}~$  & $~100\,\beta_{\rm iso}^{(1)}~$ & $~100\,\beta_{\rm iso}^{(2)}~$ &  $~n_{\mathcal{I}\mathcal{I}}~$ \\ 
     \hline
    \textbf{Uncorrelated:} {\scriptsize($n_{\mathcal{I}\mathcal{I}}$=1)} & & & & &  \\ 
    \hline
    \hspace{10pt}CDI: \textsf{Planck} & $<0.9$ & $<0.9$  & $<3.7$  & $<4.2$ & $1$  \\
    \hspace{10pt}CDI: \pact & $<1.1$ & $<1.1$  & $<4.4$  & $<4.9$ & $1$  \\
    \hspace{10pt}CDI: \pactlb & $<1.1$ & $<1.1$  & $<4.7$  & $<5.1$ & $1$  \\
    \hline
    \hspace{10pt}NDI: \textsf{Planck} & $<2.1$ & $<2.1$  & $<8.5$  & $<9.4$ & $1$  \\
    \hspace{10pt}NDI: \pact & $<1.8$ & $<1.8$  & $<7.4$  & $<8.1$ & $1$ \\
    \hspace{10pt}NDI: \pactlb & $<1.8$ & $<1.8$  & $<7.4$  & $<8.1$ & $1$  \\
     \hline
    \textbf{Uncorrelated:} {\scriptsize({\rm free} $n_{\mathcal{I}\mathcal{I}}$)} & & &  & &  \\ 
    \hline
    \hspace{10pt}CDI: \textsf{Planck} & $<0.5$ & $<59$  & $<2.0$  & $41^{+34}_{-41}$ & $2.4^{+1.1}_{-1.2}$  \\
    \hspace{10pt}CDI: \pact & $<0.7$ & $<26$  & $<3.1$  & $<55$ & $2.0^{+1.1}_{-1.0}$ \\
    \hspace{10pt}CDI: \pactlb & $<0.7$ & $<26$  & $<3.1$  & $<55$ & $2.0^{+1.1}_{-0.9}$ \\
    \hline
     \hspace{10pt}NDI: \textsf{Planck} & $<1.5$ & $4.1^{+3.8}_{-4.1}$  & $<6.3$  & $16^{+13}_{-14}$ & $1.5^{+0.6}_{-0.7}$  \\
    \hspace{10pt}NDI: \pact & $<1.6$ & $<6.1$  & $<6.7$  & $<23$ & $1.6^{+0.8}_{-1.0}$  \\
    \hspace{10pt}NDI: \pactlb & $<1.7$ & $<5.5$  & $<6.9$  & $<21$ & $1.5\pm 0.9$  \\
    \hline
     \textbf{Fully correlated:} {\scriptsize($n_{\mathcal{I}\mathcal{I}}=n_{\mathcal{R}\mathcal{R}}$)} & &  & & & \\ 
      \hline
    \hspace{10pt}CDI: \textsf{Planck} & $<0.028$ & $<0.025$  & $<0.12$  & $<0.12$ & $0.970\pm 0.010$ \\
    \hspace{10pt}CDI: \pact & $<0.025$ & $<0.023$  & $<0.11$  & $<0.11$ & $0.975^{+0.009}_{-0.010}$ \\
    \hspace{10pt}CDI: \pactlb & $<0.031$ & $<0.029$  & $<0.14$  & $<0.14$ & $0.977^{+0.007}_{-0.010}$ \\
     \hline
     \textbf{Fully anti-correlated:} {\scriptsize($n_{\mathcal{I}\mathcal{I}}=n_{\mathcal{R}\mathcal{R}}$)} & & &   & & \\
      \hline
    \hspace{10pt}CDI: \textsf{Planck} & $<0.031$ & $<0.026$  & $<0.13$  & $<0.13$ & $0.960\pm 0.010$  \\
    \hspace{10pt}CDI: \pact & $<0.027$ & $<0.023$  & $<0.11$  & $<0.11$ & $0.967^{+0.008}_{-0.010}$  \\
    \hspace{10pt}CDI: \pactlb & $<0.015$ & $<0.014$  & $<0.06$  & $<0.06$ & $0.972^{+0.007}_{-0.010}$ \\
    \hline
    \end{tabular}
    \caption{Constraints on isocurvature perturbations for the models and dataset combinations considered in~\S{\ref{sec:isocurvature}}. We report the one-tailed 95\% upper bound for parameters that are not detected; otherwise, we report the two-tailed 95\% CL (chosen to facilitate comparison with results from Table 14 of~\citealt{Planck2018_inflation}).}\label{table:isocurvature_constraints}
\end{table*}
\clearpage

\section{Additional constraints on early dark energy models}\label{app:ede_full_constraints}

\begin{figure}[htb]
    \centering
    \includegraphics[width=.4\columnwidth]{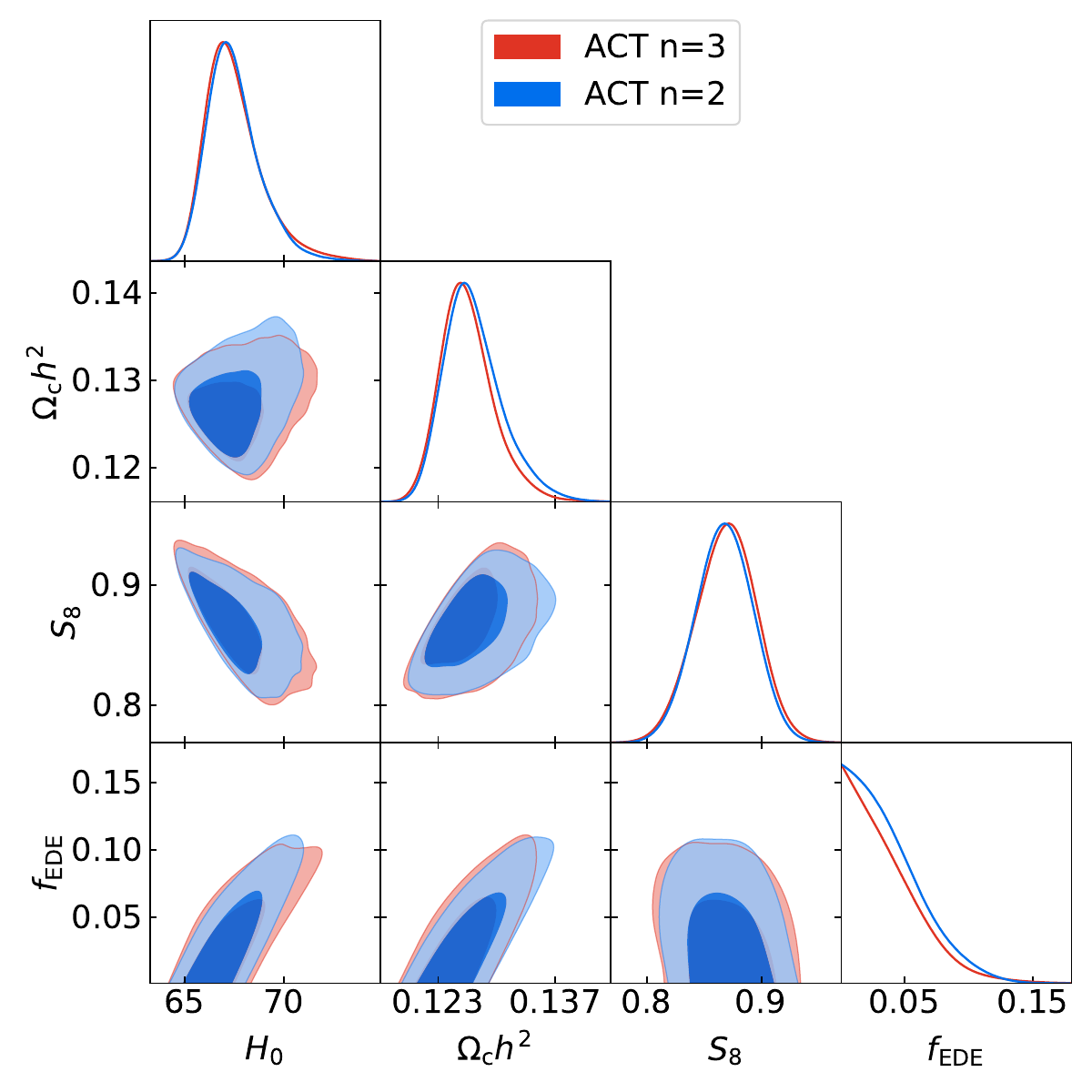}
     \includegraphics[width=.4\columnwidth]{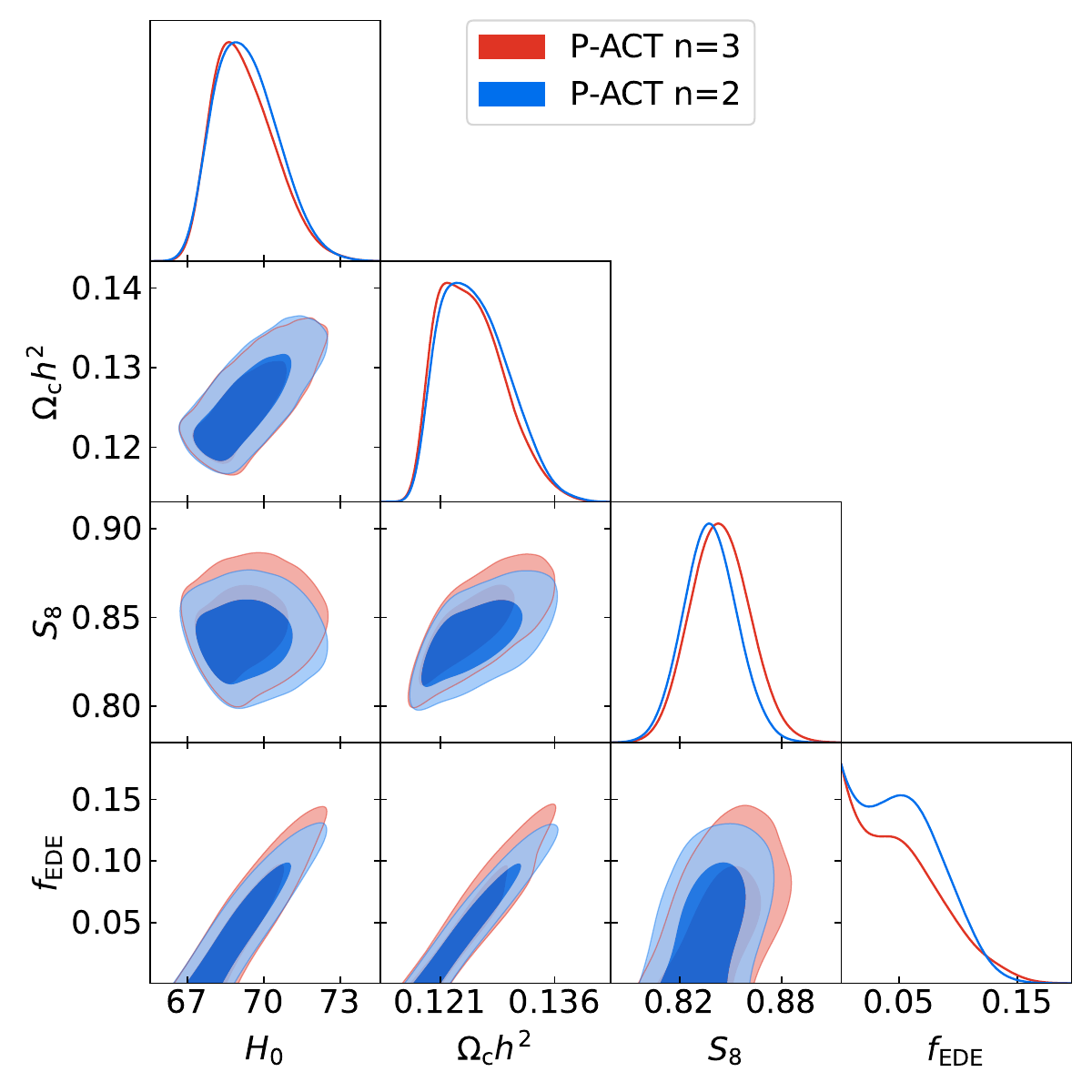}
      \includegraphics[width=.4\columnwidth]{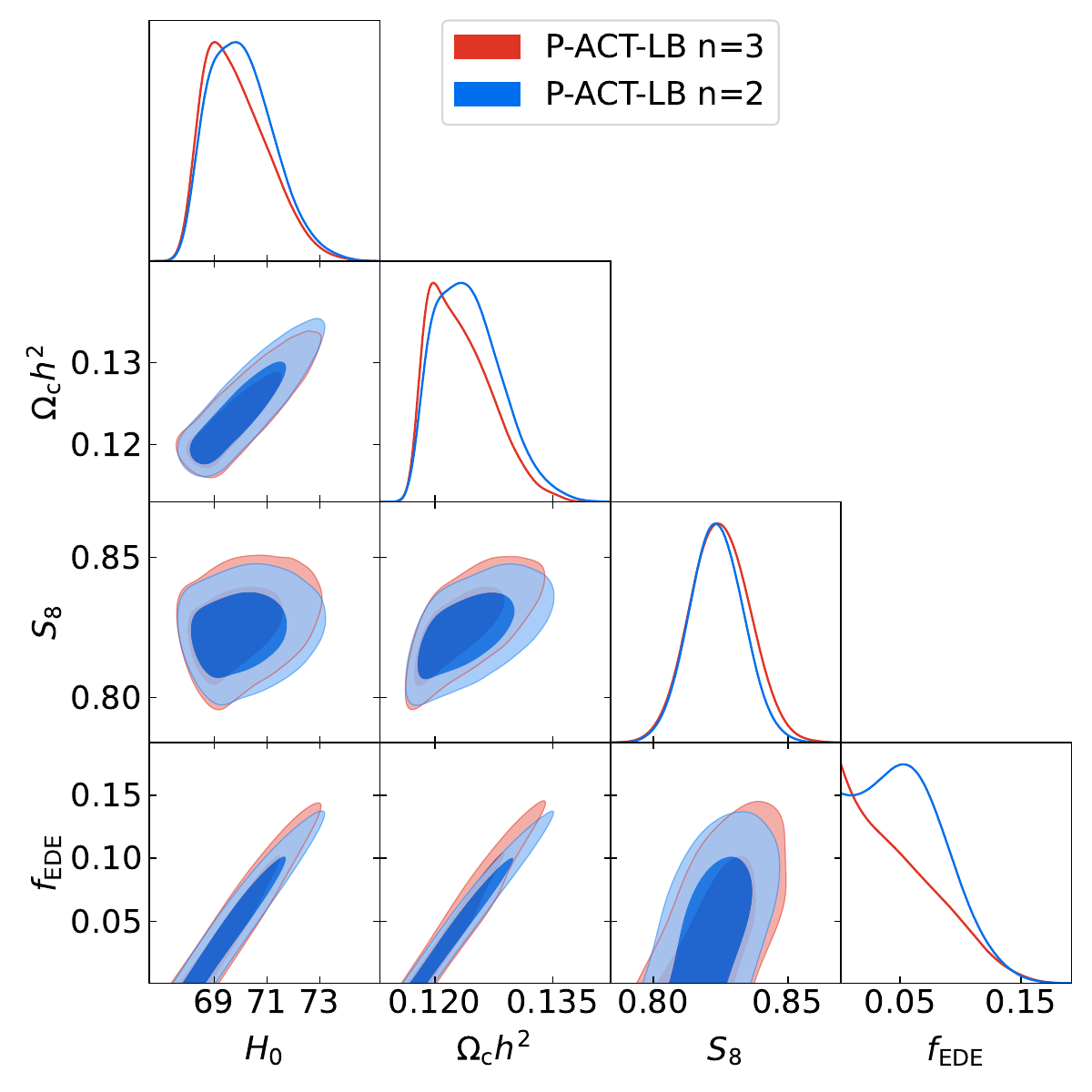}
       \includegraphics[width=.4\columnwidth]{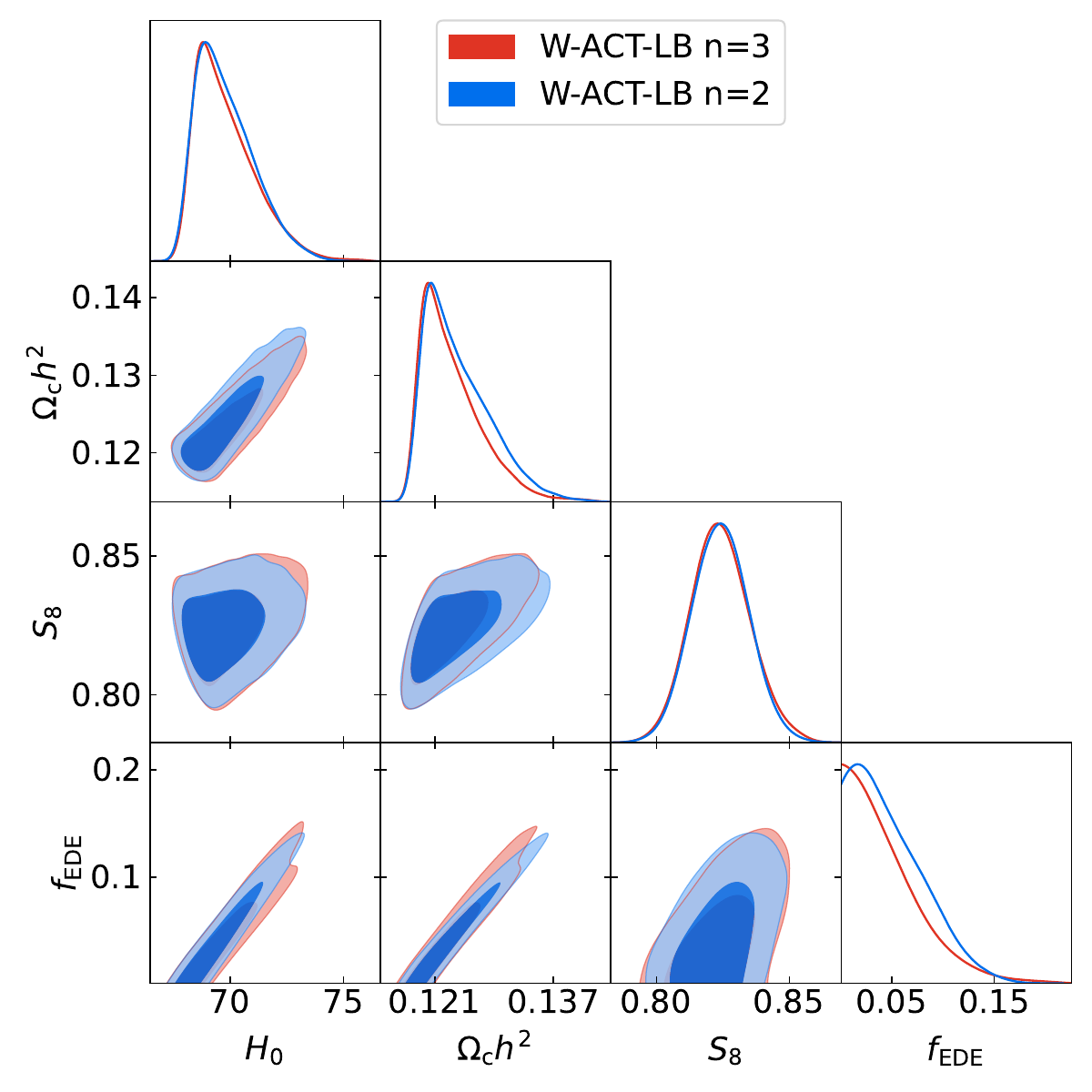}
    \includegraphics[width=.4\columnwidth]{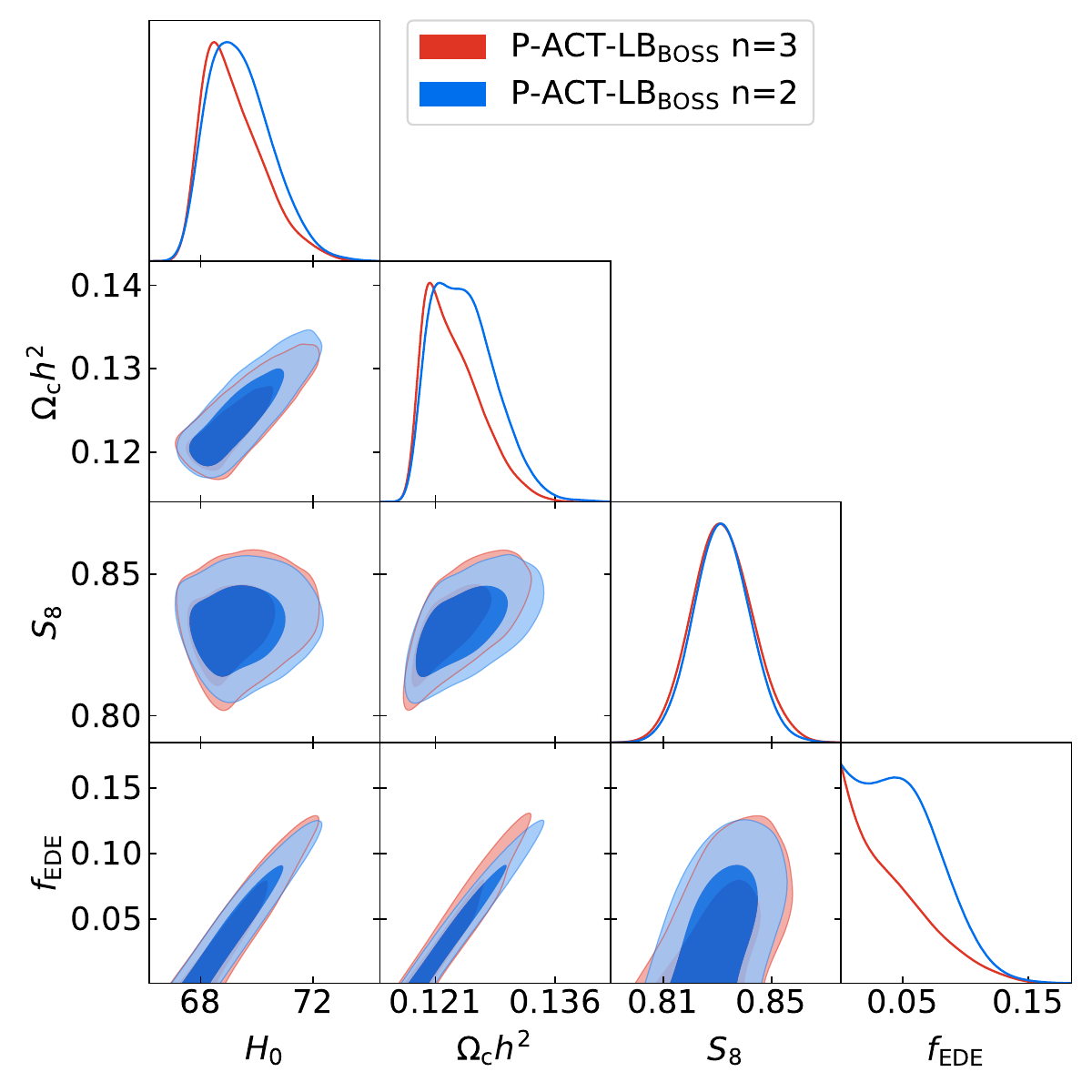}
    
    \caption{Comparison of constraints from \act\ (top left), \pact\ (top right), \pactlb\ (middle left), \wactlb\ (middle right), and \pactlbb\ (bottom) for EDE models with $n=2$ (blue) and $n=3$ (red).} \label{fig:eden2n3}  
\end{figure}

Here we compare the constraints for EDE models with $n=3$ (our baseline) and $n=2$. Figure~\ref{fig:eden2n3} shows the marginalized posterior comparisons for various dataset combinations, and Table~\ref{table:ede_n2_n3_full} provides the numerical constraints.  As for $n=3$, we find no evidence of $n=2$ EDE.  The $n=3$ model is analyzed using \texttt{CosmoPower}-based emulators of \texttt{class}, whereas the $n=2$ model is run using \texttt{camb} due to instability of \texttt{class} for this model. Note that both \texttt{class} and \texttt{camb} solve the full perturbed Klein-Gordon equation for the EDE model (in \texttt{camb} via the \texttt{EarlyQuintessence} module). 
As a validation, we check that the two codes yield very similar constraints for the $n=3$ model.

\begin{figure}[htp]
    \centering
    \includegraphics[width=.4\columnwidth]{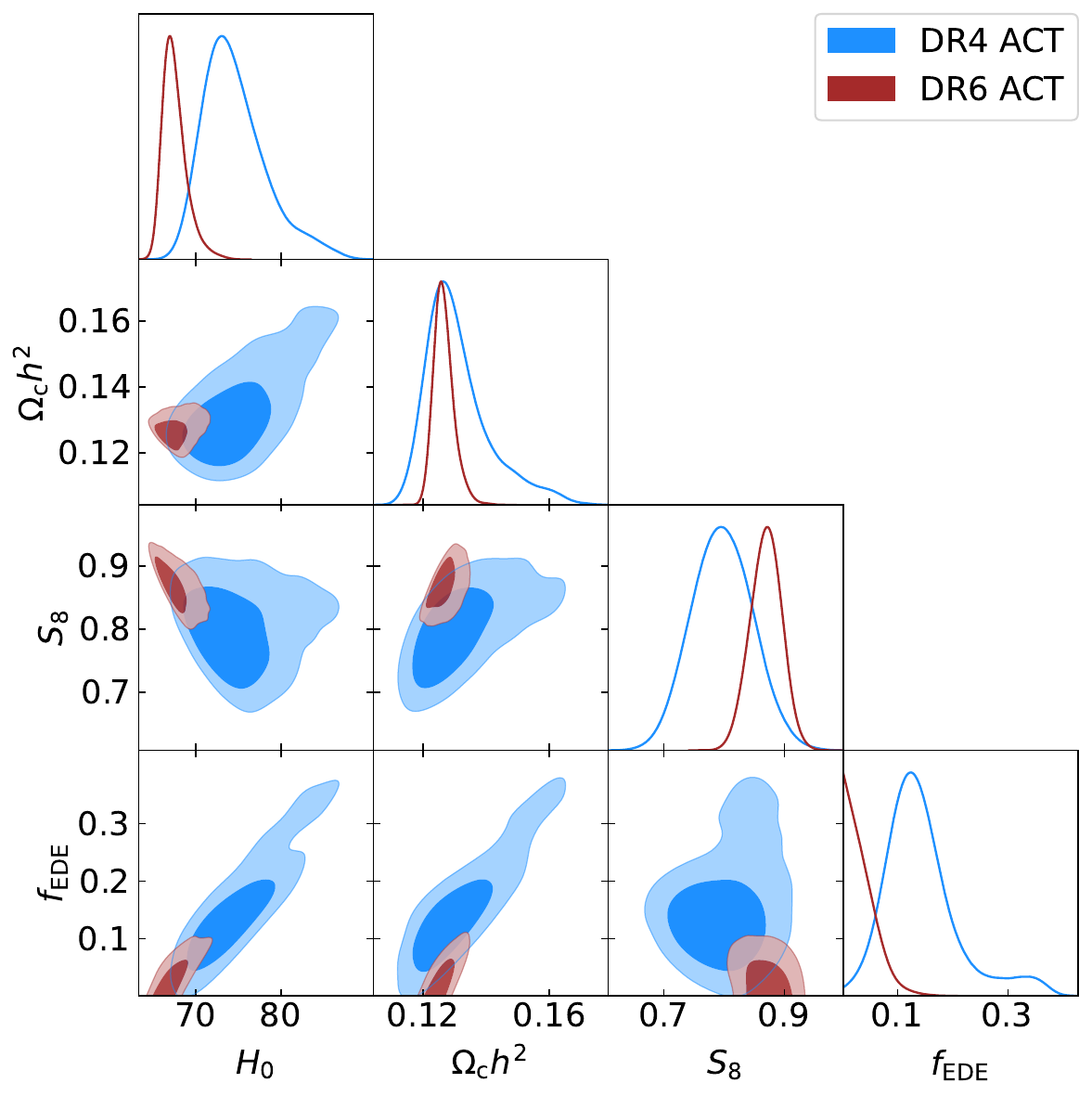}
     \includegraphics[width=.4\columnwidth]{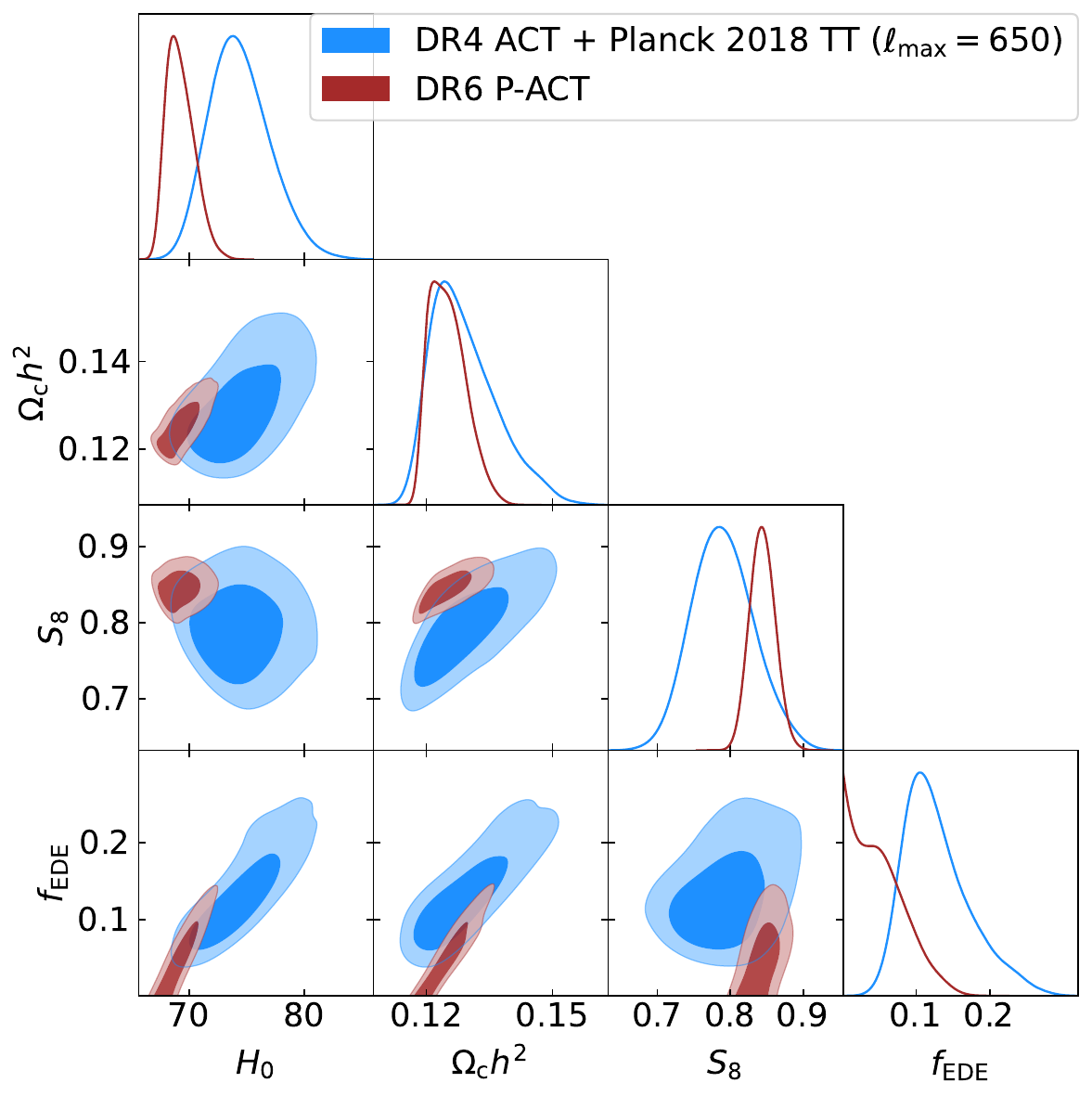}
      \includegraphics[width=.4\columnwidth]{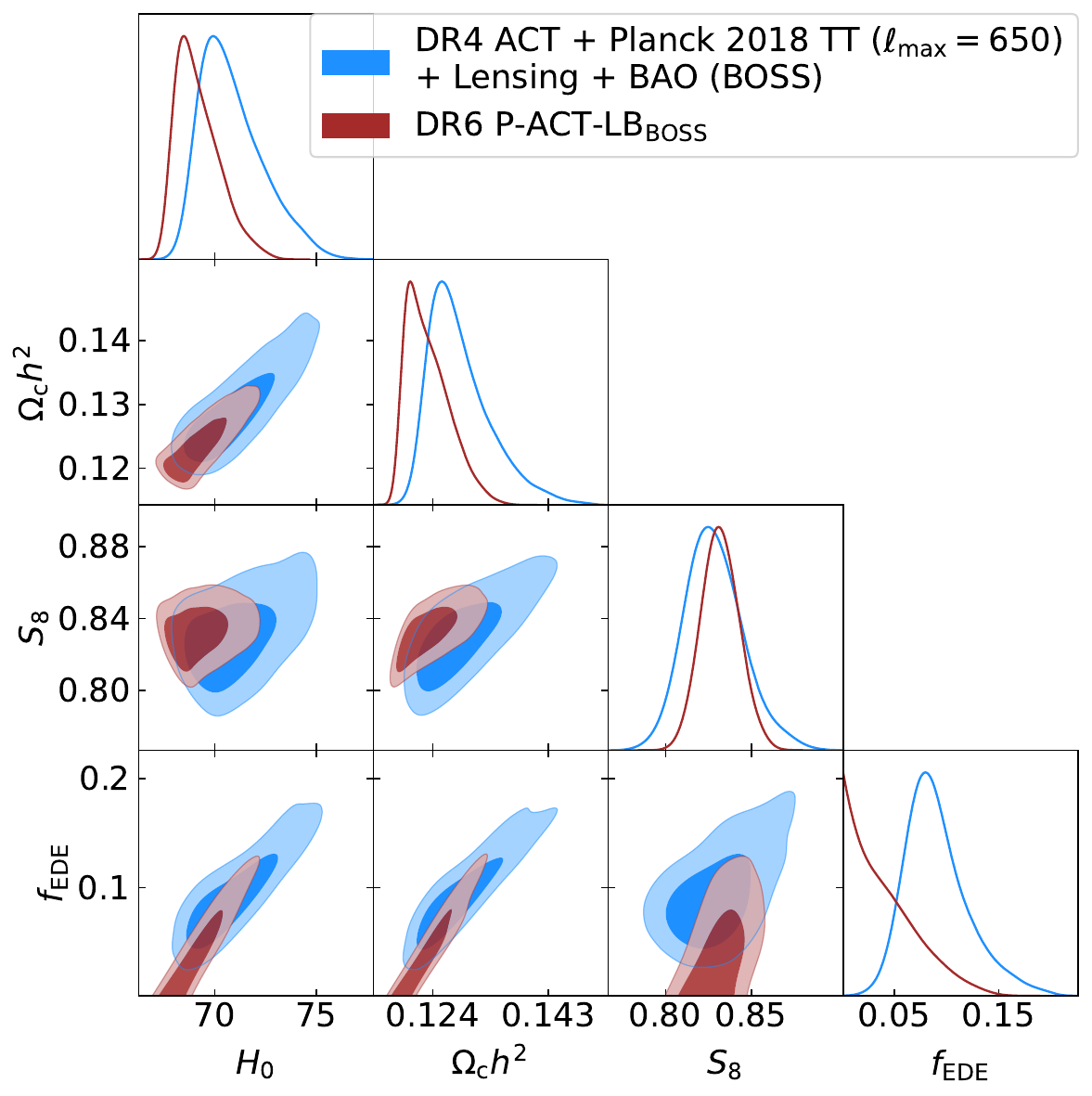}
    \caption{Comparison of constraints on the $n=3$ EDE model from \act\ (top left), \pact\ (top right), and \pactlbb\ (bottom) from the DR4 (blue) and DR6 (brown) data. The mild hint of non-zero EDE in DR4~\citep{Hill_ACT_EDE} is not seen in the more-sensitive DR6 dataset.  Beyond the significantly increased sensitivity, the DR6 dataset also benefits from improved map-making and systematic modeling compared to DR4, resulting in better-understood beams, transfer functions, and leakage corrections (see Appendix~\ref{app:dr4dr6comp} for further discussion and details).} \label{fig:ede_dr4_comparison}  
\end{figure}

To facilitate comparisons with earlier work, we also compare constraints on the $n=3$ EDE model from ACT DR4 \citep{Hill_ACT_EDE} to those obtained here from ACT DR6, as shown in Fig.~\ref{fig:ede_dr4_comparison}.  We show results for ACT alone (DR6 versus DR4), \pact\ versus the combination of ACT DR4 with \Planck\ 2018 TT data up to $\ell_{\rm max} = 650$, and \pactlbb\ versus the combination of ACT DR4 with \Planck\ 2018 TT data ($\ell_{\rm max} = 650$), \Planck\ 2018 CMB lensing data, and BAO data from BOSS (and pre-BOSS surveys).  We consider results using BOSS BAO data here to ease the DR4-DR6 comparison.  Significant portions of parameter space allowed by the DR4 data are excluded by the DR6 data, and the moderate hint of non-zero EDE in DR4 is no longer present in the more-sensitive DR6 dataset (see also Appendix~\ref{app:dr4dr6comp}). However, we also see a narrow degeneracy direction in the parameter posteriors (e.g., in the $f_{\rm EDE}$--$\Omega_c h^2$ panel) that is difficult to probe. Future experiments, such as the Simons Observatory and CMB-S4, will be needed to fully break this degeneracy and robustly detect or exclude pre-recombination EDE (see, e.g.,~\citealt{2025JCAP...01..033K}).

\begin{table}[h]
\centering
\begin{tabular}{c|c|c|c|c|c}
\hline
 &   $\Delta\chi^2$ & Pref. (in $\sigma$) & $H_0^{\rm (EDE)}$ & $f_{\rm EDE}$ & $\log_{10}{z_c}$ \\ \hline \hline
\act &$1.4$  &$0.4$  &$66.4$ &$0.032$ & 3.04   \\ \hline
\pact & $3.6$ & $1.0$ &$ 69.0$ &$0.046$ & 3.52   \\ \hline
\pactlb & $5.5$ & $1.5$ &$70.3$ &$0.064$ & 3.50 \\ \hline
\wactlb &$3.4$  &$1.0$ &$70.1$ &$0.059$ &3.49 \\ \hline
\end{tabular}
  \caption{The $\Delta\chi^2=\chi^2_{\Lambda\mathrm{CDM}}-\chi^2_{\mathrm{EDE}}$ from the multifrequency likelihood MAP points for the $n=2$ EDE model as compared to those found for $\Lambda$CDM for each dataset combination, and preference (in units of $\sigma$) for EDE using the likelihood-ratio test statistic. The MAP estimates for $H_0$, $f_{\rm EDE}$, and $\log_{10}{z_c}$ in the EDE model are also shown. As with the $n=3$ model, the data show no significant preference for non-zero EDE.}
\label{table:ede_chi2_n2}
\end{table}
\vspace{-0.2in}

\begin{table*}[h]
\centering
    \renewcommand{\arraystretch}{1.1}
    \begin{tabular}{l c c c c c c c}
    \hline
    \hline
     \multicolumn{1}{c}{Dataset}&  $\Omega_\mathrm{c}h^2$    &   $\Omega_\mathrm{b}h^2$  & $\ln 10^{10}A_s$ & $n_\mathrm{s}$ &  $H_0$ & $\tau$ &$f_\mathrm{EDE}$ \\ 
     \hline
     
     \textbf{\act}  & &  & & & \\ 
      \hline
    \hspace{10pt}$n=2$ & $0.1272^{+0.0026}_{-0.0040}$ & $0.02259\pm 0.00020$  & $3.061\pm 0.015$  & $0.962^{+0.020}_{-0.012}$ & $67.5^{+0.9}_{-1.5}$ & $0.0574\pm 0.0056$ & $< 0.091$ \\
    \hspace{10pt}$n=3$ & $0.1265^{+0.0024}_{-0.0036}$ & $0.02260\pm 0.00020$  & $3.058\pm 0.015$  & $0.964^{+0.022}_{-0.014}$ & $67.5^{+0.9}_{-1.7}$ & $0.0571\pm 0.0057$ & $< 0.088$  \\
     \hline
     
     \textbf{\pact}  & & &   & & \\ 
      \hline
    \hspace{10pt}$n=2$ & $0.1256^{+0.0033}_{-0.0055}$ & $0.02264^{+0.00015}_{-0.00018}$  & $3.066\pm 0.015$  & $0.9767\pm 0.0059$ & $69.3^{+1.0}_{-1.5}$ &$0.0607^{+0.0056}_{-0.0067}$ & $< 0.11$  \\
    \hspace{10pt}$n=3$ & $0.1251^{+0.0031}_{-0.0055}$ & $0.02267^{+0.00017}_{-0.00019}$  & $3.065\pm 0.015$  & $0.9785^{+0.0063}_{-0.0079}$ & $69.3^{+0.9}_{-1.5}$ &$0.0599^{+0.0055}_{-0.0066}$ & $< 0.12$   \\
    \hline
    
    \textbf{\pactlb}  & & &   & & \\ 
      \hline
    \hspace{10pt}$n=2$ & $0.1243^{+0.0030}_{-0.0053}$ & $0.02269^{+0.00014}_{-0.00017}$  & $3.070^{+0.012}_{-0.014}$  & $0.9796^{+0.0048}_{-0.0055}$ & $70.1^{+0.9}_{-1.5}$  &$0.0634^{+0.0057}_{-0.0068}$ & $< 0.11$ \\
    \hspace{10pt}$n=3$ & $0.1233^{+0.0025}_{-0.0052}$ & $0.02270^{+0.00015}_{-0.00018}$  & $3.065^{+0.011}_{-0.013}$  & $0.9809^{+0.0061}_{-0.0075}$ & $69.9^{+0.8}_{-1.5}$  &$0.0619^{+0.0056}_{-0.0066}$ & $< 0.12$ \\
    \hline

    \textbf{\wactlb}  & & &   & & \\ 
      \hline
    \hspace{10pt}$n=2$ & $0.1241^{+0.0024}_{-0.0056}$ & $0.02272\pm 0.00018$  & $3.067^{+0.012}_{-0.014}$  & $0.9755\pm 0.0061$ & $69.8^{+0.8}_{-1.6}$ &$0.0630^{+0.0057}_{-0.0066}$ & $<0.12$ \\
    \hspace{10pt}$n=3$ & $0.1233^{+0.0019}_{-0.0050}$ & $0.02273\pm 0.00019$  & $3.064^{+0.011}_{-0.013}$  & $0.9766^{+0.0059}_{-0.0080}$ & $69.8^{+0.7}_{-1.7}$ &$0.0614^{+0.0056}_{-0.0066}$ & $< 0.12$  \\
    \hline

    \textbf{\pactlbb}  & & &   & & \\ 
      \hline
    \hspace{10pt}$n=2$ & $0.1246^{+0.0027}_{-0.0050}$  & $0.02264^{+0.00015}_{-0.00017}$  & $3.065^{+0.011}_{-0.013}$ & $0.9769^{+0.0047}_{-0.0054}$ & $69.4^{+0.8}_{-1.4}$ &$0.0609^{+0.0054}_{-0.0064}$ & $< 0.11$  \\
    \hspace{10pt}$n=3$ & $0.1233^{+0.0021}_{-0.0045}$ & $0.02265^{+0.00015}_{-0.00018}$  & $3.061^{+0.011}_{-0.012}$  & $0.9777^{+0.0053}_{-0.0071}$ & $69.2^{+0.7}_{-1.3}$  &$0.0600^{+0.0052}_{-0.0063}$ &$< 0.10$ \\
    \hline

    \end{tabular}
    \caption{Constraints on EDE for $n=2$ and $n=3$ models, for various dataset combinations. All numbers are reported as 68\% confidence intervals, except $f_{\rm EDE}$, which is given as a 95\% upper bound.}\label{table:ede_n2_n3_full}
\end{table*}

\clearpage

\section{Additional constraints on modified recombination}\label{app:modrec_full_constraints}

\begin{figure*}[htp]
	\centering
 \includegraphics[width=\textwidth]{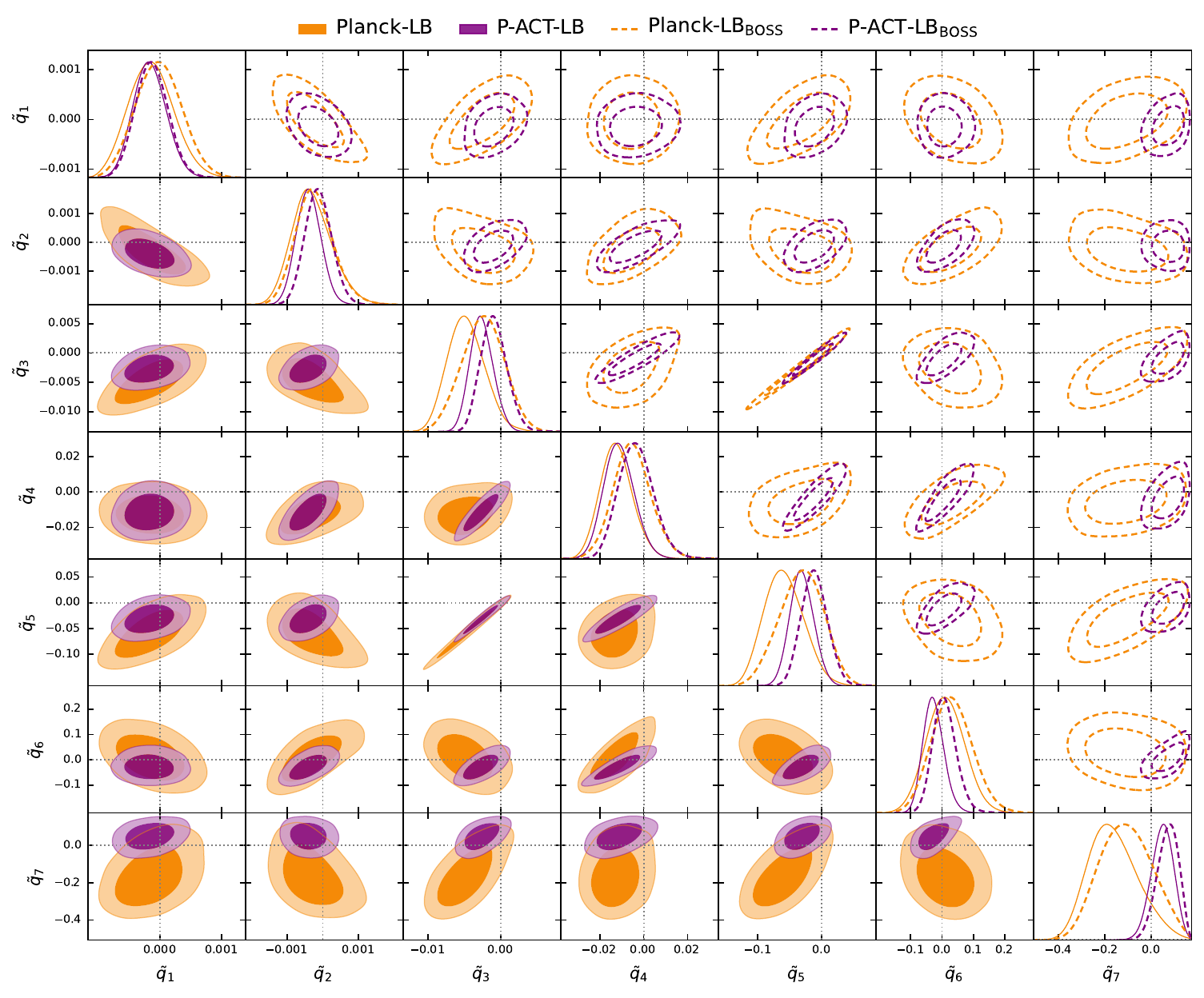}
	\caption{Marginalized parameter posteriors for the control points varied in the \texttt{ModRec} scenario analyzed in \S\ref{sec:modrec}. The bottom (top) panels use DESI (BOSS) BAO. The dotted gray lines indicate the standard recombination scenario ($\tilde{q}_i=0$).}
	\label{fig:modrec_control_points}
\end{figure*}

Figure~\ref{fig:modrec_control_points} shows the marginalized parameter posteriors for the control points\footnote{See~\cite{ Lynch:2024gmp} for the definition of $\tilde{q}_i$.} varied in the \texttt{ModRec} scenario studied in \S\ref{sec:modrec}. The lower triangle plot, shown with solid contours, compares the constraints from \plb\ (orange) and \pactlb\ (purple). The upper triangle plot, shown with dashed contours, compares the constraints from $\textsf{Planck-LB}_{\textsf{BOSS}}$ (orange) and \pactlbb\ (purple). Including ACT data significantly improves the constraints on the control points, yielding stringent bounds on the cosmological recombination history, which we find to be consistent with the standard recombination scenario ($\tilde{q}_i$=0; dotted gray lines) within 2$\sigma$ across all control points analyzed in the \texttt{ModRec} model. Note that the control points in neighboring redshift bins can be highly correlated.

\newpage 
\section{Impact of DESI DR2 BAO and optical depth measurements.} \label{app:newDESI_tau}

\subsection{DESI DR2}

\begin{figure*}[hp]
	\centering
   \includegraphics[width=0.48\textwidth]{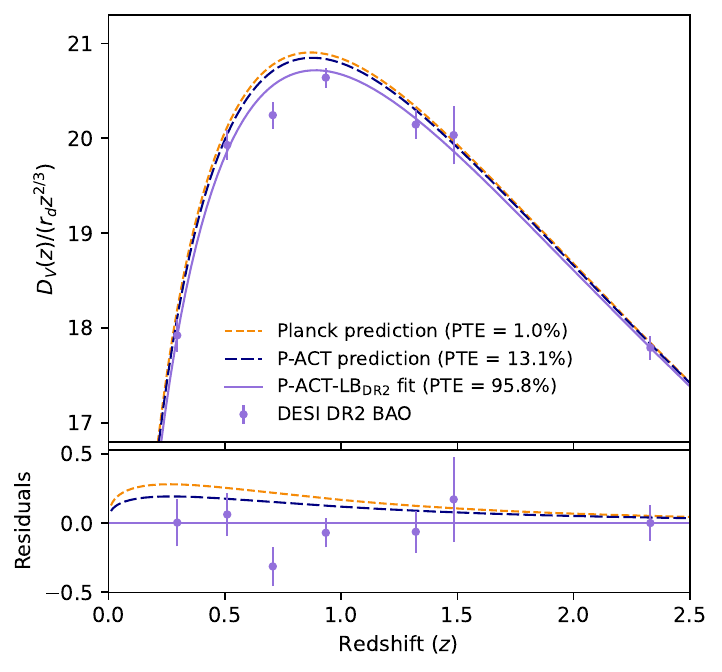}
    \includegraphics[width=0.51\textwidth]{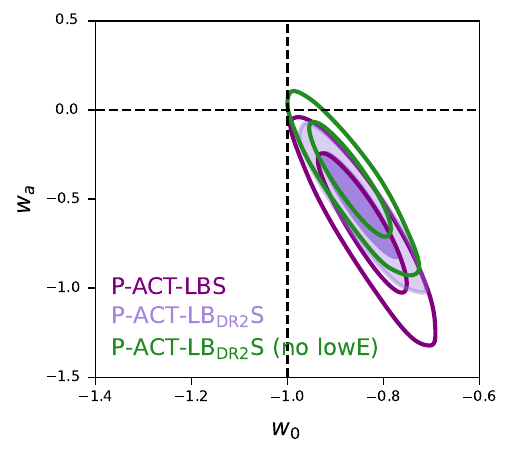}
    \vspace{-0.5cm}
	\caption{\emph{Left}: DESI DR2 BAO measurements combined in seven total data points and compared with CMB \lcdm\ predictions (from \Planck\ in orange and \pact\ in navy) and with the joint \textsf{P-ACT-LB$_{\rm DR2}$} fit (light purple), as in Fig.~\ref{fig:lensing_bao_residuals}. The \pact\ \lcdm\ best-fit model continues to provide good predictions for the DESI DR2 data. \emph{Right}: Comparison of evolving dark energy constraints using one (empty dark purple) or three years (filled light purple) of DESI BAO data, in combination with \Planck, ACT, CMB lensing, and SNIa data (\pactlbs). The evidence for evolving dark energy over \lcdm\ remains moderate, shifting from $2.2\sigma$ (one year of DESI BAO) to $2.4\sigma$ (three years of DESI BAO). Relaxing the measurement of the optical depth using a wider prior on $\tau$ does not impact these limits: a shift in the central value of $\tau$ (as for example measured in the extreme case of removing low-$\ell$ polarization data, shown in the green contours) is needed to move the constraints to within $2\sigma$ of \lcdm. }
	\label{fig:desi_dr2_lcdm}
\end{figure*}

\cite{DESI1_DR2,DESI-DR2} presented new BAO measurements from three years of observations --- DESI DR2, labeled as \textsf{B$_{\rm DR2}$} hereafter --- improving the size of the data sample and parameter sensitivity over the dataset from Year-1 exploited in the main text of this paper. In Fig.~\ref{fig:desi_dr2_lcdm}, we revisit the consistency of DESI BAO with the best-fit \lcdm\ model from \Planck\ and from \pact. We find that the \pact\ best-fit model continues to provide an accurate prediction (in fact, even more accurate than for the DESI Year-1 dataset) for these new BAO data and that the joint \lcdm\ solution is a good fit to all the datasets. In the same figure, we also compare the $w_0/w_a$ constraints presented in~\S\ref{sec:de} with the contours obtained swapping in DESI DR2 for DESI Year-1. We find that the preference for the $w_0/w_a$ model over \lcdm\ remains moderate, at the $2.4\sigma$ level with or without supernovae, i.e., for \textsf{P-ACT-LB$_{\rm DR2}$S} and \textsf{P-ACT-LB$_{\rm DR2}$}. These results are consistent with what was derived in \cite{2025arXiv250418464G}. The evidence for evolving dark energy is reduced with \pact\ compared to combinations of DESI with other CMB likelihoods (i.e., CMB spectra from \Planck\ NPIPE, \citealp{rosenberg:2022}) because the value of the matter fraction measured by \pact\ is slightly lower than the value measured by \Planck\ alone: $\Omega_m=0.3116 \pm 0.0071$ for \pact\ and $\Omega_m=0.3158 \pm 0.0085$ for \Planck\ (L25) or $\Omega_m=0.3140 \pm 0.0076$ for \Planck\ NPIPE~\citep{rosenberg:2022}, and thus lies closer to the DESI constraint in \lcdm.

\begin{figure}
\includegraphics[width=\textwidth]{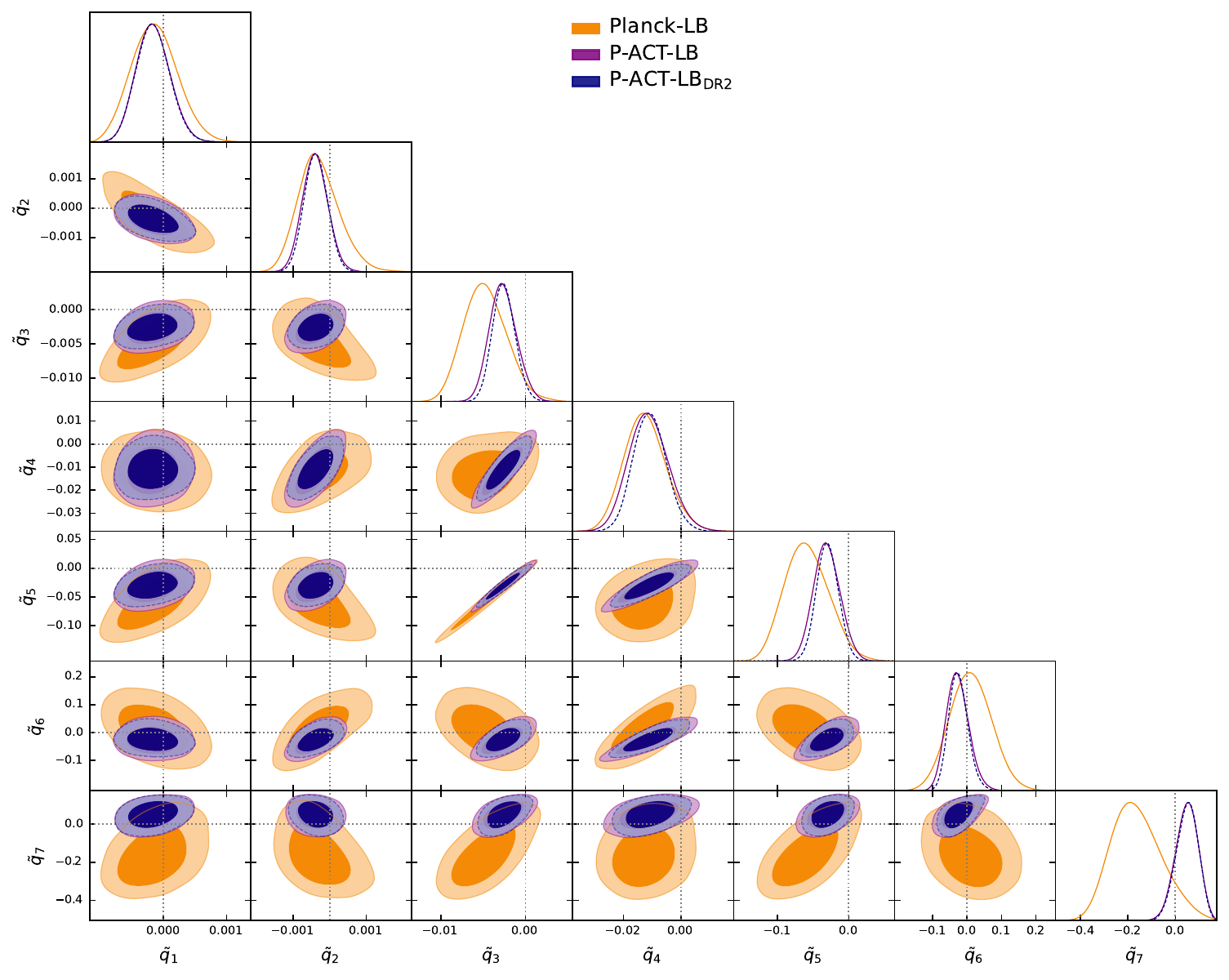}
\vspace{-0.5cm}
\caption{Marginalized parameter posteriors for the control points varied in the \texttt{ModRec} scenario analyzed in §5.5 (as in Fig.~\ref{fig:modrec_control_points}) including DESI DR2 BAO (blue). The dotted gray lines indicate the standard recombination scenario ($\tilde{q}_i = 0$).}
\label{fig:modrec_dr2}
\end{figure}

\begin{figure}
\centering
\includegraphics[width=0.5\columnwidth]{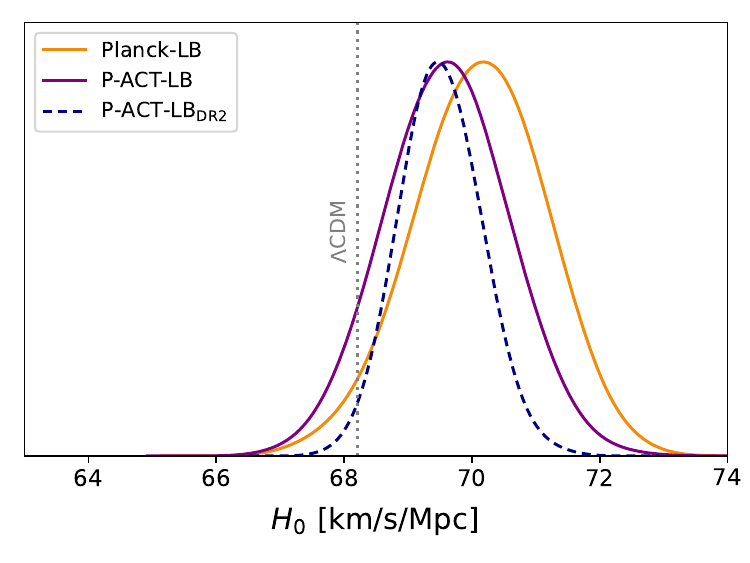}
   \vspace{-0.4cm} 
	\caption{Constraints on the Hubble constant from the \texttt{ModRec} analysis as in Fig.~\ref{fig:modrec_H0}, highlighting the impact of DESI DR2.}
	\label{fig:H0_modrec_dr2}
\end{figure}

As an example of the impact of DESI DR2 data on an extended model with new recombination-era physics, here we present updated results for the modified recombination scenario studied in \S\ref{sec:modrec}.  This example is particularly useful due to the sensitivity of the modified recombination constraints to the choice of BAO dataset (see Fig.~\ref{fig:modrec_Xe_recon}).  
Figure~\ref{fig:modrec_dr2} shows the marginalized parameter posteriors for the \texttt{ModRec} control points with DESI DR2 data (dark blue). The \textsf{P-ACT-LB$_{\rm DR2}$} constraints are consistent with the \pactlb\ constraints, as well as the standard recombination scenario. Relative to the DESI Year-1 analysis, the DR2 data yield approximately 10–15\% tighter constraints on the control points.
Figure~\ref{fig:H0_modrec_dr2} shows the marginalized posterior on $H_0$ for the \texttt{ModRec} analysis with DESI DR2 data. With the addition of DESI DR2, we find $H_0=69.5\pm 0.7 \ (68\%, \pactlb)$
which is significantly tighter than the \pactlb\ constraint ($69.6\pm1.0$~km/s/Mpc).

\begin{figure*}
\includegraphics[width=0.53\textwidth]{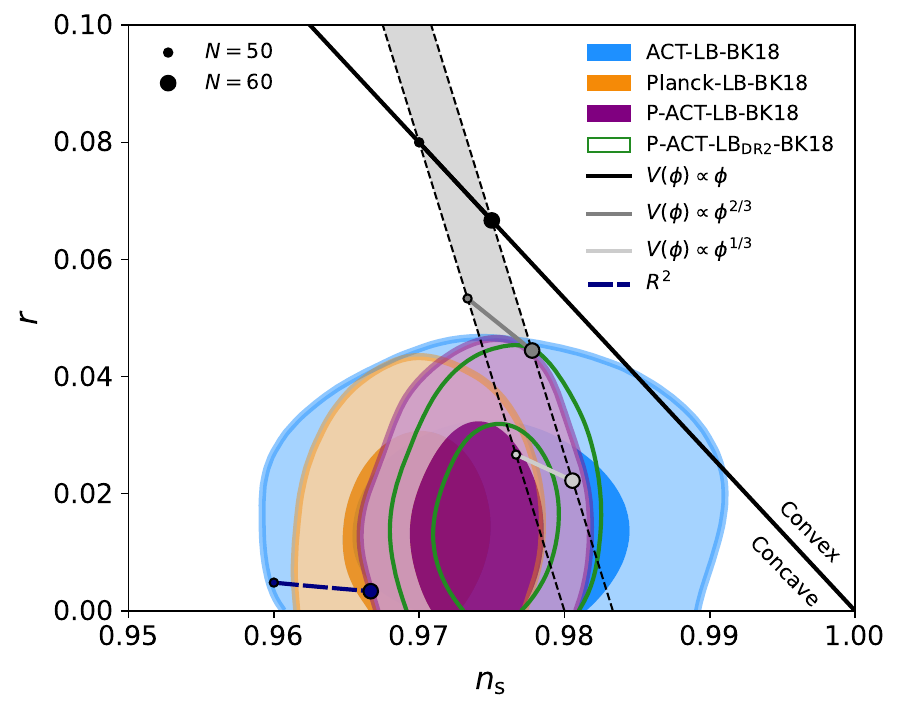}
\includegraphics[width=0.465\textwidth]{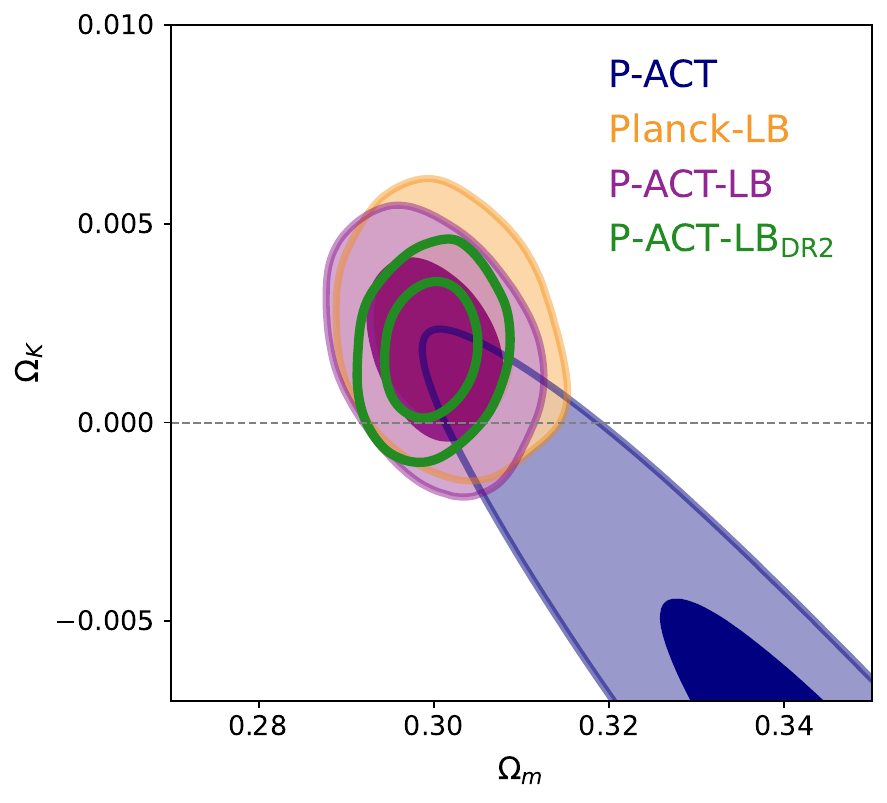}
   \vspace{-0.4cm} 
	\caption{Constraints in the $r-n_s$ plane (left, as in Fig.~\ref{fig:tensors}) and on the curvature and matter densities (right, same as Fig.~\ref{fig:omk}), with the addition of the combination \textsf{P-ACT-LB$_{\rm DR2}$} using DESI DR2 shown in green.
	\label{fig:r_ns_omk_dr2}}
\end{figure*}

The impact of DESI DR2 on most single parameter extensions studied in this paper is minimal. We report here only three cases that exhibit shifts worth noting. The measurement of $\neff$ is only marginally improved with the newer DESI data and the central value shifts towards the standard model predictions, yielding $\neff=2.92\pm0.12$ (68\%, \textsf{P-ACT-LB$_{\rm DR2}$}).  The right panel of Fig.~\ref{fig:r_ns_omk_dr2} shows that the measurement of $n_s=0.9752\pm0.0030$ obtained with \textsf{P-ACT-LB$_{\rm DR2}$} (see L25) tightens the contours in the $r-n_s$ plane, reducing further the parameter space allowed for Starobinsky-like inflation models (assuming 50--60 e-folds of inflation). A tighter measurement of the matter fraction shrinks the uncertainty on --- and reduces correlations with --- spatial curvature as shown in the left panel of Fig.~\ref{fig:r_ns_omk_dr2}, yielding $\Omega_{k} = 0.0018 \pm 0.0011$ (68\%, \textsf{P-ACT-LB$_{\rm DR2}$}), in agreement with \lcdm\ at $1.6\sigma$.

\subsection{Optical Depth}
Recent works (e.g.,~\citealp{2024arXiv241000090L,2025arXiv250305691A,2025arXiv250416932S,2025arXiv250421813J}), have shown that the contribution to tight limits on the neutrino mass sum and the evidence for evolving dark energy stemming from the matter density fraction measured by DESI BAO can be reduced by relaxing the constraint on the optical depth, $\tau$. We show the impact of $\tau$ on our \pactlb\ results by doubling the \texttt{Sroll2} uncertainty on $\tau$ and using a Gaussian prior of $\tau=0.057 \pm 0.012$.  This positions $\tau$ in between the most stringent limit from \texttt{Sroll2} and the recent measurement from CLASS~\citep{2025arXiv250111904L}. We find that a more uncertain measurement of $\tau$ at this level has no impact on the $w_0$/$w_a$ constraints. To reduce further the evidence for evolving dark energy, the central value of the prior needs to shift towards higher values of $\tau \approx 0.07$ --- this is consistent with the fact that the value of $\tau$ preferred by the combination \textsf{P-ACT-LB$_{\rm DR2}$} ($\tau=0.0643^{+0.0055}_{-0.0067}$, L25) is higher compared to the central value of the \texttt{Sroll2} measurement. We show this in Fig.~\ref{fig:desi_dr2_lcdm} by removing \texttt{Sroll2} information altogether and measuring $\tau$ from lensing effects in the power spectra, which break the $A_s$--$\tau$ degeneracy. The $w_0/w_a$ limits become broader and move closer to consistency with \lcdm, at the $2\sigma$ level for \textsf{P-ACT-LB$_{\rm DR2}$S}, and measure $\tau=0.081 \pm 0.016$. 

Given the well-known degeneracy between neutrino mass and the power spectrum amplitude, which in turn correlates with $\tau$, the neutrino mass limit is also expected to become more relaxed when considering a more uncertain $\tau$. We find that a wider prior has minimal impact, yielding $\sum m_\nu<0.10$~eV at 95\% confidence for \pactlb, which becomes $\sum m_\nu<0.19$~eV excluding \texttt{Sroll2} completely. 

\collaboration{175}{}
\allauthors
\end{document}